\documentclass[fleqn,usenatbib]{mnras}

\usepackage{newtxtext,newtxmath}

\usepackage[T1]{fontenc}

\DeclareRobustCommand{\VAN}[3]{#2}
\let\VANthebibliography\thebibliography
\def\thebibliography{\DeclareRobustCommand{\VAN}[3]{##3}\VANthebibliography}

\usepackage{graphicx}	
\usepackage{amsmath}	
\usepackage[usenames,dvipsnames]{xcolor}

\usepackage{gensymb}    
\usepackage{latexsym}
\usepackage{amsfonts}
\usepackage{pifont}
\usepackage{epstopdf}
\usepackage{rotating}
\usepackage{pdflscape}
\usepackage{changepage}
\usepackage{url}
\usepackage{color}
\usepackage{footnote}
\usepackage{cleveref}
\crefname{section}{§}{§§}
\Crefname{section}{§}{§§}
\usepackage{float}
\usepackage{subfig}
\usepackage{caption}
\usepackage{booktabs}
\usepackage{fixltx2e}
\usepackage{threeparttable}
\usepackage{longtable}
\usepackage{supertabular,booktabs}
\makesavenoteenv{itemize}
\usepackage{booktabs}
\usepackage{xspace}
\usepackage{color, colortbl}
\usepackage{listings}
\usepackage{enumerate}
\usepackage{multirow}
\usepackage{fixltx2e}
\usepackage{diagbox}
\usepackage{multirow}
\usepackage{cellspace}
\usepackage{floatpag}
\usepackage{pdfpages}

\usepackage[percent]{overpic}
\usepackage{pict2e}
\newcommand{\hst}{\textit{HST}\xspace}
\newcommand{\gaia}{\textit{Gaia}\xspace}
\newcommand{\mdot}{\dot{M}\xspace}
\newcommand{\mwd}{M_\mathrm{WD}\xspace}
\newcommand{\rwd}{R_\mathrm{WD}\xspace}
\newcommand{\msun}{\mathrm{M_\odot}\xspace}
\newcommand{\rsun}{\mathrm{R_\odot}\xspace} 
\newcommand{\porb}{\mbox{$P_\mathrm{orb}$}\xspace}

\newcommand{\teff}{\mbox{$T_\mathrm{eff}$}\xspace}

\newcommand{\xmark}{\ding{55}}%

\title[Masses and Accretion Rates of CV White Dwarfs]{
Constraining the Evolution of Cataclysmic Variables via the Masses and Accretion Rates of their Underlying White Dwarfs}

\author[A. F. Pala et al.]{A.~F.~Pala$^{1,2}$\thanks{E-mail: anna.pala@esa.int}, B.~T.~G\"ansicke$^{3}$, D.~Belloni$^{4}$, S.~G.~Parsons$^{5}$, T.~R.~Marsh$^{3}$, M.~R.~Schreiber$^{6,7}$, \newauthor
E.~Breedt$^{8}$, 
C.~Knigge$^{9}$, 
E.~M.~Sion$^{10}$, 
P.~Szkody$^{11}$, 
D.~Townsley$^{12}$, 
L.~Bildsten$^{13,14}$, 
D.~Boyd$^{15}$, \newauthor  
M.~J.~Cook$^{16,17}$, 
D.~De~Martino$^{18}$, 
P.~Godon$^{10}$, 
S.~Kafka$^{16}$,  
V.~Kouprianov$^{19}$,  
K.~S.~Long$^{20,21}$, \newauthor 
B.~Monard$^{22}$, 
G.~Myers$^{16}$, 
P.~Nelson$^{16}$, 
D.~Nogami$^{23}$,   
A.~Oksanen$^{24}$,  
R.~Pickard$^{25}$,  
G.~Poyner$^{25}$, \newauthor 
D.~E.~Reichart$^{19}$, 
D.~Rodriguez~Perez$^{16}$, 
J.~Shears$^{25}$, 
R.~Stubbings$^{16}$ and 
O.~Toloza$^{6}$\\
$^{1}$European Space Agency, European Space Astronomy Centre, Camino Bajo del Castillo s/n, 28692 Villanueva de la Ca\~nada, Madrid, Spain\\
$^{2}$European Southern Observatory, Karl Schwarzschild Stra{\ss}e 2, Garching, 85748, Germany\\
$^{3}$Department of Physics, University of Warwick, Coventry, CV4 7AL, UK\\
$^{4}$National Institute for Space Research, Av. dos Astronautas, 1758, 12227-010 S\~ao Jos\'e dos Campos, SP, Brazil\\
$^{5}$Department of Physics and Astronomy, University of Sheffield, Sheffield S3 7RH, UK\\
$^{6}$Departamento de F{\'i}sica, Universidad T{\'e}cnica Federico Santa Mar\'ia, A. Espa{\~n}a 1680, Valpara\'iso, Chile\\
$^{7}$Instituto de F{\'i}sica y Astronom{\'i}a, Millennium Nucleus for Planet Formation (NPF), Universidad de Valpara{\'i}so, 2360102 Valparaiso, Chile\\
$^{8}$Institute of Astronomy, University of Cambridge, Cambridge, CB3 0HA, UK\\
$^{9}$School of Physics and Astronomy, University of Southampton, Southampton, SO17 1BJ, UK\\
$^{10}$Astronomy \& Astrophysics, Villanova University, Villanova, PA 19085, USA\\
$^{11}$Department of Astronomy, University of Washington, Seattle, WA 98195--1580, USA\\
$^{12}$Department of Physics and Astronomy, University of Alabama, Tuscaloosa, AL 35405, USA\\
$^{13}$Department of Physics, University of California, Santa Barbara, CA 93106, USA\\
$^{14}$Kavli Institute for Theoretical Physics, University of California, Santa Barbara, CA 93106, USA\\
$^{15}$British Astronomical Association, Variable Star Section, West Challow Observatory, OX12 9TX, UK\\
$^{16}$American Association of Variable Star Observers, Cambridge, MA 02138, USA\\
$^{17}$Newcastle Observatory, Newcastle, Ontario, Canada\\
$^{18}$INAF - Osservatorio Astronomico di Capodimonte, Napoli, I--80131, Italy\\
$^{19}$Department of Physics and Astronomy, University of North Carolina at Chapel Hill, Chapel Hill, NC 27599-3255\\
$^{20}$Space Telescope Science Institute, 3700 San Martin Drive, Baltimore, MD 21218, USA\\
$^{21}$Eureka Scientific, Inc. 2452 Delmer Street, Suite 100, Oakland, CA 94602--3017, USA\\
$^{22}$CBA Kleinkaroo, Calitzdorp, South Africa\\
$^{23}$Department of Astronomy, Graduate School of Science, Kyoto University, Oiwakecho, Kitashirakawa, Sakyo-ku, Kyoto 606-8502, Japan\\
$^{24}$Hankasalmi observatory, Verkkoniementie 30, 40950 Muurame,  Finland\\
$^{25}$British Astronomical Association, Variable Star Section, Burlington House, Piccadilly, London, W1J ODU, UK\\
}
\date{Accepted 2021 November 17. Received 2021 November 17; in original form 2021 September 1.}

\pubyear{2021}

\begin{document}
\label{firstpage}
\pagerange{\pageref{firstpage}--\pageref{lastpage}}
\maketitle

\begin{abstract}
We report on the masses ($\mwd$), effective temperatures ($\teff$) and secular mean accretion rates ($\langle \mdot \rangle$) of 43 cataclysmic variable (CV) white dwarfs, 42 of which were obtained from the combined analysis of their \textit{Hubble Space Telescope} ultraviolet data with the parallaxes provided by the Early Third Data Release of the \gaia space mission, and one from the white dwarf gravitational redshift. Our results double the number of CV white dwarfs with an accurate mass measurement, bringing the total census to 89 systems.
From the study of the mass distribution, we derive $\langle \mwd \rangle = 0.81^{+0.16}_{-0.20}\,\msun$, in perfect agreement with previous results, and find no evidence of any evolution of the mass with orbital period. Moreover, we identify five systems with $\mwd < 0.5\msun$, which are most likely representative of helium-core white dwarfs, showing that these CVs are present in the overall population.
We reveal the presence of an anti-correlation between the average accretion rates and the white dwarf masses for the systems below the $2-3\,$h period gap. Since $\langle \mdot \rangle$ reflects the rate of system angular momentum loss, this correlation suggests the presence of an additional mechanism of angular momentum loss that is more efficient at low white dwarf masses. 
This is the fundamental concept of the recently proposed empirical prescription of consequential angular momentum loss (eCAML) and our results provide observational support for it, although we also highlight how its current recipe needs to be refined to better reproduce the observed scatter in $\teff$ and $\langle \mdot \rangle$, and the presence of helium-core white dwarfs.
\end{abstract}




\section{Introduction}\label{sec:intro}
Cataclysmic variables (CVs) are compact interacting binaries in which a white dwarf is accreting mass from a low-mass star via Roche lobe overflow \cite[e.g.][]{Warner1995}.
CVs descend from main-sequence binaries in which the more massive star (the primary) evolves first and leaves the main sequence. Following its expansion, the primary star fills its Roche lobe and starts unstable mass transfer on its less massive companion (the secondary), leading to the formation of a common envelope, i.e. a shared photosphere engulfing both stars \citep[e.g.][]{Paczynski+1976,Ivanova+2013}. The two stars transfer orbital energy to the envelope, which is rapidly expelled, leaving behind a post common envelope binary composed of the core of the giant primary, which evolves into a white dwarf, and a low-mass secondary star. Owing to subsequent orbital angular momentum losses, mainly via magnetic braking (which arises from a stellar wind associated with the magnetic activity of the secondary, e.g. \citealt{Mestel1968,Verbunt+1981}) and gravitational wave radiation \citep[e.g.][]{Paczynski1967}, the post common envelope binary evolves into a semi-detached configuration, becoming a CV.

During the CV phase, the white dwarf response to the mass accretion process is the subject of a long-standing debate. Many binary population studies predict an average mass of CV white dwarfs of $\langle \mwd \rangle \simeq 0.5\,\msun$ \citep[e.g.][]{deKool1992,Politano1996}, which is lower than that of single white dwarfs ($\langle \mwd \rangle \simeq 0.6\,\msun$, \citealt{Koester+1979,Liebert+2005b,Kepler+2007}). This is because it is expected that the core growth of the primary is halted by the onset of the common envelope phase. Moreover, once a CV is formed, the material piled up at the white dwarf surface is compressed by the strong gravitational field of the star, leading periodically (typically on time-scale of ten/hundred thousands of years) to the occurrence of classical nova eruptions. These are the result of the thermonuclear ignition of the accreted material on the white dwarf surface and, during these explosions, different theoretical models predict that the accreted material \citep[e.g.][]{Yaron+2005} and part of the underlying core of the white dwarf \citep{Gehrz+1998,Epelstain+2007} should be ejected in the surrounding space, thus preventing the white dwarf from growing in mass. However, white dwarfs in CVs have been found to be significantly more massive than binary population models predicted \citep[e.g.][]{deKool1992,Politano1996}, with early work showing the average mass of CV white dwarfs to lie in the range $\langle \mwd \rangle \simeq 0.8-1.2\,\mathrm{M}_\odot$ \citep{Warner1973,Ritter1987}. This result was originally interpreted as an observational bias because (i) the higher the mass of the white dwarf, the larger the accretion energy released per accreted unit mass, and (ii) for a fixed donor mass, more massive white dwarfs have larger Roche lobes that can accommodate larger, and hence brighter (especially at optically wavelengths) accretion discs. Therefore, CVs hosting massive white dwarfs were expected to be more easily discovered in magnitude limited samples \citep{Ritter+1986}.
Later on, \citet{Zorotovic+2011} reviewed the masses of CV white dwarfs available in the literature, showing that the average mass of CV white dwarfs is $\langle \mwd \rangle = 0.83 \pm 0.23\,\mathrm{M}_\odot$. Using the large and homogeneous sample of systems (\citealt{Szkody+2011}, and references therein) discovered by the Sloan Digital Sky Survey (SDSS, \citealt{York+2000}), the authors also demonstrated that there is a clear trend in disfavouring the detection of massive white dwarfs (which are smaller and less luminous than low mass white dwarfs for the same $\teff$) and therefore the high average mass of CV white dwarfs cannot be ascribed to an observational bias.

\begin{table*}
\setlength{\tabcolsep}{0.09cm}
\caption{Log of the \hst ultraviolet observations of the 43 CV white dwarfs studied in this work along with the corresponding \gaia EDR3 parallaxes (corrected for the zero point, see Section~\ref{subsec:gaia_quality}), sorted by increasing orbital period. The distances have been computed following the method described in \protect\cite{Pala+2020}, assuming the reported scale height $h$. The colour excess $E(B-V)$ have been derived using the STructuring by Inversion the Local Interstellar Medium (Stilism) reddening map \protect\citep{Lallement+2018}. Systems highlighted with a star symbol are known to be eclipsing.}\label{tab:hst_obs_log}
\begin{adjustwidth}{-3em}{0em}
\begin{tabular}{@{} lcccccccc | ccc@{}}
 \toprule
  System         & $P_\mathrm{orb}$ & $h$ & $E(B-V)$ & Instrument & Grating & Central      & Total        & Observation &\gaia EDR3 ID & $\varpi$ & $d$  \\ 
                      & (min)                     & (pc) &  (mag)    &                  &             & $\lambda$ & exposure  & date            &   & (mas)    & (pc)\\  
                      &                             &        &                &                  &             & (\AA)          & time   (s)  &                   &  &   & \\  
\midrule
\multirow{2}{*}{SDSS\,J150722.30+523039.8*} & \multirow{2}{*}{66.61} & \multirow{2}{*}{260} & \multirow{2}{*}{$0.018^{+0.006}_{-0.018}$} & COS & G140L & 1230 & 9762 & \multirow{2}{*}{2010 Feb 18}  & \multirow{2}{*}{1593140224924964864} & \multirow{2}{*}{$4.73\pm0.09$} &  \multirow{2}{*}{$211 \pm 4$}\\
& & & & STIS & G230LB & 2375 & 5226 & & & & \\[0.05cm]
SDSS\,J074531.91+453829.5 & 76.0 & 260 & $0.027\pm0.017$ & COS & G140L & 1105 & 5486 & 2012 Mar 13 & 927255749553754880 & $3.2\pm0.2$ & $310^{+23}_{-20}$\\[0.05cm]             
GW\,Lib & 76.78 & 260 & $0.022\pm0.018$ & COS & G140L & 1105 &  7417 & 2013 May 30 & 6226943645600487552 & $8.88\pm0.06$ & $112.6{\pm0.8}$\\[0.05cm]                  
SDSS\,J143544.02+233638.7 & 78.0 & 260 & $0.02\pm0.02$ & COS & G140L & 1105 & 7123 & 2013 Mar 09 &  1242828982729309952 & $4.8\pm0.2$ & $208^{+9}_{-8}$\\[0.05cm]
OT\,J213806.6+261957 & 78.1 & 260 & $0.017\pm0.015$ & COS & G140L & 1105 & 4760 & 2013 Jul 25 &  1800384942558699008 & $10.11\pm0.04$ & $98.9{\pm0.4}$\\[0.05cm]    
BW\,Scl & 78.23 & 260 & $0.002^{+0.015}_{-0.002}$ & STIS & E140M & 1425 & 1977 & 2006 Dec 27 & 2307289214897332480 & $10.71\pm0.05$ & $93.4{\pm0.5}$\\[0.05cm]      
LL\,And & 79.28 & 260 & $0.06^{+0}_{-0.02}$ & STIS & G140L & 1425 & 4499 & 2000 Dec 07 & 2809168096329043712 & $1.6\pm0.6$ & $609^{+343}_{-205}$\\[0.05cm]  
AL\,Com & 81.6 & 260 & $0.045\pm0.018$ & STIS & G140L & 1425 & 4380 & 2000 Nov 27 & 3932951035266314496 & $1.9\pm0.6$ & $523^{+252}_{-149}$\\[0.05cm]           
WZ\,Sge & 81.63 &  260 & $0.003^{+0.015}_{-0.003}$ & FOS & G130H & 1600 & 3000 & 1992 Oct 08 & 1809844934461976832 & $22.14\pm0.03$ & $45.17{\pm0.06}$\\[0.05cm]     
SW\,UMa & 81.81 & 260 & $0.008^{+0.017}_{-0.008}$ & STIS & G140L & 1425 & 4933 & 2000 Mar 26 & 1030279027003254784 & $6.23\pm0.06$ & $160.6{\pm1.6}$\\[0.05cm]       
V1108\,Her & 81.87 & 260 & $0.046\pm0.019$ & COS & G140L & 1105 & 7327 & 2013 Jun 06 & 4538504384210935424 & $6.8\pm0.1$ & $148\pm2$\\[0.05cm]            
ASAS\,J002511+1217.2 & 82.0 & 260 & $0.009^{+0.016}_{-0.009}$ & COS & G140L & 1105 & 7183 & 2012 Nov 15 & 2754909740118313344 & $6.4\pm0.1$ & $157{\pm3}$\\[0.05cm]          
HV\,Vir & 82.18 & 260 & $0.024^{+0.012}_{-0.021}$ & STIS & G140L & 1425 & 4535 & 2000 Jun 10 & 3688359000015020800 & $3.2\pm0.3$ & $317^{+29}_{-25}$\\[0.05cm] 
SDSS\,J103533.02+055158.4* & 82.22 & 260 & $0.017^{+0.018}_{-0.017}$ & COS & G140L & 1105 & 12282 & 2013 Mar 08 & 3859020040917830400 & $5.1\pm0.3$ & $195^{+12}_{-10}$\\[0.05cm]
WX\,Cet  & 83.90 & 260  & $0.013^{+0.016}_{-0.013}$ & STIS & E140M & 1425 & 7299 & 2000 Oct 30 & 2355217815809560192 & $3.97 \pm 0.13$ & $252^{+9}_{-8}$\\[0.05cm]
SDSS\,J075507.70+143547.6 & 84.76 & 260 & $0.011^{+0.015}_{-0.011}$ & COS & G140L & 1105 & 7183 & 2012 Dec 14 & 654539826068054400 & $4.2\pm0.2$ & $239^{+12}_{-11}$\\[0.05cm] 
SDSS\,J080434.20+510349.2 & 84.97 & 260 & $0.007^{+0.016}_{-0.007}$ & COS & G140L & 1105 & 5415 & 2011 Nov 03 & 935056333580267392 & $7.03\pm0.11$ & $142{\pm2}$\\[0.05cm]          
EG\,Cnc & 86.36 & 260 & $0.008^{+0.017}_{-0.008}$ & STIS & G140L & 1425 & 4579 & 2006 Dec 20 & 703580960947960576 & $5.4\pm0.2$ & $186\pm7$\\[0.05cm]  
EK\,TrA & 86.36 & 260 & $0.034\pm0.019$ & STIS & E140M & 1425 & 4302 & 1999 Jul 25 & 5825198967486003072 & $6.61\pm0.04$ & $151.4{\pm0.8}$\\[0.05cm]                 
1RXS\,J105010.8--140431 & 88.56 & 450 & $0.005^{+0.015}_{-0.005}$ & COS & G140L & 1105 & 7363 & 2013 May 10 & 3750072904055666176 & $9.19\pm0.09$ & $109{\pm1}$\\[0.05cm]   
BC\,UMa & 90.16 & 260 & $0.017^{+0}_{-0.017}$ & STIS & G140L & 1425 & 12998 & 2000 Jul 18 & 787683052032971904 & $3.41\pm0.13$ & $293^{+11}_{-10}$\\[0.05cm]    
VY\,Aqr & 90.85 & 260 & $0.008^{+0.016}_{-0.008}$ & STIS & E140M & 1425 & 7250 & 2000 Jul 10 & 6896767366186700416 & $7.08\pm0.09$ & $141.3^{+1.9}_{-1.8}$\\[0.05cm]
QZ\,Lib & 92.36 & 450 & $0.10\pm0.03$ & COS & G140L & 1105 & 7512 & 2013 Apr 26 & 6318149711371454464 & $5.0\pm0.2$ & $199^{+11}_{-10}$\\[0.05cm]            
SDSS\,J153817.35+512338.0 & 93.11 & 260 & $0.027\pm0.018$ & COS & G140L & 1105 & 4704 & 2013 May 16 & 1595085299649674240 & $1.64\pm0.12$ & $607^{+47}_{-40}$\\[0.05cm]
UV\,Per & 93.44 & 260 & $0.07\pm0.04$ & STIS & G140L & 1425 & 900 & 2002 Oct 11 & 457106501671769472 & $4.04\pm0.11$ & $248^{+7}_{-6}$\\[0.05cm]             
1RXS\,J023238.8--371812 & 95.04 & 450 & $0.005^{+0.014}_{-0.005}$ & COS & G140L & 1105 & 12556 & 2012 Nov 01 & 4953766320874344704 & $4.68\pm0.16$ & $214^{+8}_{-7}$\\[0.05cm]
RZ\,Sge & 98.32 & 260 & $0.03\pm0.3$ & STIS & G140L & 1425 & 900 & 2003 Jun 13 & 1820209309025797888 & $3.40\pm0.08$ & $294\pm7$\\[0.05cm]
CY\,UMa & 100.18 & 260 & $0.018^{+0}_{-0.015}$ & STIS & G140L & 1425 & 830 & 2002 Dec 27 & 832942871937909632 & $3.27\pm0.07$ & $306\pm6$\\[0.05cm] 
\multirow{2}{*}{GD\,552} & \multirow{2}{*}{102.73} & \multirow{2}{*}{450} & \multirow{2}{*}{$0.007^{+0.015}_{-0.007}$} & \multirow{2}{*}{STIS} & G140L & 1425 & 14580 & 2002 Oct 24 & \multirow{2}{*}{2208124536065383424} & \multirow{2}{*}{$12.41\pm0.04$} & \multirow{2}{*}{$80.6{\pm0.2}$}\\
& & & & & G230LB & 2375 & 4230 & 2002 Aug 31& & & \\[0.05cm]    
IY\,UMa* & 106.43 & 260 & $0.015^{+0}_{-0.015}$ & COS & G140L & 1105 & 4195 & 2013 Mar 30 & 855167540988615296 & $5.52\pm0.07$ & $181{\pm2}$\\[0.05cm]           
SDSS\,J100515.38+191107.9 & 107.6 & 260 & $0.021^{+0.011}_{-0.014}$ & COS & G140L & 1105 & 7093 & 2013 Jan 31 & 626719772406892288 & $2.95\pm0.17$ & $339^{+21}_{-19}$\\[0.05cm] 
RZ\,Leo & 110.17 & 260 & $0.023^{+0.015}_{-0.023}$ & COS & G140L & 1105 & 10505 & 2013 Apr 11 & 3799290858445023488 & $3.59\pm0.15$ & $279^{+12}_{-11}$\\[0.05cm] 
AX\,For & 113.04 & 260 & $0.018\pm0.018$ & COS & G140L & 1105 & 7483 & 2013 Jul 11 & 5067753236787919232 & $2.86\pm0.08$ & $349\pm10$\\[0.05cm]          
CU\,Vel & 113.04 & 260 & $0.007^{+0.015}_{-0.007}$ & COS & G140L & 1105 & 4634 & 2013 Jan 18 & 5524430207364715520 & $6.31\pm0.04$ & $158{\pm1}$\\[0.05cm]        
EF\,Peg & 120.53 & 120 & $0.048\pm0.019$ & STIS & E140M &  1425 & 6883 & 2000 Jun 18 & 1759321791033449472 & $3.5\pm0.2$ & $288^{+21}_{-18}$\\[0.05cm]           
DV\,UMa* & 123.62 & 120 & $0.02^{+0}_{-0.012}$ & STIS & G140L & 1425  & 900 & 2004 Feb 08 & 820959638305816448 & $2.60\pm0.15$ & $382^{+24}_{-21}$\\[0.05cm]    
IR\,Com* & 125.34 & 120 & $0.019^{+0.021}_{-0.019}$ & COS & G140L & 1105 & 6866 & 2013 Jul 11 & 3955313418148878080 & $4.63\pm0.06$ & $216{\pm3}$\\[0.05cm]         
AM\,Her & 185.65 & 120 & $0.017\pm0.016$ & STIS & G140L & 1425  & 10980 & 2002 Jul 11/12 & 2123837555230207744 & $11.37\pm0.03$ & $87.9{\pm0.2}$\\[0.05cm]              
DW\,UMa* & 196.71 & 120 & $0.009^{+0.018}_{-0.009}$ & STIS & G140L & 1425  & 26144 & 1999 Jan 25 & 855119196836523008 & $1.73\pm0.02$ & $579^{+7}_{-6}$\\[0.05cm]
U\,Gem & 254.74 & 120 & $0.003^{+0.015}_{-0.003}$ & FOS & G130H & 1600 & 360 & 1992 Sep 28 & 674214551557961984 & $10.75\pm0.03$ & $93.0{\pm0.3}$\\[0.05cm]
SS\,Aur & 263.23 & 120 & $0.047\pm0.033$ & STIS & G140L & 1425 & 600 & 2003 Mar 20 & 968824328534823936 & $4.02\pm0.03$ & $249{\pm2}$\\[0.05cm]
RX\,And & 302.25 & 120 & $0.02\pm0.02$ & GHRS & G140L & 1425 & 1425 & 1996 Dec 22 & 374510294830244992 & $5.08\pm0.03$ & $197{\pm1}$\\[0.05cm]
V442\,Cen & 662.4 & 120 & $0.048\pm0.015$ & STIS & G140L & 1425 & 700 & 2002 Dec 29 & 5398867830598349952 & $2.92\pm0.04$ & $343\pm5$\\[0.05cm]         
\bottomrule
\end{tabular}
\begin{tablenotes}
\item \textbf{Notes.} For several systems in our sample, \textit{IUE} spectroscopic observations covering the wavelength range $1800-3200$\,\AA\, are available. These data allow to estimate the interstellar reddening due to dust absorption from the bump at $\simeq 2175\,$\AA\, and their analysis is available in the literature (CU\,Vel, \citealt{Pala+2017}; DW\,UMa, \citealt{Szkody1987}; AM\,Her, \citealt{Raymond+1979};  UV\,Per, \citealt{Szkody1985};  SW\,UMa, WZ\,Sge, U\,Gem, RX\,And, VY\,Aqr and SS\,Aur, \citealt{LaDous1991}). The $E(B-V)$ literature measurements from \textit{IUE} are all in agreement with those from Stilism. However, since the former do not have associated uncertainties, we decided to adopt the latter in our analysis, since Stilism provides the corresponding uncertainties and allows us to properly account for them when evaluating those associate with the white dwarf parameters.
\end{tablenotes}
\end{adjustwidth}
\end{table*}

Following this result, several authors investigated the possibility that CV white dwarfs could efficiently retain the accreted material and grow in mass, either by ejecting less material than they accrete, via quasi-steady helium burning after several nova cycles \citep{Hillman+2016}, or through a phase of thermal time-scale mass transfer during the pre-CV phase \citep{Schenker+2002}. \cite{Wijnen+2015} argued that mass growth during nova cycles cannot reproduce the observed distribution. In addition, these authors found that, even though the white dwarf mass could increase during thermal time-scale mass transfer, the resulting CV population would harbour a much higher fraction of nuclear evolved donor stars than observed, and thus they concluded that mass growth does not seem to be the reason behind the high masses of CV white dwarfs.
A general consensus on the ability of the white dwarf to retain the accreted mass has not been reached yet \citep{Hillman+2020,Starrfield+2020}. Moreover, neither mechanism is able to explain the observed CV white dwarf mass distribution without creating conflicts with other observational constraints. 

A promising alternative solution proposed throughout the last couple of years assumes that the standard CV evolution model is incomplete.
\cite{Schreiber+2016} suggested that consequential angular momentum loss (CAML) is the key missing ingredient. This sort of angular momentum loss arises from the mass transfer process itself and acts in addition to magnetic braking and gravitational wave radiation. \cite{Schreiber+2016} developed an empirical model (eCAML) in which the strength of CAML is inversely proportional to the white dwarf mass, leading to dynamically unstable mass transfer in most CVs hosting low-mass white dwarfs ($\mwd \lesssim 0.6\,\msun$). The majority of these systems would not survive as semi-detached binaries but the two stellar components would instead merge into a single object \citep[see also][]{Nelemans+2016}. The main strength of the eCAML model is that it can also solve other disagreements between theory and observations, such as the observed CV space density and orbital period distribution \citep{Zorotovic+2017,Belloni+2018,Belloni+2020,Pala+2020}, without the requirement of additional fine tuning. However, the exact physical mechanism behind this additional source of angular momentum loss and the reason for its dependence on the white dwarf mass are still unclear.

Finally, it has to be considered that, while observational biases can be ruled out, it is more difficult to assess the presence of systematics affecting \cite{Zorotovic+2011}'s results, which were based on a sample of only 32 systems with accurate mass measurements, 22 of which were derived from the analysis of their eclipse light curves. Given that the previous results were mainly based on one methodology, it is necessary to increase the number of systems with an accurate mass measurement and diversify the methods employed in order to confirm the inferred high masses of CV white dwarfs, which remains one of the biggest and unresolved issues for the theoretical modelling of CV evolution.

An alternative method to measure CV white dwarf masses consists of the analysis of their ultraviolet spectra. The ultraviolet waveband is optimal for studying the underlying white dwarfs as they are relatively hot ($\teff \geq 10\,000\,$K, \citealt{Sion1999,Pala+2017}) and dominate the emission at these wavelengths, while the optical waveband is contaminated by the emission from the accretion flow and the companion star. From the knowledge of the distance to the system, the white dwarf radius ($\rwd$) can be derived from the scaling factor between the best-fitting model and the ultraviolet spectrum. Under the assumption of a mass-radius relationship, it is then possible to measure the white dwarf mass. Another possibility is to perform a dynamical study. The radial velocities of the white dwarf and the donor allow to infer the white dwarf gravitational redshift that, combined with a mass-radius relationship, provides the mass of the white dwarf.  

Over the last 30 years, the \textit{Hubble Space Telescope} (\hst) has proven to be essential for the study of CVs, delivering ultraviolet observations\footnote{It is worth mentioning that also the \textit{Far Ultraviolet Spectroscopic Explorer} \textit{(FUSE)} and the \textit{International Ultraviolet Explorer} \textit{(IUE)} provided a plethora of CV ultraviolet spectra. However, the wavelength range ($\simeq 920 - 1180\,$\AA) of the \textit{FUSE} observations is too limited for an accurate spectral fit with atmosphere models for white dwarfs with $\teff \lesssim 20\,000\,$K. This is because their spectra are characterised by a low flux level due to the broad  absorption lines from the higher orders of the Lyman series and, possibly, are complicated by heavy contamination from interstellar H$_2$. In \textit{IUE}, the 45\,cm diameter mirror only allowed good signal-to-noise ratio observations of the brightest CVs.} for 193 systems. Nonetheless, in the past, only a handful of objects had mass measurements derived from the analysis of their ultraviolet data because of the lack of accurate CV distances. In this respect, the \gaia space mission represents a milestone for CV research, delivering accurate parallaxes for a large number of these interacting binaries, finally enabling a quantitative analysis of the ultraviolet data obtained over the past decades.

We here analyse the \hst observations of 43 CVs for which the white dwarf is the dominant source of emission in the ultraviolet wavelength range, and for which \gaia parallaxes from the Third Early Data release (EDR3) are available. To provide an additional independent determination to test our results and assess the presence of systematics, we complement this data set with optical phase-resolved observations obtained with X-shooter mounted on the Very Large Telescope (VLT), which provides independent dynamical mass measurements. 
We present this large CV sample, which doubles the number of objects with accurate $\mwd$ measurements, providing new constraints on the response of the white dwarf to the mass accretion process, and for the further development of models for CV evolution.

\section{Sample selection and Observations}\label{sec:observations}
\subsection{Ultraviolet observations}\label{subsec:uv_obs}
\begin{table}
 \setlength{\tabcolsep}{0.1cm}
 \caption{Characteristics of the gratings and setup of the \hst observations used in this work.}\label{tab:gratings}
  \begin{tabular}{@{}lclccl@{}}
  \toprule
Instrument                  &     Aperture                                      & Grating   &    Central        &   Wavelength &  Resolution  \\
                                 &                                                        &              &    wavelength$^{e}$   &   coverage   &    \\
\midrule
COS                            &  PSA$^{a}$                                      & G140L   &     1105\,\AA     &   $1105-1730\,$\AA    & $\simeq 3000$\\[0.1cm]  
FOS                            &  $1"$                                              & G130H   &     1600\,\AA     &  $1150-1610\,$\AA    & $\simeq 1200$\\[0.1cm]  
GHRS                          &   LSA$^{b}$                                     & G140L	&     1425\,\AA      &  $1149-1435\,$\AA   & $\simeq 2000$\\[0.1cm]  
\multirow{2}{*}{STIS}  &  $0.2"\times0.2"$                           & E140M$^{c}$   &     1425\,\AA      & $1125-1710\,$\AA   & $\simeq 90\,000$\\ 
                                  &   \multirow{2}{*}{$52"\times0.2"$}  & G140L   &     1425\,\AA      & $1150-1700\,$\AA   & $\simeq 1000$\\
                                  &                                                        & G230L$^{d}$ &    2375\,\AA     &   $1650-3150\,$\AA    & $\simeq 800$\\
\bottomrule
\end{tabular}
\begin{tablenotes}
\item \textbf{Notes.} \emph{(a)} Primary Science Aperture (2.5"); \emph{(b)} Large Science Aperture (2.0"), \emph{(c)} We re-bin the data obtained with the E140M grating to match the resolution of the G140L observations in order to increase the signal-to-noise ratio.
\emph{(d)} Only used in the cases of SDSS\,J150722.30+523039.8 and GD\,552 to complement the G140L data, which provide the full coverage of Ly$\alpha$ from the white dwarf photosphere. 
\emph{(e)} The central wavelength is defined as the shortest wavelength recorded on the Segment A of the detector.
\end{tablenotes}
\end{table}

Among the 193 systems in the \hst ultraviolet archive, 191 have a \gaia EDR3 parallax (Table~\ref{tab:cvs_with_hst_obs}) and have been observed with either the Goddard High Resolution Spectrograph (GHRS), the Faint Object Spectrograph (FOS), the Space Telescope Imaging Spectrograph (STIS) or the Cosmic Origin Spectrograph (COS) with a setup suitable for our analysis, i.e. (at least) the full coverage of the Ly$\alpha$ ($1100-1600\,$\AA) from the white dwarf photosphere with a resolution of $R \simeq 1000-3000$. Together with a signal-to-noise ratio (SNR) of at least $\simeq 10$ and the knowledge of the distance to the systems, this setup allows an accurate determination of the white dwarf effective temperatures, chemical abundances and masses \citep{Gaensicke+2005}. 

\begin{figure}
\includegraphics[width=\columnwidth]{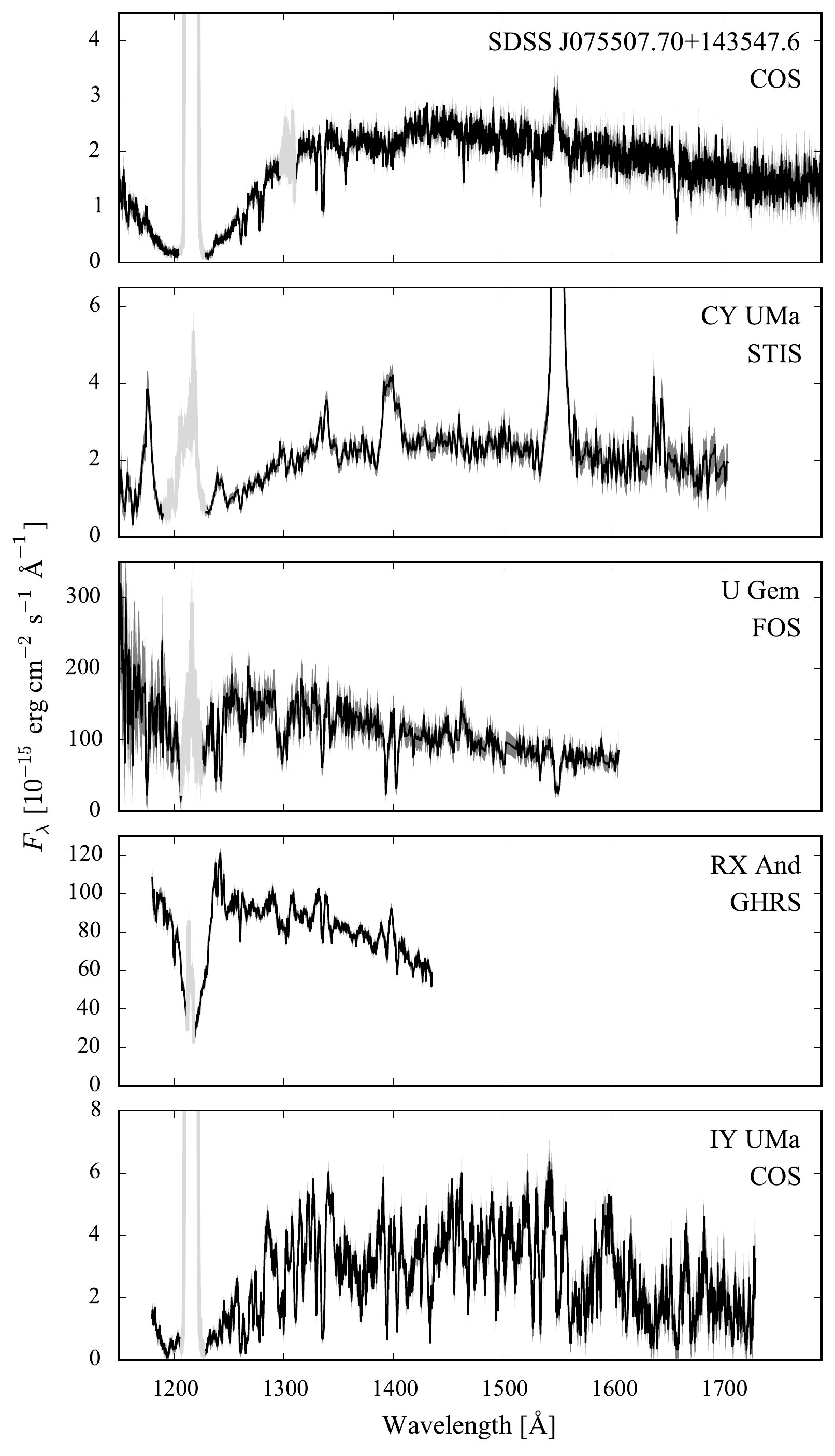}
\caption{Sample \hst spectra showing, from top to bottom, four typical quiescent CVs (SDSS\,J075507.70+143547.6, CY\,UMa, U\,Gem and RX\,And) observed with different \hst instruments, and one eclipsing system (IY\,UMa) whose spectrum is characterised by the presence of the iron-curtain. The geocoronal emission lines of Ly$\alpha$ ($1216\,$\AA) and \ion{O}{i} ($1302\,$\AA, not always detected) are plotted in grey.}\label{fig:sample_spectra}
\end{figure}

\begin{figure}
\includegraphics[width=\columnwidth]{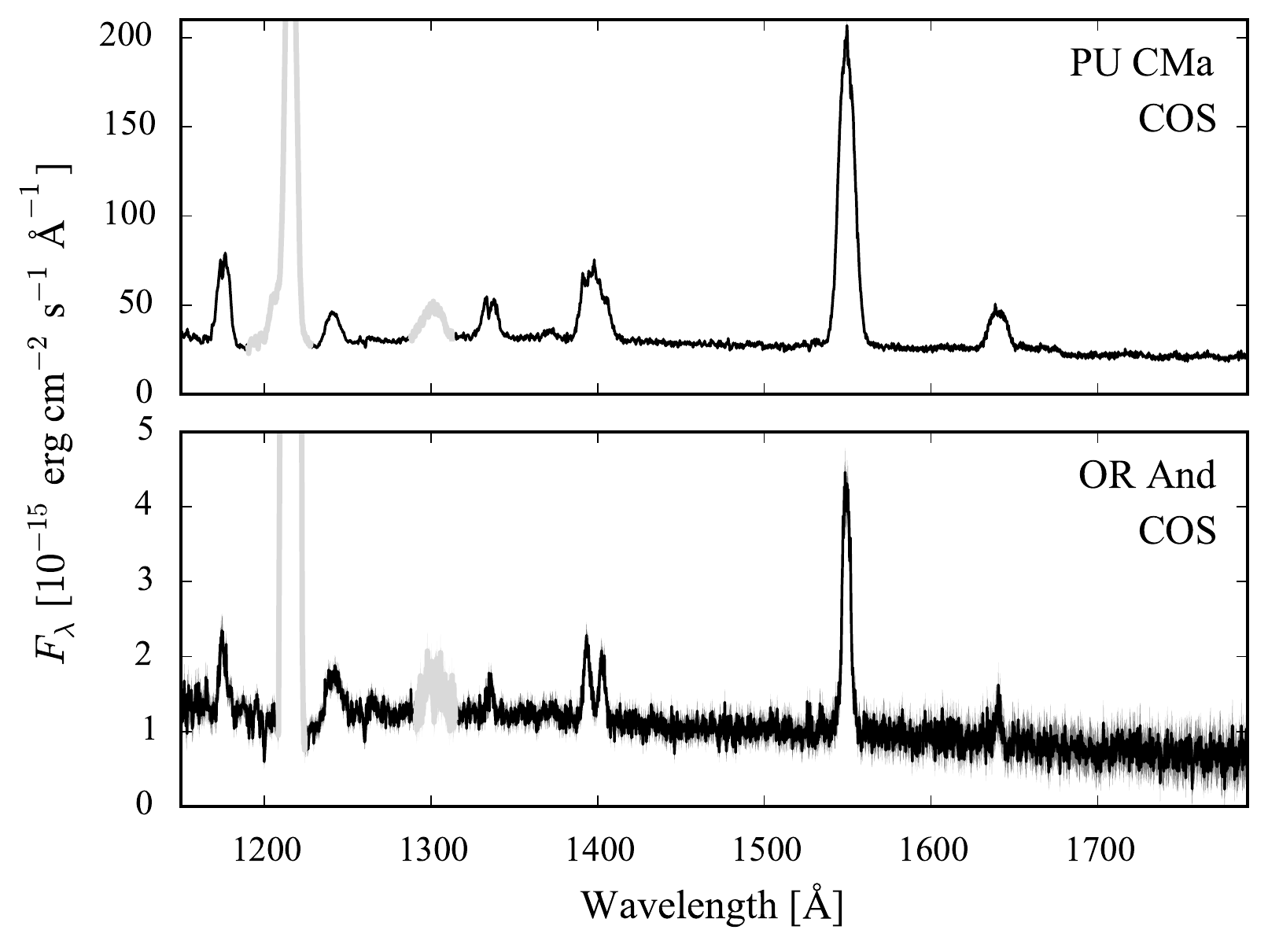}
\caption{Sample \hst spectra of a dwarf nova observed four days after a disc outburst (PU\,CMa, top panel) and a nova-like star observed in its normal high state (OR\,And, bottom panel). The spectra are dominated by the disc emission rather than the white dwarf. The geocoronal emission lines of Ly$\alpha$ ($1216\,$\AA) and \ion{O}{i} ($1302\,$\AA) are plotted in grey.}\label{fig:sample_disc}
\end{figure}

CVs are characterised by an intrinsic variable behaviour that can affect the analysis of their ultraviolet data.
A large fraction of CVs spend most of their time in a quiescent state, in which the accretion rate onto the white dwarf is very low, though the mass buffered in the accretion disc is slowly built up. During this phase, the white dwarf dominates the spectral appearance of the system and, at ultraviolet wavelengths, is recognisable from broad Ly$\alpha$ absorption centred on $1216\,$\AA. The profile of Ly$\alpha$ changes with $T_{\mathrm{eff}}$, becoming more defined and narrower in the hotter white dwarfs, while the continuum slope of the spectrum becomes steeper (Figure~\ref{fig:sample_spectra}). Periods of quiescence are interrupted by sudden brightenings with amplitudes of 2--5\,mag and, occasionally up to 9\,mag \citep{Warner1995,Maza+1983,Templeton2007}, called dwarf nova outbursts, when a fraction of the disc mass is rapidly drained onto the white dwarf. These outbursts arise from thermal-viscous instabilities in their accretion discs, causing a variation in the mass transfer rate through them \citep{Osaki1974,Hameury+1998,Meyer+1984}. Immediately after the occurrence of an outburst, the disc is hot and while it cools down it can partially or totally outshine the emission of the white dwarf even in the ultraviolet (top panel of Figure~\ref{fig:sample_disc}).

Following an outburst, the white dwarf is heated as a consequence of the increased infall of material, and this can possibly give rise to a non-homogeneous distribution of the temperature across its visible surface. Heated accretion belts or hot spots can dominate the overall ultraviolet emission of the system and the resulting temperature gradient results in the white dwarf radius being underestimated \citep[e.g][]{Toloza+2016,Pala+2019}.
The presence of a hot spot can be easily unveiled thanks to the modulation it introduces in the light curve of the system. In contrast, an equatorial accretion belt \citep{Kippenhahn+1978} can be difficult to detect since, being symmetric with respect to the rotation axis of the white dwarf, it does not cause variability on the white dwarf spin period \citep[e.g GW\,Lib,][]{Toloza+2016}.
Therefore, the analysis of the ultraviolet data obtained after a disc outburst provides only an upper limit on the white dwarf effective temperature and radius and, in turn, only a lower limit on its mass.

Additionally, members of a sub-class of CVs, known as nova-like systems, are characterised by high mean mass-transfer rates which usually keep the discs in a stable hot state, equivalent to a dwarf nova in permanent outburst. In this high state, the disc dominates the spectral appearance even in the far ultraviolet, preventing a direct detection of the white dwarf (bottom panel of Figure~\ref{fig:sample_disc}). However, occasionally, it is thought that as a consequence of starspots appearing in, or migrating into the tip of the donor star at the first Lagrangian point \citep{Livio+1994}, the mass transfer rate drops (low state) and unveils the white dwarf (e.g. \citealt{Boris+1999,Knigge+2000,Hoard+2004}). These low states typically last for days up to years \citep[e.g][]{Rodrguez-Gil+2007} and provide a window in which the white dwarf parameters can be measured (e.g. \citealt{Boris+1999a,Knigge+2000,Hoard+2004})

While some nova-likes can be magnetic, similar behaviour is observed in highly magnetic CVs. A significant number (28\,per cent, \citealt{Pala+2020}) of CVs contain strongly ($B \gtrsim 10\,\mathrm{MG}$) magnetic white dwarfs, whose magnetism suppresses the formation of an accretion disc and forces the accretion flow to follow the field lines. The strong field of the white dwarf may have a deep impact on the evolution of the system \citep{Schreiber+2021}.
Similar to nova-likes, magnetic CVs are also characterised by alternating between high and low states, on time-scales of months to years. During high states, mass accretion is stable and the ultraviolet emission is dominated by a hot polar cap close to one (or both) the magnetic pole(s) of the white dwarf. During low states, the hot-caps emission is greatly weakened and the white dwarf dominates the spectral appearance of the system. Possibly, low states in magnetic CVs are also triggered by a donor stellar spot passing around the first Lagrangian point region \citep{Hessman+2000}.
In low states, time-resolved observations are required in order to account for the possible contribution from the accretion cap and to obtain accurate white dwarf parameters.

Finally, CVs experience (at least once in their life) powerful classical nova eruptions due to thermonuclear runaways at their surface. CVs that have been observed to undergo these eruptions are known as novae and their ultraviolet spectra are dominated by the emission from a hot highly variable component \citep[e.g.][]{Cassatella+2005}, whose origin is still not clear, that prevents the direct detection of the white dwarf. 

\begin{figure}
\includegraphics[width=\columnwidth]{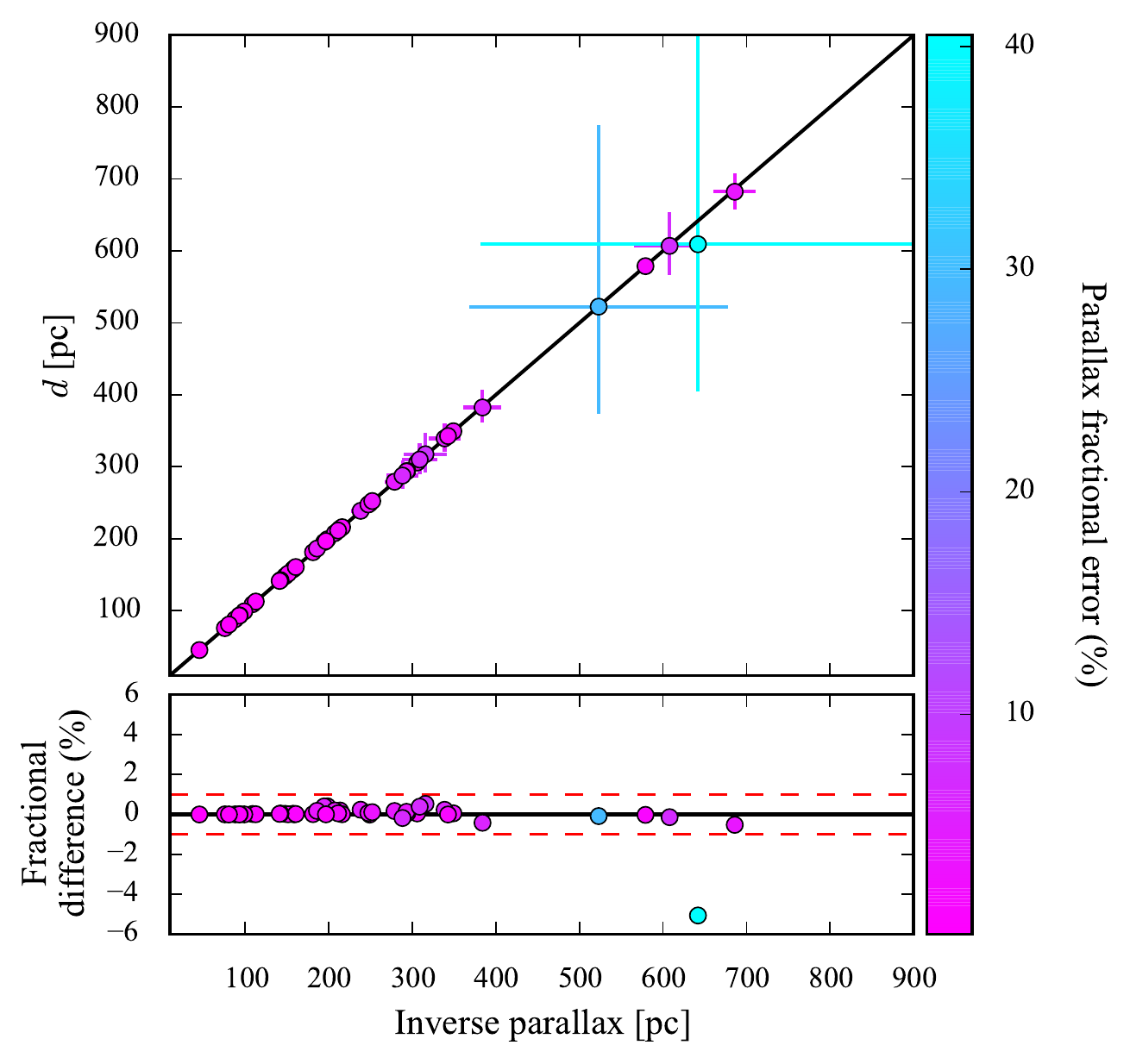}
\caption{Comparison between the distances to our targets computed as the inverse of the parallax ($\varpi^{-1}$) and using a statistical approach via the assumption of an exponentially decreasing volume density prior. The data are colour coded according to the parallax fractional error. The fractional difference, defined as ($D - \varpi^{-1})/\varpi^{-1}$, between the two methods is typically less than one per cent (red dashed lines in the bottom panel).}\label{fig:par_vs_dist}
\end{figure}

\begin{table*}
 \caption{Log of the optical observations of the three CVs observed with X-shooter in which the signatures of both the white dwarf and the secondary star were identified in their spectra. We obtained time-series of N spectra with T exposure time each.}\label{tab:xshoo_obs_log}
 \setlength{\tabcolsep}{0.17cm}
  \begin{tabular}{@{}lccccccccccccc@{}}
  \toprule
                                                &                   &  \multicolumn{3}{c}{UVB} &&\multicolumn{3}{c}{VIS} && \multicolumn{3}{c}{NIR}\\\cmidrule{3-5}\cmidrule{7-9}\cmidrule{11-13}
                                                & Observation &Exposure & Slit & \multirow{2}{*}{Resolution} && Exposure & Slit  & \multirow{2}{*}{Resolution} && Exposure & Slit & \multirow{2}{*}{Resolution}  \\
                                                & date            & time       & width  &             && time  & width    &    && time  & width  & \\
 System                                    & (YYYY-MM-DD) & N$\times$T(s)  &  (\arcsec)  & (\AA) && N$\times$T(s)  &  (\arcsec) & (\AA) && N$\times$T(s)  &  (\arcsec) & (\AA) \\
 \midrule                                                       
 \multirow{2}{*}{AX\,For}           & 2013-10-25  & $7\times 606$ & 1.0 & 0.99 && $13 \times 294$& 0.9 & 0.90 && $22 \times 200$ & 0.9 & 2.11 \\ 
                                                 &  2015-09-24  & $12\times 610$ & 1.3 & 1.10 &&  $12 \times 592$& 1.2 & 0.90 &&  $12 \times 642$ & 1.2 & 2.06\\[0.1cm]
IR\,Com                                      & 2014-03-05 & $29 \times 270$ & 1.0 & 1.01 && $21 \times 412$& 0.9 & 0.89 && $31 \times 300$ & 0.9 & 2.02 \\
V1108\,Her                               & 2015-05-12 & $12 \times 480$ & 1.0 & 1.04 && $10 \times 587$ & 0.9 &  0.92 && $12 \times 520$ & 0.9 & 2.07\\
 \bottomrule
\end{tabular}
\end{table*}

To account for the high variable nature of CVs, we inspected the archival ultraviolet \hst data and discarded (i) dwarf novae observed during or immediately after a disc outburst; (ii) nova-likes in high state and novae and (iii) polars lacking time-resolved observations with sufficient orbital phase coverage, necessary to determine a spectrum of the underlying white dwarf \citep[see e.g.][]{Gaensicke+2006}. In addition, we also discarded systems hosting a nuclear evolved donor. These CVs descend from binaries that underwent a thermal-time-scale mass transfer phase \citep{Schenker+2002} and can be easily identified from their enhanced \ion{N}{v}/\ion{C}{iv} line flux ratios \citep{Boris+2003}. In these systems, the white dwarf accretes helium rich material from its companion and this anomalously large helium abundance can cause an asymmetrical broadening of the blue wing of the Ly$\alpha$ \citep{Boris+2018}, thus affecting the estimates of the white dwarf surface gravity and temperature (Toloza et al., in preparation).

The final sample consists of 43 systems\footnote{\citet{Pala+2017} reported that AX\,For went into outburst five days before the \hst observations. The analysis of the ultraviolet data provides only a lower limit on the white dwarf mass. Nonetheless, we included this system in our sample since we measured its mass from additional optical phase-resolved observations.} and a log of their spectroscopic observations is presented in Table~\ref{tab:hst_obs_log}, while the different observational setups are listed in Table~\ref{tab:gratings}. Finally, the full list of CVs observed with \hst is provided in Appendix~\ref{ap:hst_archive}.

\subsubsection{Quality of the \gaia data}\label{subsec:gaia_quality}
\gaia astrometric solutions, and thereby parallaxes, are known to be affected by systematics arising from imperfections in the instrument and data processing methods \citep{Lindegren+2020a}. The mean value of the systematic error, the so-called parallax zero point $\varpi_\mathrm{zp}$, can be modelled according to the ecliptic latitude, magnitude and colour of each \gaia EDR3 source. We employed the python code provided by the \gaia consortium\footnote{\url{https://gitlab.com/icc-ub/public/gaiadr3_zeropoint}} to compute $\varpi_\mathrm{zp}$ for our targets, and corrected their parallaxes by subtracting the estimated zero point to the quoted \gaia EDR3 parallaxes. This correction ranges from $0.5 \mathrm{\mu}\textrm{as}$, in the case of V1108\,Her, to $-43 \mathrm{\mu}\textrm{as}$ for U\,Gem.

Together with the main kinematic parameters (positions, proper motions and magnitudes), which are used to derive the astrometric solution for each source, \gaia EDR3 also provides a series of ancillary parameters that can be used to evaluate the accuracy of this solution. Among the most relevant ones is the \texttt{astrometric\_excess\_noise}, which represents the error associated with the astrometric modelling \citep[see][]{Lindegren+2012} and that, ideally, should be zero.
Following \cite{Pala+2020}, we verified that the sources in our sample have reliable parallaxes by satisfying the condition \texttt{astrometric\_excess\_noise < 2}.

Converting parallaxes into distances is not always trivial as the mere inversion of the parallax can introduce some biases in the distance estimate, especially when the
fractional error on the parallax is larger than 20 per cent (see e.g. \citealt{Bailer-Jones2015,Luri+2018}), which is the case for two systems in our sample, AL\,Com and LL\,And. Their large uncertainties are most likely related to their intrinsic faintness ($G = 19.7$, $G = 20.1$, respectively) since  the other \gaia parameters that flag possible issues with the astronometric solution (\texttt{astrometric\_excess\_noise} and \texttt{RUWE}, described below) are within the range expected for well-behaved sources. 
We therefore computed the distance to each CV using a statistical approach, in which we assumed an exponentially decreasing volume density prior and a scale height $h$ following the method\footnote{The distance to the targets in our sample have also been computed from their \gaia EDR3 parallaxes by \citet{Bailer-Jones+2021}, with a method which also employs an exponentially decreasing volume density prior and a scale height calibrated against the stellar distribution at different Galactic latitudes. 
Nonetheless, we preferred to recompute the distances to our targets following the method described in \cite{Pala+2020}, since it employs a scale height that accounts for the age of the systems (as described in \citealt{Pretorius+2007b}) and is therefore more representative of the properties of the CV population.} described in \cite{Pala+2020}. Typically, the distances computed with the two methods differ by less than one per cent, the only exception being LL\,And with a difference of five per cent (Figure~\ref{fig:par_vs_dist}).

Another important parameter to assess the reliability of the parallaxes is the renormalised unit weight error (\texttt{RUWE}). This represents the square root of the normalised chi-square of the astrometric fit, scaled according to the source magnitude $G$, its effective wavenumber $\nu_\mathrm{eff}$ and its pseudocolour $\hat{\nu}_\mathrm{eff}$ (see \citealt{Lindegren+2020a} for more details). Ideally, for well-behaved sources\footnote{See the document ``Re-normalising the astrometric chi-square in Gaia DR2'', which can be downloaded from: \url{https://www.cosmos.esa.int/web/gaia/public-dpac-documents}}, \texttt{RUWE < 1.4}. However, we noticed that for the system in our sample with the largest \texttt{RUWE}, AM\,Her (\texttt{RUWE = 2.8}), the distance derived from its \gaia parallax ($88.1 \pm 0.4\,$pc) is consistent with the distance estimated by \citet{Thorstensen2003}, $79^{+9}_{-6}\,$pc. Similarly, in the case of U\,Gem, another system with high weight error (\texttt{RUWE = 2.4}), 
its \gaia parallax ($\varpi = 10.75 \pm 0.03$\,mas) and corresponding distance ($93.4 \pm 0.5\,$pc) are in good agreement, respectively, with the parallax measurements obtained using the \hst Fine Guidance Sensors by \citet{Harrison+2000} ($\varpi = 10.30 \pm 0.50$\,mas) and \citet{Harrison+2004} ($\varpi = 9.96 \pm 0.37$\,mas), and with the distance estimated by \citet{Beuermann2006}, $97\pm\,7$pc.
These large values of \texttt{RUWE} are most likely related to colour variations of the systems during different \gaia observations, caused either by the occurrence of low and high states in AM\,Her \citep{Gaensicke+2006}, and disc outbursts in U\,Gem. Nonetheless, since their \gaia astrometry still provides reliable distances, we decided to not apply any cut on this parameter for the remaining systems, which all have lower \texttt{RUWE} values.

\begin{figure*}
\includegraphics[width=\textwidth]{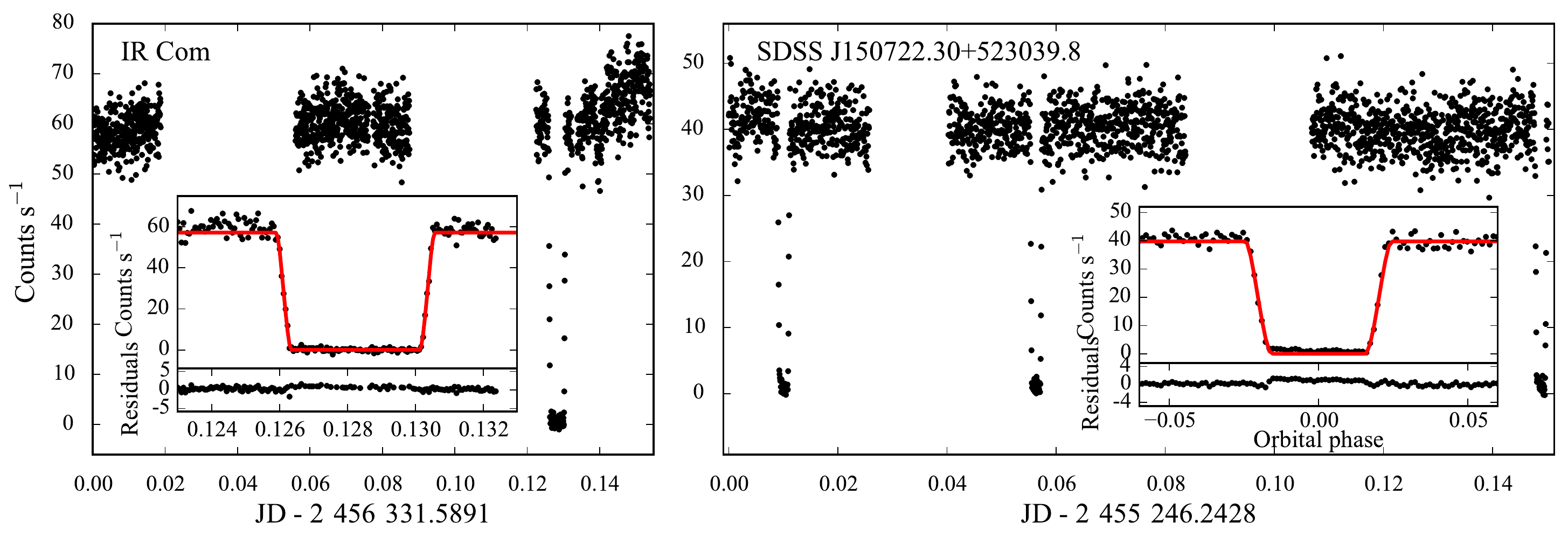}
\caption{\hst/COS light curves of IR\,Com (left) and SDSS\,J150722.30+523039.8 (right). The absence of any contamination from the bright spot allows us to fit the the eclipse light curve and measure the mass of the white dwarf in these systems. The insets show a close-up of the eclipse of IR\,Com and of the average phase-folded eclipse light curve of SDSS\,J150722.30+523039.8, along with the best-fitting models (red).}\label{fig:lc_eclipse}
\end{figure*}

\subsection{Optical spectroscopy}\label{subsec:optical_obs}
We obtained complementary phase-resolved spectroscopy with X-shooter \citep{xshooter} of the 22 targets from the \hst sample that are visible from the Southern hemisphere and in which (i) the white dwarf dominates the of emission in the ultraviolet and (ii) for which no dynamical study has been carried out before.

X-shooter is an {\'e}chelle spectrograph located at the Cassegrain focus of UT2 of the VLT at the European Southern Observatory (ESO) in Cerro Paranal (Chile). It is equipped with three arms: blue (UVB, $\lambda \simeq 3000-5595\,$\AA), visual (VIS, $\lambda \simeq 5595-10\,240\,$\AA) and near-infrared (NIR, $\lambda \simeq 10\,240-24\,800\,$\AA), with a medium spectral resolution ($R \simeq 5000-10\,000$).
For each arm, the slit width was chosen to best match the seeing and the exposure times were set with the aim to optimise the SNR and, at the same time, to minimise the orbital smearing.
At the time of the observations the atmospheric dispersion correctors of X-shooter were broken and hence the slit angle was reset to the parallactic angle position after one hour of exposures.
The data were reduced using the Reflex pipeline \citep{reflex}. To account for the well-documented wavelength shift between the three arms\footnote{A report on the wavelength shift can be found at \url{https://www.eso.org/sci/facilities/paranal/instruments/xshooter/doc/XS\_wlc\_shift\_150615.pdf}}, we used theoretical templates of sky emission lines to calculate the shift of each spectrum with respect to the expected position. We then applied this shift together with the barycentric radial velocity correction to the data. Finally, a telluric correction was performed using \texttt{molecfit} \citep{molecfit1,molecfit2}. 

In the spectra of three (AX\,For, IR\,Com and V1108\,Her) of the 22 CVs observed with X-shooter, we identified the \ion{Mg}{ii} absorption line at 4481 \AA\, that originates in the white dwarf photosphere, and several absorption features arising from the secondary photosphere, including \ion{Na}{i} ($11\,381/11\,403$ \AA), \ion{K}{i} ($11\,690/11\,769$ \AA\, and $12\,432/12\,522$ \AA). The \ion{K}{i} and \ion{Mg}{ii} lines, were used to track the motion of the two stellar components and to reconstruct their radial velocity curves from which the mass of the white dwarf can be determined. A log of the spectroscopic observations is presented in Table~\ref{tab:xshoo_obs_log}.

In the remaining 19 systems, we identified only signatures of either the white dwarf or the secondary, and in some cases of neither of them, and the analysis of these objects will be presented elsewhere. 

\section{Methods}\label{sec:methods}
\subsection{Light curve analysis}
Throughout the duration of the individual \hst observations (typically a few hours), CVs can show different types of variability, such as eclipses, modulations due to the white dwarf rotation, white dwarf pulsations, double humps and brightenings \cite[e.g.][]{Szkody+2002c,Araujo+2003,Szkody+2017,Toloza+2016,Pala+2019}.
Eclipses are particularly important since they allow a white dwarf mass measurement based only on geometrical assumptions to be obtained. In contrast, pulsations and brightenings reflect the presence of hot spots and, more generally, of a gradient in temperature over the visible white dwarf surface. When in view, the emission of the hot spots can dominate the overall ultraviolet emission of the system making the white dwarf look hotter and affecting both the temperature and mass estimates \citep{Toloza+2016,Pala+2019}. Therefore, it is important to remove the contribution of these spurious sources in order to obtain an accurate mass measurement.

The \hst \textsc{time--tag} data allow us to reconstruct a 2D image of the detector, where the dispersion direction runs along one axis and the spatial direction along the other, which can be used to reconstruct the light curve of the observed system. 
For each CV, we masked the geocoronal emission lines from Ly$\alpha$ (centred at 1216\,\AA) and \ion{O}{i} (centred at 1300\,\AA) as well as all the most prominent emission features from the accretion disc, which are not representative of the white dwarf. Using five-second bins and following the method described in \cite{Pala+2019}, we then extracted the light curve of each target in counts per second.

For the objects that did not exhibit any significant variability during their \hst observations, the data from all the orbits were summed to produce an average ultraviolet spectrum. 
We discuss, in what follows, the eclipsing systems while the remaining six CVs that showed some level of variability within the time-scale of the \hst observations are discussed in Appendix~\ref{ap:variability}.

\subsubsection{Eclipsing systems}\label{subsubsec:ircom_sdss1507} 
In eclipsing CVs, where the white dwarf is periodically obscured by its stellar companion, the duration of the ingress of the white dwarf, as well as the duration of the whole eclipse, can be used to derive the radius of both stars ($R_\mathrm{WD}$ and $R_\mathrm{donor}$ for the white dwarf and the donor, respectively) scaled by the orbital separation $a$:
\begin{eqnarray}
\frac{\rwd}{a} =\frac{1}{2} \left( \sqrt{\cos^2 i + \sin^2 i \cos^2 \Phi_1} + \sqrt{\cos^2 i + \sin^2 i \cos^2 \Phi_2} \right) \\
\frac{R_\mathrm{donor}}{a} = \frac{1}{2} \left( \sqrt{\cos^2 i + \sin^2 i \cos^2 \Phi_1} - \sqrt{\cos^2 i + \sin^2 i \cos^2 \Phi_2} \right)
\end{eqnarray}%
In the above equations, $\Phi_1$ and $\Phi_2$ are the phases of the first and second contacts and are directly measured from the light curve, while $i$ is the inclination of the system and is an additional unknown. Since the same eclipse profile can be reproduced by different inclinations and assuming different radii for the two stellar components, additional constraints are required to lift the degeneracy between these three parameters ($\rwd$, $R_\mathrm{donor}$ and $i$).

To this end, one of the most direct method consists in measuring the radial velocity amplitudes ($K_\mathrm{WD}$ and $K_\mathrm{donor}$) of the two stellar components from phase-resolved spectroscopic observations. These quantities provide the system mass ratio $q = K_\mathrm{WD}/K_\mathrm{donor} = M_\mathrm{donor}/\mwd$, which allows constraining the size of the Roche-lobe of the companion star \citep{Eggleton1983}. Under the assumption that the donor is Roche-lobe filling, the degeneracy in the three quantities can be lifted. In this case, a fit to the eclipse light curve provides $\rwd/a$ which, combined with a mass radius relationship and Kepler's third law, provides a measurement of the white dwarf mass (see e.g. \citealt{Littlefair+2006,Feline+2005,Savoury+2011,McAllister+2019}).

We detected the white dwarf eclipse in the COS light curves of four CVs, IR\,Com, IY\,UMa, SDSS\,J103533.02+055158.4 and SDSS\,J150722.30+523039.8. However, the data quality and orbital coverage allowed us to perform a fit to the eclipse light curve only in the cases of IR\,Com and SDSS\,J150722.30+523039.8. The remaining eclipsing systems are discussed in Appendix~\ref{subsubsec:eclipsing}.

We used the \texttt{lcurve} tool\footnote{\url{https://github.com/trmrsh/cpp-lcurve}} \citep[see][for a detailed description of the code]{Copperwheat+2010} to perform the light curve modelling and define the binary star model that best reproduces the observed eclipse. For IR\,Com we assumed the mass ratios derived from the radial velocity amplitudes from Section~\ref{subsec:rv}, $q=0.016 \pm 0.001$. For SDSS\,J150722.30+523039.8 we used the mass ratio from \cite{Savoury+2011}, $q=0.0647 \pm 0.0018$, which has been derived from the analysis of the optical light curve of the system. For both CVs, we assumed the white dwarf effective temperatures derived in Section~\ref{sec:uv_fit}, which are used by \texttt{lcurve} to estimate the flux contribution from the white dwarf. We kept the inclination $i$, $R_\mathrm{WD}/a$ and the time of middle eclipse $T_0$ as free parameters. The best-fitting models are shown in the insets in Figure~\ref{fig:lc_eclipse} and returned $i=80.5 \pm 0.3$ and $R_\mathrm{WD}/a = 0.00956 \pm 0.0002$ for IR\,Com and $i=83.5 \pm 0.3$ and $R_\mathrm{WD}/a = 0.0185 \pm 0.0006$ for SDSS\,J150722.30+523039.8 respectively. The $R_\mathrm{WD}/a$ ratios, combined with the white dwarf mass radius relationship\footnote{\url{http://www.astro.umontreal.ca/~bergeron/CoolingModels}} \citep{Holberg+2006,Tremblay+2011} and Kepler's third law, provide $M_\mathrm{WD} = 0.989 \pm 0.003\,\msun$ for IR\,Com and $M_\mathrm{WD} = 0.83^{+0.19}_{-0.15}\,\msun$ for SDSS\,J150722.30+523039.8 (Table~\ref{tab:comparison}).

It is worth mentioning at this point that the eclipse of the white dwarf in DW\,UMa was detected in the STIS \textsc{time--tag} data. This light curve has already been analysed by \cite{Araujo+2003}, which derived a white dwarf mass of $M_\mathrm{WD} = 0.77 \pm 0.07\,\msun$. 
Finally, DV\,UMa is also eclipsing but, since the data were acquired as snapshot, the lightcurve of the eclipse is not available.

\subsection{Ultraviolet spectral fitting}\label{sec:uv_fit}
To perform the spectral fit to the ultraviolet data, we generated a grid of white dwarf synthetic atmosphere models using \textsc{tlusty} and \textsc{synspec} \citep{tlusty,tlusty1}, covering the effective temperature range $T_\mathrm{eff} = 9\,000 -  70\,000\,$K in steps of $100\,$K, and the surface gravity range $\log(g) = 6.4 - 9.5$ in steps of $0.1$ (where $g$ is expressed in cgs units). As discussed by \citet{Pala+2017}, using a single metallicity is sufficient to account for the presence of the metal lines and possible deviations of single element abundances from the overall scaling with respect to the solar values do not affect the results of the fitting procedure. We therefore fixed the metal abundances to the values derived from the analysis of the same \hst data by previous works (see Table~\ref{tab:results_good} and references therein). 

The white dwarf effective temperature correlates with its surface gravity: strong gravitational fields translate into pressure broadening of the lines; this effect can be balanced by higher temperatures that increase the fraction of ionised hydrogen, resulting in narrower absorption lines. It is not possible to break this degeneracy from the sole analysis of the \hst data since they only provide the Ly$\alpha$ absorption profile which, in the case of cool CV white dwarfs, is limited to only the red wing of the line. 
Therefore, in the past, the analysis of CV ultraviolet \hst data has been limited mainly to accurate measurements of the white dwarf effective temperatures, obtained for a fixed $\log(g)$ for the white dwarfs. 
Nowadays, thanks to the parallaxes provided by \gaia EDR3, the knowledge of the distance $d$ to the system finally allows us to constrain the radius of the white dwarf. Under the assumption of a mass-radius relationship, the white dwarf mass can be derived from the scaling factor between the \hst data and the best-fitting model, according to the equation:
\begin{equation}\label{eq:scaling}
S = \left( \frac{R_\mathrm{WD}}{d} \right) ^2
\end{equation}
By breaking this degeneracy, we can simultaneously measure both the white dwarf effective temperature and mass. 

The \hst ultraviolet spectra are contaminated by geocoronal emission of Ly$\alpha$ and \ion{O}{i} (1302\,\AA). The first is always detected and we masked the corresponding wavelength range for our spectral analysis. In the case of \ion{O}{i}, the related wavelength range was masked only when the emission was detected in the spectrum. Moreover, CV ultraviolet spectra show the presence of an additional continuum component, which contributes $\simeq 10-30\,$per cent of the observed flux. The origin of this additional emission source is unclear and it has been suggested that it could arise from either (i) the disc, or (ii) the hot spot where the ballistic stream intersects with the disc or (iii) the interface region between the disc and the white dwarf surface \cite[e.g.][]{Long+1993,Godon+2004,Gaensicke+2005}.
In the literature, different approximations have been used to model the emission of this additional component, such as a blackbody, a power law or a constant flux (in $F_\mathrm{\lambda}$). As discussed by \cite{Pala+2017}, these assumptions represent a very simplified model of the additional continuum contribution and it is likely that none of them provides a realistic physical description of this emission component. These authors also showed that, when only a limited wavelength coverage ($1105-1800\,$\AA) is available, it is not possible to statistically discriminate among the three of them, and they all result in fits of similar quality for the white dwarf. We therefore decided to use a blackbody, which is described by two free parameters (a temperature and a scaling factor), because, in the limited wavelength here considered, its tail approximates both the power law and the constant flux cases.

\begin{figure}
\includegraphics[width=\columnwidth]{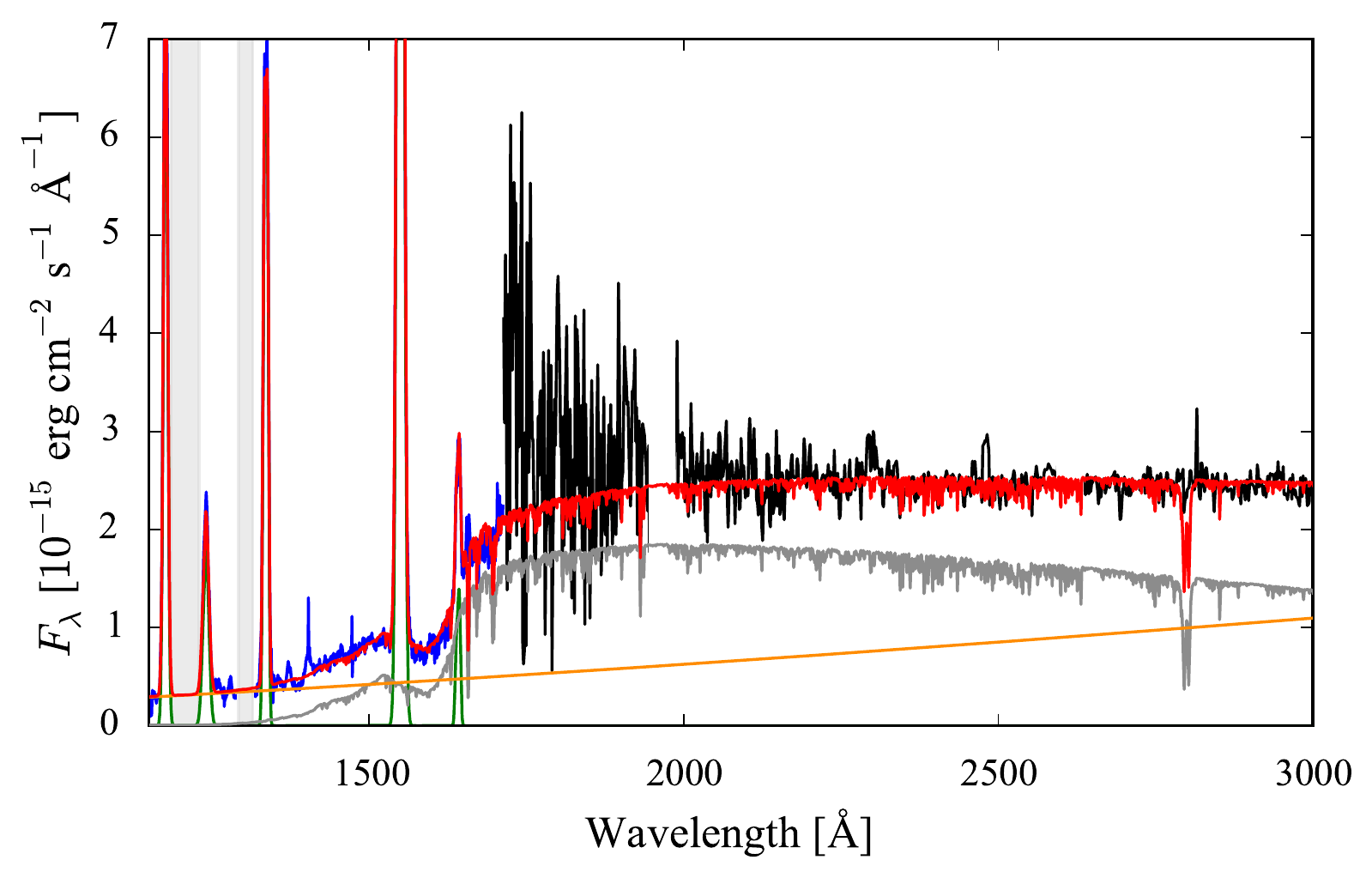}
\caption{\hst spectrum of GD\,552 obtained from the combination of the STIS/G140L far-ultraviolet (blue) and STIS/G230L (black) data (the grey bands mask the geocoronal emission lines of Ly$\alpha$ at $1216\,$\AA\, and of \ion{O}{i} at $1302\,$\AA, while the gaps at $\simeq 1950\,$\AA\, and $\simeq 2600\,$\AA\, correspond to data with bad quality flags). The best fitting model (red) is composed of the sum of the white dwarf emission (grey), the emission lines approximated as Gaussian profiles (green), and an additional second component (orange) which shows a clear dependence on wavelength. This suggests that this additional emission component in the system is either an optically thin emission region, or an optically thick accretion disc, or a bright spot. These cases are expected to display a power-law distribution and, in our fitting procedure, we assumed a power law to model the emission of this additional component in GD\,552.}\label{fig:gd552}
\end{figure}

The only exceptions are the eclipsing systems (discussed below) and GD\,552. For the latter, additional data obtained with the G230L grating, covering the wavelength range $1650-3150\,$\AA, are available. As already noticed by \citet{Unda-Sanzana2008}, the additional second component in the near-ultraviolet flux of GD\,552 shows a clear dependence on wavelength (orange line in Figure~\ref{fig:gd552}). The observed slope could arise from either an optically thin emission region, or an optically thick accretion disc, or a bright spot. All cases are expected to display a power-law distribution (which, for the optically thick disc and the bight spot, results from the approximation with a sum of blackbodies) and, given the wide wavelength coverage available for this object, a single black body would represent a poor approximation for the additional emission component. Therefore, in the case of GD\,552, we assumed a power law, which is described by two free parameters (a power law index and a scaling factor).

The analysis of the eclipsing systems is complicated by the presence of the so--called ``iron curtain'', i.e. a layer of absorbing material extending above the disc which gives rise to strong absorption features (see, for example, the spectrum of IY\,UMa in Figure~\ref{fig:sample_spectra}), mainly a forest of blended \ion{Fe}{ii} absorption lines \citep{Horne+1994}. These veil the white dwarf emission, making it difficult to establish the actual flux level (necessary to constrain the white dwarf radius). In addition, these lines modify the overall slope of the spectrum as well as the shape of the core of the Ly$\alpha$ line, which are the tracers for the white dwarf \teff.
Out of the six eclipsing systems in our sample, two are strongly affected by the veiling gas: IY\,UMa and DV\,UMa. 
Following the consideration by \cite{Pala+2017}, in the spectral fitting of these CVs, we included two homogeneous slabs, one cold ($T_\mathrm{curtain} \simeq 10\,000\,$K) and one hot ($T\mathrm{curtain} \simeq 80\,000\,$K). We generated two grids of monochromatic opacity of the slabs using \textsc{synspec}, one covering the effective temperature range $T_\mathrm{eff} = 5\,000 - 25\,000\,$K and the other the range $T_\mathrm{eff} = 25\,000 - 120\,000\,$K, both in steps of 5000\,K. Both grids covered the electron density range $n_\mathrm{e} = 10^9 - 10^{21} \mathrm{cm}^{-3}$ in steps of $10^3 \mathrm{cm}^{-3}$ and the turbulence velocity range $V_\mathrm{t} = 0 -500\,\mathrm{km\,s}^{-1}$ in steps of $100\,\mathrm{km\,s}^{-1}$. These models, combined with the column densities ($N_\mathrm{H}$, for which we assumed a flat prior in the range $10^{17} - 10^{23}\,\mathrm{cm}^{-2}$), return the absorption due to the curtain. Given the large number of free parameters involved in the spectral fitting of the eclipsing systems, we chose to use a constant flux (in $F_\mathrm{\lambda}$) to approximate the additional continuum component, as this is the simplest approximation and introduces only one more additional free parameter in the fitting procedure.
 
 \begin{figure*}
\includegraphics[width=0.31\textwidth]{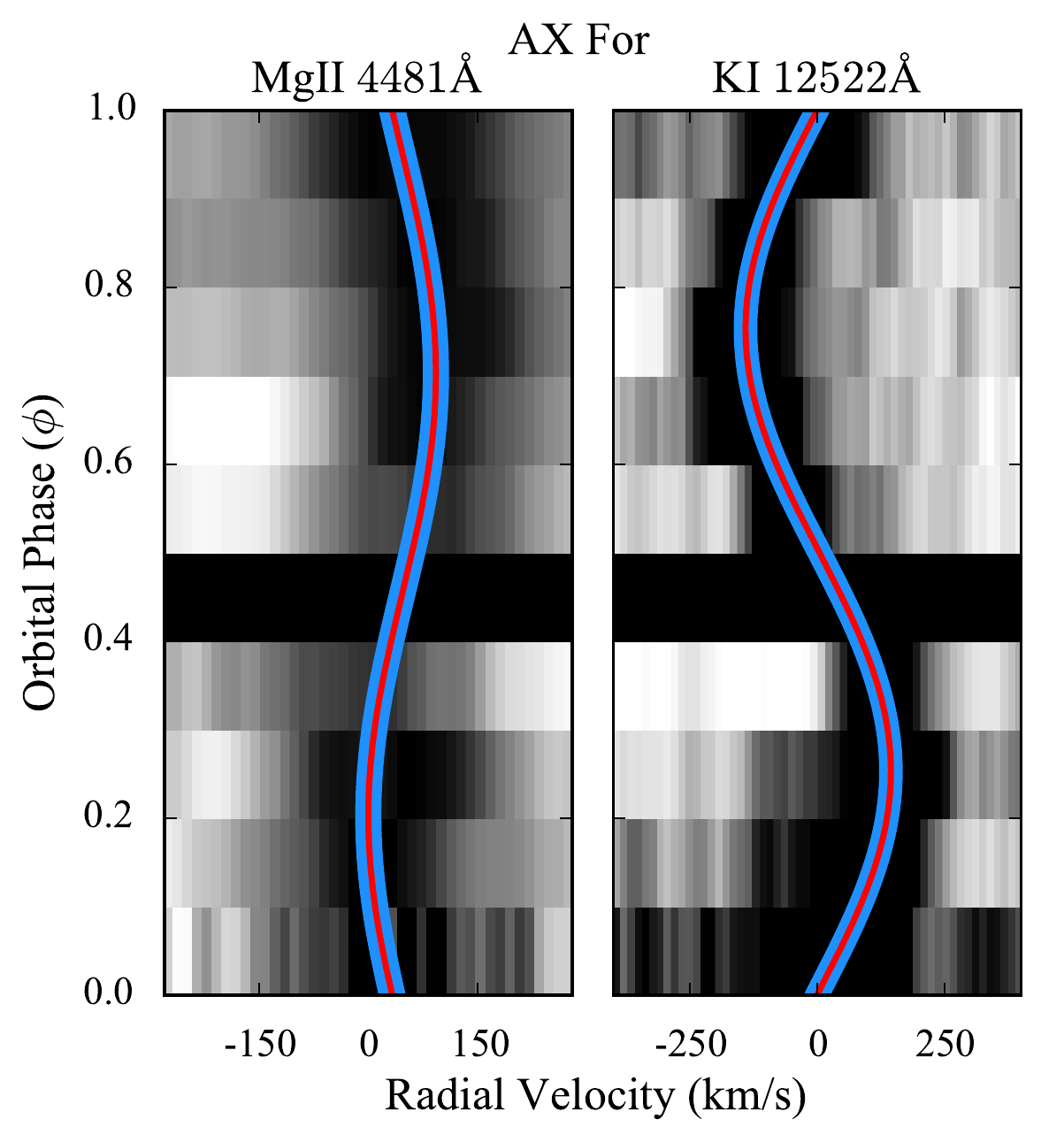}\quad\quad
\includegraphics[width=0.31\textwidth]{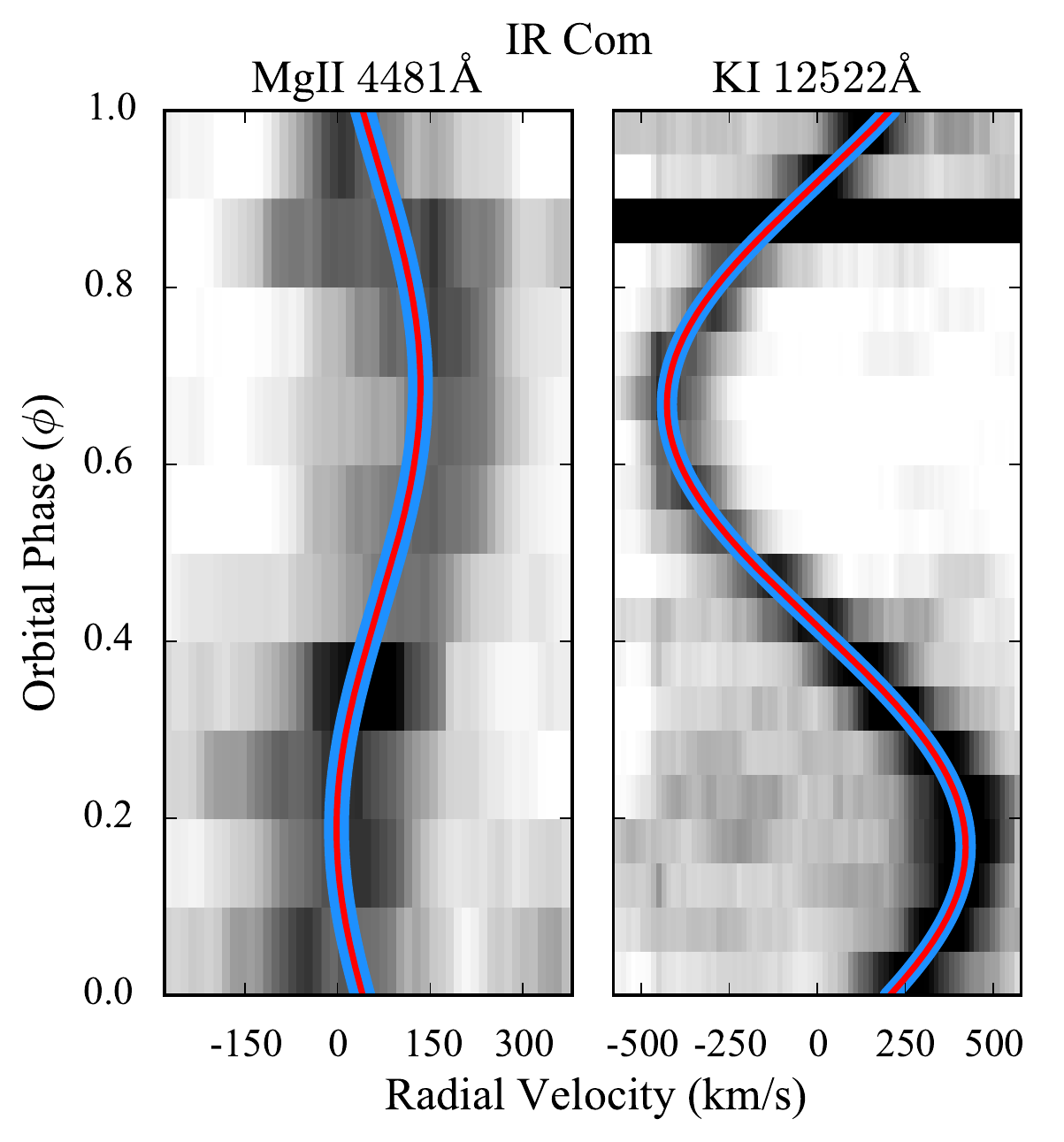}\quad\quad
\includegraphics[width=0.31\textwidth]{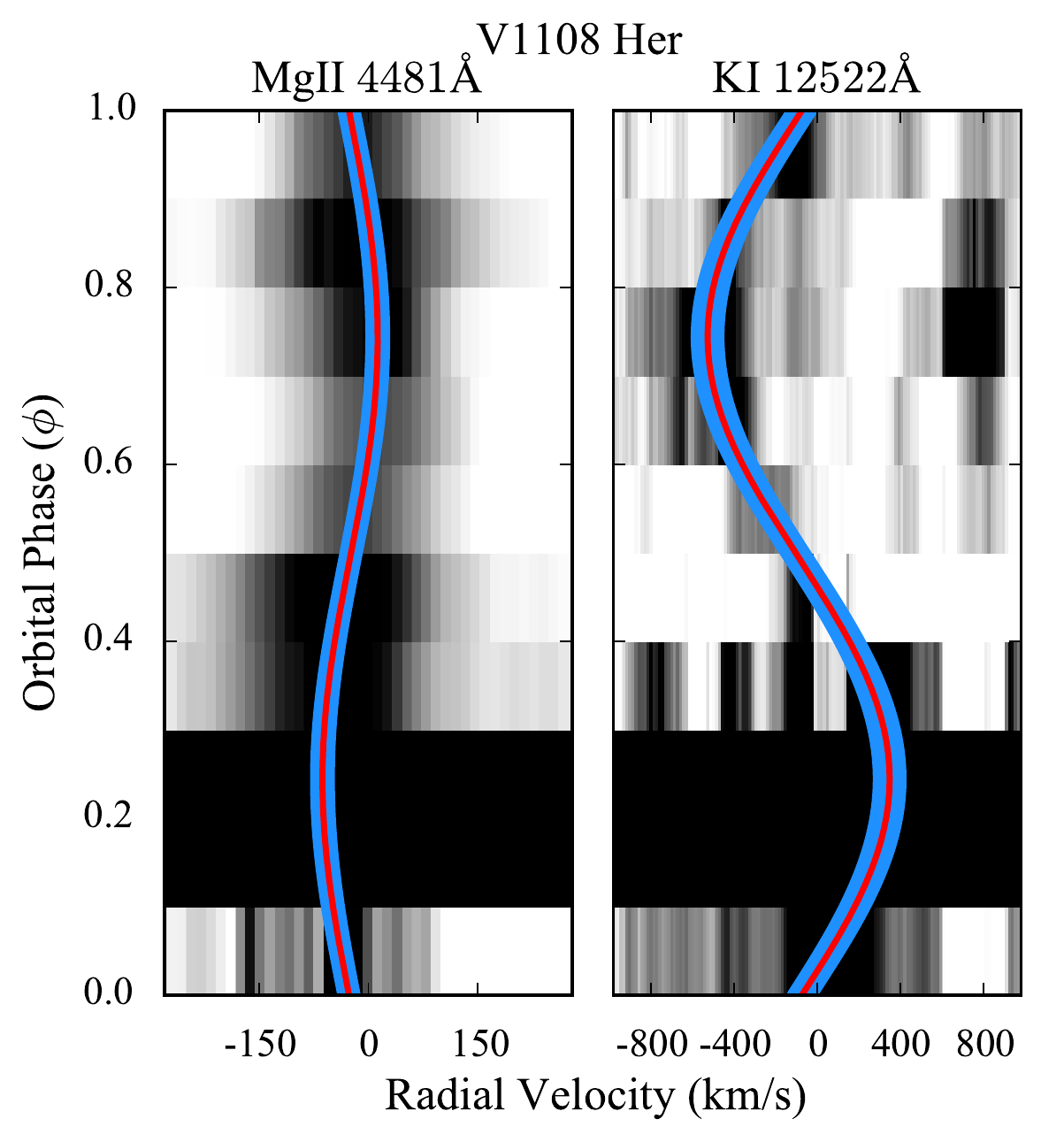}
\caption{For each system, the trailed spectra for the \ion{Mg}{ii} line ($4481$ \AA, left panel) and the \ion{K}{i} line ($12\,522$ \AA, right panel) are shown. Overplotted are the best fitting models (red) along with their uncertainties (light blue).}\label{fig:trail_spectra}
\end{figure*}

\begin{figure*}
\begin{overpic}[width=\columnwidth]{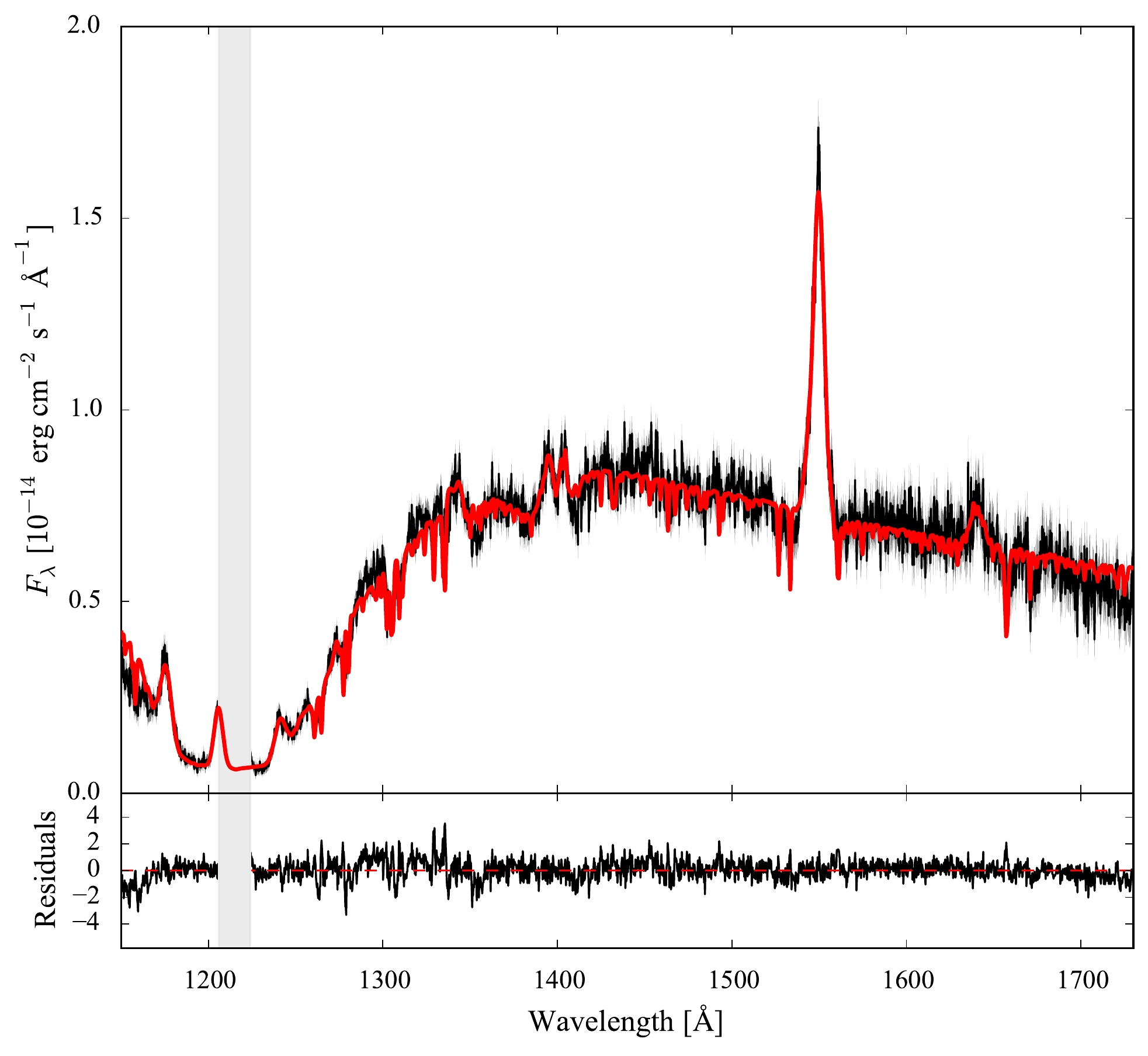}
\put (12,85) {CU\,Vel} 
\put (35,85) {$\teff = 14\,174^{+117}_{-169}\,$K}
\put (67,85) {$\mwd = 0.47^{+0.04}_{-0.05}\,\mathrm{M_\odot}$}
\end{overpic} \quad
 \begin{overpic}[width=\columnwidth]{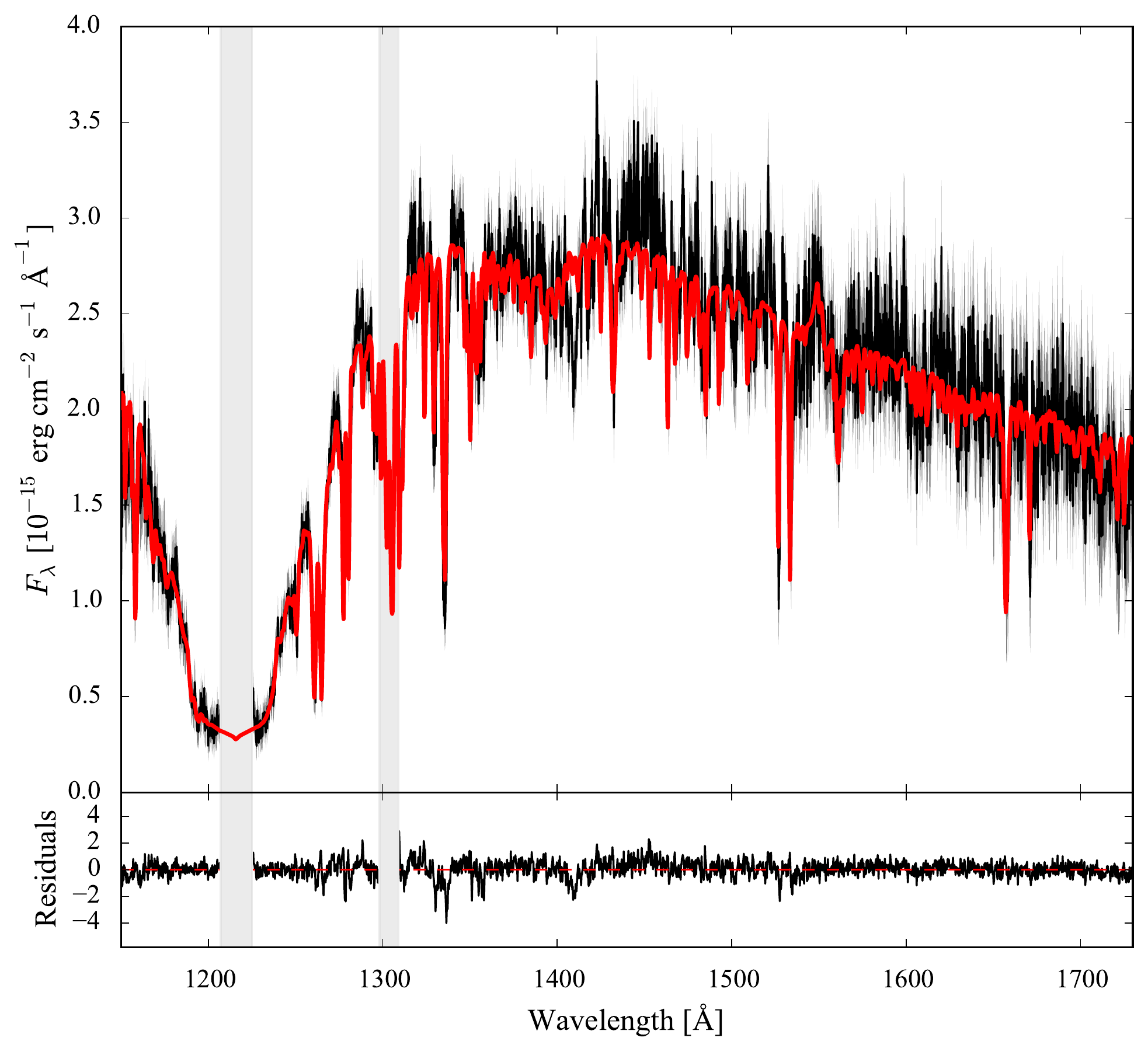}
 \put (13,85) {IR\,Com}
 \put (35,85) {$\teff = 17\,531^{+236}_{-271}\,$K} 
 \put (67,85) {$\mwd = 1.03^{+0.07}_{-0.09}\,\mathrm{M_\odot}$}  
 \end{overpic}
 \\
 \begin{overpic}[width=\columnwidth]{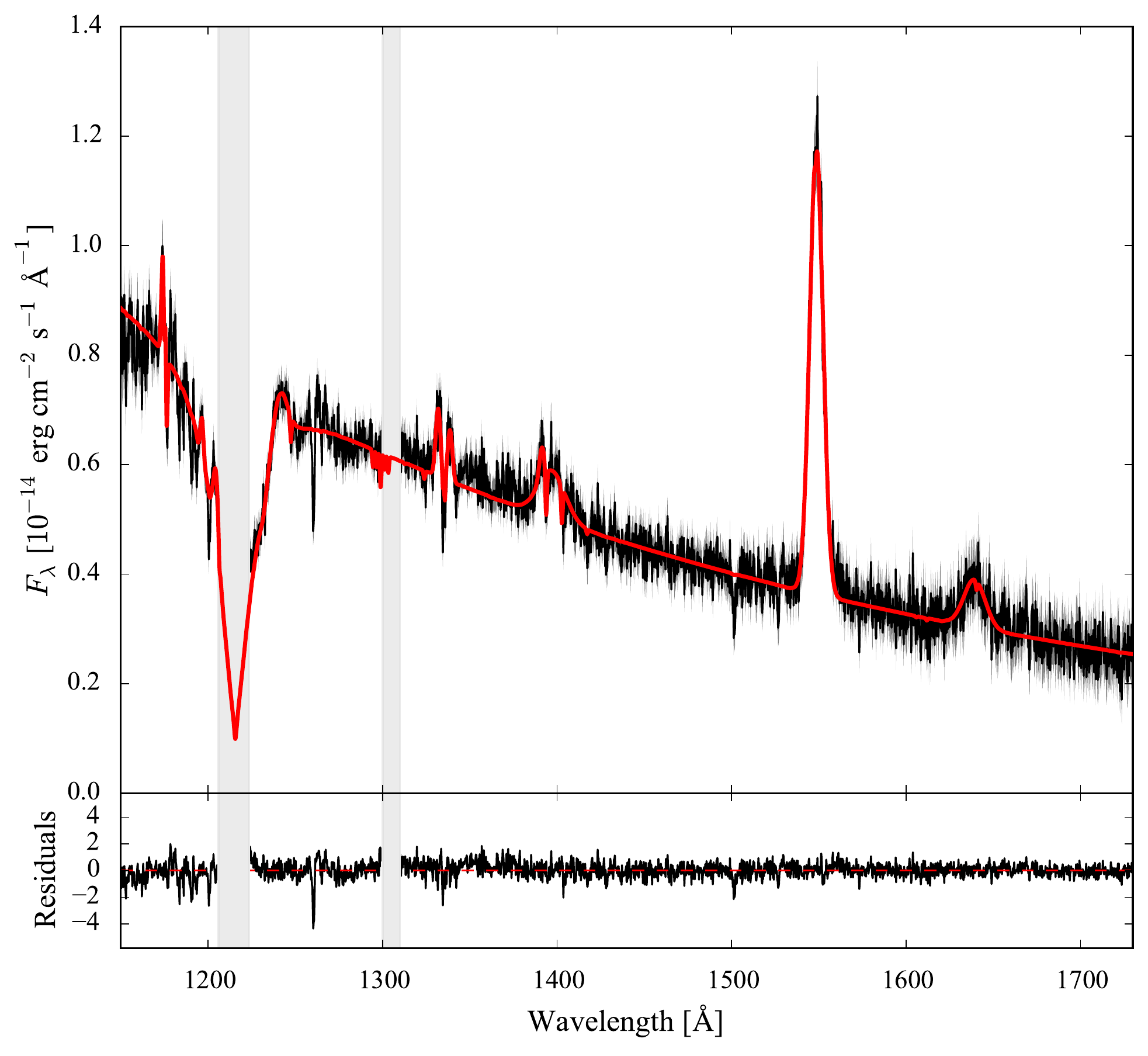}
 \put (12,85) {SDSS\,J153817.35+512338.0}
 \put (12,79) {$\teff = 35\,284^{+600}_{-688}\,$K} 
 \put (12,73) {$\mwd = 0.97^{+0.09}_{-0.11}\,\mathrm{M_\odot}$} 
 \end{overpic} \quad
 \begin{overpic}[width=\columnwidth]{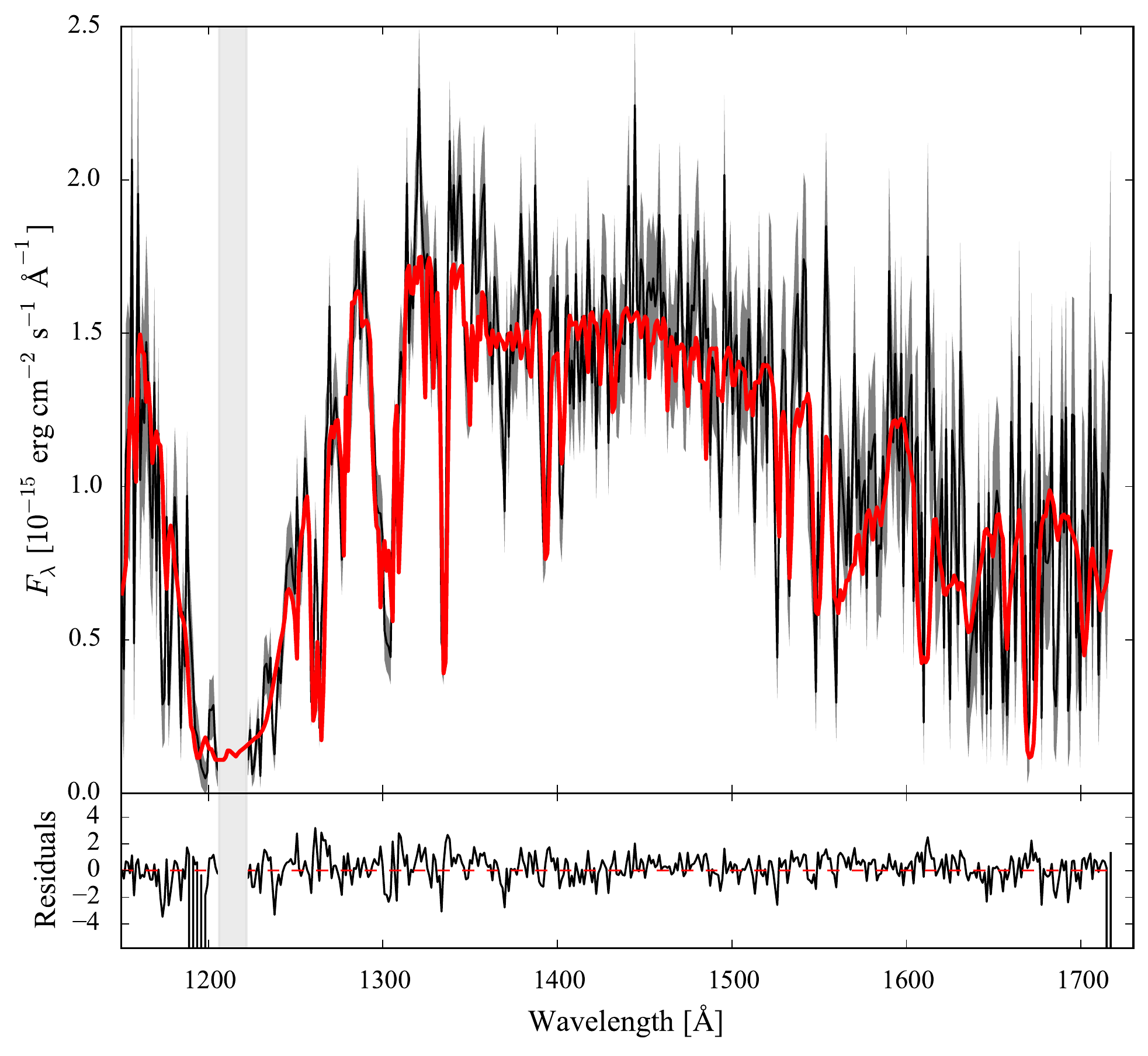}
 \put (13,85) {DV\,UMa}
 \put (35,85) {$\teff = 19\,418^{+244}_{-400}\,$K} 
 \put (67,85) {$\mwd = 0.96^{+0.07}_{-0.10}\,\mathrm{M_\odot}$}  
 \end{overpic} 
\caption{Ultraviolet spectra (black) of sample CV white dwarfs along with the best-fitting model (red), representative of cool (top left panel), warm (top right pane), hot (bottom left panel) and eclipsing (bottom right panel) systems. The best-fitting models are composed of the sum of a white dwarf synthetic atmosphere model, a second continuum emission component in the form of a blackbody, and the emission lines from the disc modelled with a Gaussian profile. In addition, the model for the eclipsing system DV\,UMa (bottom right) includes an ``absorption curtain'' component (see Section~\ref{sec:uv_fit}). The grey bands mask the geocoronal emission lines of Ly$\alpha$ ($1216\,$\AA) and, whenever present, of \ion{O}{i} ($1302\,$\AA).}\label{fig:uv_fit1}
\end{figure*}

Finally, for all systems, whenever detected, we included the emission lines arising from the accretion disc as Gaussian profiles, allowing three free parameters: amplitude ($f_\mathrm{em}$), wavelength ($\lambda_\mathrm{em}$) and width ($\sigma_\mathrm{em}$). As shown by \cite{Pala+2017}, by including or masking the disc lines has no influence on the result but their inclusion allowed us to use as much of the data as possible.

\begin{table}
\centering
\caption{Summary of the fit parameters employed in the analysis of the ultraviolet spectra described in Section~\ref{sec:uv_fit}. }\label{tab:fit_parameters}
\setlength{\tabcolsep}{0.13cm}
\begin{tabular}{lcccc}
\toprule
System component & & Parameter & Free? & Range of variation   \\
\midrule     
& & Distance & \xmark & -- \\ 
& & $E(B-V)$ & \xmark & -- \\[0.15cm]
\multirow{2}{*}{White dwarf} & & $\teff$ (K) & \checkmark & 9000 - 70\,000\\
& & $\log(g)$ & \checkmark & 6.4 - 9.5 \\[0.15cm]
\multirow{3}{*}{Emission lines}  & & $f_\mathrm{em}$ & \checkmark & > 0 \\
  & & $\lambda_\mathrm{em}$ & \checkmark & > 0 \\
  & & $\sigma_\mathrm{em}$ & \checkmark & > 0 \\[0.15cm] 
\multirow{5}{*}{Second components} & \multirow{2}{*}{BB} & $\teff$ (K) & \checkmark & > 0   \\ 
& & scaling factor & \checkmark & > 0  \\[0.1cm]
& \multirow{2}{*}{PL} & Exponent & \checkmark & $\mathbb{R}$   \\
& & scaling factor & \checkmark & > 0  \\[0.1cm]
& constant & & \checkmark & > 0  \\[0.15cm]
\multirow{8}{*}{Slabs} &
\multirow{4}{*}{Cold} & $T_\mathrm{curtain}$ (K) & \checkmark & 5000 - 25\,000 \\
& & $\log (n_\mathrm{e} \cdot \mathrm{cm}^3)$ & \checkmark &  9 - 21 \\
& & $\log (N_\mathrm{H} \cdot \mathrm{cm}^2)$ & \checkmark &  17 - 23 \\
& & $V_\mathrm{t}\,(\mathrm{km\,s}^{-1})$ & \checkmark & 0 - 500\\[0.1cm]
& \multirow{4}{*}{Hot} & $T_\mathrm{curtain}$ (K) & \checkmark & 25\,000 - 120\,000 \\
& & $\log (n_\mathrm{e} \cdot \mathrm{cm}^3)$ & \checkmark &  9 - 21 \\
& & $\log (N_\mathrm{H} \cdot \mathrm{cm}^2)$ & \checkmark &  17 - 23 \\
& & $V_\mathrm{t}\,(\mathrm{km\,s}^{-1})$ & \checkmark & 0 - 500\\
\bottomrule
\end{tabular}
\end{table}

We performed the spectral fit using the Markov chain Monte Carlo (MCMC) implementation for Python, \textsc{emcee} \citep{emcee}.

We scaled the models according to the distance to the system, computed as described in Section~\ref{subsec:gaia_quality}, and reddened them according to the extinction values reported in Table~\ref{tab:hst_obs_log}. We used the STructuring by Inversion the Local Interstellar Medium (Stilism) reddening map \citep{Lallement+2018} to derive the $E(B-V)$. For those objects not included in the Stilism reddening maps, we used instead the three-dimensional map of interstellar dust reddening based on Pan-STARRS\,1 and 2MASS photometry \citep{Green+2019}.

The free parameters of the fit and their allowed range of variations are listed in Table~\ref{tab:fit_parameters}, and we assumed a flat prior in the ranges covered by the corresponding grid of models. We assumed the mass-radius relation of \citet{Holberg+2006,Tremblay+2011} and constrained the parameters describing the black body (BB) and constant additional second components to be positive. In the case of the power-law (PL) additional second component, we only constrained its scaling factor to be positive.

The accuracy derived from the statistical uncertainties are typically two and four per cent for $\teff$ and $\rwd$, respectively.
These uncertainties together provide an accuracy on the white dwarf masses of typically $0.03\,\msun$.
However, the real uncertainties are dominated by systematic effects, which are discussed in the following Section.

\subsubsection{Uncertainty estimate}\label{subsubsec:errors}
As discussed by \citet{Pala+2017}, we can rule out the presence of systematics arising from instrument flux calibration issues as well as any noticeable contamination from additional Ly$\alpha$ absorption from the second emission component and/or interstellar gas along the line of sight.
Moreover, the uncertainties related to the unknown nature of the second component are smaller than the statistical errors from the fitting procedure \citep{Pala+2017}, therefore their effect is already accounted for in the error balance from the previous Section.

The remaining sources of uncertainties are hence those related to the precision on the \gaia parallaxes and the reddening measurements.
The reddening due to interstellar dust along the line of sight affects the overall slope and flux level of the observed spectrum, thus influencing both the radius and \teff measurements. Similarly, the precision on the \gaia parallax directly reflects that in the radius and hence in the white dwarf mass.
 
Allowing a $3\,\sigma$ variation for both the parallax and the reddening, the observed flux level of a system could be reproduced by a combination of high reddening (i.e. more absorption along the line of sight) and a large parallax (i.e. moving the object closer to the observer) and vice versa.
This degeneracy cannot be broken from the sole analysis of the \hst data since they do not extend to the wavelength range where the signature of interstellar dust absorption could be detected as a bump at $\simeq 2175\,$\AA, from which the colour excess can be estimated.

Ideally, the spectroscopic fitting procedure described in the previous section could account for this correlation by allowing both the distance and the reddening to vary according to suitable priors. In practice, this approach would require including in our fitting procedure the probability density function we used to compute the distance to each system in Section~\ref{subsec:gaia_quality}. Since this is not straightforward, we preferred to use an alternative method and estimate the systematic uncertainty on our mass measurements employing a Monte Carlo approach.
We used a $\chi^2$ minimisation routine to fit each spectrum 5000 times. During each execution, the models were scaled assuming a distance drawn randomly from the probability density function used to compute the distance to each system in Section~\ref{subsec:gaia_quality}. Similarly, the models were also corrected for the reddening, assuming a colour excess drawn randomly from a normal distribution, centred on the $E(B-V)$ reported in Table~\ref{tab:hst_obs_log} and weighted according to its uncertainty.
Keeping the distance and the reddening fixed to the values drawn from these distributions, we allowed as free parameters the white dwarf $\teff$ and $\log(g)$ and the second continuum component (as defined in the previous Section). To speed up the calculation, we masked the emission lines.

From the best-fitting parameters obtained from each of the 5000 executions, we derived the posterior distribution of $\teff$ and $\log(g)$, which provided the estimates of the systematic uncertainties related to the accuracy on the distance and the reddening. We compared these systematics with the statistical uncertainties derived in the previous Section and assumed as final value the maximum of the two.

In the next few years, the upcoming \gaia data releases will improve the accuracy on the parallaxes, and thereby that on the distances to CVs, and will allow reconstruction of more detailed three-dimensional reddening maps. Eventually, these will make it possible to reduce the uncertainties on the white dwarf parameters to the level of the statistical ones.

\subsection{Radial velocity measurements from optical spectra}\label{subsec:rv}
Among the different lines arising from the secondary photosphere listed in Section~\ref{subsec:optical_obs}, the \ion{K}{i} ($12\,432/12\,522$ \AA) lines are the strongest and the only ones visible in the spectra of all three systems. The \ion{Na}{i} doublet is only visible in the spectra of IR\,Com but it is contaminated by the residual of the telluric line removal. Therefore, we decided not to include it in the following analysis. We used the \ion{Mg}{ii} absorption feature and the \ion{K}{i} lines to track the reflex motion of the white dwarf and the donor, respectively.
Our data were characterised by a relatively low SNR ($\simeq 10-20$ in the UVB and $\simeq 5$ in the NIR) and, moreover, the NIR spectra showed strong contamination from the residuals from the sky lines subtraction. Therefore, to achieve more robust radial velocity measurements, we fit all the spectra of each object simultaneously (as done, for example, by \citealt{steven_sdss0857} and \citealt{Pala+2019}).

We first fitted the \ion{K}{i} absorption lines using a combination of a constant and a double Gaussian of fixed separation. We allowed the wavelength of the Gaussians to change according to the following Equation:
\begin{equation}
V = \gamma + K \sin[2\pi(\phi-\phi_0)]
\end{equation}
where $V$ is the radial velocity, $\gamma$ is the systemic velocity, $K$ is the velocity amplitude, $\phi$ is the orbital phase and $\phi_0$ is the zero point of the ephemeris.
We fitted the \ion{Mg}{ii} absorption lines using a combination of a constant and a single Gaussian. From these fitting procedures, we derived the systemic velocities of the white dwarf and the donor ($\gamma_\mathrm{WD}$ and $\gamma_\mathrm{donor}$, Figure~\ref{fig:trail_spectra}). Their difference provides a direct measurement of the gravitational redshift of the white dwarf and thereby of its surface gravity\citep{Greenstein+1967}:
\begin{equation}\label{eq:gr}
v_\mathrm{grav}(\mathrm{WD}) = \gamma_\mathrm{WD} - \gamma_\mathrm{donor} = 0.635 \, \frac{M_\mathrm{WD}}{M_\odot} \, \frac{\rsun}{\rwd}\,\mathrm{km\,s}^{-1}
\end{equation} 
At the surface of both the white dwarf and the secondary, a contribution from the gravitational field of the other star is present.
For $\porb < 5\,$h, CV secondary stars have typically $M_\mathrm{donor} \lesssim 0.6\,\mathrm{M}_\odot$ and $\log(g)_\mathrm{donor} \lesssim 5$ \citep[e.g.][]{Knigge+2011}, therefore their contribution only introduces a small correction near the white dwarf surface, $\simeq 0.1\, \mathrm{km s^{-1}}$. Similarly, the influence of the gravitational potential of the white dwarf introduces a correction of $\simeq 1\, \mathrm{km s^{-1}}$ near the donor star surface. Both these effects are negligible compared to the typical uncertainties ($\simeq 5-10\, \mathrm{km s^{-1}}$) on $v_\mathrm{grav}(\mathrm{WD})$ and can be safely ignored.

By assuming the same mass-radius relationship as in Section~\ref{subsubsec:ircom_sdss1507} and using Equation~\ref{eq:gr}, we measured the masses of the three CV white dwarfs, which result in $0.76^{+0.06}_{-0.07}\,\msun$, $0.95\pm0.04 \,\msun$ and $0.91^{+0.14}_{-0.20}\,\msun$, for AX\,For, IR\,Com and V1108\,Her, respectively (Table~\ref{tab:comparison}).
In Section~\ref{subsubsec:rv_uv_comparison}, we discuss these results in comparison with the masses obtained from the analysis of the ultraviolet spectra.

The accuracy we achieved on these measurements is directly related to the quality of the data. An optimal sampling of the orbital period is crucial to precisely measure radial velocities. This can be seen by comparing the results for IR\,Com with those for AX\,For and V1108\,Her. The data of the former are almost evenly distributed along the orbital period and allow us to obtain accuracy of the order $\simeq 7$ per cent. However, the observations of AX\,For were affected by clouds while those of V1108\,Her were contaminated by the presence of a close background star. In both cases, we were forced to reject some spectra, resulting in poor orbital sampling and larger uncertainties.

\begin{table*}
\thisfloatpagestyle{empty}
\vspace{-0.7cm}
 \caption{White dwarf parameters and mass accretion rates for the 42 CVs observed with \hst during their quiescent state and for AX\,For, whose radius and mass have been obtained from the white dwarf gravitational redshift.}\label{tab:results_good}
\setlength{\tabcolsep}{0.08cm}
\begin{tabular}{@{} lcccc|cccccc @{}}
\toprule
System & $\porb$ & $d$ & $Z$             & References & $\teff$ & $\rwd$ & $\mwd$ & $\log(g)$ & $\langle \dot{M} \rangle $ & White dwarf \\
           &  (min)     & (pc) & $Z_\odot$  &                   & (K)       & ($0.01\,\mathrm{R}_\odot$) & ($\mathrm{M}_\odot$) &  & $(10^{-10}\,\mathrm{M}_\odot$ yr$^\mathrm{-1}$) & contribution ($\%$)\\ 
\midrule
SDSS\,J150722.30+523039.8 & 66.61 & $211 \pm 4$ & 0.1 & 1 & $14\,207^{+356}_{-403}$ & $0.93^{+0.17}_{-0.12}$ & $0.90^{+0.10}_{-0.14}$ & $8.45^{+0.16}_{-0.22}$ & $0.53^{+0.18}_{-0.11}$ & 78 \\ [0.13cm]
SDSS\,J074531.91+453829.5 & 76.0 & $310^{+23}_{-20}$ & 0.1 & 2 & $15\,447^{+556}_{-654}$ & $1.1^{+0.3}_{-0.2}$ & $0.75^{+0.18}_{-0.20}$ & $8.2^{+0.3}_{-0.4}$ & $1.2^{+0.7}_{-0.4}$ & 88 \\ [0.13cm]
GW\,Lib & 76.78 & $112.6 \pm 0.8$ & 0.2 & 3 & $16\,166^{+253}_{-350}$ & $1.03^{+0.15}_{-0.1}$ & $0.83^{+0.08}_{-0.12}$ & $8.33^{+0.13}_{-0.18}$ & $1.12^{+0.32}_{-0.19}$ & 86 \\ [0.13cm]
SDSS\,J143544.02+233638.7 & 78.0 & $208^{+9}_{-8}$ & 0.01 & 3 & $11\,997^{+99}_{-160}$ & $1.00^{+0.08}_{-0.09}$ & $0.84^{+0.07}_{-0.06}$ & $8.36^{+0.11}_{-0.1}$ & $0.32^{+0.08}_{-0.05}$ & 90 \\ [0.13cm]
OT\,J213806.6+261957 & 78.1 & $98.9 \pm 0.4$ & 0.2 & 3 & $15\,317^{+216}_{-228}$ & $1.39^{+0.15}_{-0.12}$ & $0.57^{+0.06}_{-0.07}$ & $7.91^{+0.12}_{-0.15}$ & $1.9^{+0.4}_{-0.3}$ & 73 \\ [0.13cm]
BW\,Scl & 78.23 & $93.4 \pm 0.5$ & 0.5 & 4 & $15\,145^{+51}_{-57}$ & $0.8^{+0.014}_{-0.011}$ & $1.007^{+0.010}_{-0.012}$ & $8.635^{+0.017}_{-0.02}$ & $0.483^{+0.014}_{-0.012}$ & 87 \\ [0.13cm]
LL\,And & 79.28 & $609^{+343}_{-205}$ & 1.0 & 5 & $14\,353^{+1210}_{-743}$ & $1.2^{+0.8}_{-0.6}$ & $0.7^{+0.4}_{-0.3}$ & $8.2 \pm 0.7$ & $0.9^{+2.4}_{-0.9}$ & 80 \\ [0.13cm]
AL\,Com & 81.6 & $523^{+252}_{-149}$ & 0.2 & 6 & $15\,840^{+1453}_{-1243}$ & $1.0^{+0.9}_{-0.6}$ & $0.9^{+0.5}_{-0.3}$ & $8.4 \pm 0.8$ & $0.9^{+3.1}_{-1.0}$ & 79 \\ [0.13cm]
WZ\,Sge & 81.63 & $45.17 \pm 0.06$ & 0.01 & 7 & $13\,190^{+115}_{-105}$ & $1.05 \pm 0.03$ & $0.8 \pm 0.02$ & $8.3 \pm 0.04$ & $0.52^{+0.014}_{-0.01}$ & 78 \\ [0.13cm]
SW\,UMa & 81.81 & $160.6^{+1.6}_{-1.5}$ & 0.2 & 4 & $13\,854^{+189}_{-131}$ & $1.29^{+0.09}_{-0.10}$ & $0.61^{+0.06}_{-0.04}$ & $8.01^{+0.11}_{-0.09}$ & $1.07^{+0.15}_{-0.14}$ & 61 \\ [0.13cm]
V1108\,Her & 81.87 & $148 \pm 2$ & 1.0 & 3 & $13\,943^{+185}_{-226}$ & $0.95^{+0.14}_{-0.11}$ & $0.88^{+0.09}_{-0.11}$ & $8.42^{+0.15}_{-0.17}$ & $0.52^{+0.15}_{-0.11}$ & 75 \\ [0.13cm]
ASAS\,J002511+1217.2 & 82.0 & $157 \pm 3$ & 0.1 & 3 & $13\,208^{+154}_{-112}$ & $0.78 \pm 0.04$ & $1.02 \pm 0.04$ & $8.66 \pm 0.07$ & $0.27^{+0.03}_{-0.02}$ & 74 \\ [0.13cm]
HV\,Vir & 82.18 & $317^{+29}_{-25}$ & 0.2 & 8 & $12\,958^{+257}_{-291}$ & $0.96^{+0.21}_{-0.16}$ & $0.87^{+0.13}_{-0.17}$ & $8.4^{+0.2}_{-0.3}$ & $0.4^{+0.2}_{-0.12}$ & 93 \\ [0.13cm]
SDSS\,J103533.02+055158.4 & 82.22 & $195^{+12}_{-10}$ & 0.01 & 3 & $11\,876^{+108}_{-115}$ & $0.8^{+0.11}_{-0.09}$ & $1.00^{+0.08}_{-0.10}$ & $8.63^{+0.14}_{-0.16}$ & $0.18^{+0.06}_{-0.04}$ & 91 \\ [0.13cm]
WX\,Cet & 83.9 & $252^{+9}_{-8}$ & 0.1 & 9 & $15\,186^{+1460}_{-1468}$ & $0.63^{+0.28}_{-0.18}$ & $1.10^{+0.17}_{-0.21}$ & $8.8 \pm 0.4$ & $0.3^{+0.07}_{-0.06}$ & 24 \\ [0.13cm]
SDSS\,J075507.70+143547.6 & 84.76 & $239^{+12}_{-11}$ & 0.5 & 3 & $16\,193^{+280}_{-357}$ & $0.92^{+0.15}_{-0.11}$ & $0.91^{+0.09}_{-0.12}$ & $8.47^{+0.15}_{-0.19}$ & $0.87^{+0.27}_{-0.18}$ & 87 \\ [0.13cm]
SDSS\,J080434.20+510349.2 & 84.97 & $142 \pm 2$ & 0.5 & 10 & $13\,715^{+55}_{-79}$ & $0.80^{+0.04}_{-0.03}$ & $1.01^{+0.03}_{-0.04}$ & $8.64^{+0.05}_{-0.06}$ & $0.32 \pm 0.03$ & 87 \\ [0.13cm]
EK\,TrA & 86.36 & $151.4 \pm 0.8$ & 0.5 & 11 & $17\,608^{+269}_{-481}$ & $0.97^{+0.17}_{-0.11}$ & $0.87^{+0.09}_{-0.13}$ & $8.4^{+0.14}_{-0.21}$ & $1.4^{+0.4}_{-0.3}$ & 58 \\ [0.13cm]
EG\,Cnc & 86.36 & $186 \pm 7$ & 0.2 & 8 & $12\,295^{+56}_{-57}$ & $0.77^{+0.06}_{-0.05}$ & $1.03^{+0.04}_{-0.05}$ & $8.67^{+0.07}_{-0.08}$ & $0.2^{+0.03}_{-0.02}$ & 100 \\ [0.13cm]
1RXS\,J105010.8--140431 & 88.56 & $108.9^{+1.1}_{-1.0}$ & 0.1 & 3 & $11\,523^{+29}_{-47}$ & $1.08^{+0.04}_{-0.03}$ & $0.77^{+0.02}_{-0.03}$ & $8.25^{+0.04}_{-0.05}$ & $0.332^{+0.028}_{-0.018}$ & 87 \\ [0.13cm]
BC\,UMa & 90.16 & $293^{+11}_{-10}$ & 0.2 & 4 & $14\,378^{+272}_{-327}$ & $1.57^{+0.19}_{-0.16}$ & $0.48^{+0.08}_{-0.09}$ & $7.73^{+0.15}_{-0.19}$ & $2.0^{+0.5}_{-0.3}$ & 85 \\ [0.13cm]
VY\,Aqr & 90.85 & $141.3^{+1.9}_{-1.8}$ & 0.5 & 9 & $14\,453^{+316}_{-366}$ & $0.74^{+0.07}_{-0.06}$ & $1.06 \pm 0.06$ & $8.73^{+0.12}_{-0.11}$ & $0.33^{+0.04}_{-0.03}$ & 45 \\ [0.13cm]
QZ\,Lib & 92.36 & $199^{+11}_{-10}$ & 0.01 & 3 & $11\,419^{+175}_{-229}$ & $1.01^{+0.23}_{-0.18}$ & $0.82^{+0.14}_{-0.19}$ & $8.3^{+0.2}_{-0.3}$ & $0.27^{+0.15}_{-0.09}$ & 74 \\ [0.13cm]
SDSS\,J153817.35+512338.0 & 93.11 & $607^{+47}_{-40}$ & 0.01 & 3 & $35\,284^{+600}_{-688}$ & $0.89^{+0.16}_{-0.12}$ & $0.97^{+0.09}_{-0.11}$ & $8.53^{+0.17}_{-0.2}$ & $18.0^{+7.0}_{-4.0}$ & 100 \\ [0.13cm]
UV\,Per & 93.44 & $248^{+7}_{-6}$ & 0.2 & 3 & $14\,040^{+539}_{-645}$ & $1.2^{+0.4}_{-0.3}$ & $0.7 \pm 0.2$ & $8.1 \pm 0.4$ & $0.9^{+0.7}_{-0.4}$ & 69 \\ [0.13cm]
1RXS\,J023238.8--371812 & 95.04 & $214^{+8}_{-7}$ & 0.2 & 3 & $14\,457^{+118}_{-135}$ & $0.63^{+0.05}_{-0.04}$ & $1.15 \pm 0.04$ & $8.9^{+0.07}_{-0.08}$ & $0.23 \pm 0.03$ & 71 \\ [0.13cm]
RZ\,Sge & 98.32 & $294 \pm 7$ & 0.5 & 3 & $15\,197^{+419}_{-507}$ & $1.09^{+0.27}_{-0.17}$ & $0.78^{+0.13}_{-0.18}$ & $8.3^{+0.2}_{-0.3}$ & $1.0^{+0.5}_{-0.3}$ & 51 \\ [0.13cm]
CY\,UMa & 100.18 & $306 \pm 6$ & 0.1 & 3 & $14\,692^{+471}_{-394}$ & $1.40 \pm 0.13$ & $0.57^{+0.08}_{-0.06}$ & $7.91^{+0.14}_{-0.13}$ & $1.6 \pm 0.2$ & 63 \\ [0.13cm]
GD\,552 & 102.73 & $80.6 \pm 0.2$ & 0.1 & 12 & $10\,761^{+37}_{-43}$ & $1.07^{+0.05}_{-0.04}$ & $0.78^{+0.03}_{-0.04}$ & $8.27^{+0.05}_{-0.06}$ & $0.243^{+0.023}_{-0.018}$ & 55 \\ [0.13cm]
IY\,UMa* & 106.43 & $181 \pm 2$ & 1.0 & 3 & $17\,057^{+179}_{-79}$ & $0.83^{+0.04}_{-0.05}$ & $0.99^{+0.04}_{-0.03}$ & $8.59^{+0.07}_{-0.05}$ & $0.85 \pm 0.07$ & 79 \\ [0.13cm]
SDSS\,J100515.38+191107.9 & 107.6 & $339^{+21}_{-19}$ & 0.2 & 3 & $14\,483^{+520}_{-430}$ & $1.7^{+0.4}_{-0.3}$ & $0.44^{+0.15}_{-0.09}$ & $7.6 \pm 0.3$ & $2.4^{+1.2}_{-0.8}$ & 76 \\ [0.13cm]
RZ\,Leo & 110.17 & $279^{+12}_{-11}$ & 0.5 & 3 & $15\,573^{+437}_{-424}$ & $0.85^{+0.2}_{-0.16}$ & $0.97^{+0.13}_{-0.16}$ & $8.6^{+0.2}_{-0.3}$ & $0.62^{+0.3}_{-0.19}$ & 81 \\ [0.13cm]
CU\,Vel & 113.04 & $158.5 \pm 1.1$ & 0.1 & 3 & $14\,174^{+117}_{-169}$ & $1.58^{+0.11}_{-0.08}$ & $0.47^{+0.04}_{-0.05}$ & $7.71^{+0.08}_{-0.11}$ & $1.97^{+0.28}_{-0.19}$ & 90 \\ [0.13cm]
AX\,For & 113.04 & $349 \pm 10$ & 1.0 & 3 & -- & $1.09^{+0.08}_{-0.09}$ & $0.76 \pm 0.07$ & $8.24 \pm 0.08$ & -- & -- \\ [0.13cm]
EF\,Peg & 120.53 & $288^{+21}_{-18}$ & 0.2 & 5 & $16\,644^{+448}_{-570}$ & $1.1 \pm 0.2$ & $0.8^{+0.16}_{-0.17}$ & $8.3 \pm 0.3$ & $1.3^{+0.5}_{-0.4}$ & 90 \\ [0.13cm]
DV\,UMa* & 123.62 & $382^{+24}_{-21}$ & 1.0 & 3 & $19\,410^{+244}_{-400}$ & $0.86^{+0.12}_{-0.09}$ & $0.96^{+0.07}_{-0.1}$ & $8.55^{+0.13}_{-0.16}$ & $1.5^{+0.4}_{-0.3}$ & 89 \\ [0.13cm]
IR\,Com & 125.34 & $216 \pm 3$ & 1.0 & 3 & $17\,531^{+236}_{-271}$ & $0.78^{+0.11}_{-0.08}$ & $1.03^{+0.07}_{-0.09}$ & $8.67^{+0.12}_{-0.15}$ & $0.82^{+0.21}_{-0.15}$ & 87 \\ [0.13cm]
AM\,Her & 185.65 & $87.9 \pm 0.2$ & 0.001 & 13 & $19\,248^{+486}_{-450}$ & $1.3^{+0.17}_{-0.14}$ & $0.63^{+0.1}_{-0.08}$ & $8.01^{+0.17}_{-0.16}$ & $4.0^{+0.9}_{-0.7}$ & 92 \\ [0.13cm]
DW\,UMa & 196.71 & $579^{+7}_{-6}$ & 0.71 & 14 & $56\,760^{+146}_{-259}$ & $1.19^{+0.07}_{-0.05}$ & $0.82^{+0.03}_{-0.04}$ & $8.2^{+0.05}_{-0.07}$ & $228.0 \pm 24.0$ & 97 \\ [0.13cm]
U\,Gem & 254.74 & $93.0 \pm 0.3$ & 1.0 & 15 & $33\,070^{+648}_{-616}$ & $0.634 \pm 0.016$ & $1.160 \pm 0.013$ & $8.90 \pm 0.03$ & $6.53^{+0.22}_{-0.17}$ & 100 \\ [0.13cm]
SS\,Aur & 263.23 & $249.0^{+1.8}_{-1.7}$ & 0.1 & 16 & $28\,627^{+190}_{-269}$ & $0.86^{+0.15}_{-0.11}$ & $0.98^{+0.09}_{-0.11}$ & $8.56^{+0.16}_{-0.19}$ & $7.3^{+3.0}_{-1.9}$ & 100 \\ [0.13cm]
RX\,And & 302.25 & $196.7 \pm 1.0$ & 0.5 & 17 & $33\,900^{+634}_{-995}$ & $1.12^{+0.12}_{-0.08}$ & $0.81^{+0.06}_{-0.08}$ & $8.25^{+0.1}_{-0.13}$ & $26.0^{+6.0}_{-4.0}$ & 100 \\ [0.13cm]
V442\,Cen & 662.4 & $343 \pm 5$ & 0.01 & 16 & $29\,802^{+211}_{-247}$ & $1.35^{+0.14}_{-0.12}$ & $0.64^{+0.06}_{-0.05}$ & $7.98 \pm 0.12$ & $25.0^{+5.0}_{-4.0}$ & 100 \\ [0.13cm]
\bottomrule
\end{tabular}
\begin{tablenotes}
\item \textbf{Notes.} For each object, its orbital period and metallicity are compiled from the literature. The last five columns report the results from this work. The two systems highlighted with a star are those for which the curtain of veiling gas has been detected in their ultraviolet spectra.\\
\item \textbf{References.} 
(1) \cite{Uthas+2011}, 
(2) \cite{Mukadam+2013}, 
(3) \cite{Pala+2017}, 
(4) \cite{Gaensicke+2005}, 
(5) \cite{Howell+2002}, 
(6) \cite{Szkody+2003b}, 
(7) \cite{Sion+1995}, 
(8) \cite{Szkody+2002a}, 
(9) \cite{Sion+2003}, 
(10) \cite{Szkody+2013}, 
(11) \cite{Gaensicke+2001a}, 
(12) \cite{Unda-Sanzana2008}, 
(13) \cite{Gaensicke+2006}, 
(14) \cite{Araujo+2003}, 
(15) \cite{Cheng+1997}, 
(16) \cite{Sion+2008}, 
(17) \cite{Sion+2001}. 
\end{tablenotes}
\end{table*}

\subsection{Summary on the mass measurements}
The results of our fitting procedures are summarised in Table~\ref{tab:results_good}. 
In the case of the white dwarf in AX\,For, from the analysis of the ultraviolet data, we derived a smaller radius ($\simeq 0.0142\,\rsun$) than the one estimated from the gravitational redshift  ($\simeq 0.0190\,\rsun$). We can thus roughly estimate that at the time of the \hst observations $\simeq 77\,$per cent of the white dwarf surface was still heated by the recent outburst and, therefore, we assumed as final measurements for its mass and radius those obtained from the gravitational redshift in Section~\ref{subsec:rv}.

We show in Figure~\ref{fig:uv_fit1} some examples of best-fitting models for three non-eclipsing systems in different temperature regimes, and for an eclipsing CV. All spectra, along with their best-fitting models, are available in Appendix~\ref{ap:best-fit}.

\begin{table*}
\centering
\caption{Summary of the CV white dwarf masses derived in this work from the analysis of the ultraviolet data and those obtained employing other methods. The systems are sorted by increasing masses.}\label{tab:comparison}
\begin{tabular}{@{} lccccccl@{}}
 \toprule
                                       & \multicolumn{6}{c}{$M_\mathrm{WD}$ ($\mathrm{M_\odot}$)}  \\ 
\cmidrule{2-7}
\multirow{3}{*}{System}   & ultraviolet & ultraviolet & ultraviolet & optical & gravitational & radial   & \multirow{3}{*}{References} \\
                                       & spectral fit & spectral fit & light curve  & light curve & redshift & velocities   & \\
                                       & (this work) & (literature${^a}$)  &                 &              &               &             & \\
\midrule 
BC\,UMa         & $0.48^{+0.08}_{-0.09}$ & $0.53^{+0.07}_{-0.09}$ & -- & -- & -- & -- &\cite{Gaensicke+2005} \\[0.1cm]
AX\,For          &  $0.56 \uparrow$             & -- & -- & -- & $0.76^{+0.06}_{-0.07}$ & -- &This work \\[0.1cm]
SW\,UMa        &  $0.61^{+0.06}_{-0.04}$ & $0.61\pm 0.6$  & -- & -- & -- &  -- &\cite{Gaensicke+2005} \\[0.1cm]
AM\,Her         & $0.63^{+0.10}_{-0.08}$   & $0.53^{+0.10}_{-0.08}$ & -- & -- & -- & -- &\citet{Gaensicke+2006} \\[0.1cm]
LL\,And          & $0.7^{+0.4}_{-0.3}$        & $0.7^{+0.4}_{-0.3}$   & -- & -- & -- & -- &\cite{Howell+2002}\\[0.1cm]
EF\,Peg          & $0.80^{+0.16}_{-0.17}$  & $0.88^{+0.16}_{-0.17}$ & -- & -- & -- & -- &\cite{Howell+2002}\\[0.1cm]
WZ\,Sge         & $0.80 \pm 0.02$             & -- & -- & -- & $0.85 \pm 0.04$ & -- &\citet{Steeghs+2007} \\[0.1cm]
DW\,UMa        & $0.82^{+0.03}_{-0.04}$ & -- & $0.77\pm0.07$ & -- & -- & -- &\citet{Araujo+2003}\\[0.1cm]
GW\,Lib          & $0.83^{+0.08}_{-0.12}$ & -- & -- & -- & $0.84 \pm 0.02$ & -- & \citet{vanSpaandonk+2010}\\[0.1cm]
\multirow{2}{*}{\,J150722.30+523039.8}      & \multirow{2}{*}{$0.90^{+0.10}_{-0.14}$} &\multirow{2}{*}{--} & \multirow{2}{*}{$0.83^{+0.19}_{-0.15}$} & \multirow{2}{*}{$0.89\pm0.01$} & \multirow{2}{*}{--} & \multirow{2}{*}{--} & Ultraviolet light curve, this work; \\
                      &                                        &    &     &    &   & & optical light curve \citet{Savoury+2011}\\[0.1cm]
V1108\,Her    & $0.88^{+0.09}_{-0.11}$  & -- & -- & -- & $0.91^{+0.14}_{-0.20}$ &-- & This work \\[0.1cm]
DV\,UMa         & $0.96^{+0.07}_{-0.10}$ & -- & -- &$1.09\pm0.03$ & -- & -- &\citet{McAllister+2019}\\[0.1cm]
IY\,UMa          & $0.99^{+0.04}_{-0.03}$ & -- & --   & $0.955^{+0.013}_{-0.028}$& -- & -- &\citet{McAllister+2019} \\[0.1cm]
SDSS\,J103533.02+055158.4      & $1.00^{+0.08}_{-0.10}$ & -- & --  & $0.835\pm0.009$& -- & -- &\citet{Savoury+2011} \\[0.1cm]
BW\,Scl           & $1.007^{+0.01}_{-0.012}$ & $1.10^{+0.03}_{-0.06}$ & -- & -- & -- & -- &\cite{Gaensicke+2005} \\[0.1cm]
IR\,Com          & $1.03^{+0.07}_{-0.09}$ & -- & $0.989\pm0.003$ & -- & $0.95\pm0.04$ & -- & This work\\[0.1cm]
U\,Gem           & $1.16 \pm 0.013$ & -- & -- & -- & -- & $1.2 \pm 0.05$ & \citet{Echevarria+2007} \\[0.1cm]
HV\,Vir           & $0.87^{+0.13}_{-0.17}$ & $1.27^{+0.13}_{-0.17}$ & -- & -- & -- & -- &\cite{Szkody+2002a}\\[0.1cm]
EG\,Cnc          & $1.03^{+0.04}_{-0.05}$ &  $1.28^{+0.04}_{-0.05}$  & -- & -- & -- & -- &\cite{Szkody+2002a}\\[0.1cm]
\bottomrule
\end{tabular}
\begin{tablenotes}
\item \textbf{Notes.} (a) The reported values have been corrected accounting for the temperature dependency of the mass-radius relationship and the reddening, as discussed in Section~\ref{subsec:comparison_uv_lit}\\
\end{tablenotes}
\end{table*}

\begin{figure}
\includegraphics[width=\columnwidth]{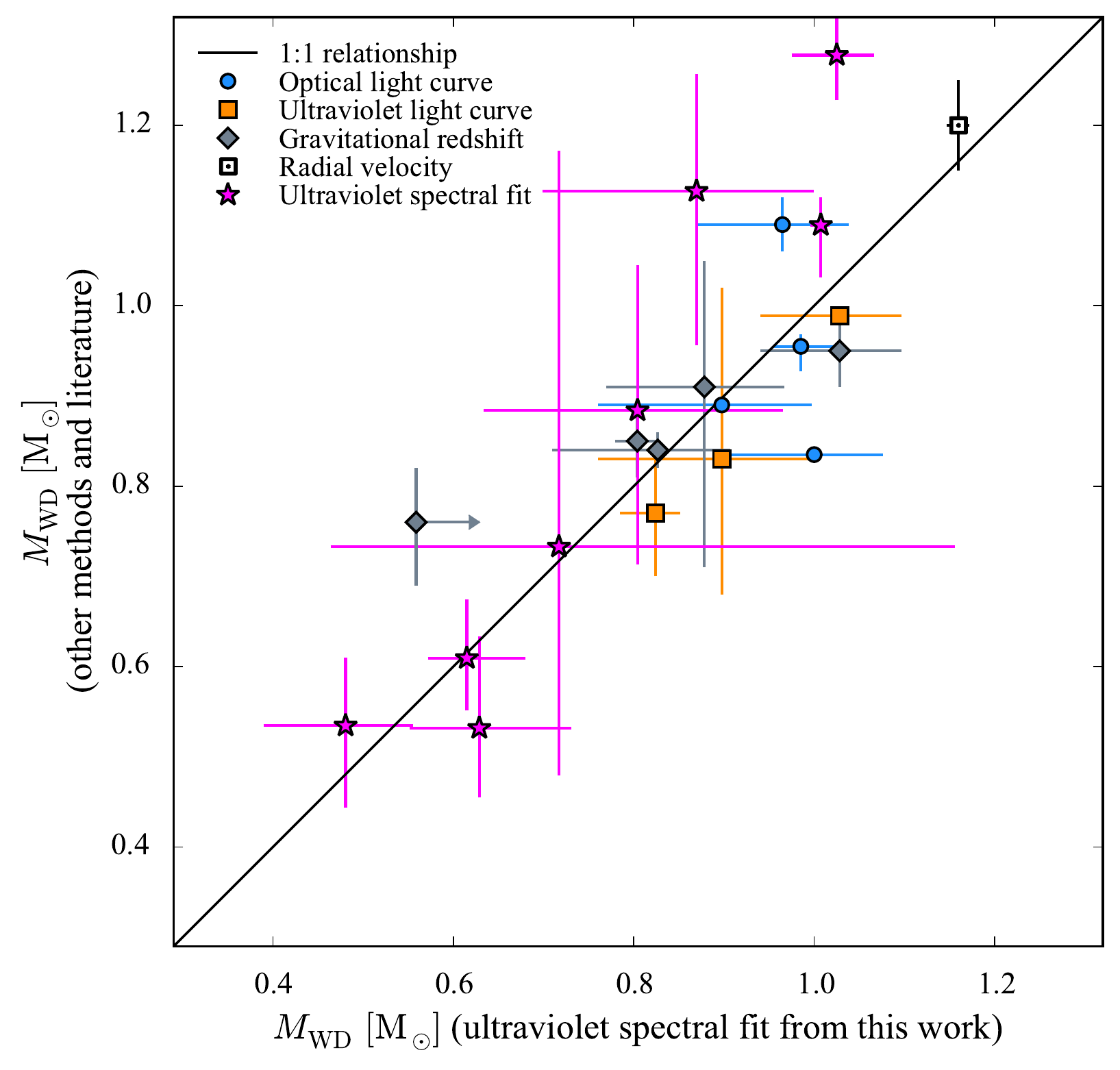}
\caption{Comparison between the CV white dwarf masses derived in this work from the analysis of the ultraviolet \hst spectra and those obtained employing other methods and from the analysis of the same ultraviolet data in the literature. The error bars show the corresponding 1$\sigma$ uncertainties.}\label{fig:comparison}
\end{figure}

\begin{figure*}
\includegraphics[width=\textwidth]{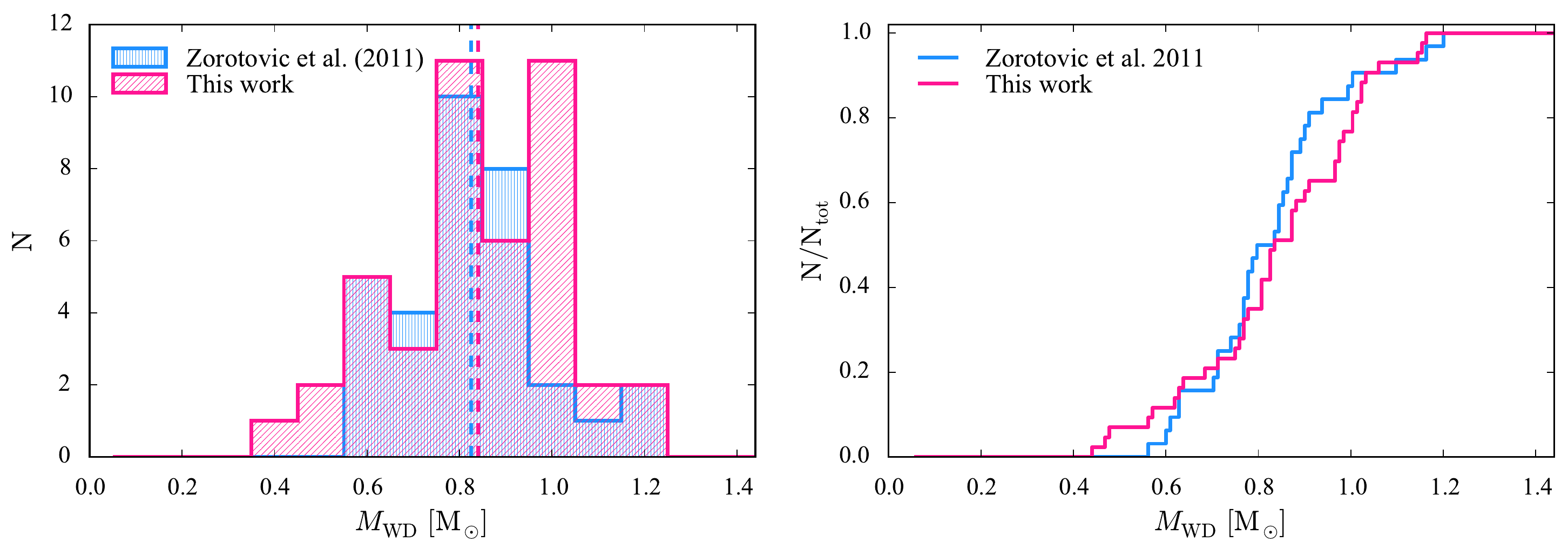}
\caption{Comparison between the mass distributions (left) and the cumulative distributions (right) of the 43 white dwarfs from this work (pink) and the 32 measurements compiled by \citeauthor{Zorotovic+2011} (\citeyear{Zorotovic+2011}, blue). The dashed vertical lines correspond to the average masses. Our results show the presence of a tail extending to low masses corresponding to helium-core white dwarfs, which is not detected in the sample studied by \citet{Zorotovic+2011}.}\label{fig:mass_distributions}
\end{figure*}

\section{Discussion}
\subsection{Comparison with previous mass measurements and different techniques}
\subsubsection{Eclipsing systems}\label{subsec:comparison_eclipsing}
Previous mass measurements from the analysis of the eclipse light curve of the white dwarf were obtained from optical observations for DV\,UMa,  IY\,UMa, SDSS\,J103533.02+055158.4 and SDSS\,J150722.30+523039.8 \citep{Savoury+2011,McAllister+2019}, and from ultraviolet observations for DW\,UMa \citep{Araujo+2003}. Moreover, we here derived a mass measurement for IR\,Com and SDSS\,J150722.30+523039.8 from the analysis of their ultraviolet light curves (Section~\ref{subsubsec:ircom_sdss1507}). We find a good agreement (within 3$\sigma$, Table~\ref{tab:comparison} and Figure~\ref{fig:comparison}) between the masses derived employing the two different methods (ultraviolet spectral fit vs. analysis of the white dwarf eclipse).

\subsubsection{Systems with previous mass measurements from ultraviolet analysis}\label{subsec:comparison_uv_lit}
For several systems in our sample (AM\,Her, BC\,UMa, BW\,Scl, EF\,Peg, EG\,Cnc,  HV\,Vir, LL\,And and SW\,UMa), a mass estimate derived from the analysis of their ultraviolet \hst data is available in the literature.
However these studies lacked the knowledge of the distance to the systems and therefore did not provide a single mass estimate but a range of possible values, computed assuming different distances. The method employed is described in detail by \citet{Gaensicke+2005} and consists of fitting the ultraviolet data by stepping through a grid of atmosphere models with fixed values for $\log(g)$, leaving the temperature and scaling factor free. In this way, it is possible to investigate the correlation between the assumed $\log(g)$ (i.e. the white dwarf mass under the assumption of a mass-radius relationship) and the best-fit value for \teff (see e.g. figure~3 from \citealt{Gaensicke+2005}) for different distances.

We retrieved the $\log(g)$-distance correlations for AM\,Her, BC\,UMa, BW\,Scl, EF\,Peg, EG\,Cnc,  HV\,Vir, LL\,And and SW\,UMa from the works by \cite{Howell+2002,Szkody+2002a,Gaensicke+2005,Gaensicke+2006}.
We applied a correction to account for the dependency of the mass-radius relationship on the white dwarf \teff and for the reddening (see Appendix~\ref{ap:correction} for the details), which were not accounted for by these studies, and estimated the mass of the white dwarf assuming their distances from \gaia EDR3 (as described in Section~\ref{subsec:gaia_quality}). We computed the associated error bars by assuming a typical statistical uncertainty on the white dwarf temperature (for a fixed $\log(g)$) of $\simeq 200\,$K (B.T.~G{\"a}nsicke, private communications). For each object, we compared them with the systematic uncertainties derived in Section~\ref{subsubsec:errors} (which are representative of the typical uncertainties related to the accuracy of the reddening and distance) and assumed as final uncertainties the larger between the two values. The final results are listed in Table~\ref{tab:comparison} and shown in Figure~\ref{fig:comparison}. 

With the exception of EG\,Cnc, the results from the literature are in good agreement with ours within the uncertainties.
The origin of the disagreement in the case of EG\,Cnc is not easy to unveil and could possibly be related to the different atmosphere models used by \citet{Szkody+2002a}, which were generated with an older version of \textsc{tlusty} (\# 195) than the one that we used (\# 204n). The  most relevant differences between the two versions are an improved treatment of H$_2^{+}$ quasi-molecular absorption lines (dominant in this cool CV) and of the Stark-broadening profiles using the calculation by \citet{Tremblay+2009}.
Therefore, we consider our results more reliable and representative of the observed flux emission of EG\,Cnc. 

\subsubsection{Systems with radial velocity measurements}\label{subsubsec:rv_uv_comparison}
For three systems in our sample, AX\,For, IR\,Com and V1108\,Her, we derived an independent mass measurement from the white dwarf gravitational redshift (Section~\ref{subsec:rv}). Moreover, additional mass measurements, which have been determined from the gravitational redshift of the white dwarfs in GW\,Lib \citep{vanSpaandonk+2010} and WZ\,Sge \citep{Steeghs+2007}, and from the radial velocities of the two stellar components in U\,Gem \citep{Echevarria+2007}, are available in the literature (see also Table~\ref{tab:masses_lit} and references therein). These masses are all in good agreement with those we derived from the spectral fit to the ultraviolet data (Table~\ref{tab:comparison} and Figure~\ref{fig:comparison}).

\subsection{Comparison with \citet{Zorotovic+2011}}\label{subsec:zorotovic}
Figure~\ref{fig:mass_distributions} shows the comparison between the 43 mass measurement from this work and the compilation from \citet{Zorotovic+2011}, which includes 22 masses derived from the analysis of the white dwarf eclipses and 10 measurements from spectroscopic studies. The two samples have eight systems in common (AM\,Her, DV\,UMa, DW\,UMa, IY\,UMa, SDSS\,J103533.02+055158.4, SDSS\,J150722.30+523039.8, U\,Gem and WZ\,Sge, see Table~\ref{tab:comparison}) and this partial overlap highlights possible differences associated with the different methods employed to measure the masses of the white dwarfs.

The mass distribution we derived presents a tail extending towards low masses, consisting of three systems (BC\,UMa, CU\,Vel and SDSS\,J100515.38+191107.9) with $\mwd < 0.5\,\msun$. Such low masses are consistent with either He core or, possibly, hybrid CO/He core white dwarfs. In contrast, the sample studied by \citet{Zorotovic+2011} does not contain any white dwarf with $\mwd < 0.5\,\msun$ although, from evolutionary considerations, they estimated that CV helium-core white dwarfs should represent $\lesssim 10\,$per cent of the systems.

\begin{figure}
\includegraphics[width=\columnwidth]{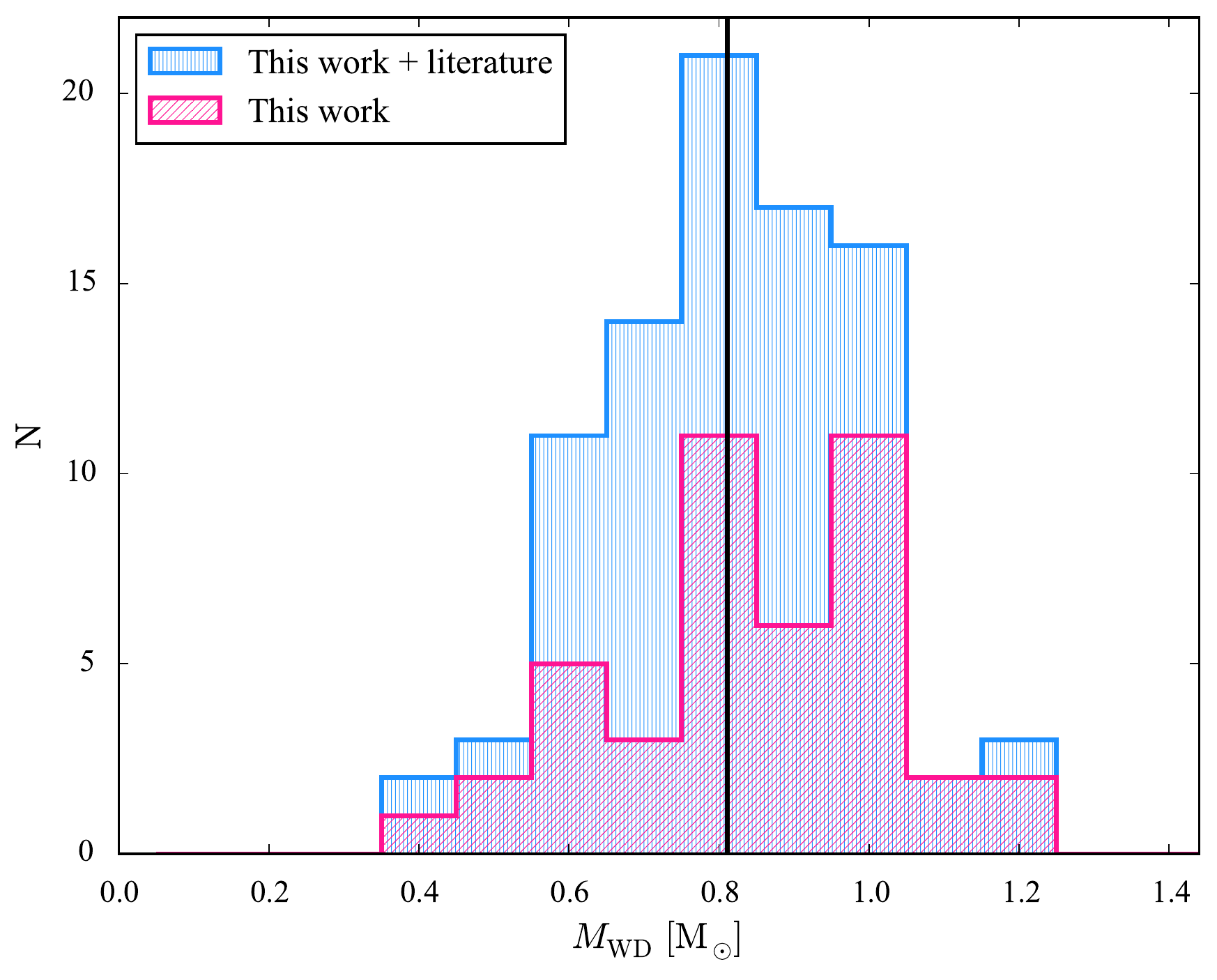}
\caption{Mass distribution for the sample of 89 CV white dwarfs obtained combining the 43 mass measurements from this work (shown in pink) with the 46 from the literature. The black vertical line corresponds to the average mass of the total sample of 89 systems, $\langle \mwd \rangle = 0.81^{+0.16}_{-0.20}\,\msun$.}\label{fig:mass_all}
\end{figure}

For both distributions, we determined the average white dwarf mass $\langle \mwd \rangle$ and the corresponding uncertainties as the 16th and the 84th percentiles, obtaining $\langle \mwd \rangle = 0.82 \pm 0.12\,\msun$ for the values from \citet{Zorotovic+2011} and $\langle \mwd \rangle = 0.84^{+0.18}_{-0.23}$ for the masses here derived. The agreement between these values allows us to rule out the presence of any systematics affecting the masses derived from the analysis of eclipse light curves. 

\subsection{The mass distribution of CV white dwarfs}\label{subsec:mass_distribution}
Since the work by \citet{Zorotovic+2011}, more mass measurements of CV white dwarfs have been made available in the literature, mainly thanks to the systematic observations of eclipsing systems (e.g. \citealt{Littlefair+2006,Feline+2005,Savoury+2011,McAllister+2019}) with the fast triple-beam camera ULTRACAM \citep{Dhillon+2007}. A more comprehensive and up-to-date census of masses in the literature consists of 54 measurements (see Table~\ref{tab:masses_lit} and reference therein), eight of which (AM\,Her, DV\,UMa, DW\,UMa, SDSS\,J103533.02+055158.4, SDSS\,J150722.30+523039.8, U Gem WZ\,Sge and IY\,UMa, see Table~\ref{tab:comparison}) are in common with our sample. For these systems, we here assume the white dwarf parameters obtained in this work. Of the remaining 46 CV white dwarf masses from the literature, 34 have been derived from the analysis of the eclipse light curves of the white dwarf and 12 from spectroscopic studies. 

Combining these 46 measurements with our results, we obtained a sample of 89 CV white dwarfs with an accurate mass, and the corresponding average mass results $\langle \mwd \rangle = 0.81^{+0.16}_{-0.20}\,\msun$ (Figure~\ref{fig:mass_all}). As demonstrated by \citet{Zorotovic+2011}, this high average mass of CV white dwarfs cannot be ascribed to an observational bias since the detection of massive white dwarfs is disfavoured by the fact that they have smaller radii and are less luminous than low mass white dwarfs for the same $\teff$. This confirms the earlier results that CV white dwarfs are genuinely more massive than predicted by most models of CV evolution, which only account for orbital angular momentum losses arising from magnetic braking and gravitational wave radiation.

Among the systems from the literature, two (HY\,Eri and KIC 5608384) have $\mwd < 0.5\,\msun$. These, combined with the  three low-mass primaries likely representative of He-core white dwarfs we have identified, bring the total census to five systems.
On the one hand, compared to the standard models for CV evolution, the number of observed He-core white dwarfs is still far lower than predicted (e.g. $\simeq 53\,$per cent, \citealt{Politano1996} and $\simeq 30\,$per cent, \citealt{Goliasch-Nelson2015} of the present-day CV population). On the other hand, more modern scenarios that take into account a mass dependent consequential angular momentum loss and that can reproduce the overall white dwarf mass distribution in CVs do not predict any He-core white dwarfs \citep[e.g.][]{Schreiber+2016}, which is also in contrast with the observations.
Our result shows that CVs hosting low mass white dwarfs contribute to the overall CV population and that their non-zero fraction should be properly taken into account in the modelling of CV evolution.

\subsection{Orbital period dependecy}\label{subsec:mass_porb}
\begin{figure}
\includegraphics[width=\columnwidth]{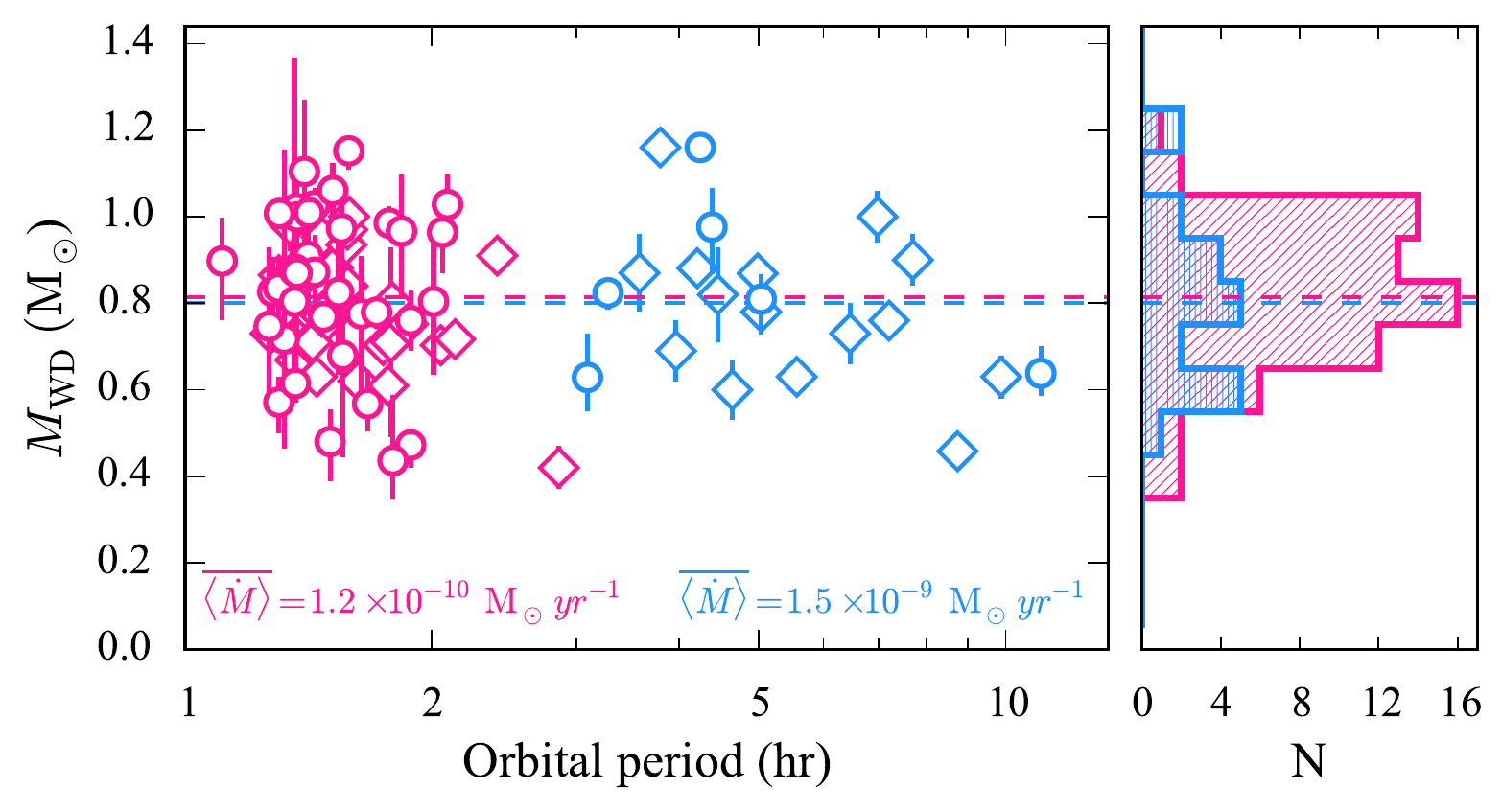}
\caption{White dwarf masses from this work (circles) and from the literature (diamonds), as a function of their orbital periods. Long period ($\porb > 3\,$hr) and short ($\porb < 3\,$hr) period CVs are shown in blue and pink, respectively. The right panel shows the corresponding distributions and the corresponding average masses (dashed lines), colour coded as in the left panel. Note that long period CVs have, on average, secular mass accretion rates about one order of magnitude higher compared to those of short period CVs, as reported in the figure.}\label{fig:porb}
\end{figure}

While losing orbital angular momentum, CVs evolve from long to short orbital periods and, by comparing the average mass of the white dwarfs in long ($\porb > 3\,$hr) and short ($\porb < 3\,$hr) period CVs, we can investigate possible overall variations (due to either mass growth or mass erosion) with time. 
We do not find any difference between the average white dwarfs mass of long ($\langle \mwd \rangle = 0.80^{+0.17}_{-0.19}\,\msun$) and short ($\langle \mwd \rangle = 0.81^{+0.17}_{-0.16}\,\msun$) period CVs. Moreover, by performing an F-test on the best linear fit to the masses as a function of the orbital period, we derive an F-statistic $F_0 = 2.22$ with p-value = 0.23 and therefore we do not find any evidence for a clear dependency on the white dwarf mass with the orbital period (Figure~\ref{fig:porb}). 
Nonetheless, only 21 long period CVs have mass determinations (in contrast to the 68 systems at short orbital periods) and additional measurements at long orbital periods are required to further constrain any correlation.

\begin{figure}
\includegraphics[width=1.05\columnwidth]{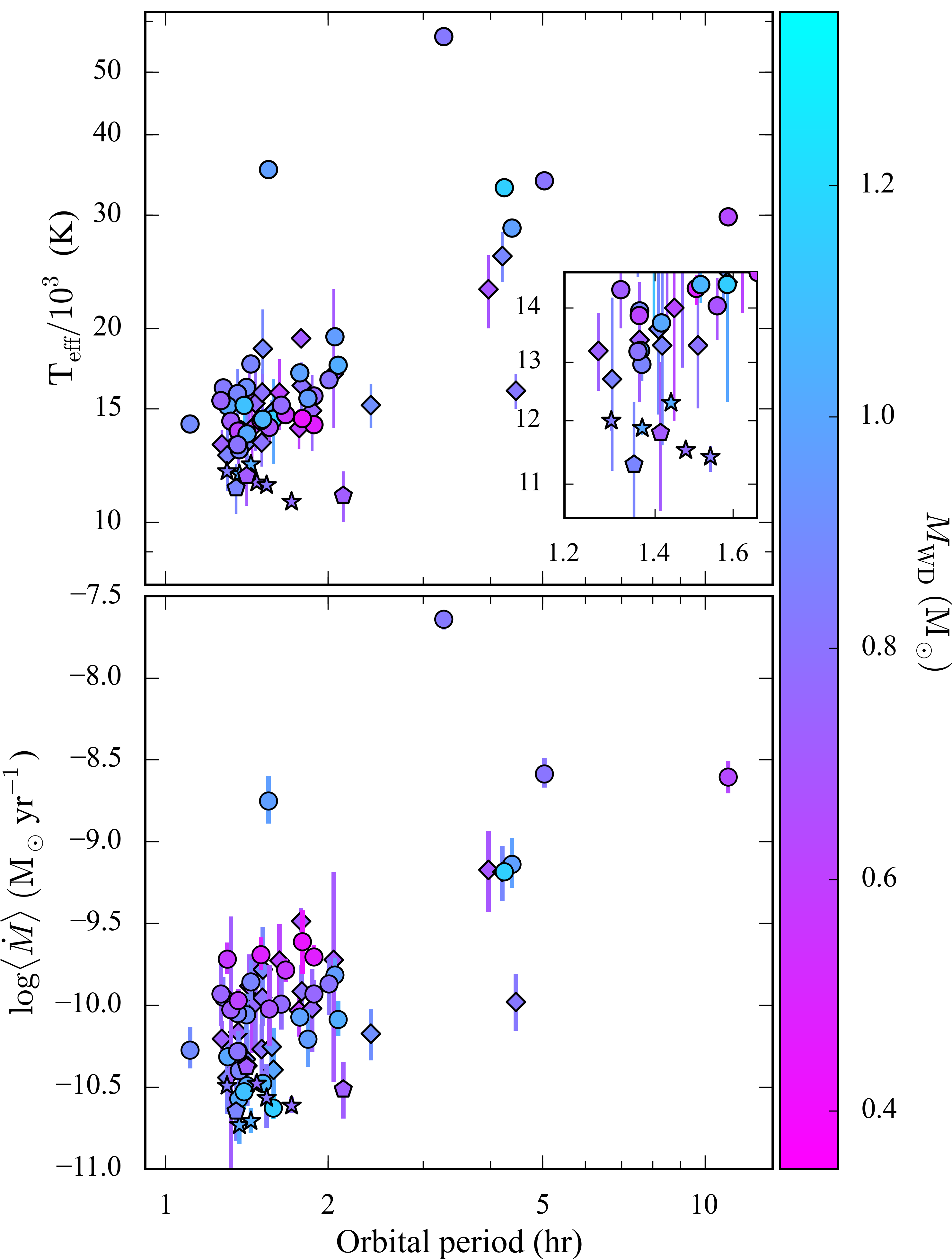}
\caption{Effective temperatures (top) and mass accretion rates (bottom) as a function of the orbital period, for the systems in our \hst sample (circles for pre-bounce and stars for period bounce CVs) and those from the literature (diamonds for pre-bounce and pentagons for period bounce CVs). The inset shows a closeup of the period bounce systems.}\label{fig:results}
\end{figure}

\begin{figure*}
\includegraphics[width=1.05\textwidth]{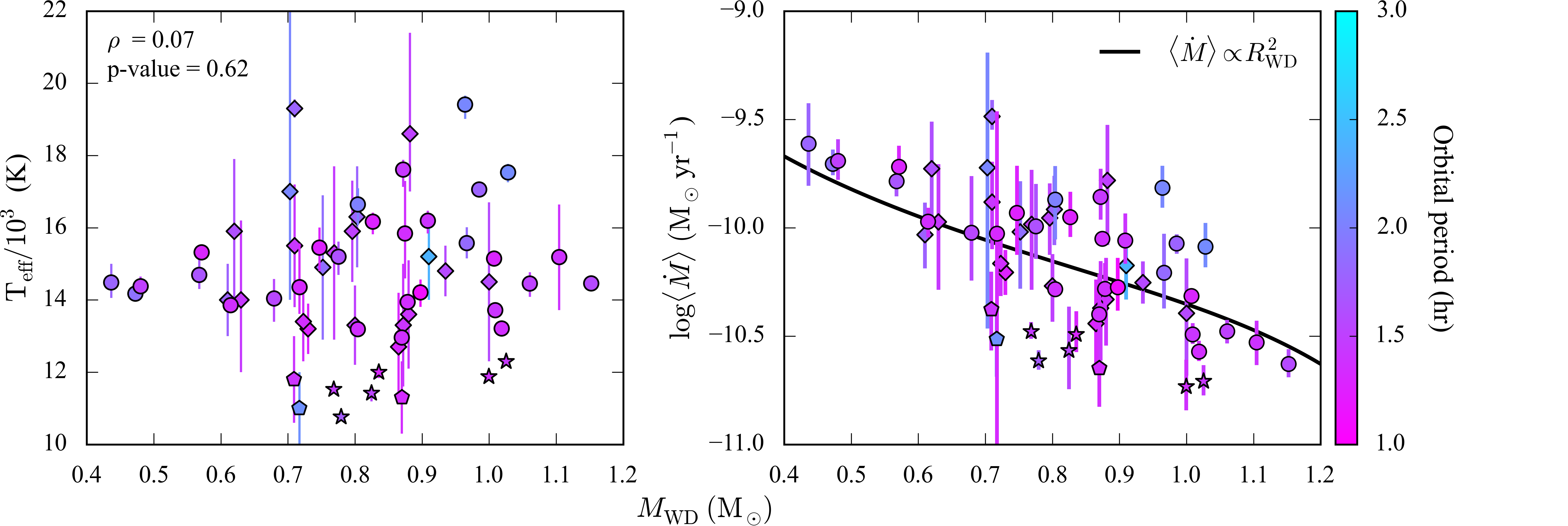}
\caption{For systems at short orbital periods ($\porb < 3\,$hr), the white dwarf effective temperatures show no clear dependency on the white dwarf mass (left). Consequently, the relation between the mass accretion rates and the mass of the white dwarf (right) is dominated by the mass-radius relationship ($\langle \dot{M} \rangle \propto \rwd^2$, solid black line, arbitrarily normalised so that $\langle \dot{M} \rangle = 7\times10^{-11} \mathrm{M_\odot} \mathrm{yr^{-1}}$ for $\mwd = 0.8\,\msun$). The data are colour-coded according to the orbital periods and the symbol convention is the same as in Figure~\ref{fig:results}.}\label{fig:correlations}
\end{figure*}

\subsection{Mass accretion rates and CV evolution}\label{subsec:evolution}
The evolution of CVs is dictated by orbital angular momentum losses, which continuously shrink the orbit and keep the secondary in touch with its Roche lobe, ensuring the stability of the mass transfer process.
For systems with $\porb \gtrsim 3\,$h, the main mechanisms of angular momentum loss are magnetic wind braking and gravitational wave radiation.
As the system loses orbital angular momentum, the two stellar components spiral inwards and the system evolves towards short orbital periods, while the donor star is constantly stripped of more and more mass. At $\porb \simeq 3\,\mathrm{h}$, the donor star has become fully convective and, in the frequently referenced interrupted magnetic braking scenario \citep{Rappaport+1983,Paczynski+1983, Spruit+1983},  
a re-configuration of the magnetic fields takes place on the donor, leading to a great reduction in the efficiency of magnetic braking. As a consequence, the secondary star detaches from its Roche-lobe and, in the period range $2\,\mathrm{h} \lesssim \porb \lesssim 3\,$h (the so-called period gap) the system evolves as a detached binary whilst still losing angular momentum through gravitational wave radiation.
Accretion then resumes at $\porb \simeq 2\,$h, when the orbital separation brings again the donor in contact with its Roche-lobe, and the system keep evolving towards shorter orbital periods.
When the system reaches the ``period minimum'' at $\porb \simeq 80\,\mathrm{min}$, the time-scale on which the secondary star loses mass becomes much shorter than its thermal time-scale and the secondary stops shrinking in response to the mass loss. Consequently, systems that have passed the period minimum evolve back towards longer orbital periods and, for this reason, are called ``period bouncers''.

The different efficiencies of magnetic braking and gravitational wave radiation in removing angular momentum from the binary orbit cause long period CVs to have $\langle\dot{M}\rangle$ about one order of magnitude higher compared to those of short period CVs. Determining the rate of angular momentum loss is therefore important in order to test the models of CV evolution. However, a direct measurement via detection of orbital period changes is impossible on human time scales. A very good proxy for the angular momentum loss rate is the white dwarf effective temperature \citep{Townsley+2003}, as it is determined by the compressional heating of the accreted material \citep{Sion1995,Townsley+2004}. Therefore, $\teff$ provides a constraint on the mean mass-accretion rate $\langle \mdot \rangle$, averaged over the thermal time-scale of the white dwarf envelope ($10^3-10^5\,$yr), which is a direct measurement of the angular momentum loss rate in the system \citep{Townsley+2009}.

An accurate determination of the mass accretion rate requires the knowledge of both $\teff$ and $\mwd$:
\begin{equation}\label{eq:mdot}
L_\mathrm{WD} = 4 \pi \rwd^{2} \sigma \teff^4 =  6 \times 10^{-3} \mathrm{L_\odot} \left( \frac{\langle \dot{M} \rangle}{10^{-10} \mathrm{M_\odot}\mathrm{yr^{-1}}} \right)  \left( \frac{\mwd}{0.9 \mathrm{M_\odot}} \right)^{0.4}
\end{equation}
where $\mathrm{L}$ is the luminosity and $\sigma$ is the Stefan-Boltzmann constant (eq.~1 from \citealt{Townsley+2009}). 
Thanks to the accurate parallaxes provided by \gaia, we have been able to measure both parameters and the mass accretion rates reported in Tables~\ref{tab:results_good} and \ref{tab:masses_lit} can then be used to test and constrain the current models of CV evolution.

Figure~\ref{fig:results} shows the effective temperatures (top) and the mass accretion rates (bottom) as a function of the orbital period for 65 systems, 41 from our \hst sample\footnote{The evolution of CVs hosting magnetic white dwarfs is different compared to that of non-magnetic CVs, as there are evidence of the reduction of magnetic braking efficiency due to the coupling of the secondary and the white dwarf magnetic fields \citep{Belloni+2020}. Our \hst sample contains one strongly magnetised CV white dwarf (AM\,Her) which we do not include in the discussion.} (circles and stars) and 34 from the literature (diamonds and pentagons), for which both $\teff$ and $\mwd$ are available.
We find that systems above the gap are hotter and accrete at higher rates than systems below the gap, reflecting the different rates of angular momentum loss driving the evolution of these binaries in different orbital period regimes.

Two outliers clearly stand out: SDSS\,J153817.35+512338.0 ($\porb = 93.11\,$min) and DW\,UMa ($\porb = 196.71\,$min), which are both much hotter than the other CV white dwarfs at similar orbital periods. As already suggested by \cite{Pala+2017}, SDSS\,J153817.35+512338.0 could be a young CV which just formed at this orbital period and is undergoing a phase of high mass accretion rate that is expected to occur at the onset of the mass transfer \citep{Dantona+1989}. An alternative possibility is a CV that recently experienced a nova eruption and the white dwarf has not cooled down yet. DW\,UMa is a member of the nova-like CV subclass\footnote{Effective temperatures obtained during a low state are available for other two nova-like CVs, TT\,Ari ($\porb = 198.07$\,min, $\teff=39\,000$\,K, \citealt{Boris+1999a}) and MV\,Lyr ($\porb = 191.38\,$min, $\teff=47\,000$\,K, \citealt{Hoard+2004}). Similarly to DW\,UMa, both systems are much hotter than other CV white dwarfs at similar orbital periods. However, TT\,Ari and MV\,Lyr lack of an accurate mass measurement and therefore they have not been included in this analysis.}, which dominates the population of CVs in the  $3-4$ hours period range. Possibly, also the high temperature and mass accretion rate of DW\,UMa, and those of nova-likes in general, could be related to their young ages if CVs are preferentially formed in the $3-4$\,h $\porb$ range \citep{Townsley+2009}. This could be the case if the initial mass ratio distribution of main sequence binaries peaks toward equal masses \citep{deKool1992}. Alternatively, nova-likes could arise naturally from systems close to the regime of unstable mass transfer \citep{Goliasch-Nelson2015}, where the mass of the donor star is similar to the white dwarf mass. These two outliers are not considered in the following discussion since their effective temperatures possibly reflect peculiar stages of their evolution. 

Below the period gap, two branches are visible. One is composed of systems with $\teff \geq 12\,500\,$K, whose temperatures and mass accretion accretion rates decrease as the systems evolve towards the period minimum. The second branch consists of the period bounce CVs, which are evolving towards longer orbital periods and can be easily recognised as such thanks to their effective temperatures being $\simeq 3000 - 4000\,$K lower than those of the pre-bounce CVs at similar orbital periods (see the inset in the top panel of Figure~\ref{fig:results}). We identify seven previously known period bouncers (EG\,Cnc, \citealt{Patterson2011}; GD\,552, \citealt{Unda-Sanzana2008}; SDSS\,J103533.02+055158.4, \citealt{Littlefair+2006}; SDSS\,J150240.98+333423.9, \citealt{McAllister+2017}; 1RXS\,J105010.8--140431, \citealt{Patterson2011,Pala+2017}, QZ\,Lib, \citealt{Pala+2018} and V455\,And \citealt{Patterson2011}) and two new period bounce CVs, SDSS\,J143544.02+233638.7 and CTCV\,J1300--3052. 
The fraction of period bouncers is thus $(13 \pm 4)\,$per cent, consistent with that derived by \cite{Pala+2020} from the analysis of a volume-limited sample of CVs ($7-14\,$per cent). 

We also noticed three additional weaker candidates (WZ\,Sge, SDSS\,J080434.20+510349.2 and SDSS\,J123813.73--033932.9), which are all known to host brown-dwarf companions \citep{Howell+2004,Zharikov+2013,Pala+2019}. However, these white dwarfs are slightly hotter ($\teff \simeq 13\,000\,$K) than other confirmed period bouncers and, since they are located right at the period minimum, it is difficult to asses whether they have already bounced back or not.

For the short period systems, we found that the white dwarf effective temperatures show a very weak dependence on the masses, i.e. systems hosting white dwarfs spanning a wide range in masses ($0.4-1.2\,\msun$) all have very similar temperatures (left panel in Figure~\ref{fig:correlations}). From the statistical point of view, this is confirmed by the Pearson coefficient $\rho = 0.07$ and p-value = 0.62 of the distribution. In contrast, the mass accretion rates appear to be anti-correlated with the white dwarf mass. 
Note that, for a given $\langle \mdot \rangle$, $\teff$ increases as the layer of accreted material builds up to the next classical nova eruption \citep{Townsley+2004}. 
Below the period gap, the expected range of variation of $\teff$ is of the order of $\pm 1000\,$K and implies that the combination of $\teff$ and $\mwd$ via Equation~\ref{eq:mdot} does not provide a single value for $\langle \mdot \rangle$, but rather a range of possible mass accretion rates with a flat probability distribution. Nonetheless, this effect is not large enough (see e.g. figure~10 from \citealt{Townsley+2004}) to explain the scatter observed in the left panel of Figure~\ref{fig:correlations}. In contrast, the $\langle \mdot \rangle - \mwd$ dependency directly descends from the white dwarf mass-radius relationship. Given that the quiescence luminosity is very weakly dependent on the white dwarf mass (Equation~\ref{eq:mdot}), it follows that $L \propto \langle \mdot \rangle \propto \rwd^{2} \teff^{4}$. Therefore $\langle \mdot \rangle \propto \rwd^{2}$, since $\teff$ is observed to have no clear dependence on the mass. For comparison, this dependency is plotted as the solid black line in the left panel in Figure~\ref{fig:correlations}, which has been computed using the mass-radius relationship of \citet{Holberg+2006,Tremblay+2011} and has been arbitrarily normalised so that $\langle \dot{M} \rangle = 7\times10^{-11} \mathrm{M_\odot} \mathrm{yr^{-1}}$ for $\mwd = 0.8\,\msun$.

\begin{figure*}
\includegraphics[width=1.05\textwidth]{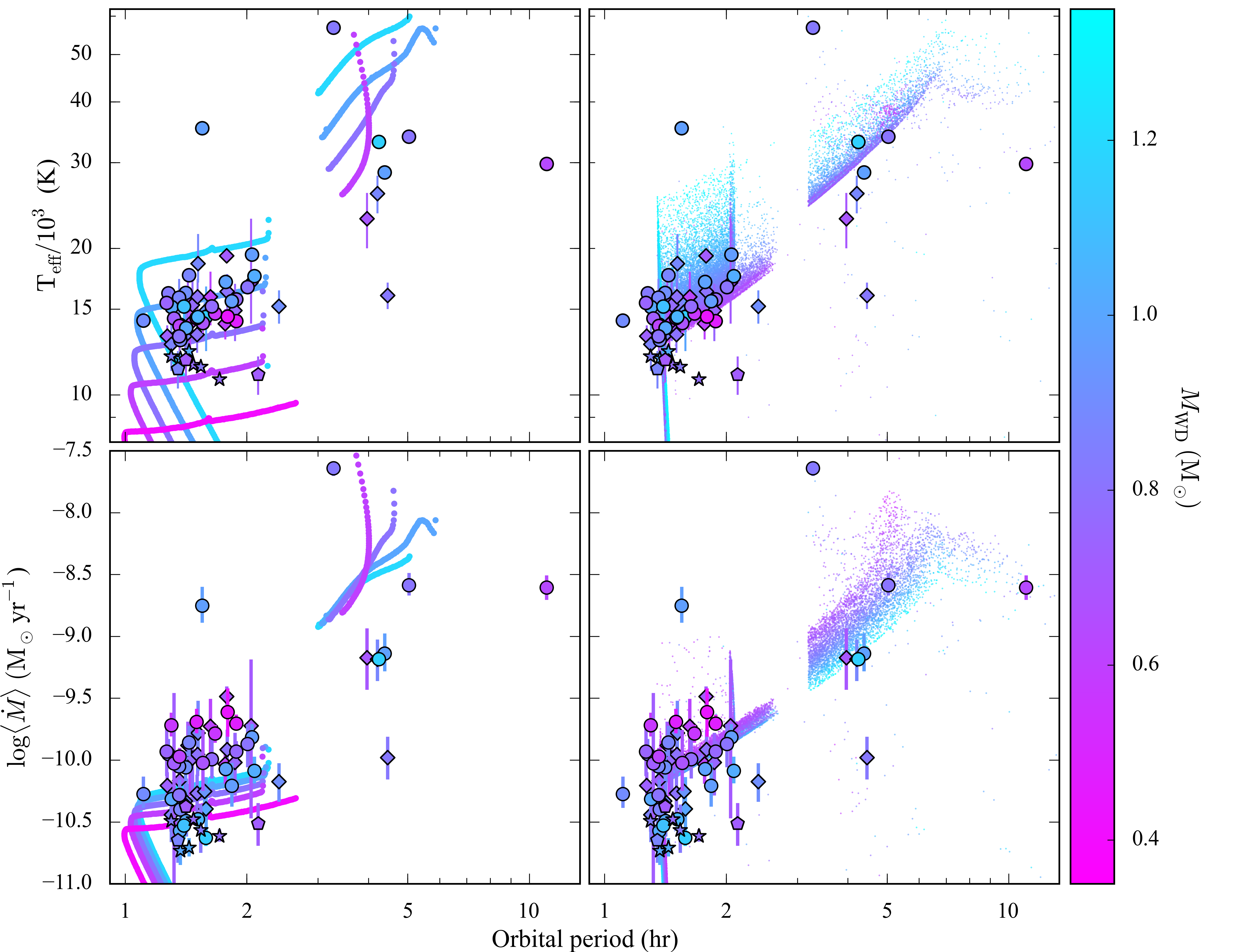}
\caption{Effective temperatures (top) and mass accretion rates (bottom) as a function of the orbital period, for the systems in our \hst sample (circles for pre-bounce and stars for period bounce CVs) and those from the literature (diamonds for pre-bounce and pentagons for period bounce CVs). For a comparison with the models, we show the theoretically predicted values for the classical recipe for CV evolution (left panels, see the text for more details) and for eCAML (right panels, small dots), as computed by \protect\cite{Belloni+2020}. Observations and theoretical values are colour-coded according to the white dwarf mass. The \textsc{MESA} tracks have been computed assuming the following combination for the masses of the white dwarf and the donor (top to bottom): $M_\mathrm{WD}=1.2\,\mathrm{M}_\odot$ and $M_\mathrm{2}=0.8\,\mathrm{M}_\odot$; $M_\mathrm{WD}=1.0\,\mathrm{M}_\odot$ and $M_\mathrm{2}=0.8\,\mathrm{M}_\odot$; $M_\mathrm{WD}=0.8\,\mathrm{M}_\odot$ and $M_\mathrm{2}=0.6\,\mathrm{M}_\odot$; $M_\mathrm{WD}=0.6\,\mathrm{M}_\odot$ and $M_\mathrm{2}=0.5\,\mathrm{M}_\odot$; $M_\mathrm{WD}=0.4\,\mathrm{M}_\odot$ and $M_\mathrm{2}=0.3\,\mathrm{M}_\odot$.}\label{fig:comparison_evolution}
\end{figure*}

This behaviour differs significantly from the predictions of the classical models of CV evolution, which assume that, in this period range, angular momentum is mainly removed by gravitational wave radiation. This mechanism implies higher accretion rates (and hence higher effective temperatures) for larger white dwarf masses because the (i) the rate of gravitational wave radiation is directly proportional to the masses of the two stellar components in the system, and (ii) $\rwd$ is smaller for more massive white dwarf while the surface luminosity is not strongly dependent on $\mwd$ and thus $\teff \propto \rwd^{-2}$ \citep{Townsley+2004}. For comparison with this model, in the left panels of Figure~\ref{fig:comparison_evolution}, we show our results against various evolutionary tracks computed with \textsc{MESAbinary} (revision 15140) assuming magnetic wind braking and gravitational wave radiation above the period gap, and gravitational wave radiation only below the period gap, for different initial white dwarf masses. For completeness, these models also account for the loss of angular momentum associated with mass ejection following a classical nova eruption (assuming that all the accreted material is ejected), which is, however, negligible compared to magnetic wind braking and gravitational wave radiation. The models cover a much wider range in temperatures than the observations and predict a strong correlation between the white dwarf effective temperature and mass, with more massive CV white dwarfs being hotter than their less massive counterparts. 
They also imply lower accretion rates than observed and a direct correlation between the accretion rates and the white dwarf mass, i.e. CVs hosting more massive white dwarfs have higher accretion rates. Moreover, the evolutionary tracks underestimate the location of the period minimum and are not able to reproduce the observed temperatures of period bounce CVs. 
This is also the case for similar evolutionary models available in the literature (see e.g. figure~3 from \citealt{Howell+2001} or figure~2 from \citealt{Goliasch-Nelson2015}), which in general predict that all systems, soon after the onset of mass transfer, will converge into a narrow track in the $\langle \mdot \rangle - \porb$ plane, with a very weak dependence on their masses. 

These theoretical results are in contrast with our findings, which suggest that the classical recipe of CV evolution needs to be revised in order to explain the absence of a clear dependency of the effective temperature on the white dwarf mass.
The effective temperature is set by the secular mass accretion rate onto the white dwarf which, in turn, reflects the rate of angular momentum loss in the system. Therefore, our results suggest that the missing ingredient of the theoretical modelling causing low mass CV white dwarfs to be hotter than predicted by the classical recipes (and to have similar effective temperature to their more massive counterparts, left panel of Figure~\ref{fig:correlations}) is an additional source of angular momentum loss which is more efficient the lower the mass of the white dwarf (right panel of Figure~\ref{fig:correlations}).

Such a dependency of the angular momentum loss rate on the white dwarf mass is the fundamental concept of the eCAML prescription developed by \cite{Schreiber+2016} and \cite{Belloni+2018,Belloni+2020} and already discussed in Section~\ref{sec:intro}. This model includes an empirically mass-dependent additional source of angular momentum loss that better accounts for the observed dependency of the accretion rates on the white dwarf mass and better reproduces the observed temperatures and mass accretion rates (right panels in Figure~\ref{fig:comparison_evolution}) than the classical models. The enhanced angular momentum loss leads to a faster erosion of the donor star, causing the systems to bounce at longer orbital periods. Consequently, the period minimum predicted by eCAML is anticipated compared to the standard prediction and agrees better with the observations.
Moreover, without requiring any additional fine-tuning, eCAML is also able to solve other disagreements that, for long time, have been found between the standard model of CV evolution and the observed properties of the CV population (such as their space density, and orbital period and mass distributions \citealt{Schreiber+2016,Belloni+2020,Pala+2020}). 

Despite the significant progress that is provided by the eCAML prescription, our observations provide reasons for improving this model in order to account for (i) the observed scatter in the parameters, (ii) the period bounce systems, for which the models predict a steep decrease in their effective temperatures that is not observed in data (which, instead, suggest the presence of enhanced angular momentum loss also in the post-bounce regime, as previously discussed also by \citealt{Pala+2017,Pala+2020}), and (iii) the presence of helium-core white dwarfs.
The difference between the null fraction of helium white dwarfs predicted by eCAML and the observations is not as dramatic as in the case of the former estimates from more classical evolutionary models (see also Section~\ref{subsec:mass_distribution}). Our results suggest that some low mass systems can survive (at least for some time) in a semi-detached configuration, thus providing valuable observational constraints to further refine this model and thereby help to understand the physical mechanism that is driving the additional angular momentum loss.

Different authors have suggested that an additional source of angular momentum loss that is more efficient the lower the mass of the white dwarf could arise from friction between the binary and the shell of ejected material following a nova eruptions \citep{Schreiber+2016,Nelemans+2016,Sparks+2021}.
To further investigate this possibility, our observational and theoretical understanding of classical novae and their impact on the secular evolution of CVs need to be improved.

Finally, additional issues are observed for systems above the period gap. The 3-4\,h period range is mainly populated by the unexpectedly hot nova-likes CVs (which have already been discussed in the first part of this Section) while, at $\porb \gtrsim 4\,$h, CV white dwarfs are found to be systematically colder than predicted. This is a long-standing issue in our understanding of compact binary evolution \citep{Knigge+2011,Pala+2017} and, as discussed in great detail by \cite{Townsley+2009} and \cite{Belloni+2020}, could be related to either inaccurate modelling of magnetic wind braking (the dominant angular momentum loss mechanism driving the evolution of the systems in this period range) or incomplete understanding of the heating of the white dwarf as a consequence of the mass accretion process (for example, owing to the presence of long-term mass-transfer rate fluctuations associated with the secondary, to which long period CVs would be more susceptible than short-period systems, \citealt{Knigge+2011}).
However, the limited number of available measurements (eight) severely limits any conclusions we can draw in this period range and more observations are needed to increase the number of systems with accurate parameters above the period gap. 

\section{Conclusions}
We analysed high quality \hst ultraviolet spectra for 42 CVs. Making use of the astrometry delivered by the ESA \gaia space mission in its EDR3, we accurately measured the white dwarf effective temperatures and masses. We complemented this sample with an additional mass measurement for the white dwarf in AX\,For, obtained from its gravitational redshift. Our results are in good agreement with independent measurements obtained from analysis of the white dwarf eclipses and from radial velocity studies.

Combining our results with the effective temperatures and masses for 46 CV white dwarfs from the literature, we assembled the largest sample of systems with accurate white dwarf parameters. We derived an average white dwarf mass of $\langle \mwd \rangle = 0.81^{+0.16}_{-0.20}\,\msun$, in perfect agreement with former results, which allows us to definitively rule out any systematics affecting the masses derived from the analysis of eclipse light curves. 
In our mass distribution, we identify a tail extending towards low masses, consisting of five systems with $\mwd < 0.5\,\msun$. Such low masses are consistent with either He core or, possibly, hybrid CO/He core white dwarfs. 

The white dwarf response to the mass accretion process and its capability to retain the accreted mass are of key interest in the context of Type Ia Supernova (SNe\,Ia) progenitors. If the mass transfer process can lead to the mass growth of the white dwarf, CVs could represent a possible channel for SN\,Ia explosions. By comparing the average mass of the white dwarfs in long ($\porb > 3\,$hr) and short ($\porb < 3\,$hr) period CVs, we do not find any evidence for a clear dependency on the white dwarf mass with the orbital period. However, additional measurements at long orbital periods are required to further constrain any correlation.

Thanks to the accurate parallaxes provided by \gaia, we have been able to measure both the white dwarf masses and temperatures. The combination of these parameters allows us to derive the secular mean of the mass accretion rates onto the white dwarf, which can be used to test and constrain the current models of CV evolution.
For CVs at short orbital periods ($\porb < 3\,$hr), we show an anti-correlation between the mass accretion rates and the mass of the white dwarf, which implies the presence of an additional mechanism of angular momentum loss that is more efficient the lower the mass of the white dwarf. 
This finding is in very good agreement with the predictions of the recently proposed eCAML prescription. Including an empirically mass-dependent additional source of angular momentum loss, eCAML is able to explain the observed high average mass of CV white dwarfs and also to solve other disagreements between theory and observations, including the CV space density and orbital period distribution.
The eCAML model provides an improved understanding of the observational properties of CVs and our results provide observational support for it. Some disagreement between eCAML and the observations still need to be addressed, like the observed scatter in the parameters and the presence of helium-core white dwarfs. Nonetheless, we highlight that the difference between the null fraction predicted by eCAML and the observations is not as dramatic as in the case of the former estimates from more classical evolutionary models. Our results suggest that some low mass systems can survive (at least for some time) in a semi-detached configuration, thus providing valuable observational constraints to understand the physical mechanism that is driving the additional angular momentum loss.

Finally, an additional discrepancy between theory and observations is noticeable for the period bounce systems, for which the models predict a steep decrease in their $\teff$ and $\langle \mdot \rangle$ that is not observed in data, which, instead, suggest the presence of enhanced angular momentum loss also in the post-bounce regime.

\section*{Acknowledgements}
This work has made use of data from the European Space Agency (ESA) mission {\it Gaia} (\url{https://www.cosmos.esa.int/gaia}), processed by the {\it Gaia} Data Processing and Analysis Consortium (DPAC, \url{https://www.cosmos.esa.int/web/gaia/dpac/consortium}). Funding for the DPAC has been provided by national institutions, in particular the institutions participating in the {\it Gaia} Multilateral Agreement.

The research leading to these results has received funding from the European Research Council under the European Union's Seventh Framework Programme (FP/2007--2013) / ERC Grant Agreement n. 320964 (WDTracer). 

The work presented in this article made large use of \textsc{TOPCAT} and \textsc{STILTS} Table/VOTable Processing Software \citep{Topcat}.

B.T.G. was supported by the UK  Science and Technology Facilities Council (STFC) grant ST/P000495 and ST/T000406/1. T.R.M. acknowledges support from STFC grants ST/T000406/1 and from a Leverhulme Research Fellowship.

D.B. was supported by the grant {\#2017/14289-3}, S\~ao Paulo Research Foundation (FAPESP) and ESO/Gobierno de Chile.

M.R.S. acknowledges support from Fondecyt (grant 1181404) and ANID, -- Millennium Science Initiative Program -- NCN19\_171.

P.S. acknowledges support from NSF grant AST-1514737 and NASA grant HST GO-15703.

D.D.M. acknowledges support from  the Italian Space Agency (ASI) and National Institute for Astrophysics (INAF) under agreements I/037/12/0 and 2017-14-H.0 and from INAF projects funded with Presidential Decrees N.43/2018 and  N.70/2016.

\section*{Data Availability}
All data underlying this article is publicly available from the relevant observatory archive or will be shared on reasonable request to the corresponding author.



\bibliographystyle{mnras}
\bibliography{main} 




\clearpage
\appendix
\section{Variable systems}\label{ap:variability}
\subsubsection{Eclipsing systems}\label{subsubsec:eclipsing} 
\begin{figure}
\includegraphics[width=\columnwidth]{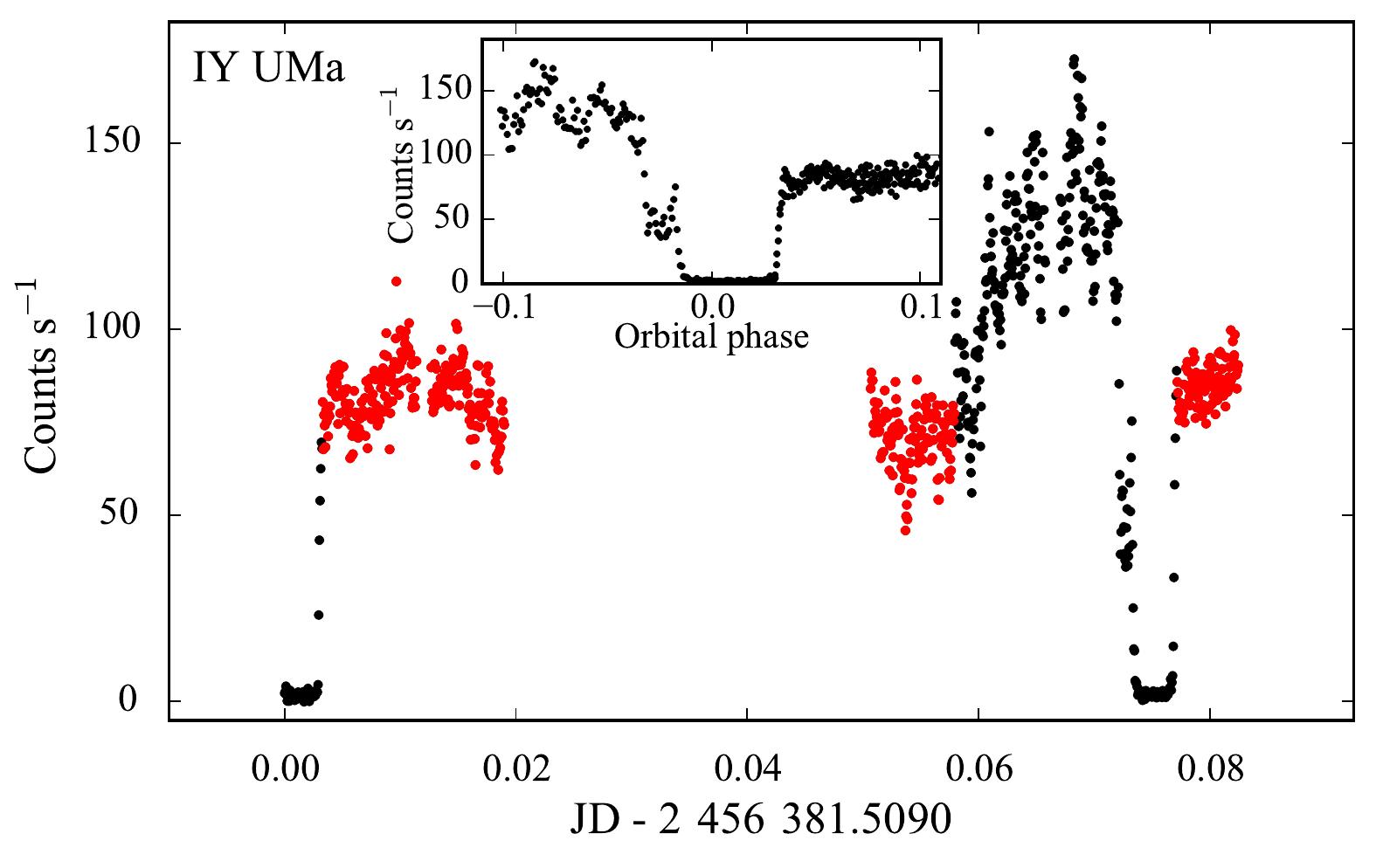}\\%
\includegraphics[width=\columnwidth]{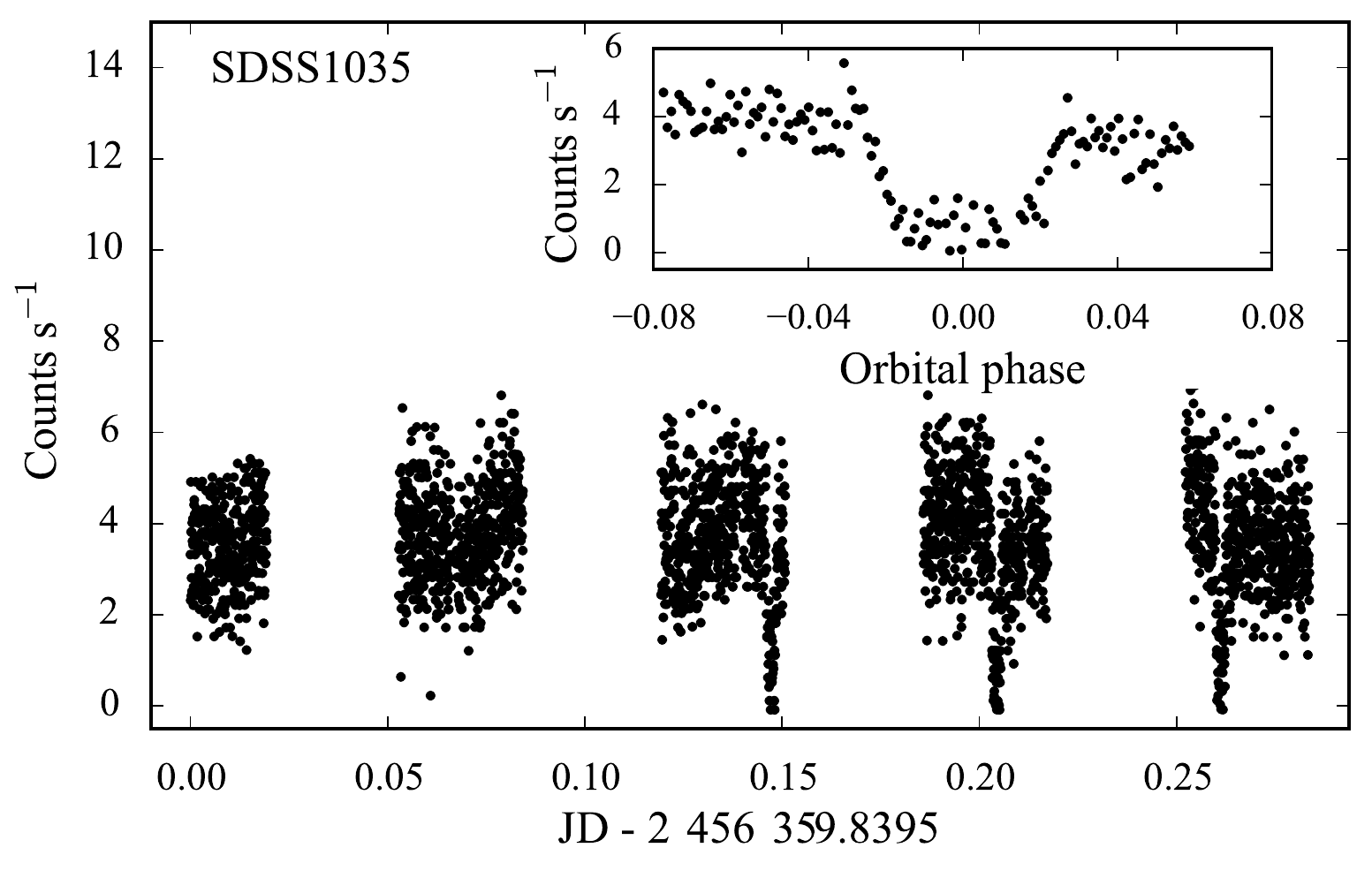}
\caption{\hst/COS light curves of some of the two eclipsing systems in our sample for which a fit to the eclipse light curve was not possible. In the case of IY\,UMa (top panel), the red dots highlight the regions that have been used to extract the spectrum of the system, when the emission was not dominated by the bright spot and the white dwarf was not eclipsed. The inset show the phase-folded light curve. Three white dwarf eclipses were detected in the light curve of SDSS\,J103533.02+055158.4 (bottom panel). Their average light curve is shown in the inset. However, the data quality is too poor to allow for an accurate mass determination of the white dwarf.}\label{fig:lc_eclipse_bad}
\end{figure}

Among the four eclipsing CVs, a fit to the eclipse light curve was only possible in the case of IR\,Com and SDSS\,J150722.30+523039.8 (Section~\ref{subsubsec:ircom_sdss1507}).
Although three eclipses of the white dwarf in SDSS\,J103533.02+055158.4 were observed in its COS light curve, their count level is too low (bottom panel of Figure~\ref{fig:lc_eclipse_bad}) to accurately resolve the ingress and egress of the white dwarf, thus preventing an accurate mass determination. 

Ideally, in systems in which the bright spot strongly contributes, multiple light curve eclipses are required in order to correctly time the ingress and egress of the white dwarf. Unfortunately, in the case of IY\,UMa, only one full eclipse was observed during the \hst observations and, therefore, an accurate mass determination is not possible from the \textsc{time--tag} data of this system. In the light curves of IY\,UMa, the hump observed just before the eclipse arises from the bright spot coming into view. To minimise the contamination from the bright spot emission for the analysis in Section~\ref{sec:uv_fit}, we extracted the spectrum out of eclipse and when the bright spot is not dominant (top panel of Figure~\ref{fig:lc_eclipse_bad}).

\subsubsection{Brightenings}\label{subsubsec:brightenings} 
\begin{figure}
\includegraphics[width=\columnwidth]{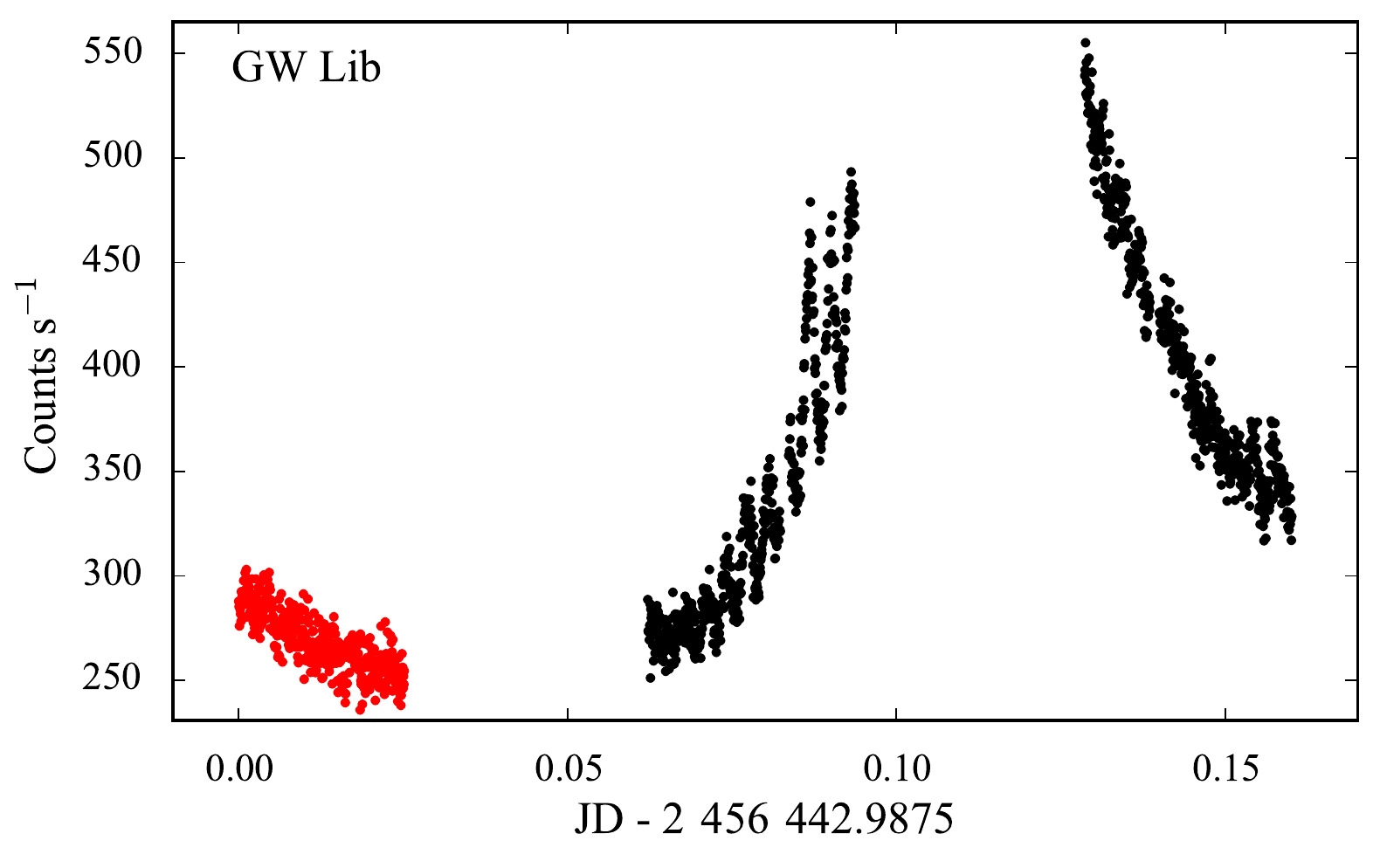}\\%
\caption{\hst/COS light curve of GW\,Lib obtained over three consecutive spacecraft orbits, the gaps result from the target being occulted by the Earth. The red dots highlights the exposures from which the spectrum of GW\,Lib has been extracted.}\label{fig:lc_bright}
\end{figure}
During the COS observations, GW\,Lib underwent a brightening event (Figure~\ref{fig:lc_bright}). This phenomenon has been studied in detailed by \cite{Toloza+2016}, who suggested that it could be associated with a fluctuation in the mass accretion rate. The brightening is related to the increase in temperature in a region covering up to $\simeq 30\,$per cent of the white dwarf surface \citep{Toloza+2016}. Since the presence of this hot spot could alter our mass measurement, we used the \textit{splitcorr} and \textit{x1dcorr} routine to extract the spectrum corresponding to the first spacecraft orbit, when the hot region covers only $\simeq 2\,$per cent of the white dwarf surface (see fig.~6 from \citealt{Toloza+2016}). This spectrum is used later in this work to measure the mass of the white dwarf as it is the most representative of the quiescent phase of the system.

\subsubsection{AM\,Her}\label{subsubsec:amher} 
\begin{figure}
\includegraphics[width=\columnwidth]{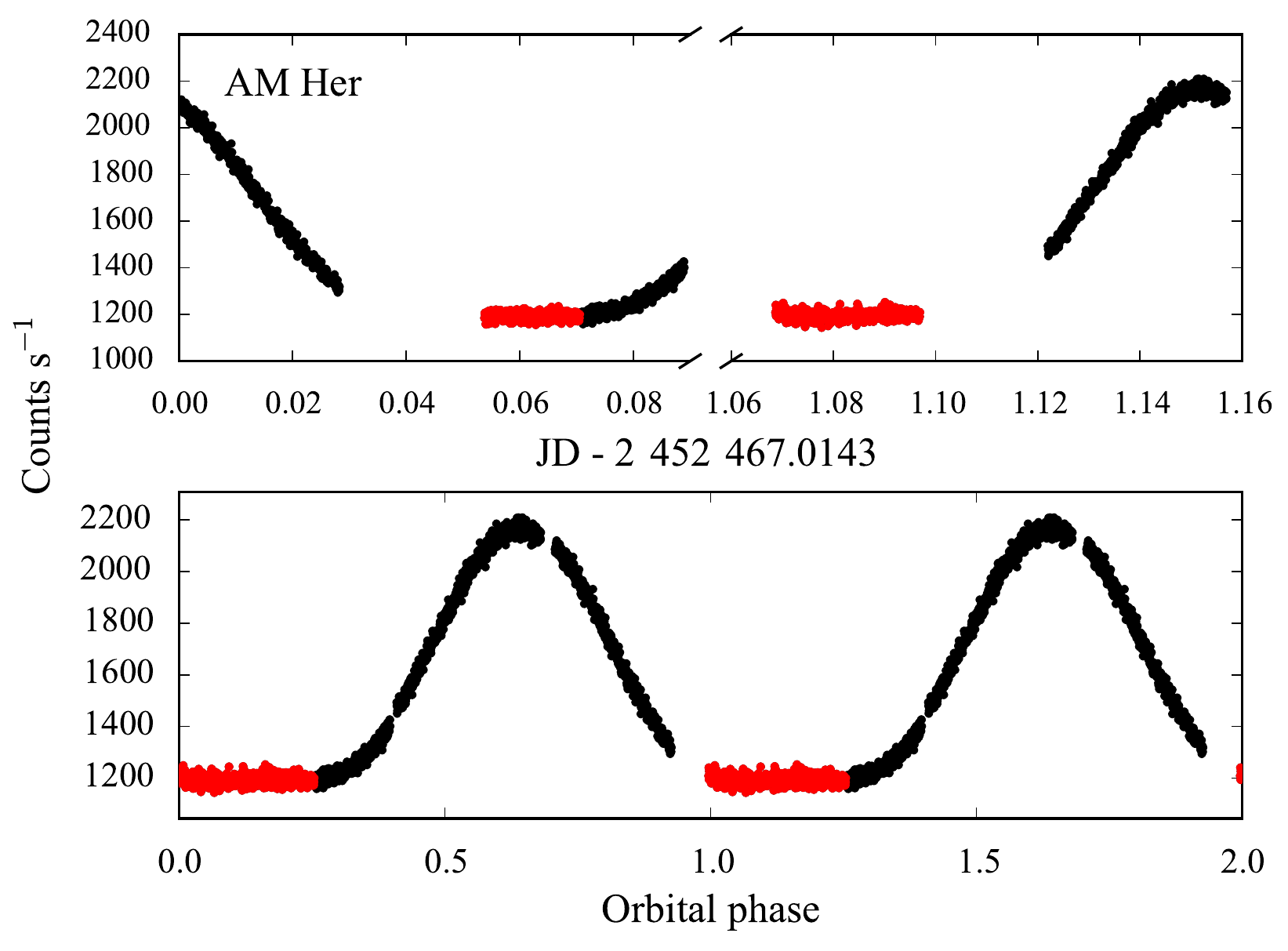}
\caption{\hst/STIS light curve of AM\,Her from the \textsc{time-tag} data (top panel). The bottom panel shows the data phase-folded using the ephemeris reported by \citet{Heise+1988}. Two cycles are plotted for clarity. The red dots highlights the exposures from which the spectrum of AM\,Her has been extracted.}\label{fig:lc_amher}
\end{figure}
AM\,Her is the prototype for the class of CVs known as polars. The \hst/STIS light curve of AM\,Her is characterised by a strong quasi-sinusoidal orbital flux modulation which, as discussed in great detail by \citet{Gaensicke+2006}, is associated with the hot polar cap on the rotating white dwarf coming into view (Figure~\ref{fig:lc_amher}). The flat part of the light curve corresponds to the times when the heated pole cap is self-eclipsed by the body of the white dwarf and we used the \textit{inttag} routine from the \textit{calstis} pipeline to split the observations into shorter exposures. Using the \textit{calstis} \textit{x1d} routine, we extracted the spectrum corresponding to the minimum of the modulation (red dots in Figure~\ref{fig:lc_amher}), which was then analysed to in Section~\ref{sec:uv_fit}.

\subsubsection{CVs hosting pulsating white dwarfs}\label{subsubsec:pulsators} 
1RXS\,J023238.8--371812 and SDSS\,J075507.70+143547.6 both contain a white dwarf pulsating with a period of  $\simeq 267.5\,$s \citep{Mukadam+2017}.
The pulsation is visible as a modulation in the COS light curve with an amplitude of $\simeq 5\,\mathrm{counts/s}$, too low to allow the extraction of a trough spectrum in a similar fashion as done for GW\,Lib. Therefore, we analysed their averaged ultraviolet spectrum obtained from summing the data from all orbits.

\section{Fit results}\label{ap:best-fit}
In the following, for each object, we present a table summarising the best-fitting parameters and a plot showing the best-fitting model (red) to the corresponding \hst data (black). The grey bands mask the geocoronal emission lines of Ly$\alpha$ ($1216\,$\AA) and, whenever present, of \ion{O}{i} ($1302\,$\AA). The systems are sorted according to their orbital periods.
\clearpage

\thispagestyle{empty}
\begin{table*}
  \centering
  \caption{Best-fitting parameters for SDSS\,J150722.30+523039.8.}
\begin{tabular}{ l r }
    \toprule
     Instrument & COS/STIS \\ [0.13cm]
     Observation date & 2010 Feb 18 \\ [0.13cm]
     $E(B-V)$ (mag) & $0.018^{+0.006}_{-0.018}$ \\ [0.13cm]
     $P_\mathrm{orb}$ (min) & 66.61 \\ [0.13cm]
     $\mathit{Gaia}$ EDR3 ID & 1593140224924964864 \\ [0.13cm]
     $\varpi$ (mas) & $4.73\pm0.09$ \\ [0.13cm]
     $d$ (pc) & $211 \pm 4$ \\ [0.13cm]
     $Z$ (Z$_\mathrm{\odot}$) & 0.1 \\ [0.13cm]
     $T_\mathrm{eff}$ (K) & $14207^{+356}_{-403}$ \\ [0.13cm]
     $R_\mathrm{WD}$ ($0.01\,\mathrm{R}_\odot$) & $0.93^{+0.17}_{-0.12}$ \\ [0.13cm]
     $M_\mathrm{WD}$ ($\mathrm{M}_\odot$) & $0.90^{+0.10}_{-0.14}$ \\ [0.13cm]
     $\log(g)$ & $8.45^{+0.16}_{-0.22}$ \\ [0.13cm]
     $\langle \dot{M} \rangle$ $(10^{-10}\,\mathrm{M}_\odot$ yr$^\mathrm{-1}$) & $0.53^{+0.18}_{-0.11}$ \\ [0.13cm]
     WD contribution & 78\% \\ [0.13cm]
     \bottomrule
    \end{tabular}
\end{table*}
\vspace{0.8cm}
\begin{figure*}
    \centering
\begin{overpic}[width=0.8\textwidth]{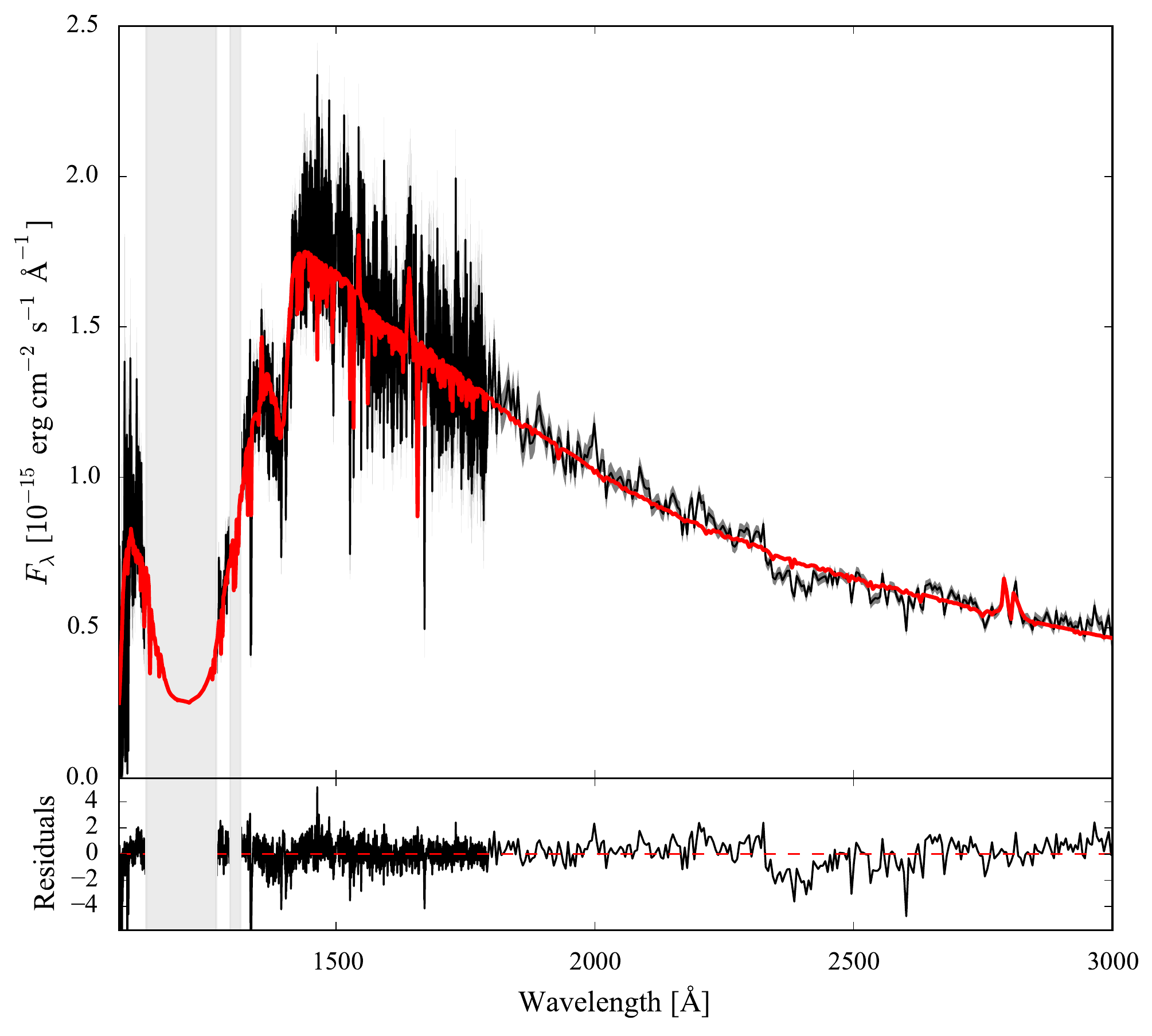}
\put (15,84) {SDSS\,J150722.30+523039.8}
\end{overpic}
  \caption{Ultraviolet \hst spectra (black) of  SDSS\,J150722.30+523039.8 along with the best-fitting model (red), computed as described in Section~\ref{sec:uv_fit}.}
\end{figure*}
\newpage
\clearpage

\thispagestyle{empty}
\begin{table*}
  \centering
  \caption{Best-fitting parameters for SDSS\,J074531.91+453829.5.}
  \begin{tabular}{ l r }
    \toprule
     Instrument & COS \\ [0.13cm]
     Observation date & 2012 Mar 13 \\ [0.13cm]
     $E(B-V)$ (mag) & $0.027\pm0.017$ \\ [0.13cm]
     $P_\mathrm{orb}$ (min) & 76.0 \\ [0.13cm]
     $\mathit{Gaia}$ EDR3 ID & 927255749553754880 \\ [0.13cm]
     $\varpi$ (mas) & $3.2\pm0.2$ \\ [0.13cm]
     $d$ (pc) & $310^{+23}_{-20}$ \\ [0.13cm]
     $Z$ (Z$_\mathrm{\odot}$) & 0.1 \\ [0.13cm]
     $T_\mathrm{eff}$ (K) & $15447^{+556}_{-654}$ \\ [0.13cm]
     $R_\mathrm{WD}$ ($0.01\,\mathrm{R}_\odot$) & $1.1^{+0.3}_{-0.2}$ \\ [0.13cm]
     $M_\mathrm{WD}$ ($\mathrm{M}_\odot$) & $0.75^{+0.18}_{-0.20}$ \\ [0.13cm]
     $\log(g)$ & $8.2^{+0.3}_{-0.4}$ \\ [0.13cm]
     $\langle \dot{M} \rangle$ $(10^{-10}\,\mathrm{M}_\odot$ yr$^\mathrm{-1}$) & $1.2^{+0.7}_{-0.4}$ \\ [0.13cm]
     WD contribution & 88\% \\ [0.13cm]
     \bottomrule
    \end{tabular}
\end{table*}
\vspace{0.8cm}
\begin{figure*}
    \centering
\begin{overpic}[width=0.8\textwidth]{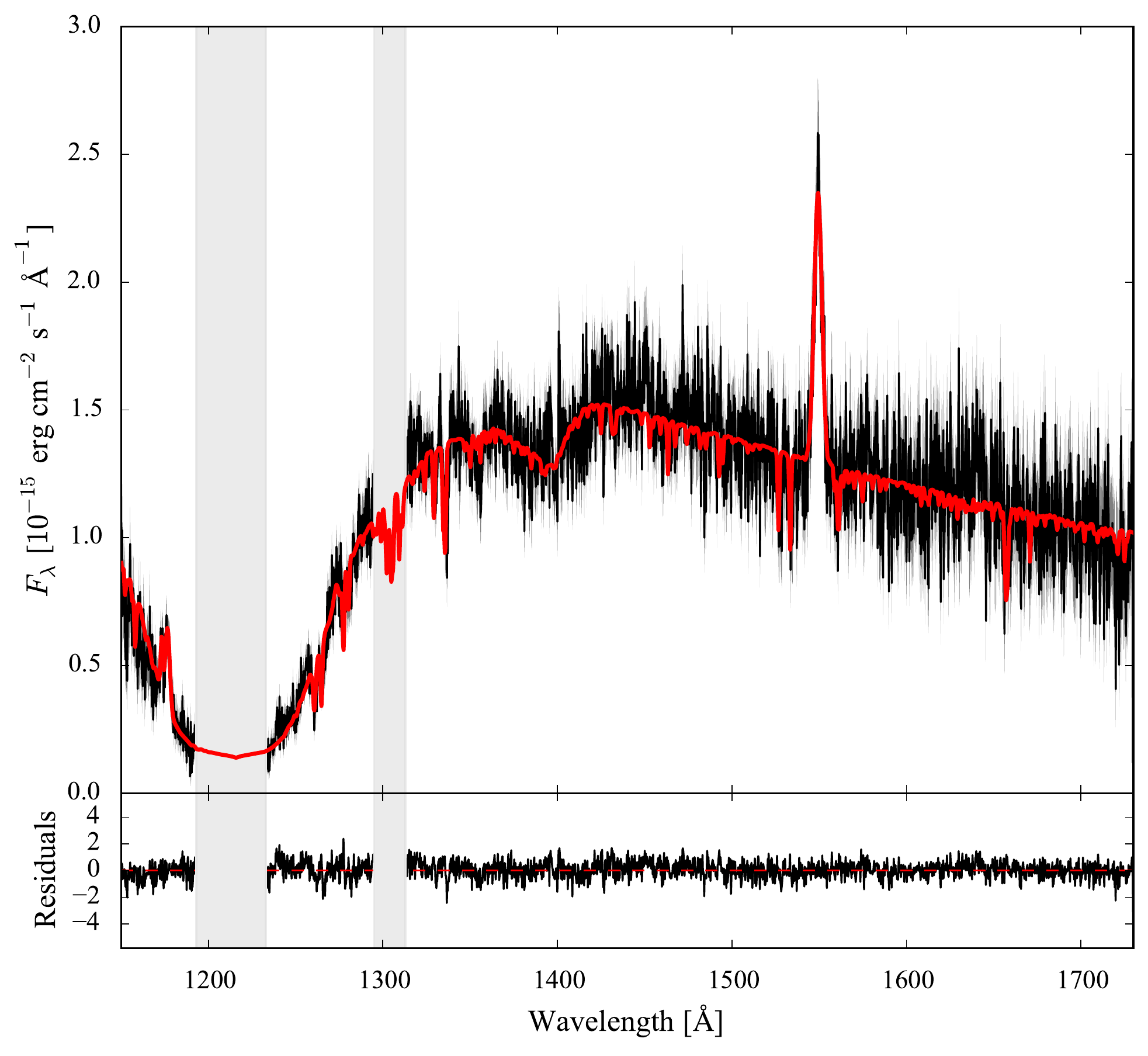}
\put (15,84) {SDSS\,J074531.91+453829.5}
\end{overpic}
  \caption{Ultraviolet \hst spectra (black) of  SDSS\,J074531.91+453829.5 along with the best-fitting model (red), computed as described in Section~\ref{sec:uv_fit}.}
\end{figure*}
\newpage
\clearpage

\thispagestyle{empty}
\begin{table*}
  \centering
  \caption{Best-fitting parameters for GW\,Lib.}
  \begin{tabular}{ l r }
    \toprule
     Instrument & COS \\ [0.13cm]
     Observation date & 2013 May 30 \\ [0.13cm]
     $E(B-V)$ (mag) & $0.022\pm0.018$ \\ [0.13cm]
     $P_\mathrm{orb}$ (min) & 76.78 \\ [0.13cm]
     $\mathit{Gaia}$ EDR3 ID & 6226943645600487552 \\ [0.13cm]
     $\varpi$ (mas) & $8.88\pm0.06$ \\ [0.13cm]
     $d$ (pc) & $112.6{\pm0.8}$ \\ [0.13cm]
     $Z$ (Z$_\mathrm{\odot}$) & 0.2 \\ [0.13cm]
     $T_\mathrm{eff}$ (K) & $16166^{+253}_{-350}$ \\ [0.13cm]
     $R_\mathrm{WD}$ ($0.01\,\mathrm{R}_\odot$) & $1.03^{+0.15}_{-0.1}$ \\ [0.13cm]
     $M_\mathrm{WD}$ ($\mathrm{M}_\odot$) & $0.83^{+0.08}_{-0.12}$ \\ [0.13cm]
     $\log(g)$ & $8.33^{+0.13}_{-0.18}$ \\ [0.13cm]
     $\langle \dot{M} \rangle$ $(10^{-10}\,\mathrm{M}_\odot$ yr$^\mathrm{-1}$) & $1.12^{+0.32}_{-0.19}$ \\ [0.13cm]
     WD contribution & 86\% \\ [0.13cm]
     \bottomrule
    \end{tabular}
\end{table*}
\vspace{0.8cm}
\begin{figure*}
    \centering
\begin{overpic}[width=0.8\textwidth]{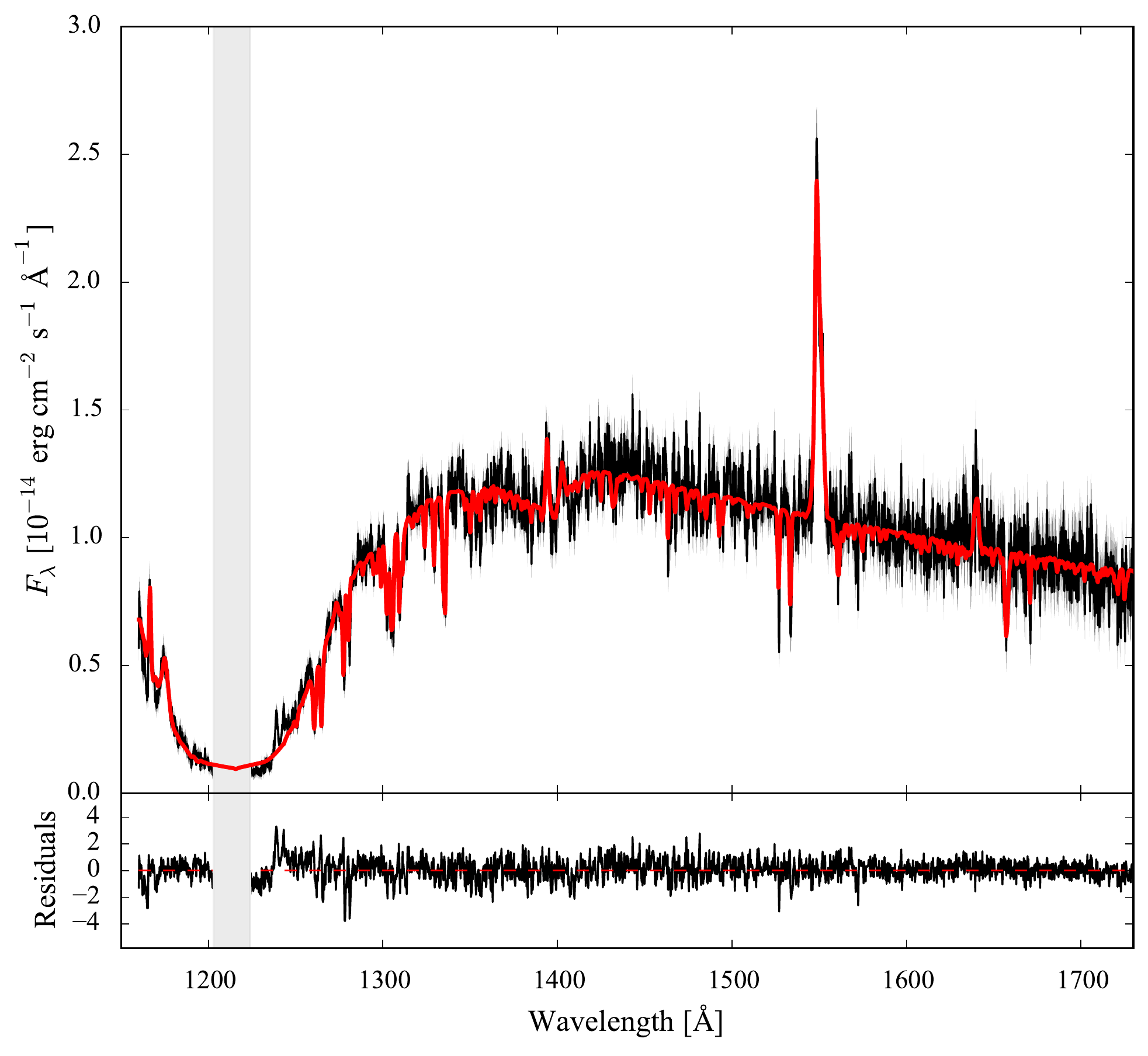}
\put (15,84) {GW\,Lib}
\end{overpic}
  \caption{Ultraviolet \hst spectra (black) of  GW\,Lib along with the best-fitting model (red), computed as described in Section~\ref{sec:uv_fit}.}
\end{figure*}
\newpage
\clearpage

\thispagestyle{empty}
\begin{table*}
  \centering
  \caption{Best-fitting parameters for SDSS\,J143544.02+233638.7.}
  \begin{tabular}{ l r }
    \toprule
     Instrument & COS \\ [0.13cm]
     Observation date & 2013 Mar 09 \\ [0.13cm]
     $E(B-V)$ (mag) & $0.02\pm0.02$ \\ [0.13cm]
     $P_\mathrm{orb}$ (min) & 78.0 \\ [0.13cm]
     $\mathit{Gaia}$ EDR3 ID & 1242828982729309952 \\ [0.13cm]
     $\varpi$ (mas) & $4.8\pm0.2$ \\ [0.13cm]
     $d$ (pc) & $208^{+9}_{-8}$ \\ [0.13cm]
     $Z$ (Z$_\mathrm{\odot}$) & 0.01 \\ [0.13cm]
     $T_\mathrm{eff}$ (K) & $11997^{+99}_{-160}$ \\ [0.13cm]
     $R_\mathrm{WD}$ ($0.01\,\mathrm{R}_\odot$) & $1.00^{+0.08}_{-0.09}$ \\ [0.13cm]
     $M_\mathrm{WD}$ ($\mathrm{M}_\odot$) & $0.84^{+0.07}_{-0.06}$ \\ [0.13cm]
     $\log(g)$ & $8.36^{+0.11}_{-0.1}$ \\ [0.13cm]
     $\langle \dot{M} \rangle$ $(10^{-10}\,\mathrm{M}_\odot$ yr$^\mathrm{-1}$) & $0.32^{+0.08}_{-0.05}$ \\ [0.13cm]
     WD contribution & 90\% \\ [0.13cm]
     \bottomrule
    \end{tabular}
\end{table*}
\vspace{0.8cm}
\begin{figure*}
    \centering
\begin{overpic}[width=0.8\textwidth]{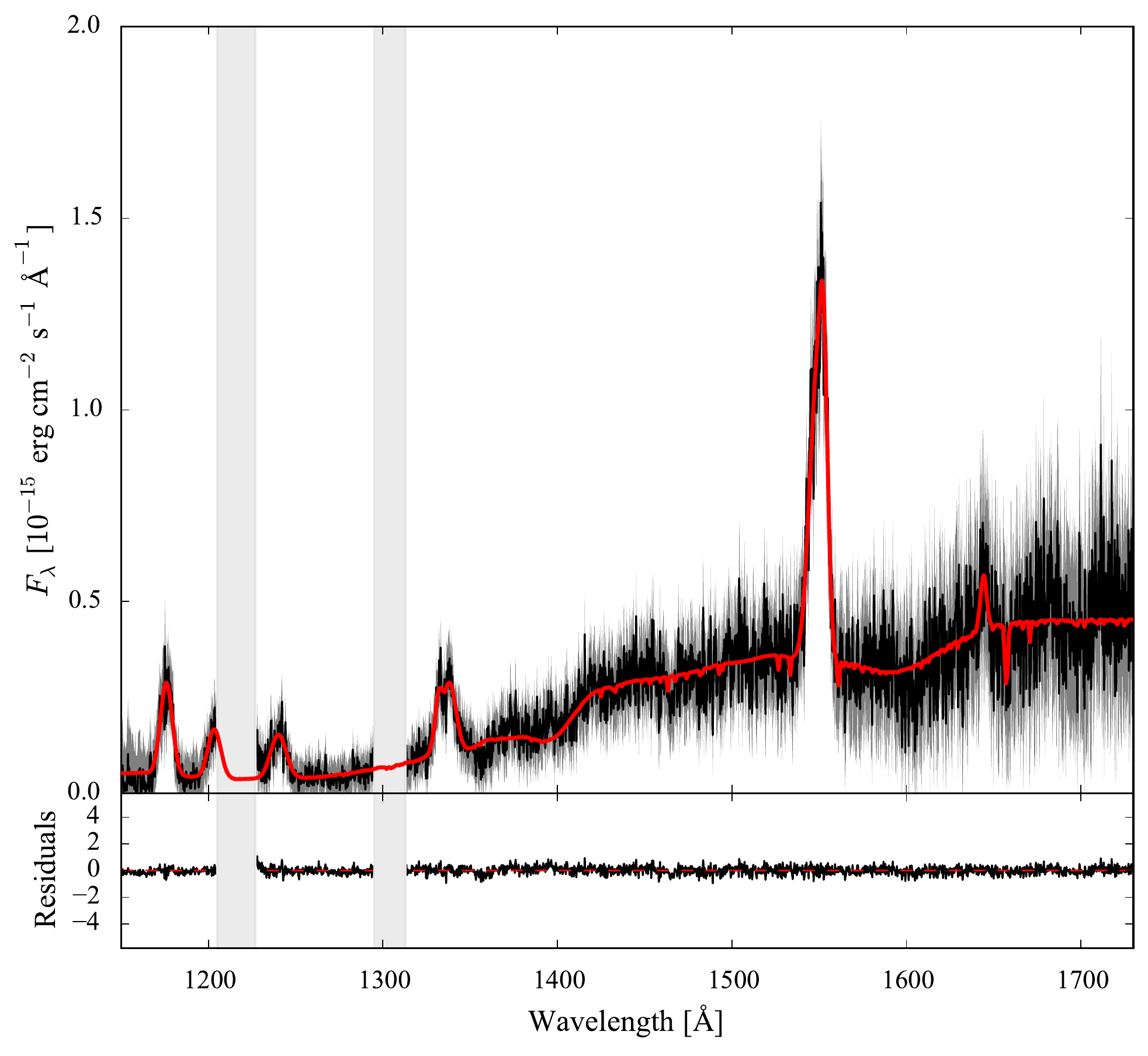}
\put (15,84) {SDSS\,J143544.02+233638.7}
\end{overpic}
  \caption{Ultraviolet \hst spectra (black) of  SDSS\,J143544.02+233638.7 along with the best-fitting model (red), computed as described in Section~\ref{sec:uv_fit}.}
\end{figure*}
\newpage
\clearpage

\thispagestyle{empty}
\begin{table*}
  \centering
  \caption{Best-fitting parameters for OT\,J213806.6+261957.}
  \begin{tabular}{ l r }
    \toprule
     Instrument & COS \\ [0.13cm]
     Observation date & 2013 Jul 25 \\ [0.13cm]
     $E(B-V)$ (mag) & $0.017\pm0.015$ \\ [0.13cm]
     $P_\mathrm{orb}$ (min) & 78.1 \\ [0.13cm]
     $\mathit{Gaia}$ EDR3 ID & 1800384942558699008 \\ [0.13cm]
     $\varpi$ (mas) & $10.11\pm0.04$ \\ [0.13cm]
     $d$ (pc) & $98.9{\pm0.4}$ \\ [0.13cm]
     $Z$ (Z$_\mathrm{\odot}$) & 0.2 \\ [0.13cm]
     $T_\mathrm{eff}$ (K) & $15317^{+216}_{-228}$ \\ [0.13cm]
     $R_\mathrm{WD}$ ($0.01\,\mathrm{R}_\odot$) & $1.39^{+0.15}_{-0.12}$ \\ [0.13cm]
     $M_\mathrm{WD}$ ($\mathrm{M}_\odot$) & $0.57^{+0.06}_{-0.07}$ \\ [0.13cm]
     $\log(g)$ & $7.91^{+0.12}_{-0.15}$ \\ [0.13cm]
     $\langle \dot{M} \rangle$ $(10^{-10}\,\mathrm{M}_\odot$ yr$^\mathrm{-1}$) & $1.9^{+0.4}_{-0.3}$ \\ [0.13cm]
     WD contribution & 73\% \\ [0.13cm]
     \bottomrule
    \end{tabular}
\end{table*}
\vspace{0.8cm}
\begin{figure*}
    \centering
\begin{overpic}[width=0.8\textwidth]{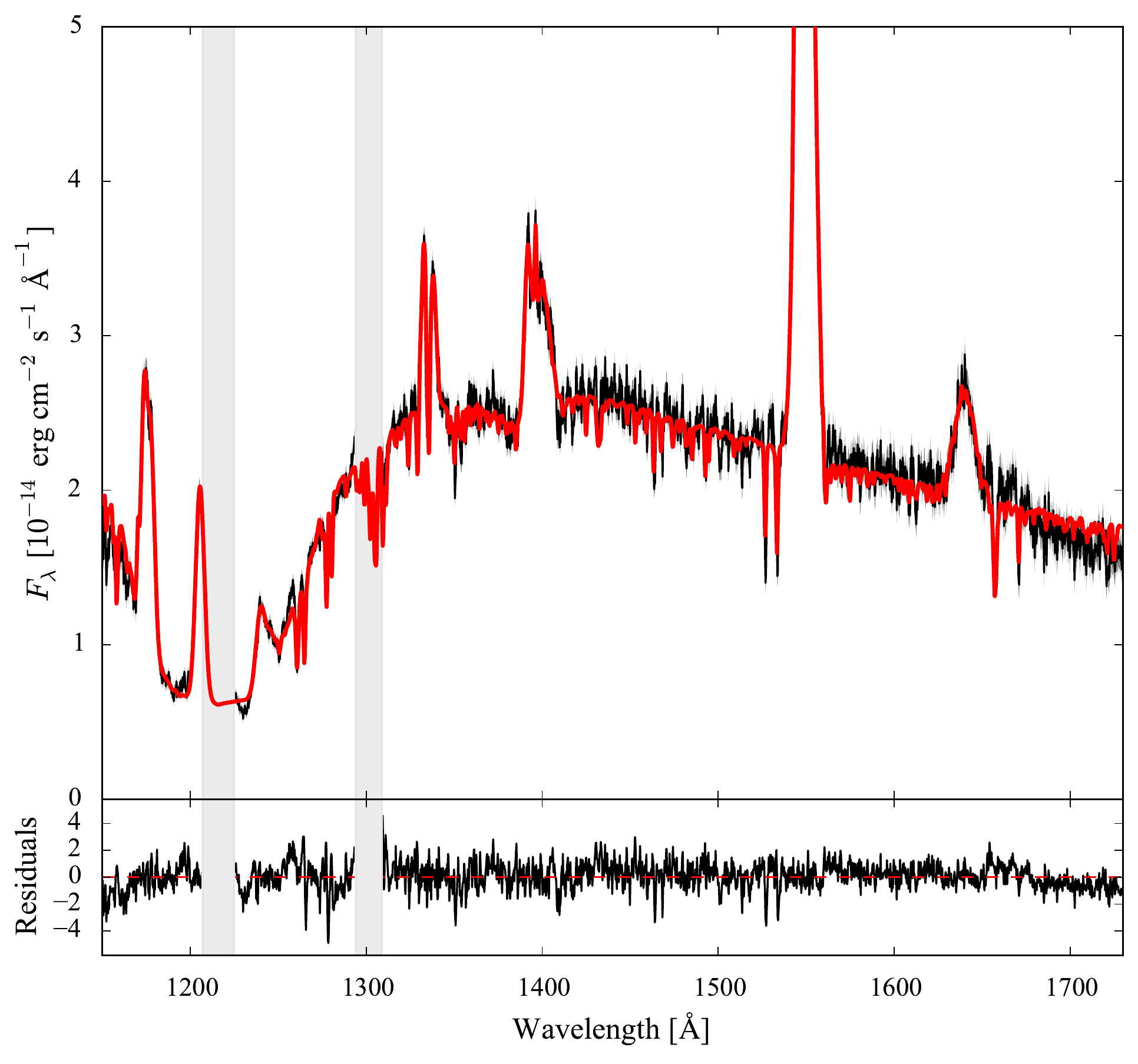}
\put (15,84) {OT\,J213806.6+261957}
\end{overpic}
  \caption{Ultraviolet \hst spectra (black) of  OT\,J213806.6+261957 along with the best-fitting model (red), computed as described in Section~\ref{sec:uv_fit}.}
\end{figure*}
\newpage
\clearpage

\thispagestyle{empty}
\begin{table*}
  \centering
  \caption{Best-fitting parameters for BW\,Scl.}
  \begin{tabular}{ l r }
    \toprule
     Instrument & STIS \\ [0.13cm]
     Observation date & 2006 Dec 27 \\ [0.13cm]
     $E(B-V)$ (mag) & $0.002^{+0.015}_{-0.002}$ \\ [0.13cm]
     $P_\mathrm{orb}$ (min) & 78.23 \\ [0.13cm]
     $\mathit{Gaia}$ EDR3 ID & 2307289214897332480 \\ [0.13cm]
     $\varpi$ (mas) & $10.71\pm0.05$ \\ [0.13cm]
     $d$ (pc) & $93.4{\pm0.5}$ \\ [0.13cm]
     $Z$ (Z$_\mathrm{\odot}$) & 0.5 \\ [0.13cm]
     $T_\mathrm{eff}$ (K) & $15145^{+51}_{-57}$ \\ [0.13cm]
     $R_\mathrm{WD}$ ($0.01\,\mathrm{R}_\odot$) & $0.8^{+0.014}_{-0.011}$ \\ [0.13cm]
     $M_\mathrm{WD}$ ($\mathrm{M}_\odot$) & $1.007^{+0.010}_{-0.012}$ \\ [0.13cm]
     $\log(g)$ & $8.635^{+0.017}_{-0.02}$ \\ [0.13cm]
     $\langle \dot{M} \rangle$ $(10^{-10}\,\mathrm{M}_\odot$ yr$^\mathrm{-1}$) & $0.483^{+0.014}_{-0.012}$ \\ [0.13cm]
     WD contribution & 87\% \\ [0.13cm]
     \bottomrule
    \end{tabular}
\end{table*}
\vspace{0.8cm}
\begin{figure*}
    \centering
\begin{overpic}[width=0.8\textwidth]{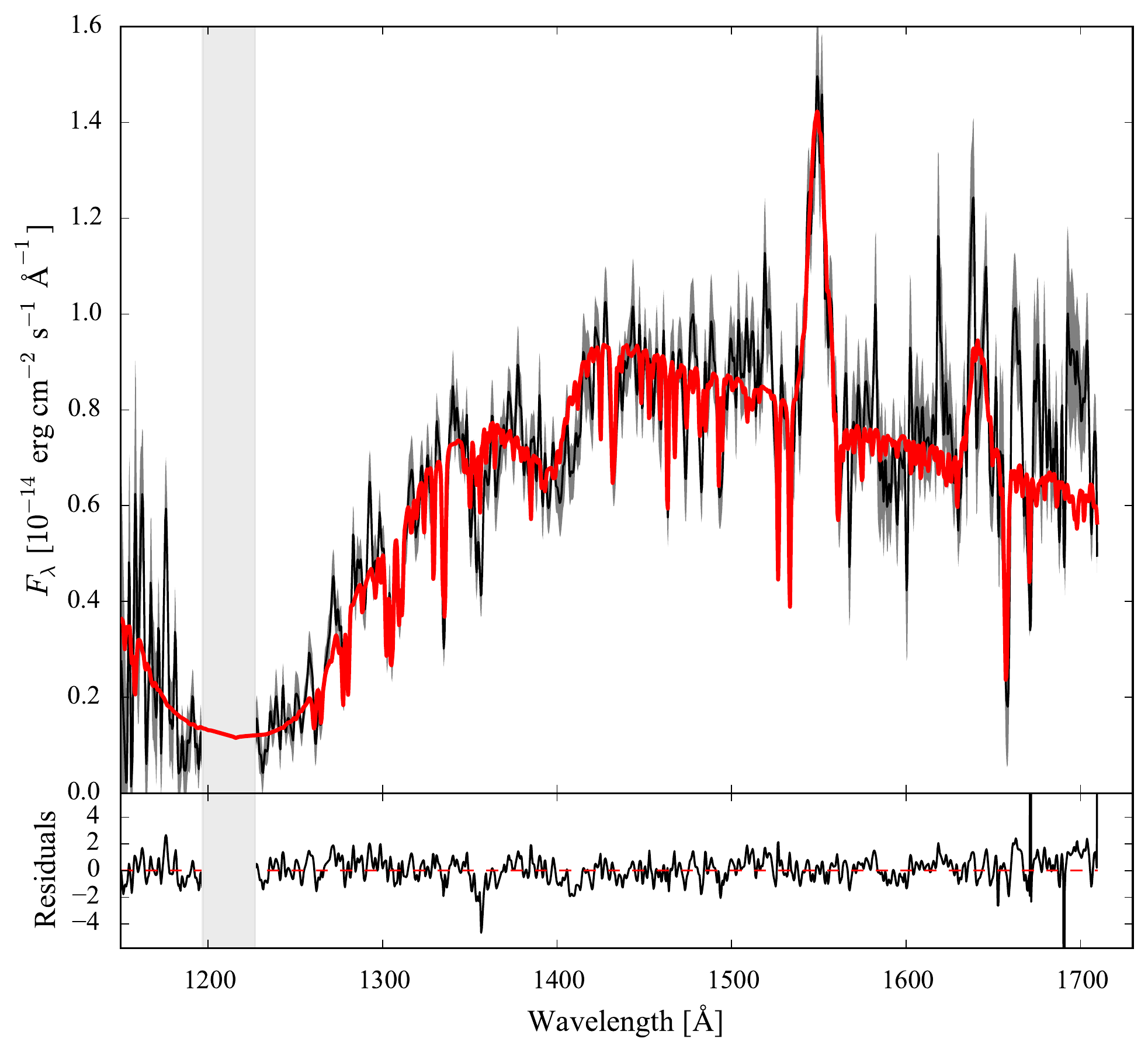}
\put (15,84) {BW\,Scl}
\end{overpic}
  \caption{Ultraviolet \hst spectra (black) of  BW\,Scl along with the best-fitting model (red), computed as described in Section~\ref{sec:uv_fit}.}
\end{figure*}
\newpage
\clearpage

\thispagestyle{empty}
\begin{table*}
  \centering
  \caption{Best-fitting parameters for LL\,And.}
  \begin{tabular}{ l r }
    \toprule
     Instrument & STIS \\ [0.13cm]
     Observation date & 2000 Dec 07 \\ [0.13cm]
     $E(B-V)$ (mag) & $0.06^{+0}_{-0.02}$ \\ [0.13cm]
     $P_\mathrm{orb}$ (min) & 79.28 \\ [0.13cm]
     $\mathit{Gaia}$ EDR3 ID & 2809168096329043712 \\ [0.13cm]
     $\varpi$ (mas) & $1.6\pm0.6$ \\ [0.13cm]
     $d$ (pc) & $609^{+343}_{-205}$ \\ [0.13cm]
     $Z$ (Z$_\mathrm{\odot}$) & 1.0 \\ [0.13cm]
     $T_\mathrm{eff}$ (K) & $14353^{+1210}_{-743}$ \\ [0.13cm]
     $R_\mathrm{WD}$ ($0.01\,\mathrm{R}_\odot$) & $1.2^{+0.8}_{-0.6}$ \\ [0.13cm]
     $M_\mathrm{WD}$ ($\mathrm{M}_\odot$) & $0.7^{+0.4}_{-0.3}$ \\ [0.13cm]
     $\log(g)$ & $8.2 \pm 0.7$ \\ [0.13cm]
     $\langle \dot{M} \rangle$ $(10^{-10}\,\mathrm{M}_\odot$ yr$^\mathrm{-1}$) & $0.9^{+2.4}_{-0.9}$ \\ [0.13cm]
     WD contribution & 80\% \\ [0.13cm]
     \bottomrule
    \end{tabular}
\end{table*}
\vspace{0.8cm}
\begin{figure*}
    \centering
\begin{overpic}[width=0.8\textwidth]{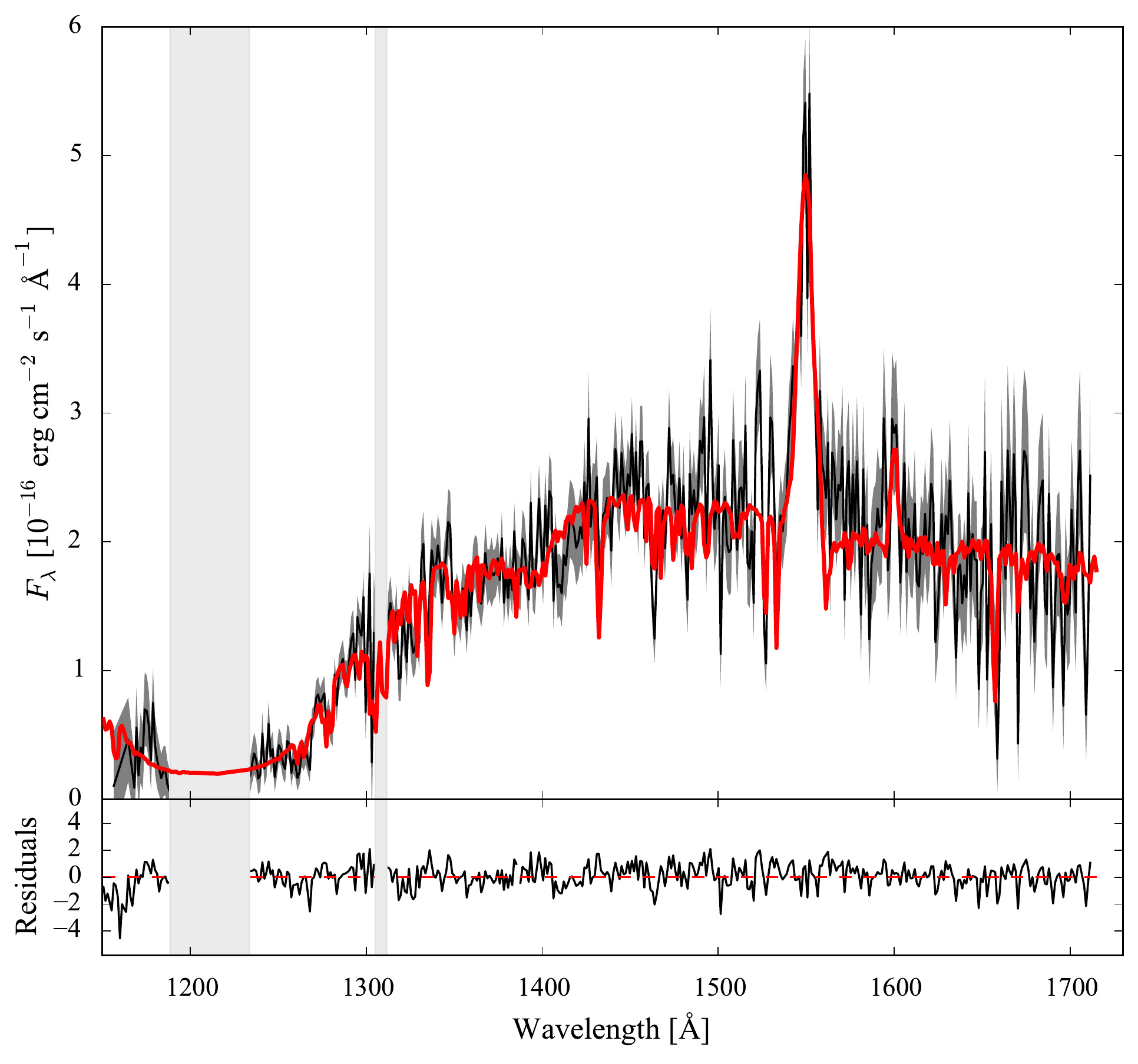}
\put (15,84) {LL\,And}
\end{overpic}
  \caption{Ultraviolet \hst spectra (black) of  LL\,And along with the best-fitting model (red), computed as described in Section~\ref{sec:uv_fit}.}
\end{figure*}
\newpage
\clearpage

\thispagestyle{empty}
\begin{table*}
  \centering
  \caption{Best-fitting parameters for AL\,Com.}
  \begin{tabular}{ l r }
    \toprule
     Instrument & STIS \\ [0.13cm]
     Observation date & 2000 Nov 27 \\ [0.13cm]
     $E(B-V)$ (mag) & $0.045\pm0.018$ \\ [0.13cm]
     $P_\mathrm{orb}$ (min) & 81.6 \\ [0.13cm]
     $\mathit{Gaia}$ EDR3 ID & 3932951035266314496 \\ [0.13cm]
     $\varpi$ (mas) & $1.9\pm0.6$ \\ [0.13cm]
     $d$ (pc) & $523^{+252}_{-149}$ \\ [0.13cm]
     $Z$ (Z$_\mathrm{\odot}$) & 0.2 \\ [0.13cm]
     $T_\mathrm{eff}$ (K) & $15840^{+1453}_{-1243}$ \\ [0.13cm]
     $R_\mathrm{WD}$ ($0.01\,\mathrm{R}_\odot$) & $1.0^{+0.9}_{-0.6}$ \\ [0.13cm]
     $M_\mathrm{WD}$ ($\mathrm{M}_\odot$) & $0.9^{+0.5}_{-0.3}$ \\ [0.13cm]
     $\log(g)$ & $8.4 \pm 0.8$ \\ [0.13cm]
     $\langle \dot{M} \rangle$ $(10^{-10}\,\mathrm{M}_\odot$ yr$^\mathrm{-1}$) & $0.9^{+3.1}_{-1.0}$ \\ [0.13cm]
     WD contribution & 79\% \\ [0.13cm]
     \bottomrule
    \end{tabular}
\end{table*}
\vspace{0.8cm}
\begin{figure*}
    \centering
\begin{overpic}[width=0.8\textwidth]{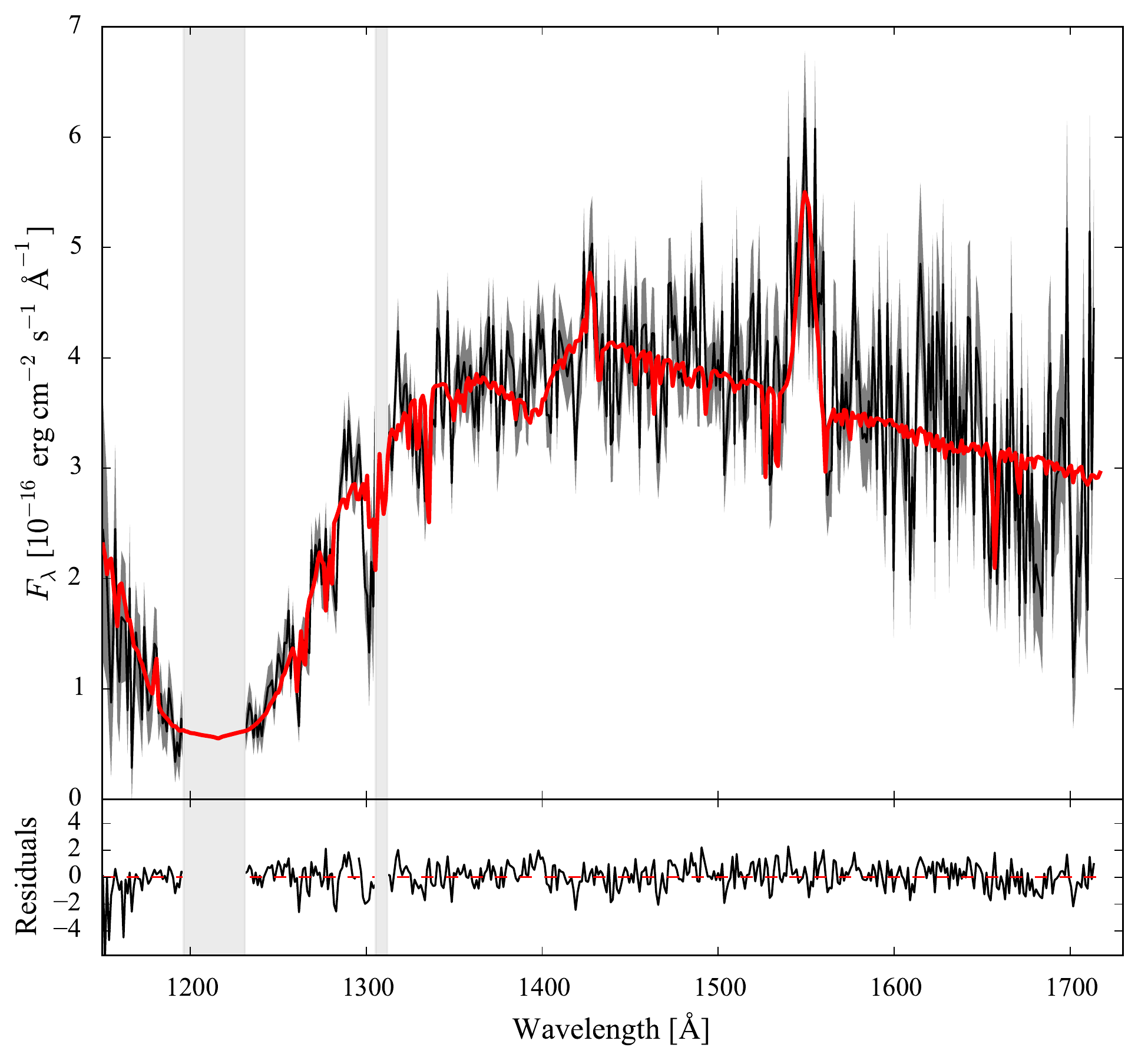}
\put (15,84) {AL\,Com}
\end{overpic}
  \caption{Ultraviolet \hst spectra (black) of  AL\,Com along with the best-fitting model (red), computed as described in Section~\ref{sec:uv_fit}.}
\end{figure*}
\newpage
\clearpage

\thispagestyle{empty}
\begin{table*}
  \centering
  \caption{Best-fitting parameters for WZ\,Sge.}
  \begin{tabular}{ l r }
    \toprule
     Instrument & FOS \\ [0.13cm]
     Observation date & 1992 Oct 08 \\ [0.13cm]
     $E(B-V)$ (mag) & $0.003^{+0.015}_{-0.003}$ \\ [0.13cm]
     $P_\mathrm{orb}$ (min) & 81.63 \\ [0.13cm]
     $\mathit{Gaia}$ EDR3 ID & 1809844934461976832 \\ [0.13cm]
     $\varpi$ (mas) & $22.14\pm0.03$ \\ [0.13cm]
     $d$ (pc) & $45.17{\pm0.06}$ \\ [0.13cm]
     $Z$ (Z$_\mathrm{\odot}$) & 0.01 \\ [0.13cm]
     $T_\mathrm{eff}$ (K) & $13190^{+115}_{-105}$ \\ [0.13cm]
     $R_\mathrm{WD}$ ($0.01\,\mathrm{R}_\odot$) & $1.05 \pm 0.03$ \\ [0.13cm]
     $M_\mathrm{WD}$ ($\mathrm{M}_\odot$) & $0.8 \pm 0.02$ \\ [0.13cm]
     $\log(g)$ & $8.3 \pm 0.04$ \\ [0.13cm]
     $\langle \dot{M} \rangle$ $(10^{-10}\,\mathrm{M}_\odot$ yr$^\mathrm{-1}$) & $0.52^{+0.014}_{-0.01}$ \\ [0.13cm]
     WD contribution & 78\% \\ [0.13cm]
     \bottomrule
    \end{tabular}
\end{table*}
\vspace{0.8cm}
\begin{figure*}
    \centering
\begin{overpic}[width=0.8\textwidth]{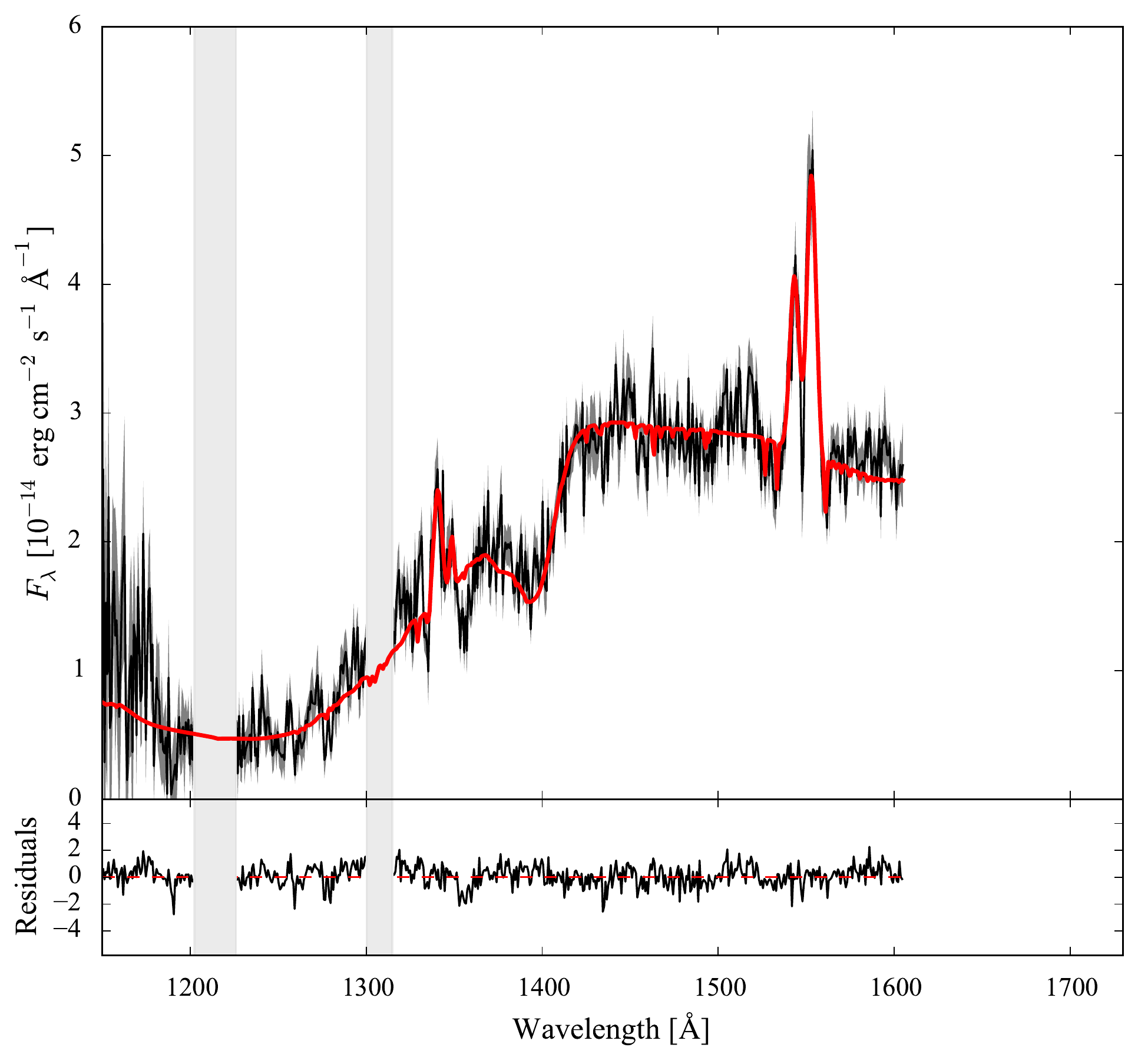}
\put (15,84) {WZ\,Sge}
\end{overpic}
  \caption{Ultraviolet \hst spectra (black) of  WZ\,Sge along with the best-fitting model (red), computed as described in Section~\ref{sec:uv_fit}.}
\end{figure*}
\newpage
\clearpage

\thispagestyle{empty}
\begin{table*}
  \centering
  \caption{Best-fitting parameters for SW\,UMa.}
  \begin{tabular}{ l r }
    \toprule
     Instrument & STIS \\ [0.13cm]
     Observation date & 2000 Mar 26 \\ [0.13cm]
     $E(B-V)$ (mag) & $0.008^{+0.017}_{-0.008}$ \\ [0.13cm]
     $P_\mathrm{orb}$ (min) & 81.81 \\ [0.13cm]
     $\mathit{Gaia}$ EDR3 ID & 1030279027003254784 \\ [0.13cm]
     $\varpi$ (mas) & $6.23\pm0.06$ \\ [0.13cm]
     $d$ (pc) & $160.6{\pm1.6}$ \\ [0.13cm]
     $Z$ (Z$_\mathrm{\odot}$) & 0.2 \\ [0.13cm]
     $T_\mathrm{eff}$ (K) & $13854^{+189}_{-131}$ \\ [0.13cm]
     $R_\mathrm{WD}$ ($0.01\,\mathrm{R}_\odot$) & $1.29^{+0.09}_{-0.10}$ \\ [0.13cm]
     $M_\mathrm{WD}$ ($\mathrm{M}_\odot$) & $0.61^{+0.06}_{-0.04}$ \\ [0.13cm]
     $\log(g)$ & $8.01^{+0.11}_{-0.09}$ \\ [0.13cm]
     $\langle \dot{M} \rangle$ $(10^{-10}\,\mathrm{M}_\odot$ yr$^\mathrm{-1}$) & $1.07^{+0.15}_{-0.14}$ \\ [0.13cm]
     WD contribution & 61\% \\ [0.13cm]
     \bottomrule
    \end{tabular}
\end{table*}
\vspace{0.8cm}
\begin{figure*}
    \centering
\begin{overpic}[width=0.8\textwidth]{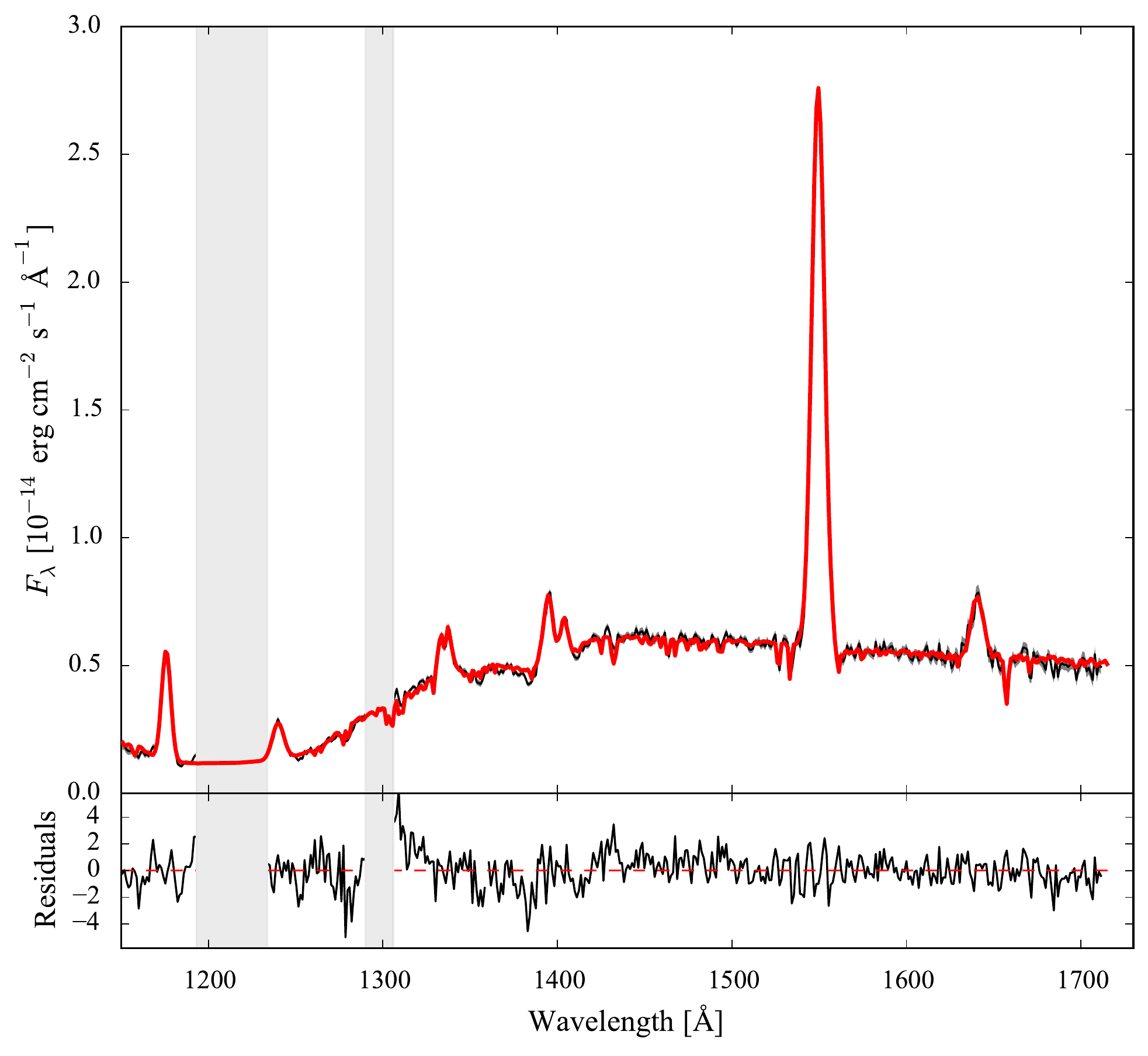}
\put (15,84) {SW\,UMa}
\end{overpic}
  \caption{Ultraviolet \hst spectra (black) of  SW\,UMa along with the best-fitting model (red), computed as described in Section~\ref{sec:uv_fit}.}
\end{figure*}
\newpage
\clearpage

\thispagestyle{empty}
\begin{table*}
  \centering
  \caption{Best-fitting parameters for V1108\,Her.}
  \begin{tabular}{ l r }
    \toprule
     Instrument & COS \\ [0.13cm]
     Observation date & 2013 Jun 06 \\ [0.13cm]
     $E(B-V)$ (mag) & $0.046\pm0.019$ \\ [0.13cm]
     $P_\mathrm{orb}$ (min) & 81.87 \\ [0.13cm]
     $\mathit{Gaia}$ EDR3 ID & 4538504384210935424 \\ [0.13cm]
     $\varpi$ (mas) & $6.8\pm0.1$ \\ [0.13cm]
     $d$ (pc) & $148\pm2$ \\ [0.13cm]
     $Z$ (Z$_\mathrm{\odot}$) & 1.0 \\ [0.13cm]
     $T_\mathrm{eff}$ (K) & $13943^{+185}_{-226}$ \\ [0.13cm]
     $R_\mathrm{WD}$ ($0.01\,\mathrm{R}_\odot$) & $0.95^{+0.14}_{-0.11}$ \\ [0.13cm]
     $M_\mathrm{WD}$ ($\mathrm{M}_\odot$) & $0.88^{+0.09}_{-0.11}$ \\ [0.13cm]
     $\log(g)$ & $8.42^{+0.15}_{-0.17}$ \\ [0.13cm]
     $\langle \dot{M} \rangle$ $(10^{-10}\,\mathrm{M}_\odot$ yr$^\mathrm{-1}$) & $0.52^{+0.15}_{-0.11}$ \\ [0.13cm]
     WD contribution & 75\% \\ [0.13cm]
     \bottomrule
    \end{tabular}
\end{table*}
\vspace{0.8cm}
\begin{figure*}
    \centering
\begin{overpic}[width=0.8\textwidth]{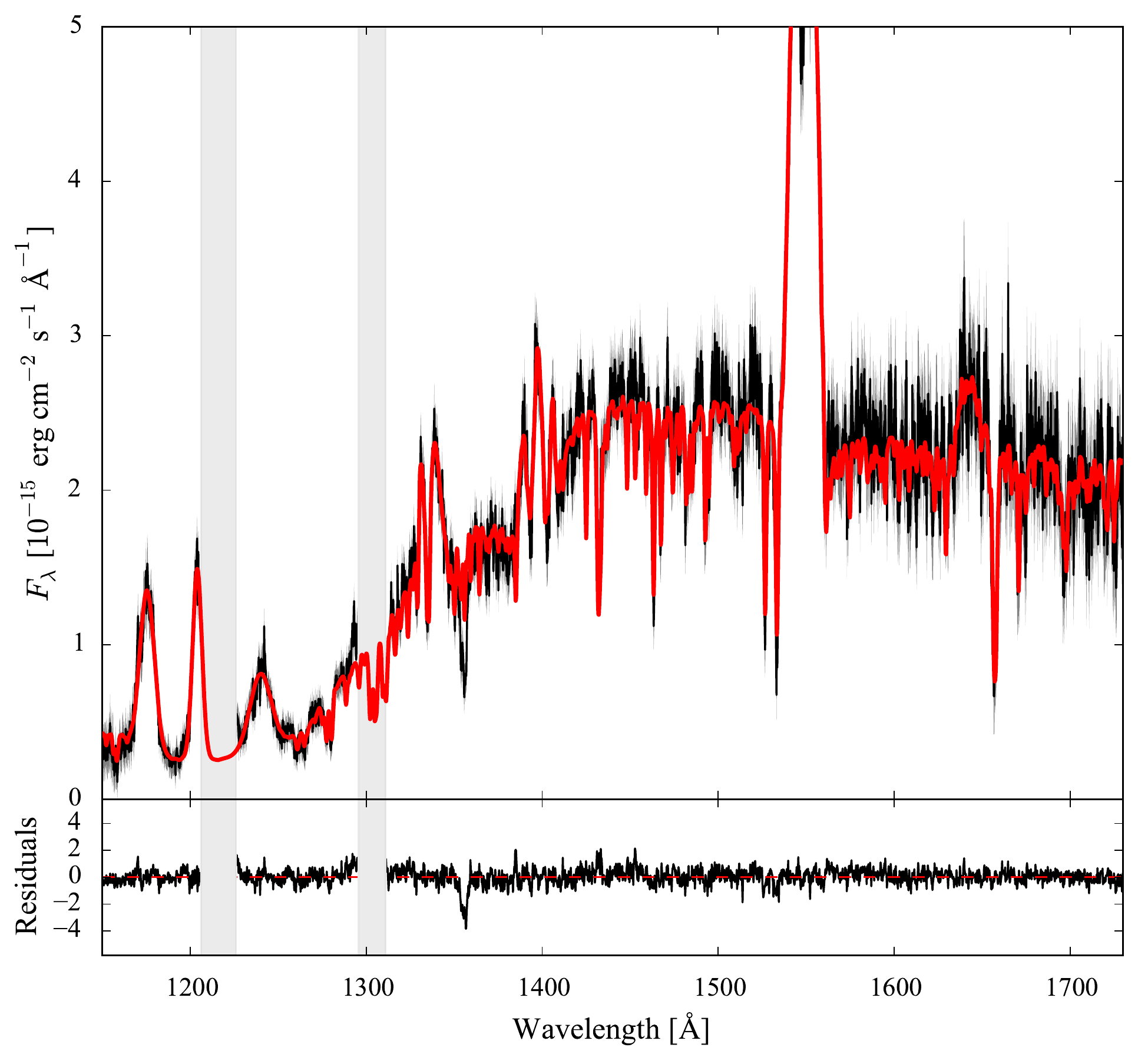}
\put (15,84) {V1108\,Her}
\end{overpic}
  \caption{Ultraviolet \hst spectra (black) of  V1108\,Her along with the best-fitting model (red), computed as described in Section~\ref{sec:uv_fit}.}
\end{figure*}
\newpage
\clearpage

\thispagestyle{empty}
\begin{table*}
  \centering
  \caption{Best-fitting parameters for ASAS\,J002511+1217.2.}
  \begin{tabular}{ l r }
    \toprule
     Instrument & COS \\ [0.13cm]
     Observation date & 2012 Nov 15 \\ [0.13cm]
     $E(B-V)$ (mag) & $0.009^{+0.016}_{-0.009}$ \\ [0.13cm]
     $P_\mathrm{orb}$ (min) & 82.0 \\ [0.13cm]
     $\mathit{Gaia}$ EDR3 ID & 2754909740118313344 \\ [0.13cm]
     $\varpi$ (mas) & $6.4\pm0.1$ \\ [0.13cm]
     $d$ (pc) & $157{\pm3}$ \\ [0.13cm]
     $Z$ (Z$_\mathrm{\odot}$) & 0.1 \\ [0.13cm]
     $T_\mathrm{eff}$ (K) & $13208^{+154}_{-112}$ \\ [0.13cm]
     $R_\mathrm{WD}$ ($0.01\,\mathrm{R}_\odot$) & $0.78 \pm 0.04$ \\ [0.13cm]
     $M_\mathrm{WD}$ ($\mathrm{M}_\odot$) & $1.02 \pm 0.04$ \\ [0.13cm]
     $\log(g)$ & $8.66 \pm 0.07$ \\ [0.13cm]
     $\langle \dot{M} \rangle$ $(10^{-10}\,\mathrm{M}_\odot$ yr$^\mathrm{-1}$) & $0.27^{+0.03}_{-0.02}$ \\ [0.13cm]
     WD contribution & 74\% \\ [0.13cm]
     \bottomrule
    \end{tabular}
\end{table*}
\vspace{0.8cm}
\begin{figure*}
    \centering
\begin{overpic}[width=0.8\textwidth]{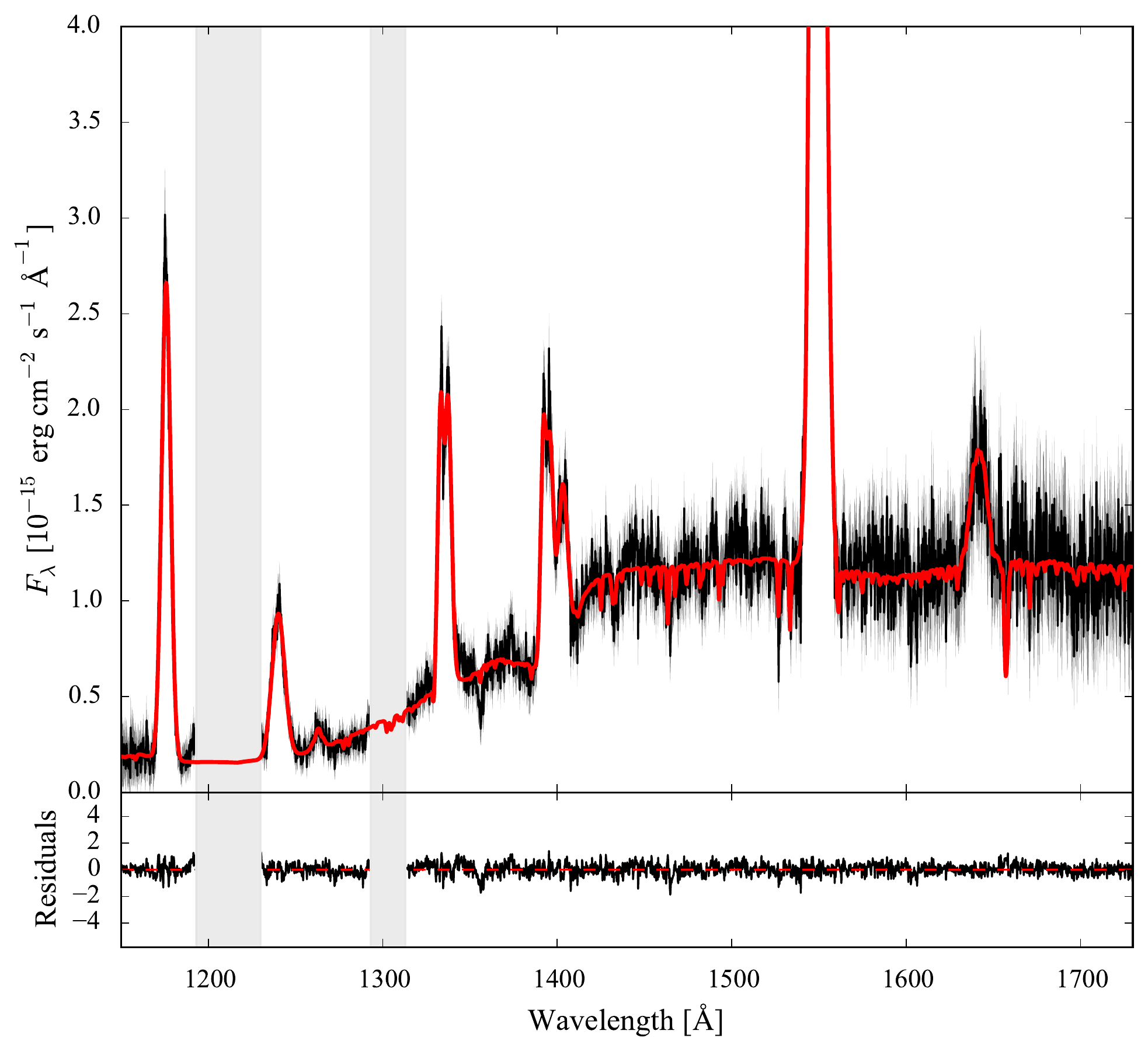}
\put (15,84) {ASAS\,J002511+1217.2}
\end{overpic}
  \caption{Ultraviolet \hst spectra (black) of  ASAS\,J002511+1217.2 along with the best-fitting model (red), computed as described in Section~\ref{sec:uv_fit}.}
\end{figure*}
\newpage
\clearpage

\thispagestyle{empty}
\begin{table*}
  \centering
  \caption{Best-fitting parameters for HV\,Vir.}
  \begin{tabular}{ l r }
    \toprule
     Instrument & STIS \\ [0.13cm]
     Observation date & 2000 Jun 10 \\ [0.13cm]
     $E(B-V)$ (mag) & $0.024^{+0.012}_{-0.021}$ \\ [0.13cm]
     $P_\mathrm{orb}$ (min) & 82.18 \\ [0.13cm]
     $\mathit{Gaia}$ EDR3 ID & 3688359000015020800 \\ [0.13cm]
     $\varpi$ (mas) & $3.2\pm0.3$ \\ [0.13cm]
     $d$ (pc) & $317^{+29}_{-25}$ \\ [0.13cm]
     $Z$ (Z$_\mathrm{\odot}$) & 0.2 \\ [0.13cm]
     $T_\mathrm{eff}$ (K) & $12958^{+257}_{-291}$ \\ [0.13cm]
     $R_\mathrm{WD}$ ($0.01\,\mathrm{R}_\odot$) & $0.96^{+0.21}_{-0.16}$ \\ [0.13cm]
     $M_\mathrm{WD}$ ($\mathrm{M}_\odot$) & $0.87^{+0.13}_{-0.17}$ \\ [0.13cm]
     $\log(g)$ & $8.4^{+0.2}_{-0.3}$ \\ [0.13cm]
     $\langle \dot{M} \rangle$ $(10^{-10}\,\mathrm{M}_\odot$ yr$^\mathrm{-1}$) & $0.4^{+0.2}_{-0.12}$ \\ [0.13cm]
     WD contribution & 93\% \\ [0.13cm]
     \bottomrule
    \end{tabular}
\end{table*}
\vspace{0.8cm}
\begin{figure*}
    \centering
\begin{overpic}[width=0.8\textwidth]{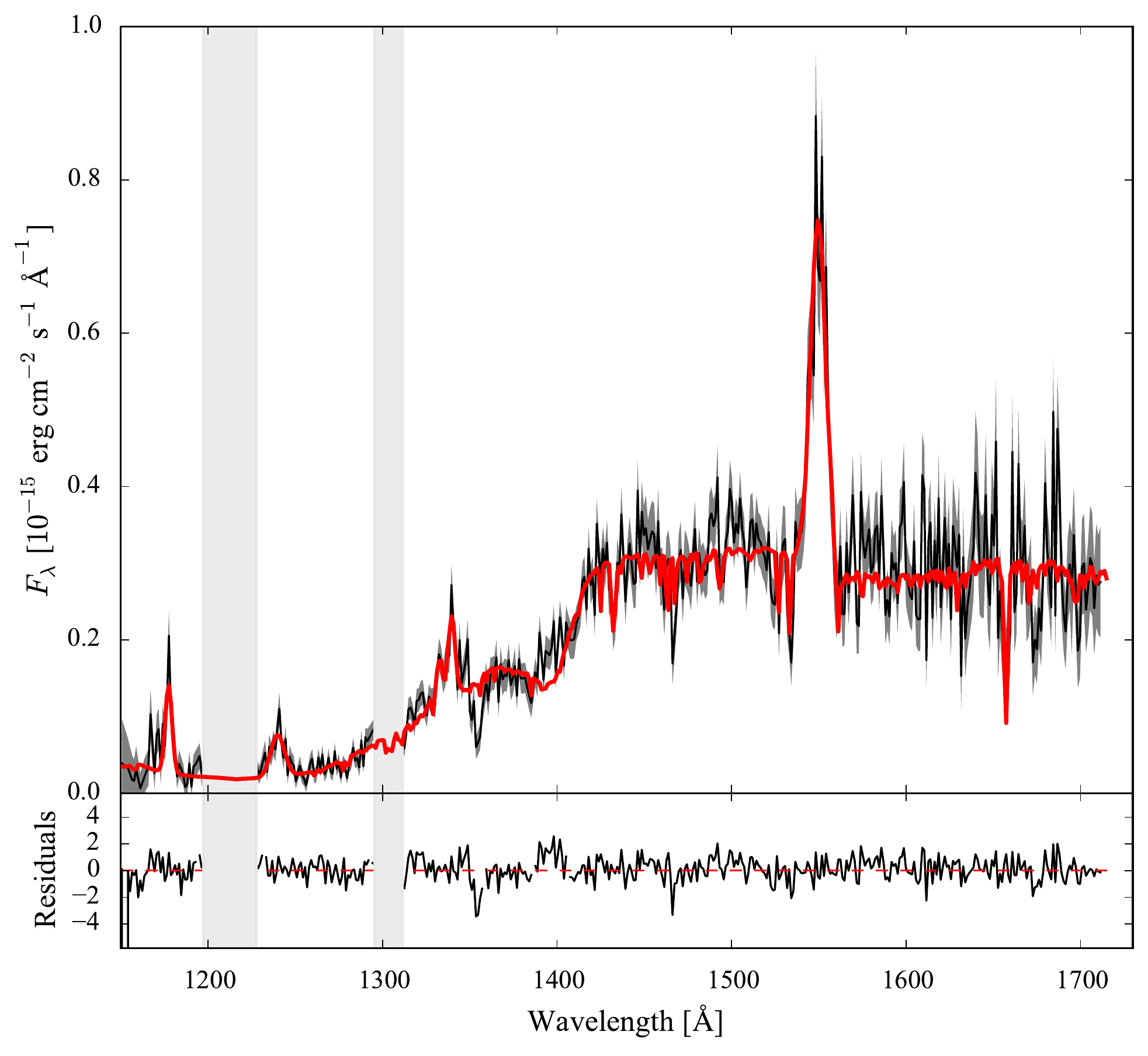}
\put (15,84) {HV\,Vir}
\end{overpic}
  \caption{Ultraviolet \hst spectra (black) of  HV\,Vir along with the best-fitting model (red), computed as described in Section~\ref{sec:uv_fit}.}
\end{figure*}
\newpage
\clearpage

\thispagestyle{empty}
\begin{table*}
  \centering
  \caption{Best-fitting parameters for SDSS\,J103533.02+055158.4.}
  \begin{tabular}{ l r }
    \toprule
     Instrument & COS \\ [0.13cm]
     Observation date & 2013 Mar 08 \\ [0.13cm]
     $E(B-V)$ (mag) & $0.017^{+0.018}_{-0.017}$ \\ [0.13cm]
     $P_\mathrm{orb}$ (min) & 82.22 \\ [0.13cm]
     $\mathit{Gaia}$ EDR3 ID & 3859020040917830400 \\ [0.13cm]
     $\varpi$ (mas) & $5.1\pm0.3$ \\ [0.13cm]
     $d$ (pc) & $195^{+12}_{-10}$ \\ [0.13cm]
     $Z$ (Z$_\mathrm{\odot}$) & 0.01 \\ [0.13cm]
     $T_\mathrm{eff}$ (K) & $11876^{+108}_{-115}$ \\ [0.13cm]
     $R_\mathrm{WD}$ ($0.01\,\mathrm{R}_\odot$) & $0.8^{+0.11}_{-0.09}$ \\ [0.13cm]
     $M_\mathrm{WD}$ ($\mathrm{M}_\odot$) & $1.00^{+0.08}_{-0.10}$ \\ [0.13cm]
     $\log(g)$ & $8.63^{+0.14}_{-0.16}$ \\ [0.13cm]
     $\langle \dot{M} \rangle$ $(10^{-10}\,\mathrm{M}_\odot$ yr$^\mathrm{-1}$) & $0.18^{+0.06}_{-0.04}$ \\ [0.13cm]
     WD contribution & 91\% \\ [0.13cm]
     \bottomrule
    \end{tabular}
\end{table*}
\vspace{0.8cm}
\begin{figure*}
    \centering
\begin{overpic}[width=0.8\textwidth]{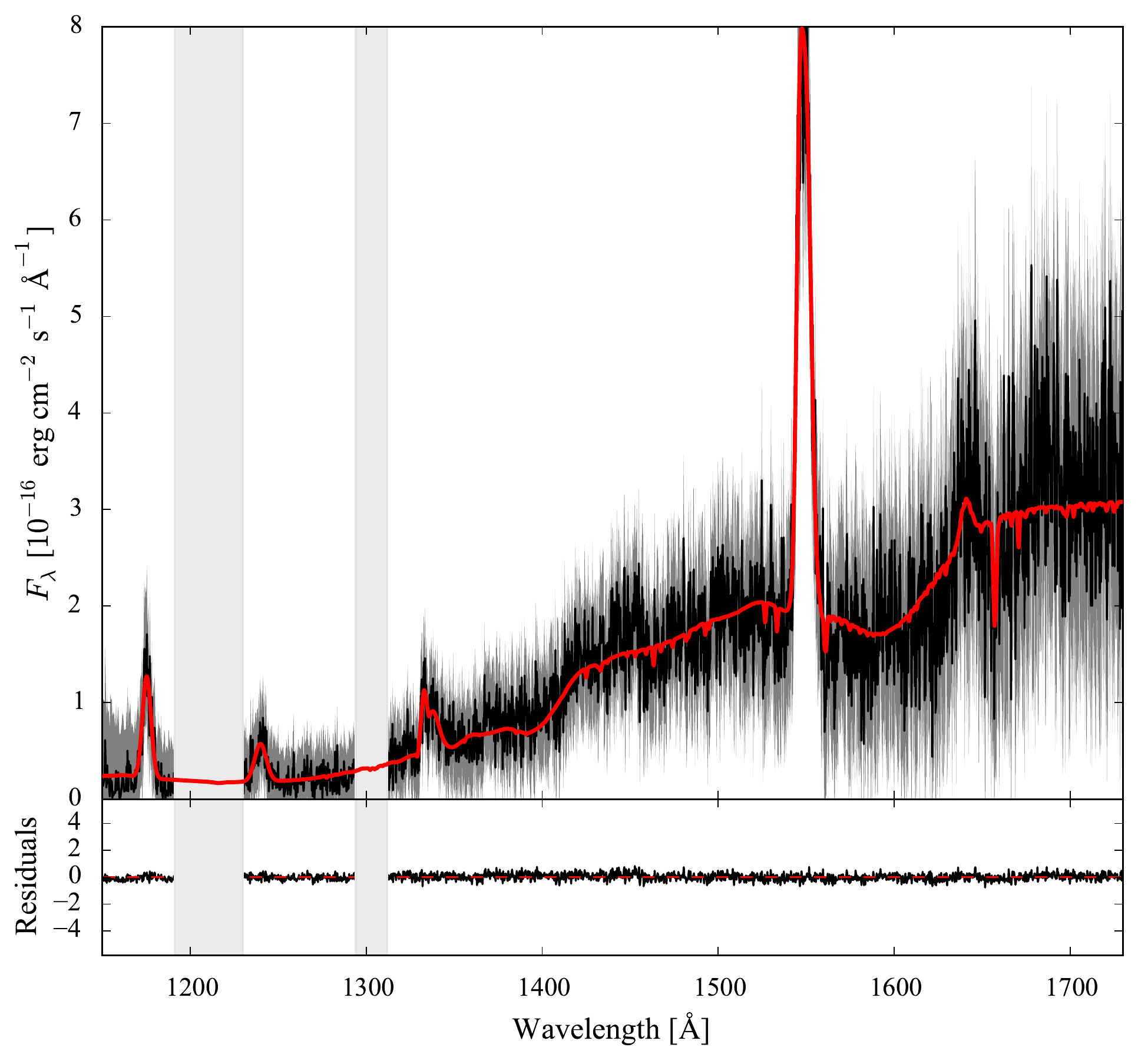}
\put (15,84) {SDSS\,J103533.02+055158.4}
\end{overpic}
  \caption{Ultraviolet \hst spectra (black) of  SDSS\,J103533.02+055158.4 along with the best-fitting model (red), computed as described in Section~\ref{sec:uv_fit}.}
\end{figure*}
\newpage
\clearpage

\thispagestyle{empty}
\begin{table*}
  \centering
  \caption{Best-fitting parameters for WX\,Cet.}
  \begin{tabular}{ l r }
    \toprule
     Instrument & STIS \\ [0.13cm]
     Observation date & 2000 Oct 30 \\ [0.13cm]
     $E(B-V)$ (mag) & $0.013^{+0.016}_{-0.013}$ \\ [0.13cm]
     $P_\mathrm{orb}$ (min) & 83.9 \\ [0.13cm]
     $\mathit{Gaia}$ EDR3 ID & 2355217815809560192 \\ [0.13cm]
     $\varpi$ (mas) & $3.97 \pm 0.13$ \\ [0.13cm]
     $d$ (pc) & $252^{+9}_{-8}$ \\ [0.13cm]
     $Z$ (Z$_\mathrm{\odot}$) & 0.1 \\ [0.13cm]
     $T_\mathrm{eff}$ (K) & $15186^{+1460}_{-1468}$ \\ [0.13cm]
     $R_\mathrm{WD}$ ($0.01\,\mathrm{R}_\odot$) & $0.63^{+0.28}_{-0.18}$ \\ [0.13cm]
     $M_\mathrm{WD}$ ($\mathrm{M}_\odot$) & $1.10^{+0.17}_{-0.21}$ \\ [0.13cm]
     $\log(g)$ & $8.8 \pm 0.4$ \\ [0.13cm]
     $\langle \dot{M} \rangle$ $(10^{-10}\,\mathrm{M}_\odot$ yr$^\mathrm{-1}$) & $0.3^{+0.07}_{-0.06}$ \\ [0.13cm]
     WD contribution & 24\% \\ [0.13cm]
     \bottomrule
    \end{tabular}
\end{table*}
\vspace{0.8cm}
\begin{figure*}
    \centering
\begin{overpic}[width=0.8\textwidth]{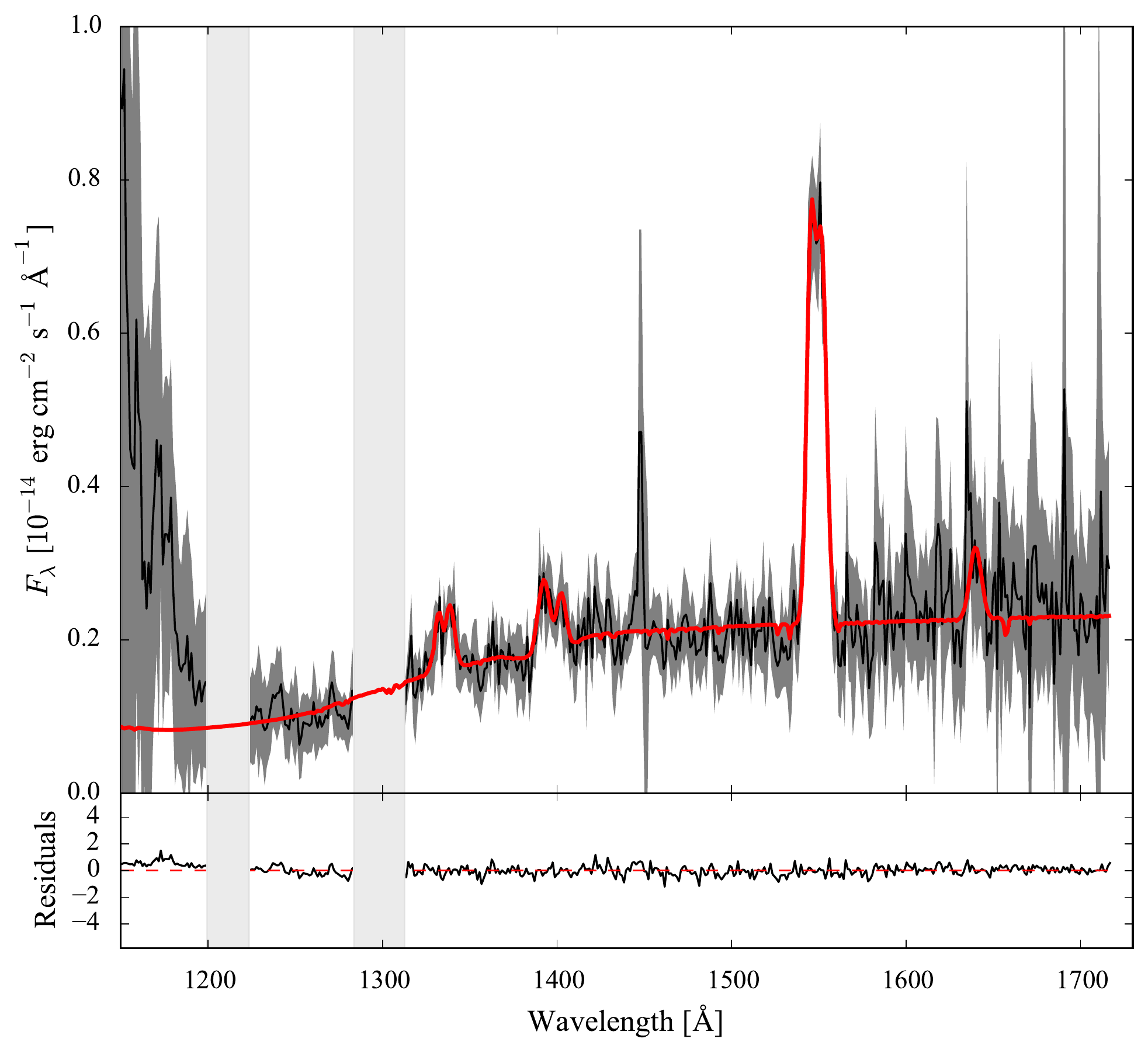}
\put (15,84) {WX\,Cet}
\end{overpic}
  \caption{Ultraviolet \hst spectra (black) of  WX\,Cet along with the best-fitting model (red), computed as described in Section~\ref{sec:uv_fit}.}
\end{figure*}
\newpage
\clearpage

\thispagestyle{empty}
\begin{table*}
  \centering
  \caption{Best-fitting parameters for SDSS\,J075507.70+143547.6.}
  \begin{tabular}{ l r }
    \toprule
     Instrument & COS \\ [0.13cm]
     Observation date & 2012 Dec 14 \\ [0.13cm]
     $E(B-V)$ (mag) & $0.011^{+0.015}_{-0.011}$ \\ [0.13cm]
     $P_\mathrm{orb}$ (min) & 84.76 \\ [0.13cm]
     $\mathit{Gaia}$ EDR3 ID & 654539826068054400 \\ [0.13cm]
     $\varpi$ (mas) & $4.2\pm0.2$ \\ [0.13cm]
     $d$ (pc) & $239^{+12}_{-11}$ \\ [0.13cm]
     $Z$ (Z$_\mathrm{\odot}$) & 0.5 \\ [0.13cm]
     $T_\mathrm{eff}$ (K) & $16193^{+280}_{-357}$ \\ [0.13cm]
     $R_\mathrm{WD}$ ($0.01\,\mathrm{R}_\odot$) & $0.92^{+0.15}_{-0.11}$ \\ [0.13cm]
     $M_\mathrm{WD}$ ($\mathrm{M}_\odot$) & $0.91^{+0.09}_{-0.12}$ \\ [0.13cm]
     $\log(g)$ & $8.47^{+0.15}_{-0.19}$ \\ [0.13cm]
     $\langle \dot{M} \rangle$ $(10^{-10}\,\mathrm{M}_\odot$ yr$^\mathrm{-1}$) & $0.87^{+0.27}_{-0.18}$ \\ [0.13cm]
     WD contribution & 87\% \\ [0.13cm]
     \bottomrule
    \end{tabular}
\end{table*}
\vspace{0.8cm}
\begin{figure*}
    \centering
\begin{overpic}[width=0.8\textwidth]{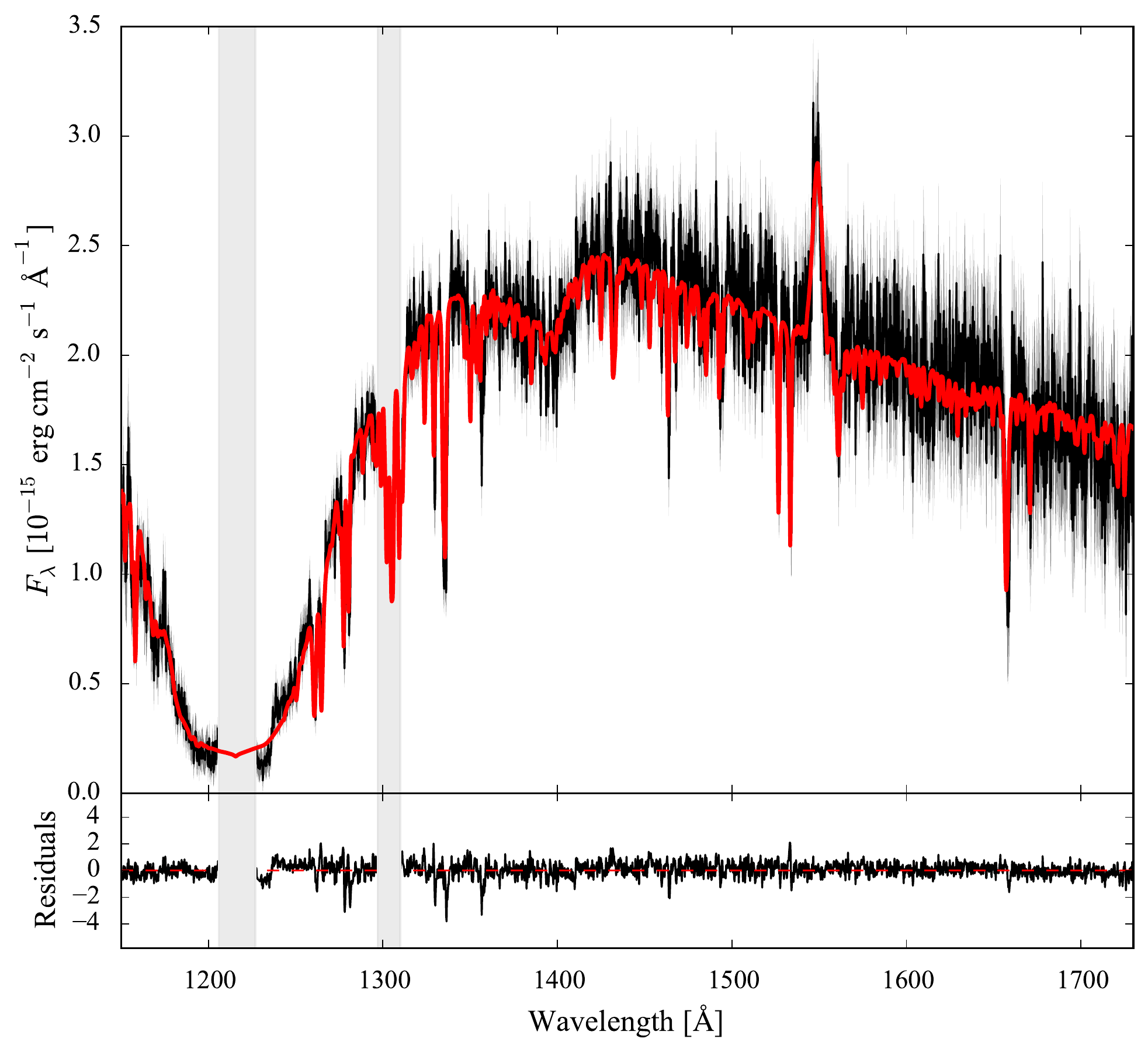}
\put (15,84) {SDSS\,J075507.70+143547.6}
\end{overpic}
  \caption{Ultraviolet \hst spectra (black) of  SDSS\,J075507.70+143547.6 along with the best-fitting model (red), computed as described in Section~\ref{sec:uv_fit}.}
\end{figure*}
\newpage
\clearpage

\thispagestyle{empty}
\begin{table*}
  \centering
  \caption{Best-fitting parameters for SDSS\,J080434.20+510349.2.}
  \begin{tabular}{ l r }
    \toprule
     Instrument & COS \\ [0.13cm]
     Observation date & 2011 Nov 03 \\ [0.13cm]
     $E(B-V)$ (mag) & $0.007^{+0.016}_{-0.007}$ \\ [0.13cm]
     $P_\mathrm{orb}$ (min) & 84.97 \\ [0.13cm]
     $\mathit{Gaia}$ EDR3 ID & 935056333580267392 \\ [0.13cm]
     $\varpi$ (mas) & $7.03\pm0.11$ \\ [0.13cm]
     $d$ (pc) & $142{\pm2}$ \\ [0.13cm]
     $Z$ (Z$_\mathrm{\odot}$) & 0.5 \\ [0.13cm]
     $T_\mathrm{eff}$ (K) & $13715^{+55}_{-79}$ \\ [0.13cm]
     $R_\mathrm{WD}$ ($0.01\,\mathrm{R}_\odot$) & $0.80^{+0.04}_{-0.03}$ \\ [0.13cm]
     $M_\mathrm{WD}$ ($\mathrm{M}_\odot$) & $1.01^{+0.03}_{-0.04}$ \\ [0.13cm]
     $\log(g)$ & $8.64^{+0.05}_{-0.06}$ \\ [0.13cm]
     $\langle \dot{M} \rangle$ $(10^{-10}\,\mathrm{M}_\odot$ yr$^\mathrm{-1}$) & $0.32 \pm 0.03$ \\ [0.13cm]
     WD contribution & 87\% \\ [0.13cm]
     \bottomrule
    \end{tabular}
\end{table*}
\vspace{0.8cm}
\begin{figure*}
    \centering
\begin{overpic}[width=0.8\textwidth]{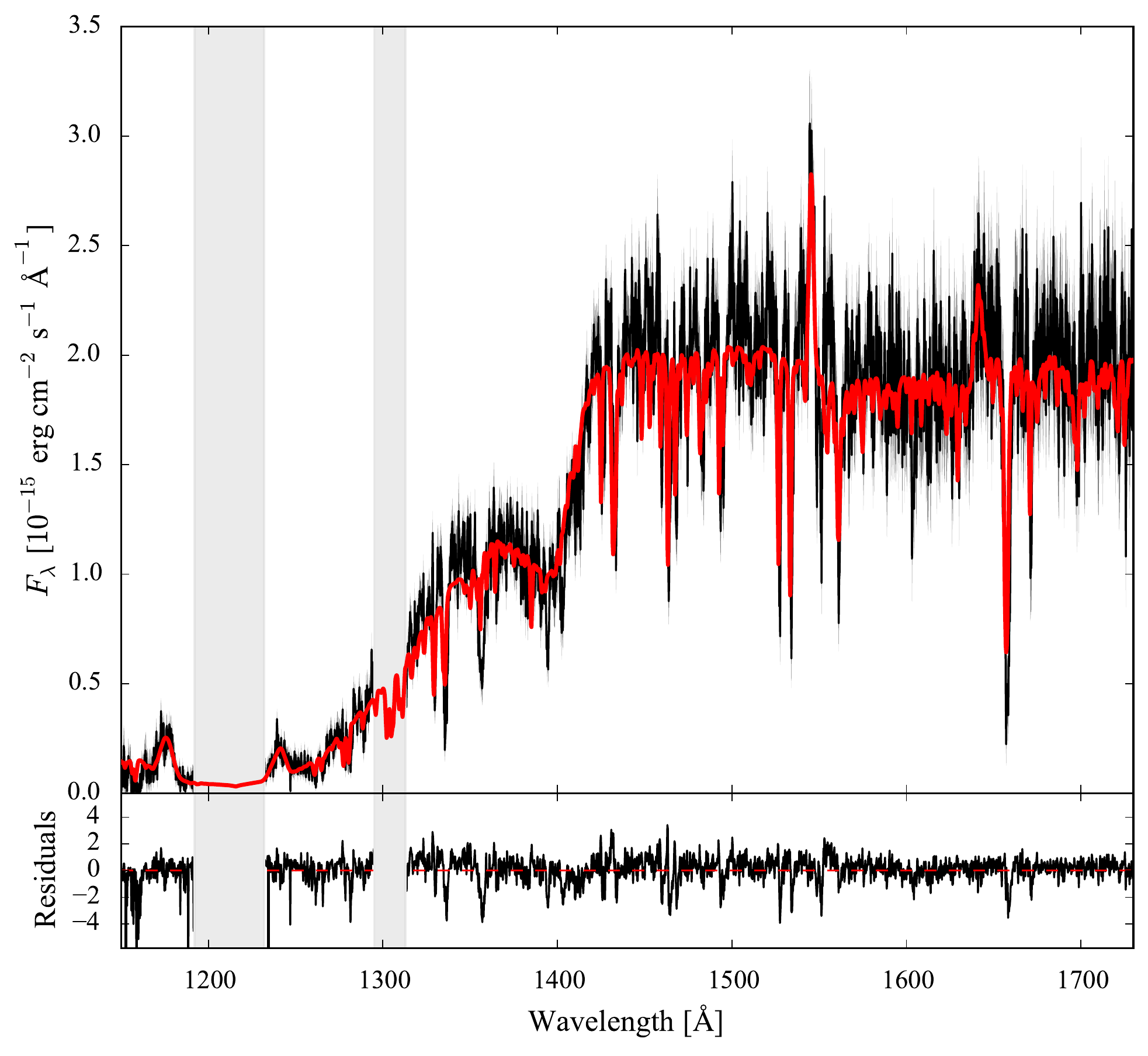}
\put (15,84) {SDSS\,J080434.20+510349.2}
\end{overpic}
  \caption{Ultraviolet \hst spectra (black) of  SDSS\,J080434.20+510349.2 along with the best-fitting model (red), computed as described in Section~\ref{sec:uv_fit}.}
\end{figure*}
\newpage
\clearpage

\thispagestyle{empty}
\begin{table*}
  \centering
  \caption{Best-fitting parameters for EG\,Cnc.}
  \begin{tabular}{ l r }
    \toprule
     Instrument & STIS \\ [0.13cm]
     Observation date & 2006 Dec 20 \\ [0.13cm]
     $E(B-V)$ (mag) & $0.008^{+0.017}_{-0.008}$ \\ [0.13cm]
     $P_\mathrm{orb}$ (min) & 86.36 \\ [0.13cm]
     $\mathit{Gaia}$ EDR3 ID & 703580960947960576 \\ [0.13cm]
     $\varpi$ (mas) & $5.4\pm0.2$ \\ [0.13cm]
     $d$ (pc) & $186\pm7$ \\ [0.13cm]
     $Z$ (Z$_\mathrm{\odot}$) & 0.2 \\ [0.13cm]
     $T_\mathrm{eff}$ (K) & $12295^{+56}_{-57}$ \\ [0.13cm]
     $R_\mathrm{WD}$ ($0.01\,\mathrm{R}_\odot$) & $0.77^{+0.06}_{-0.05}$ \\ [0.13cm]
     $M_\mathrm{WD}$ ($\mathrm{M}_\odot$) & $1.03^{+0.04}_{-0.05}$ \\ [0.13cm]
     $\log(g)$ & $8.67^{+0.07}_{-0.08}$ \\ [0.13cm]
     $\langle \dot{M} \rangle$ $(10^{-10}\,\mathrm{M}_\odot$ yr$^\mathrm{-1}$) & $0.2^{+0.03}_{-0.02}$ \\ [0.13cm]
     WD contribution & 100\% \\ [0.13cm]
     \bottomrule
    \end{tabular}
\end{table*}
\vspace{0.8cm}
\begin{figure*}
    \centering
\begin{overpic}[width=0.8\textwidth]{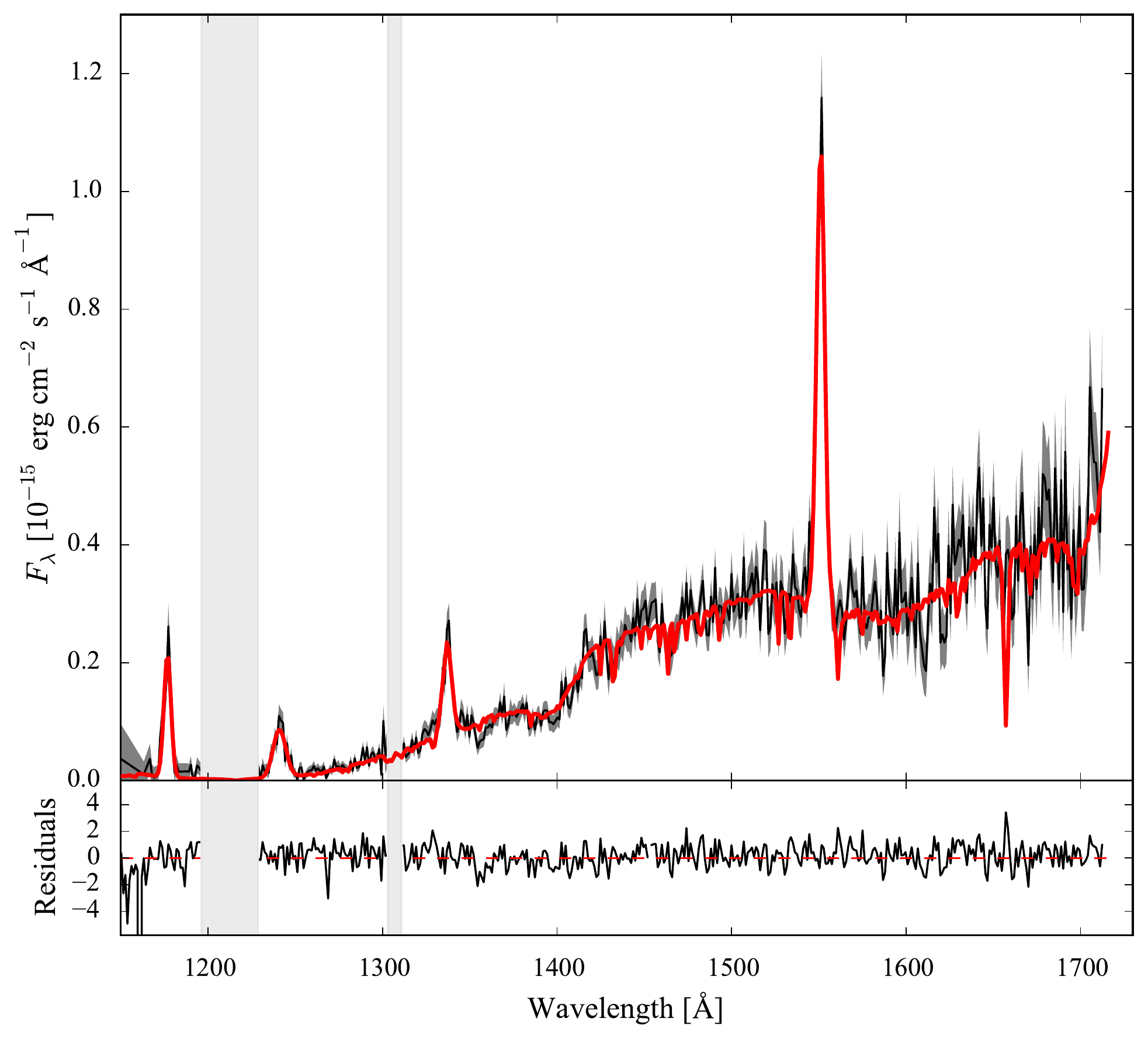}
\put (15,84) {EG\,Cnc}
\end{overpic}
  \caption{Ultraviolet \hst spectra (black) of  EG\,Cnc along with the best-fitting model (red), computed as described in Section~\ref{sec:uv_fit}.}
\end{figure*}
\newpage
\clearpage

\thispagestyle{empty}
\begin{table*}
  \centering
  \caption{Best-fitting parameters for EK\,TrA.}
  \begin{tabular}{ l r }
    \toprule
     Instrument & STIS \\ [0.13cm]
     Observation date & 1999 Jul 25 \\ [0.13cm]
     $E(B-V)$ (mag) & $0.034\pm0.019$ \\ [0.13cm]
     $P_\mathrm{orb}$ (min) & 86.36 \\ [0.13cm]
     $\mathit{Gaia}$ EDR3 ID & 5825198967486003072 \\ [0.13cm]
     $\varpi$ (mas) & $6.61\pm0.04$ \\ [0.13cm]
     $d$ (pc) & $151.4{\pm0.8}$ \\ [0.13cm]
     $Z$ (Z$_\mathrm{\odot}$) & 0.5 \\ [0.13cm]
     $T_\mathrm{eff}$ (K) & $17608^{+269}_{-481}$ \\ [0.13cm]
     $R_\mathrm{WD}$ ($0.01\,\mathrm{R}_\odot$) & $0.97^{+0.17}_{-0.11}$ \\ [0.13cm]
     $M_\mathrm{WD}$ ($\mathrm{M}_\odot$) & $0.87^{+0.09}_{-0.13}$ \\ [0.13cm]
     $\log(g)$ & $8.4^{+0.14}_{-0.21}$ \\ [0.13cm]
     $\langle \dot{M} \rangle$ $(10^{-10}\,\mathrm{M}_\odot$ yr$^\mathrm{-1}$) & $1.4^{+0.4}_{-0.3}$ \\ [0.13cm]
     WD contribution & 58\% \\ [0.13cm]
     \bottomrule
    \end{tabular}
\end{table*}
\vspace{0.8cm}
\begin{figure*}
    \centering
\begin{overpic}[width=0.8\textwidth]{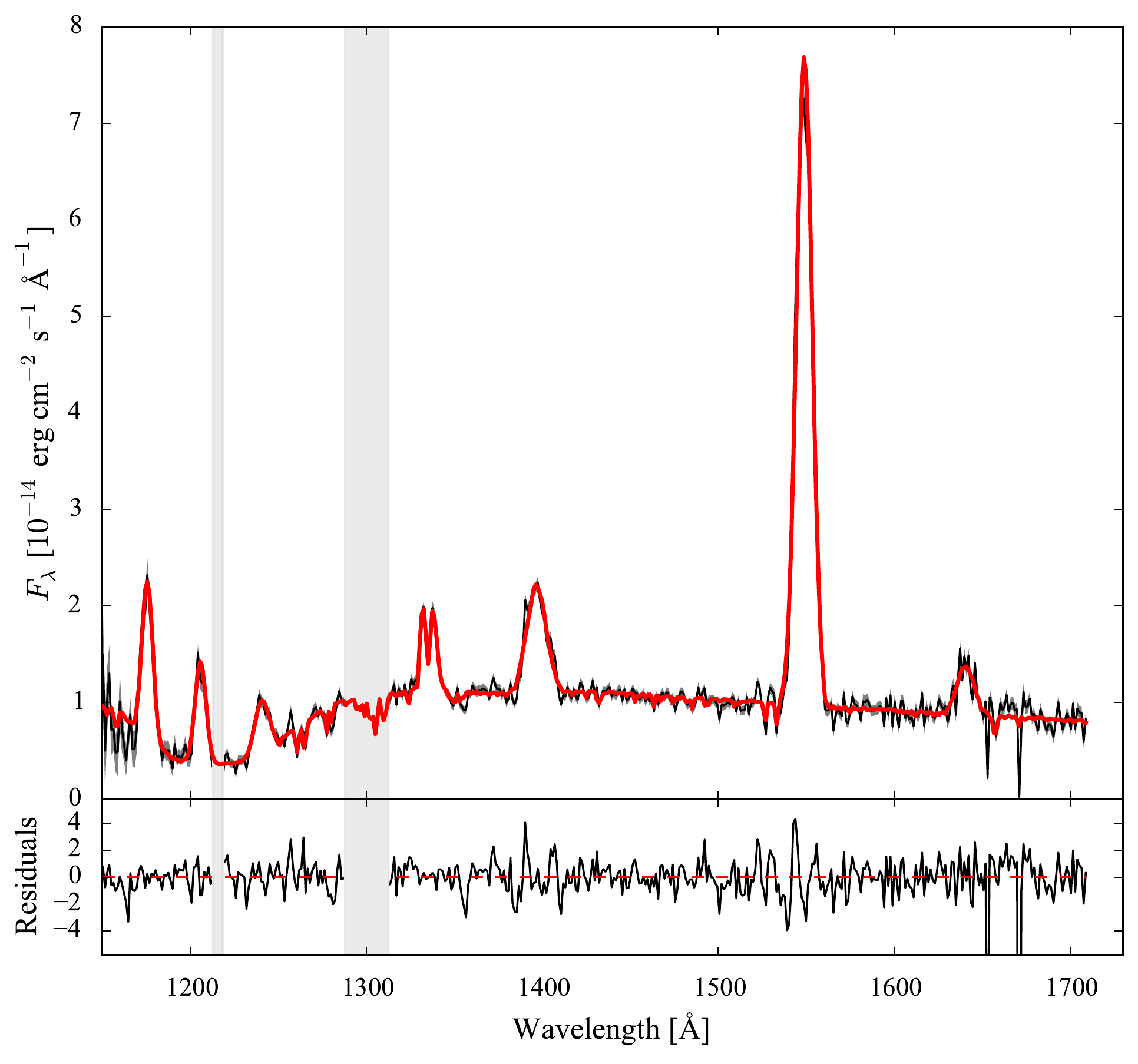}
\put (15,84) {EK\,TrA}
\end{overpic}
  \caption{Ultraviolet \hst spectra (black) of  EK\,TrA along with the best-fitting model (red), computed as described in Section~\ref{sec:uv_fit}.}
\end{figure*}
\newpage
\clearpage

\thispagestyle{empty}
\begin{table*}
  \centering
  \caption{Best-fitting parameters for 1RXS\,J105010.8--140431.}
  \begin{tabular}{ l r }
    \toprule
     Instrument & COS \\ [0.13cm]
     Observation date & 2013 May 10 \\ [0.13cm]
     $E(B-V)$ (mag) & $0.005^{+0.015}_{-0.005}$ \\ [0.13cm]
     $P_\mathrm{orb}$ (min) & 88.56 \\ [0.13cm]
     $\mathit{Gaia}$ EDR3 ID & 3750072904055666176 \\ [0.13cm]
     $\varpi$ (mas) & $9.19\pm0.09$ \\ [0.13cm]
     $d$ (pc) & $109{\pm1}$ \\ [0.13cm]
     $Z$ (Z$_\mathrm{\odot}$) & 0.1 \\ [0.13cm]
     $T_\mathrm{eff}$ (K) & $11523^{+29}_{-47}$ \\ [0.13cm]
     $R_\mathrm{WD}$ ($0.01\,\mathrm{R}_\odot$) & $1.08^{+0.04}_{-0.03}$ \\ [0.13cm]
     $M_\mathrm{WD}$ ($\mathrm{M}_\odot$) & $0.77^{+0.02}_{-0.03}$ \\ [0.13cm]
     $\log(g)$ & $8.25^{+0.04}_{-0.05}$ \\ [0.13cm]
     $\langle \dot{M} \rangle$ $(10^{-10}\,\mathrm{M}_\odot$ yr$^\mathrm{-1}$) & $0.332^{+0.028}_{-0.018}$ \\ [0.13cm]
     WD contribution & 87\% \\ [0.13cm]
     \bottomrule
    \end{tabular}
\end{table*}
\vspace{0.8cm}
\begin{figure*}
    \centering
\begin{overpic}[width=0.8\textwidth]{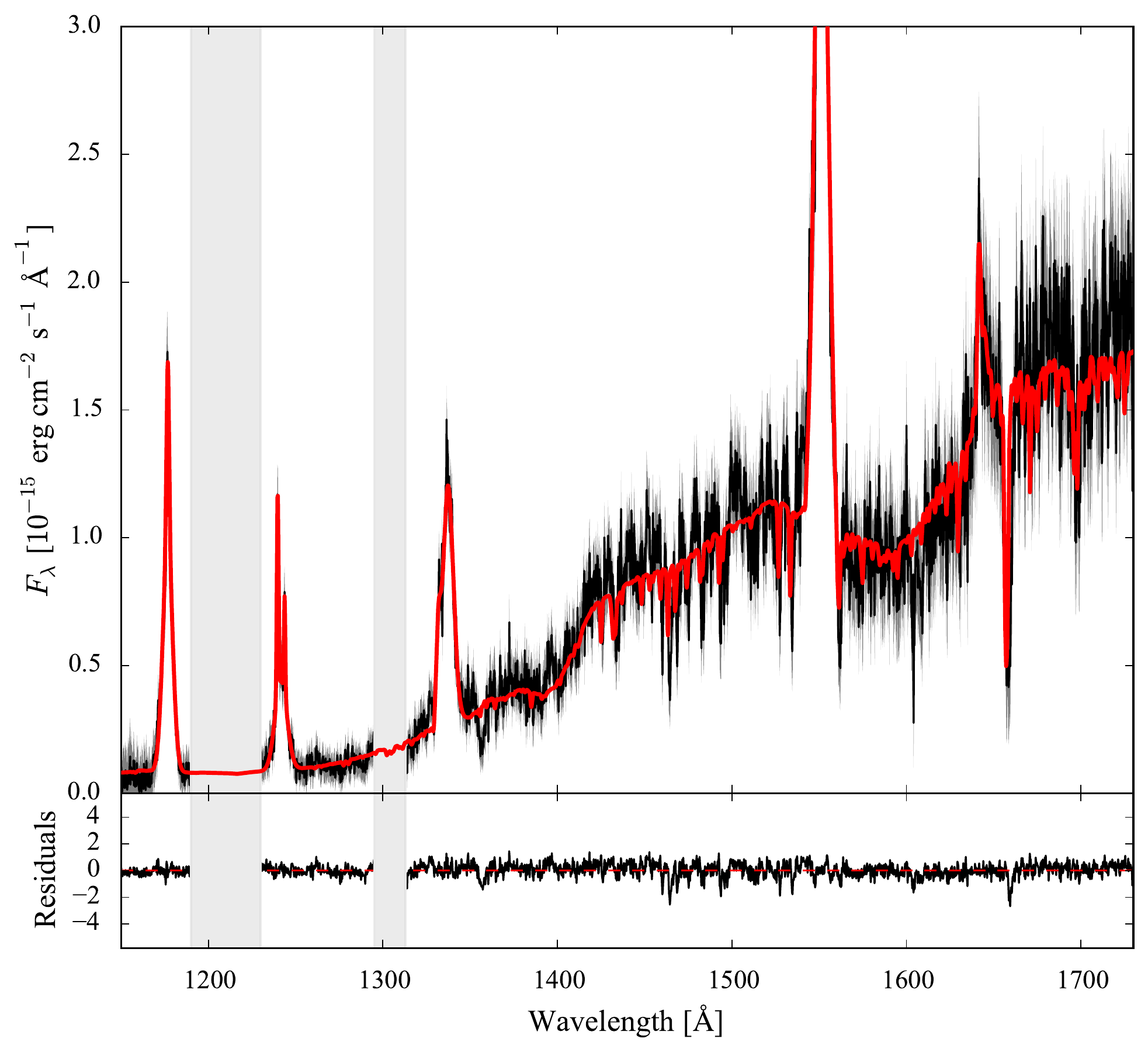}
\put (15,84) {1RXS\,J105010.8--140431}
\end{overpic}
  \caption{Ultraviolet \hst spectra (black) of  1RXS\,J105010.8--140431 along with the best-fitting model (red), computed as described in Section~\ref{sec:uv_fit}.}
\end{figure*}
\newpage
\clearpage

\thispagestyle{empty}
\begin{table*}
  \centering
  \caption{Best-fitting parameters for BC\,UMa.}
  \begin{tabular}{ l r }
    \toprule
     Instrument & STIS \\ [0.13cm]
     Observation date & 2000 Jul 18 \\ [0.13cm]
     $E(B-V)$ (mag) & $0.017^{+0}_{-0.017}$ \\ [0.13cm]
     $P_\mathrm{orb}$ (min) & 90.16 \\ [0.13cm]
     $\mathit{Gaia}$ EDR3 ID & 787683052032971904 \\ [0.13cm]
     $\varpi$ (mas) & $3.41\pm0.13$ \\ [0.13cm]
     $d$ (pc) & $293^{+11}_{-10}$ \\ [0.13cm]
     $Z$ (Z$_\mathrm{\odot}$) & 0.2 \\ [0.13cm]
     $T_\mathrm{eff}$ (K) & $14378^{+272}_{-327}$ \\ [0.13cm]
     $R_\mathrm{WD}$ ($0.01\,\mathrm{R}_\odot$) & $1.57^{+0.19}_{-0.16}$ \\ [0.13cm]
     $M_\mathrm{WD}$ ($\mathrm{M}_\odot$) & $0.48^{+0.08}_{-0.09}$ \\ [0.13cm]
     $\log(g)$ & $7.73^{+0.15}_{-0.19}$ \\ [0.13cm]
     $\langle \dot{M} \rangle$ $(10^{-10}\,\mathrm{M}_\odot$ yr$^\mathrm{-1}$) & $2.0^{+0.5}_{-0.3}$ \\ [0.13cm]
     WD contribution & 85\% \\ [0.13cm]
     \bottomrule
    \end{tabular}
\end{table*}
\vspace{0.8cm}
\begin{figure*}
    \centering
\begin{overpic}[width=0.8\textwidth]{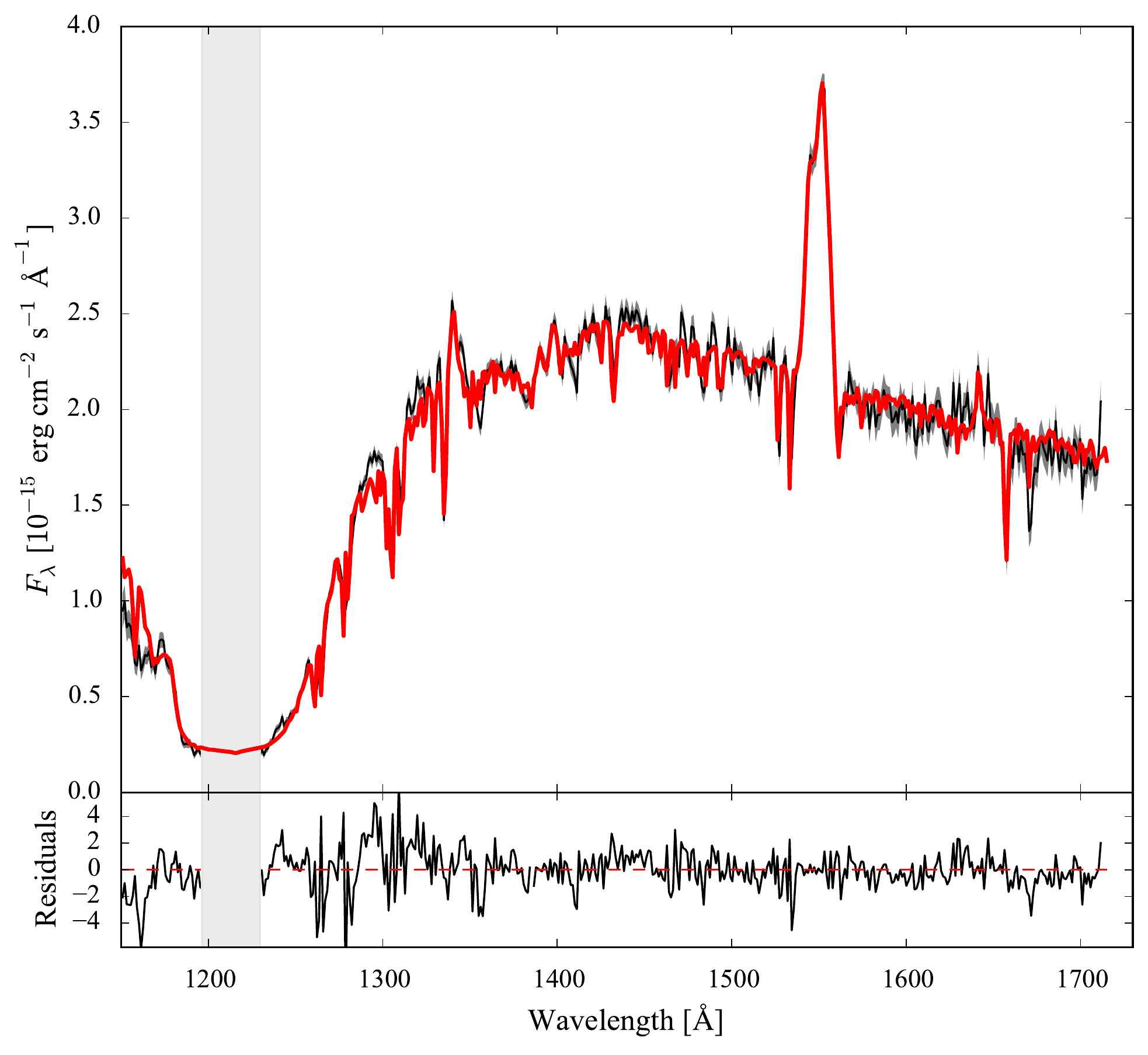}
\put (15,84) {BC\,UMa}
\end{overpic}
  \caption{Ultraviolet \hst spectra (black) of  BC\,UMa along with the best-fitting model (red), computed as described in Section~\ref{sec:uv_fit}.}
\end{figure*}
\newpage
\clearpage

\thispagestyle{empty}
\begin{table*}
  \centering
  \caption{Best-fitting parameters for VY\,Aqr.}
  \begin{tabular}{ l r }
    \toprule
     Instrument & STIS \\ [0.13cm]
     Observation date & 2000 Jul 10 \\ [0.13cm]
     $E(B-V)$ (mag) & $0.008^{+0.016}_{-0.008}$ \\ [0.13cm]
     $P_\mathrm{orb}$ (min) & 90.85 \\ [0.13cm]
     $\mathit{Gaia}$ EDR3 ID & 6896767366186700416 \\ [0.13cm]
     $\varpi$ (mas) & $7.08\pm0.09$ \\ [0.13cm]
     $d$ (pc) & $141.3^{+1.9}_{-1.8}$ \\ [0.13cm]
     $Z$ (Z$_\mathrm{\odot}$) & 0.5 \\ [0.13cm]
     $T_\mathrm{eff}$ (K) & $14453^{+316}_{-366}$ \\ [0.13cm]
     $R_\mathrm{WD}$ ($0.01\,\mathrm{R}_\odot$) & $0.74^{+0.07}_{-0.06}$ \\ [0.13cm]
     $M_\mathrm{WD}$ ($\mathrm{M}_\odot$) & $1.06 \pm 0.06$ \\ [0.13cm]
     $\log(g)$ & $8.73^{+0.12}_{-0.11}$ \\ [0.13cm]
     $\langle \dot{M} \rangle$ $(10^{-10}\,\mathrm{M}_\odot$ yr$^\mathrm{-1}$) & $0.33^{+0.04}_{-0.03}$ \\ [0.13cm]
     WD contribution & 45\% \\ [0.13cm]
     \bottomrule
    \end{tabular}
\end{table*}
\vspace{0.8cm}
\begin{figure*}
    \centering
\begin{overpic}[width=0.8\textwidth]{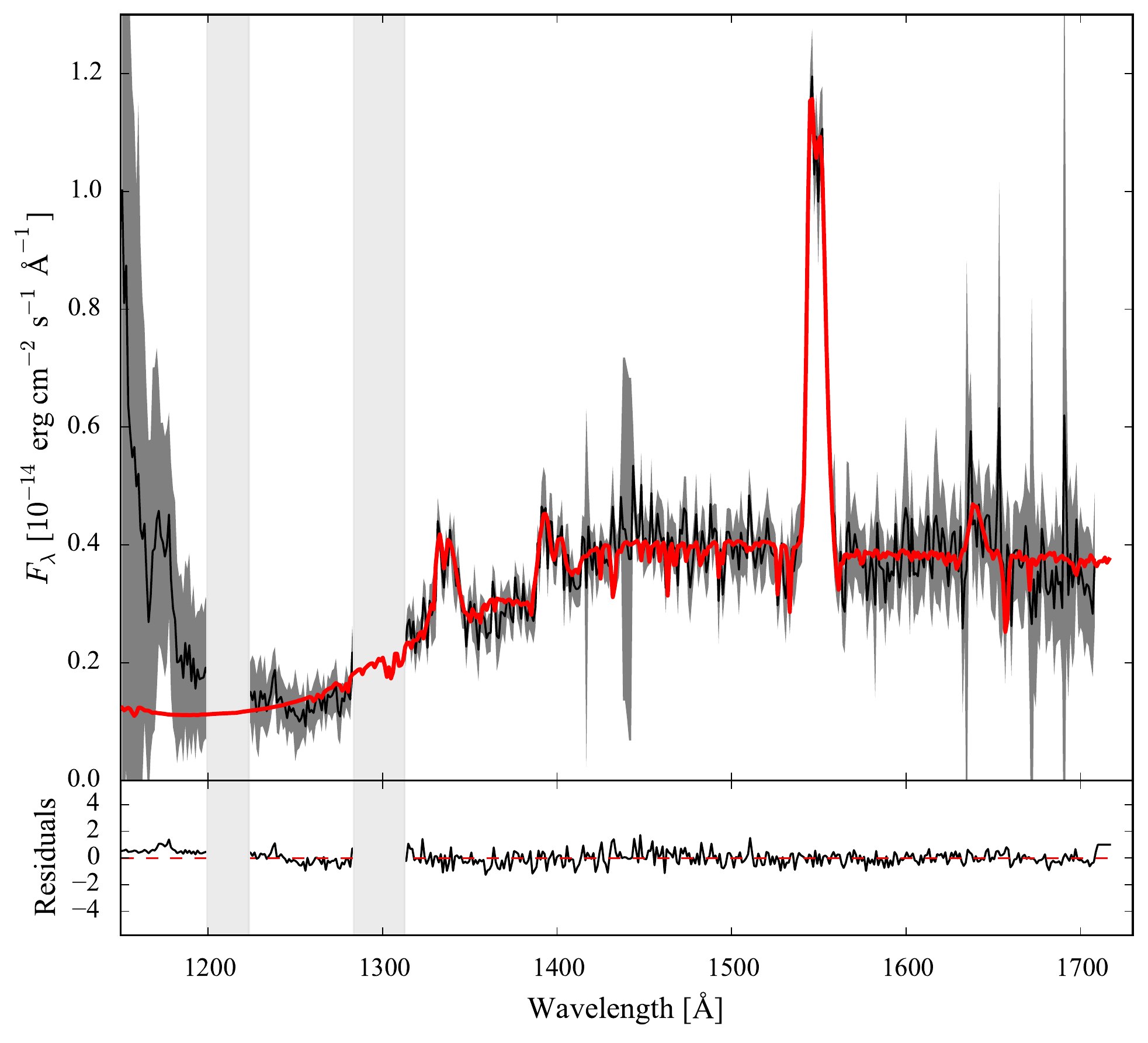}
\put (15,84) {VY\,Aqr}
\end{overpic}
  \caption{Ultraviolet \hst spectra (black) of  VY\,Aqr along with the best-fitting model (red), computed as described in Section~\ref{sec:uv_fit}.}
\end{figure*}
\newpage
\clearpage

\thispagestyle{empty}
\begin{table*}
  \centering
  \caption{Best-fitting parameters for QZ\,Lib.}
  \begin{tabular}{ l r }
    \toprule
     Instrument & COS \\ [0.13cm]
     Observation date & 2013 Apr 26 \\ [0.13cm]
     $E(B-V)$ (mag) & $0.10\pm0.03$ \\ [0.13cm]
     $P_\mathrm{orb}$ (min) & 92.36 \\ [0.13cm]
     $\mathit{Gaia}$ EDR3 ID & 6318149711371454464 \\ [0.13cm]
     $\varpi$ (mas) & $5.0\pm0.2$ \\ [0.13cm]
     $d$ (pc) & $199^{+11}_{-10}$ \\ [0.13cm]
     $Z$ (Z$_\mathrm{\odot}$) & 0.01 \\ [0.13cm]
     $T_\mathrm{eff}$ (K) & $11419^{+175}_{-229}$ \\ [0.13cm]
     $R_\mathrm{WD}$ ($0.01\,\mathrm{R}_\odot$) & $1.01^{+0.23}_{-0.18}$ \\ [0.13cm]
     $M_\mathrm{WD}$ ($\mathrm{M}_\odot$) & $0.82^{+0.14}_{-0.19}$ \\ [0.13cm]
     $\log(g)$ & $8.3^{+0.2}_{-0.3}$ \\ [0.13cm]
     $\langle \dot{M} \rangle$ $(10^{-10}\,\mathrm{M}_\odot$ yr$^\mathrm{-1}$) & $0.27^{+0.15}_{-0.09}$ \\ [0.13cm]
     WD contribution & 74\% \\ [0.13cm]
     \bottomrule
    \end{tabular}
\end{table*}
\vspace{0.8cm}
\begin{figure*}
    \centering
\begin{overpic}[width=0.8\textwidth]{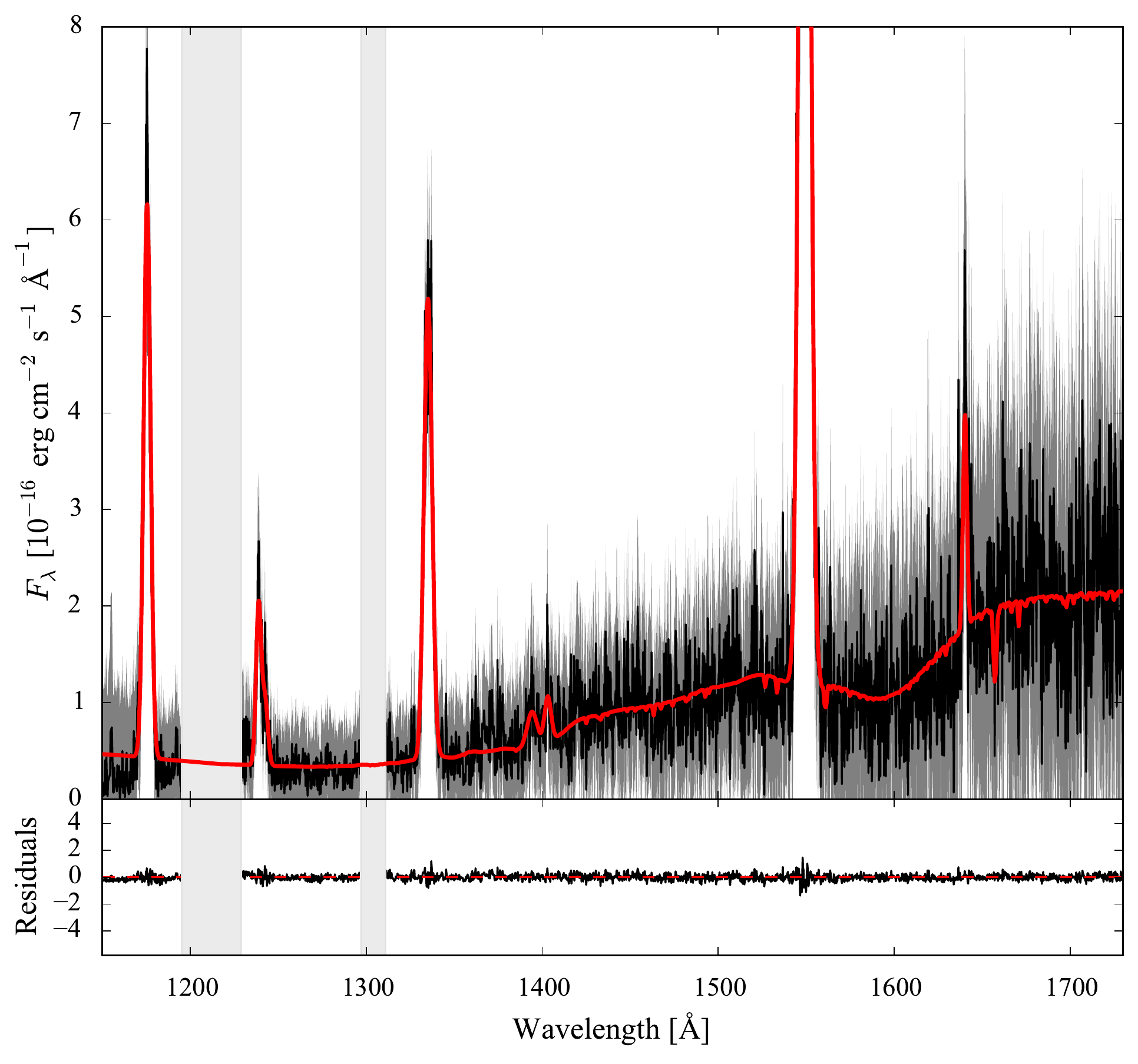}
\put (15,84) {QZ\,Lib}
\end{overpic}
  \caption{Ultraviolet \hst spectra (black) of  QZ\,Lib along with the best-fitting model (red), computed as described in Section~\ref{sec:uv_fit}.}
\end{figure*}
\newpage
\clearpage

\thispagestyle{empty}
\begin{table*}
  \centering
  \caption{Best-fitting parameters for SDSS\,J153817.35+512338.0.}
  \begin{tabular}{ l r }
    \toprule
     Instrument & COS \\ [0.13cm]
     Observation date & 2013 May 16 \\ [0.13cm]
     $E(B-V)$ (mag) & $0.027\pm0.018$ \\ [0.13cm]
     $P_\mathrm{orb}$ (min) & 93.11 \\ [0.13cm]
     $\mathit{Gaia}$ EDR3 ID & 1595085299649674240 \\ [0.13cm]
     $\varpi$ (mas) & $1.64\pm0.12$ \\ [0.13cm]
     $d$ (pc) & $607^{+47}_{-40}$ \\ [0.13cm]
     $Z$ (Z$_\mathrm{\odot}$) & 0.01 \\ [0.13cm]
     $T_\mathrm{eff}$ (K) & $35284^{+600}_{-688}$ \\ [0.13cm]
     $R_\mathrm{WD}$ ($0.01\,\mathrm{R}_\odot$) & $0.89^{+0.16}_{-0.12}$ \\ [0.13cm]
     $M_\mathrm{WD}$ ($\mathrm{M}_\odot$) & $0.97^{+0.09}_{-0.11}$ \\ [0.13cm]
     $\log(g)$ & $8.53^{+0.17}_{-0.2}$ \\ [0.13cm]
     $\langle \dot{M} \rangle$ $(10^{-10}\,\mathrm{M}_\odot$ yr$^\mathrm{-1}$) & $18.0^{+7.0}_{-4.0}$ \\ [0.13cm]
     WD contribution & 100\% \\ [0.13cm]
     \bottomrule
    \end{tabular}
\end{table*}
\vspace{0.8cm}
\begin{figure*}
    \centering
\begin{overpic}[width=0.8\textwidth]{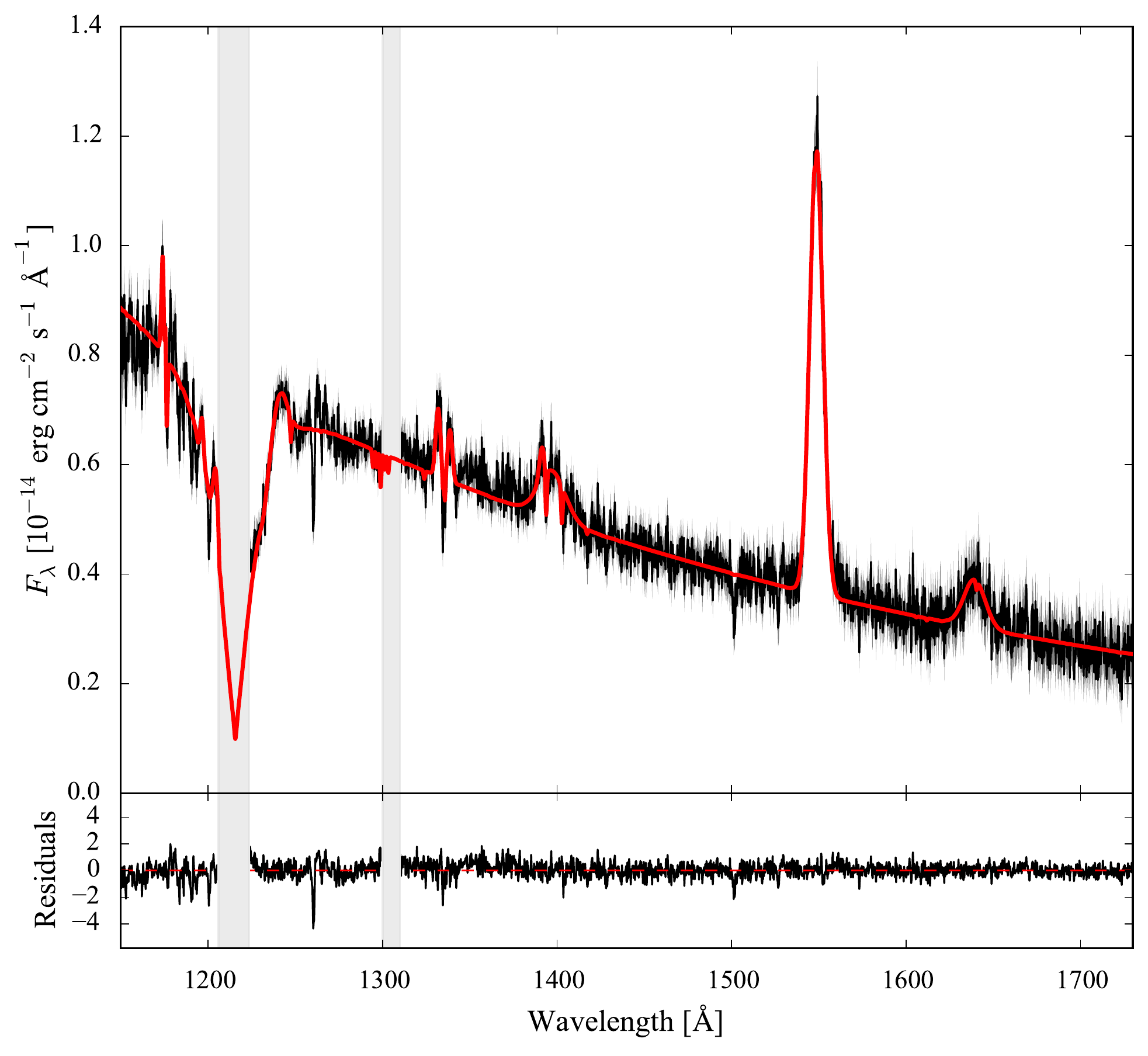}
\put (15,84) {SDSS\,J153817.35+512338.0}
\end{overpic}
  \caption{Ultraviolet \hst spectra (black) of  SDSS\,J153817.35+512338.0 along with the best-fitting model (red), computed as described in Section~\ref{sec:uv_fit}.}
\end{figure*}
\newpage
\clearpage

\thispagestyle{empty}
\begin{table*}
  \centering
  \caption{Best-fitting parameters for UV\,Per.}
  \begin{tabular}{ l r }
    \toprule
     Instrument & STIS \\ [0.13cm]
     Observation date & 2002 Oct 11 \\ [0.13cm]
     $E(B-V)$ (mag) & $0.07\pm0.04$ \\ [0.13cm]
     $P_\mathrm{orb}$ (min) & 93.44 \\ [0.13cm]
     $\mathit{Gaia}$ EDR3 ID & 457106501671769472 \\ [0.13cm]
     $\varpi$ (mas) & $4.04\pm0.11$ \\ [0.13cm]
     $d$ (pc) & $248^{+7}_{-6}$ \\ [0.13cm]
     $Z$ (Z$_\mathrm{\odot}$) & 0.2 \\ [0.13cm]
     $T_\mathrm{eff}$ (K) & $14040^{+539}_{-645}$ \\ [0.13cm]
     $R_\mathrm{WD}$ ($0.01\,\mathrm{R}_\odot$) & $1.2^{+0.4}_{-0.3}$ \\ [0.13cm]
     $M_\mathrm{WD}$ ($\mathrm{M}_\odot$) & $0.7 \pm 0.2$ \\ [0.13cm]
     $\log(g)$ & $8.1 \pm 0.4$ \\ [0.13cm]
     $\langle \dot{M} \rangle$ $(10^{-10}\,\mathrm{M}_\odot$ yr$^\mathrm{-1}$) & $0.9^{+0.7}_{-0.4}$ \\ [0.13cm]
     WD contribution & 69\% \\ [0.13cm]
     \bottomrule
    \end{tabular}
\end{table*}
\vspace{0.8cm}
\begin{figure*}
    \centering
\begin{overpic}[width=0.8\textwidth]{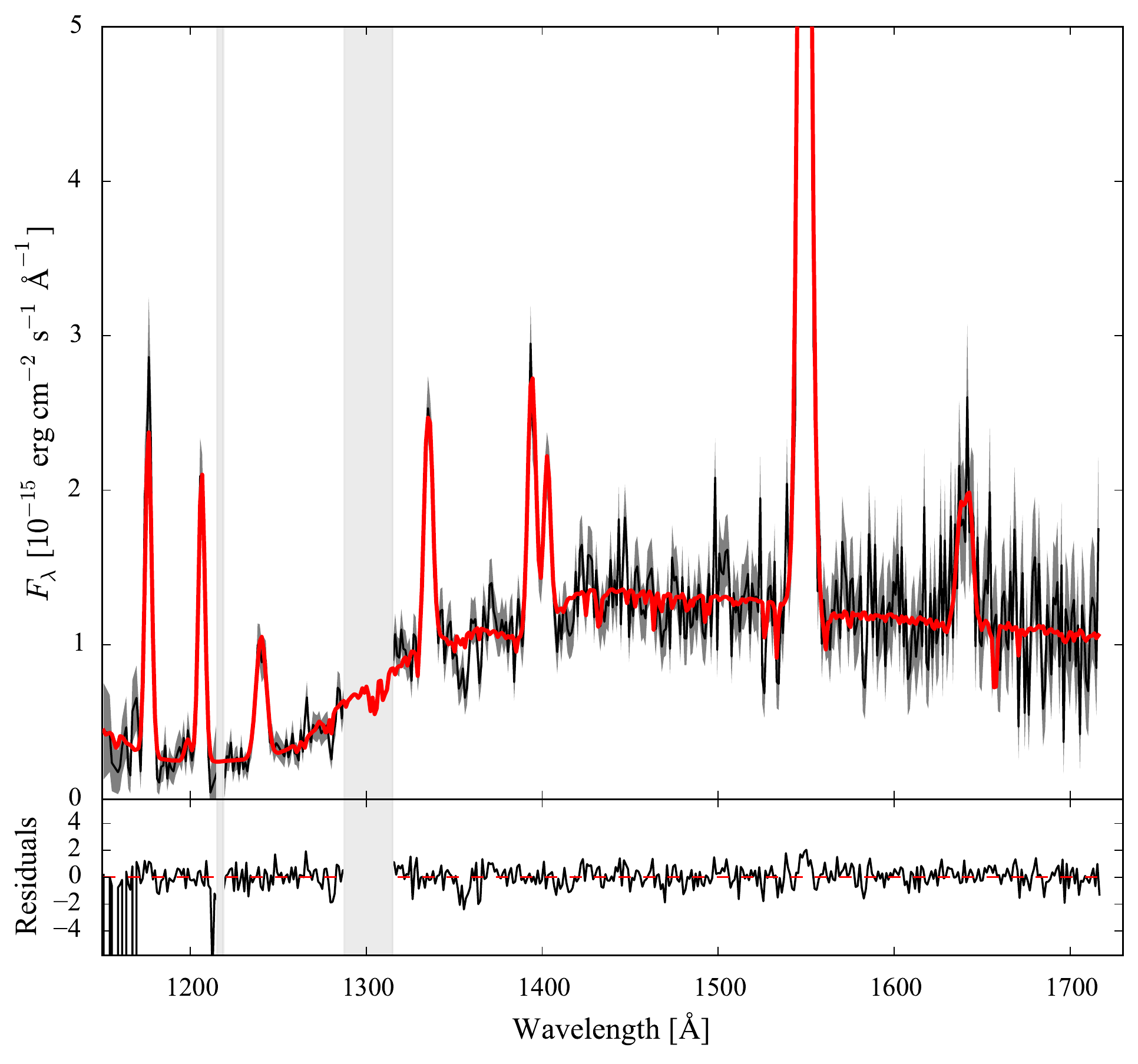}
\put (15,84) {UV\,Per}
\end{overpic}
  \caption{Ultraviolet \hst spectra (black) of  UV\,Per along with the best-fitting model (red), computed as described in Section~\ref{sec:uv_fit}.}
\end{figure*}
\newpage
\clearpage

\thispagestyle{empty}
\begin{table*}
  \centering
  \caption{Best-fitting parameters for 1RXS\,J023238.8--371812.}
  \begin{tabular}{ l r }
    \toprule
     Instrument & COS \\ [0.13cm]
     Observation date & 2012 Nov 01 \\ [0.13cm]
     $E(B-V)$ (mag) & $0.005^{+0.014}_{-0.005}$ \\ [0.13cm]
     $P_\mathrm{orb}$ (min) & 95.04 \\ [0.13cm]
     $\mathit{Gaia}$ EDR3 ID & 4953766320874344704 \\ [0.13cm]
     $\varpi$ (mas) & $4.68\pm0.16$ \\ [0.13cm]
     $d$ (pc) & $214^{+8}_{-7}$ \\ [0.13cm]
     $Z$ (Z$_\mathrm{\odot}$) & 0.2 \\ [0.13cm]
     $T_\mathrm{eff}$ (K) & $14457^{+118}_{-135}$ \\ [0.13cm]
     $R_\mathrm{WD}$ ($0.01\,\mathrm{R}_\odot$) & $0.63^{+0.05}_{-0.04}$ \\ [0.13cm]
     $M_\mathrm{WD}$ ($\mathrm{M}_\odot$) & $1.15 \pm 0.04$ \\ [0.13cm]
     $\log(g)$ & $8.9^{+0.07}_{-0.08}$ \\ [0.13cm]
     $\langle \dot{M} \rangle$ $(10^{-10}\,\mathrm{M}_\odot$ yr$^\mathrm{-1}$) & $0.23 \pm 0.03$ \\ [0.13cm]
     WD contribution & 71\% \\ [0.13cm]
     \bottomrule
    \end{tabular}
\end{table*}
\vspace{0.8cm}
\begin{figure*}
    \centering
\begin{overpic}[width=0.8\textwidth]{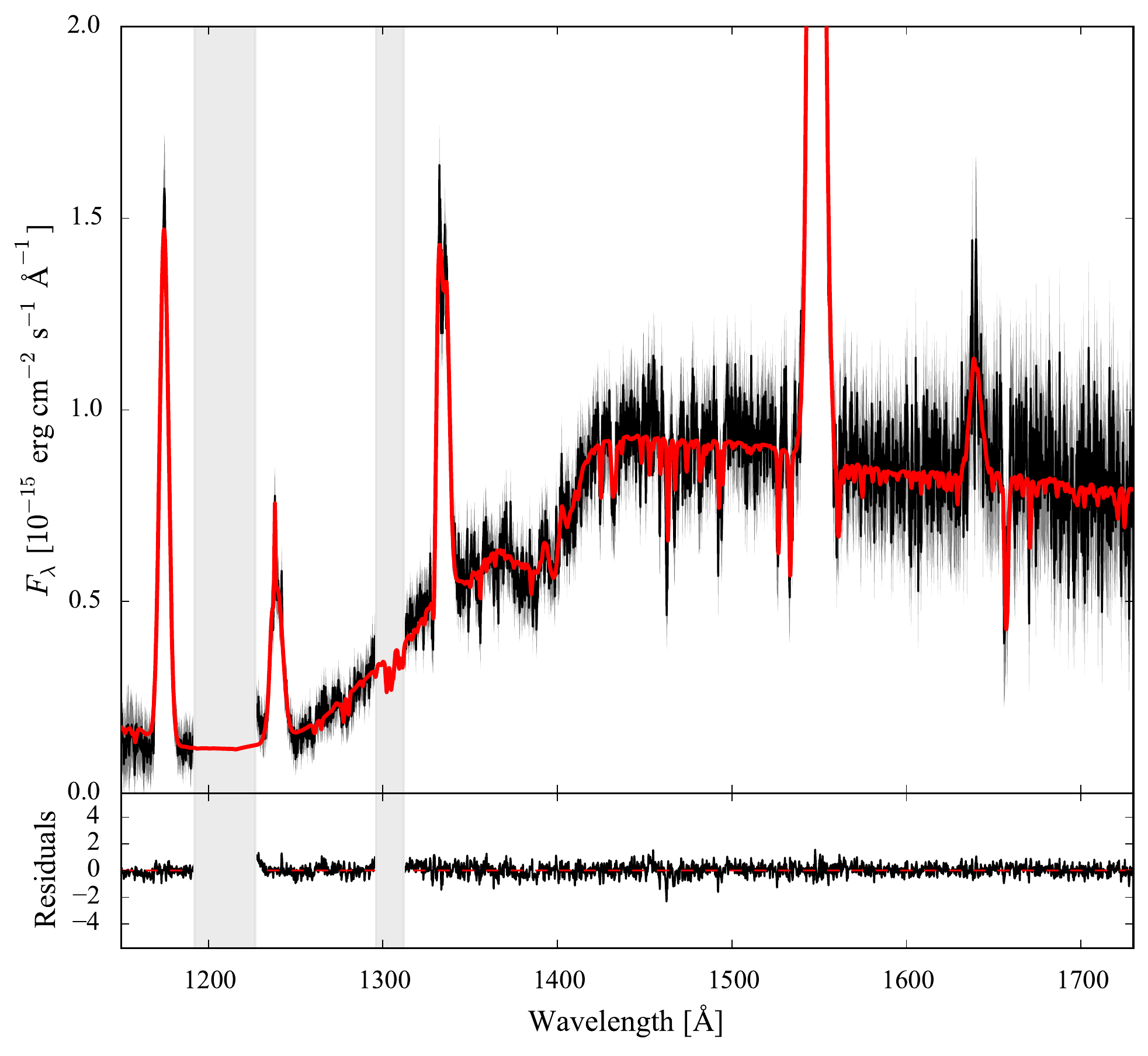}
\put (15,84) {1RXS\,J023238.8--371812}
\end{overpic}
  \caption{Ultraviolet \hst spectra (black) of  1RXS\,J023238.8--371812 along with the best-fitting model (red), computed as described in Section~\ref{sec:uv_fit}.}
\end{figure*}
\newpage
\clearpage

\thispagestyle{empty}
\begin{table*}
  \centering
  \caption{Best-fitting parameters for RZ\,Sge.}
  \begin{tabular}{ l r }
    \toprule
     Instrument & STIS \\ [0.13cm]
     Observation date & 2003 Jun 13 \\ [0.13cm]
     $E(B-V)$ (mag) & $0.03\pm0.3$ \\ [0.13cm]
     $P_\mathrm{orb}$ (min) & 98.32 \\ [0.13cm]
     $\mathit{Gaia}$ EDR3 ID & 1820209309025797888 \\ [0.13cm]
     $\varpi$ (mas) & $3.40\pm0.08$ \\ [0.13cm]
     $d$ (pc) & $294\pm7$ \\ [0.13cm]
     $Z$ (Z$_\mathrm{\odot}$) & 0.5 \\ [0.13cm]
     $T_\mathrm{eff}$ (K) & $15197^{+419}_{-507}$ \\ [0.13cm]
     $R_\mathrm{WD}$ ($0.01\,\mathrm{R}_\odot$) & $1.09^{+0.27}_{-0.17}$ \\ [0.13cm]
     $M_\mathrm{WD}$ ($\mathrm{M}_\odot$) & $0.78^{+0.13}_{-0.18}$ \\ [0.13cm]
     $\log(g)$ & $8.3^{+0.2}_{-0.3}$ \\ [0.13cm]
     $\langle \dot{M} \rangle$ $(10^{-10}\,\mathrm{M}_\odot$ yr$^\mathrm{-1}$) & $1.0^{+0.5}_{-0.3}$ \\ [0.13cm]
     WD contribution & 51\% \\ [0.13cm]
     \bottomrule
    \end{tabular}
\end{table*}
\vspace{0.8cm}
\begin{figure*}
    \centering
\begin{overpic}[width=0.8\textwidth]{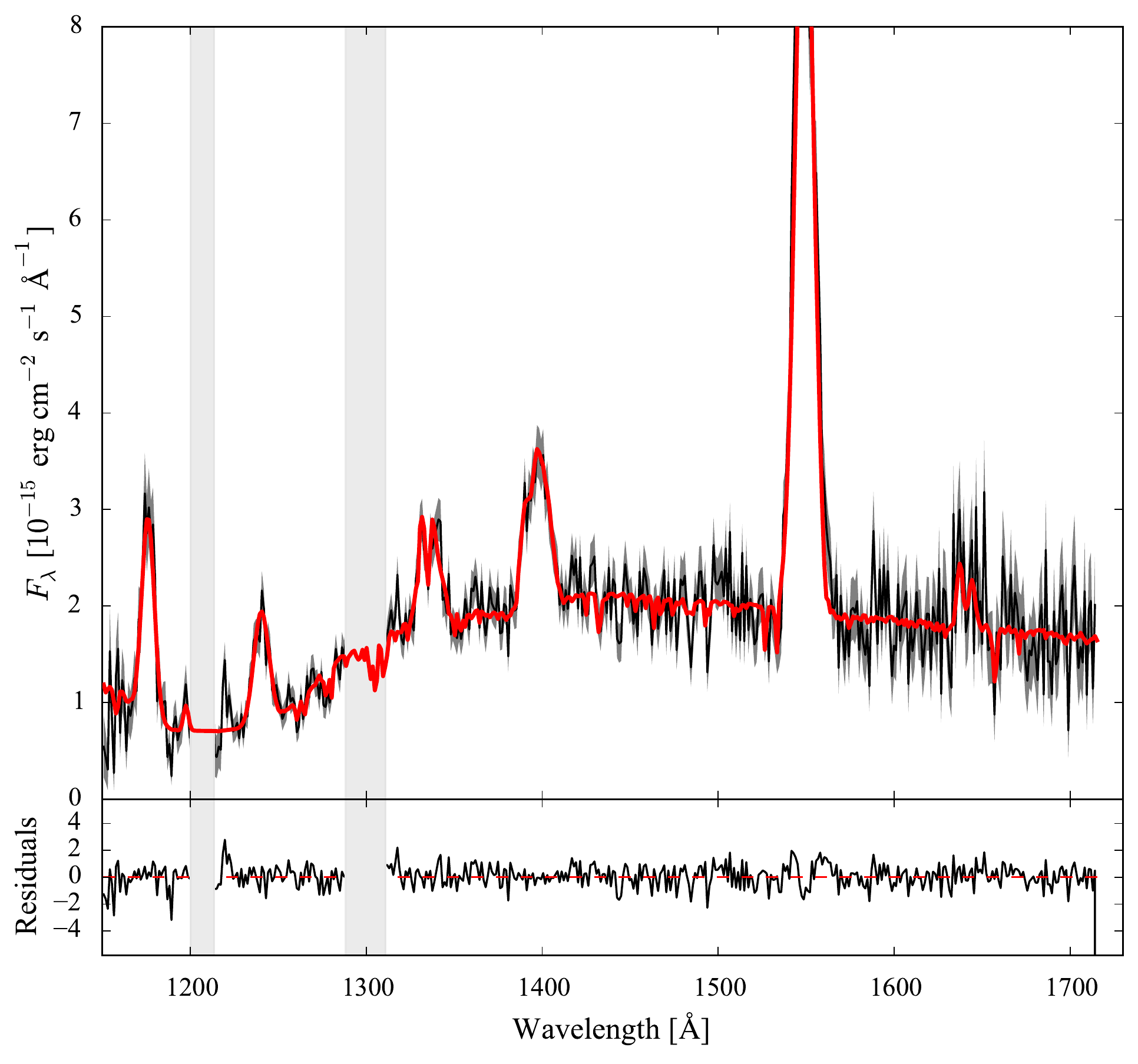}
\put (15,84) {RZ\,Sge}
\end{overpic}
  \caption{Ultraviolet \hst spectra (black) of  RZ\,Sge along with the best-fitting model (red), computed as described in Section~\ref{sec:uv_fit}.}
\end{figure*}
\newpage
\clearpage

\thispagestyle{empty}
\begin{table*}
  \centering
  \caption{Best-fitting parameters for CY\,UMa.}
  \begin{tabular}{ l r }
    \toprule
     Instrument & STIS \\ [0.13cm]
     Observation date & 2002 Dec 27 \\ [0.13cm]
     $E(B-V)$ (mag) & $0.018^{+0}_{-0.015}$ \\ [0.13cm]
     $P_\mathrm{orb}$ (min) & 100.18 \\ [0.13cm]
     $\mathit{Gaia}$ EDR3 ID & 832942871937909632 \\ [0.13cm]
     $\varpi$ (mas) & $3.27\pm0.07$ \\ [0.13cm]
     $d$ (pc) & $306\pm6$ \\ [0.13cm]
     $Z$ (Z$_\mathrm{\odot}$) & 0.1 \\ [0.13cm]
     $T_\mathrm{eff}$ (K) & $14692^{+471}_{-394}$ \\ [0.13cm]
     $R_\mathrm{WD}$ ($0.01\,\mathrm{R}_\odot$) & $1.40 \pm 0.13$ \\ [0.13cm]
     $M_\mathrm{WD}$ ($\mathrm{M}_\odot$) & $0.57^{+0.08}_{-0.06}$ \\ [0.13cm]
     $\log(g)$ & $7.91^{+0.14}_{-0.13}$ \\ [0.13cm]
     $\langle \dot{M} \rangle$ $(10^{-10}\,\mathrm{M}_\odot$ yr$^\mathrm{-1}$) & $1.6 \pm 0.2$ \\ [0.13cm]
     WD contribution & 63\% \\ [0.13cm]
     \bottomrule
    \end{tabular}
\end{table*}
\vspace{0.8cm}
\begin{figure*}
    \centering
\begin{overpic}[width=0.8\textwidth]{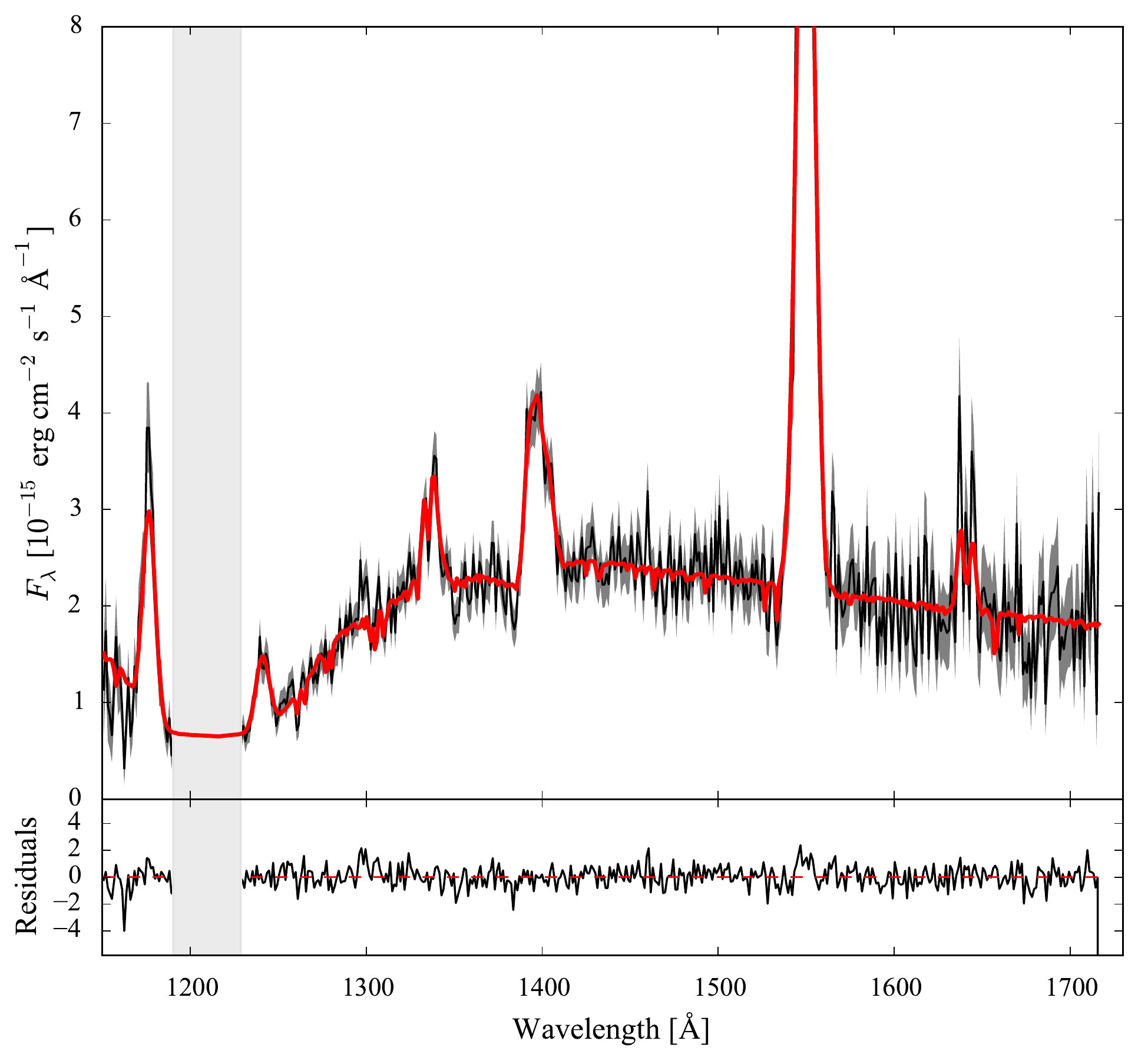}
\put (15,84) {CY\,UMa}
\end{overpic}
  \caption{Ultraviolet \hst spectra (black) of  CY\,UMa along with the best-fitting model (red), computed as described in Section~\ref{sec:uv_fit}.}
\end{figure*}
\newpage
\clearpage

\thispagestyle{empty}
\begin{table*}
  \centering
  \caption{Best-fitting parameters for GD\,552.}
  \begin{tabular}{ l r }
    \toprule
     Instrument & STIS \\ [0.13cm]
     Observation date & 2002 Oct 24/2002 Aug 31 \\ [0.13cm]
     $E(B-V)$ (mag) & $0.007^{+0.015}_{-0.007}$ \\ [0.13cm]
     $P_\mathrm{orb}$ (min) & 102.73 \\ [0.13cm]
     $\mathit{Gaia}$ EDR3 ID & 2208124536065383424 \\ [0.13cm]
     $\varpi$ (mas) & $12.41\pm0.04$ \\ [0.13cm]
     $d$ (pc) & $80.6{\pm0.2}$ \\ [0.13cm]
     $Z$ (Z$_\mathrm{\odot}$) & 0.1 \\ [0.13cm]
     $T_\mathrm{eff}$ (K) & $10761^{+37}_{-43}$ \\ [0.13cm]
     $R_\mathrm{WD}$ ($0.01\,\mathrm{R}_\odot$) & $1.07^{+0.05}_{-0.04}$ \\ [0.13cm]
     $M_\mathrm{WD}$ ($\mathrm{M}_\odot$) & $0.78^{+0.03}_{-0.04}$ \\ [0.13cm]
     $\log(g)$ & $8.27^{+0.05}_{-0.06}$ \\ [0.13cm]
     $\langle \dot{M} \rangle$ $(10^{-10}\,\mathrm{M}_\odot$ yr$^\mathrm{-1}$) & $0.243^{+0.023}_{-0.018}$ \\ [0.13cm]
     WD contribution & 55\% \\ [0.13cm]
     \bottomrule
    \end{tabular}
\end{table*}
\vspace{0.8cm}
\begin{figure*}
    \centering
\begin{overpic}[width=0.8\textwidth]{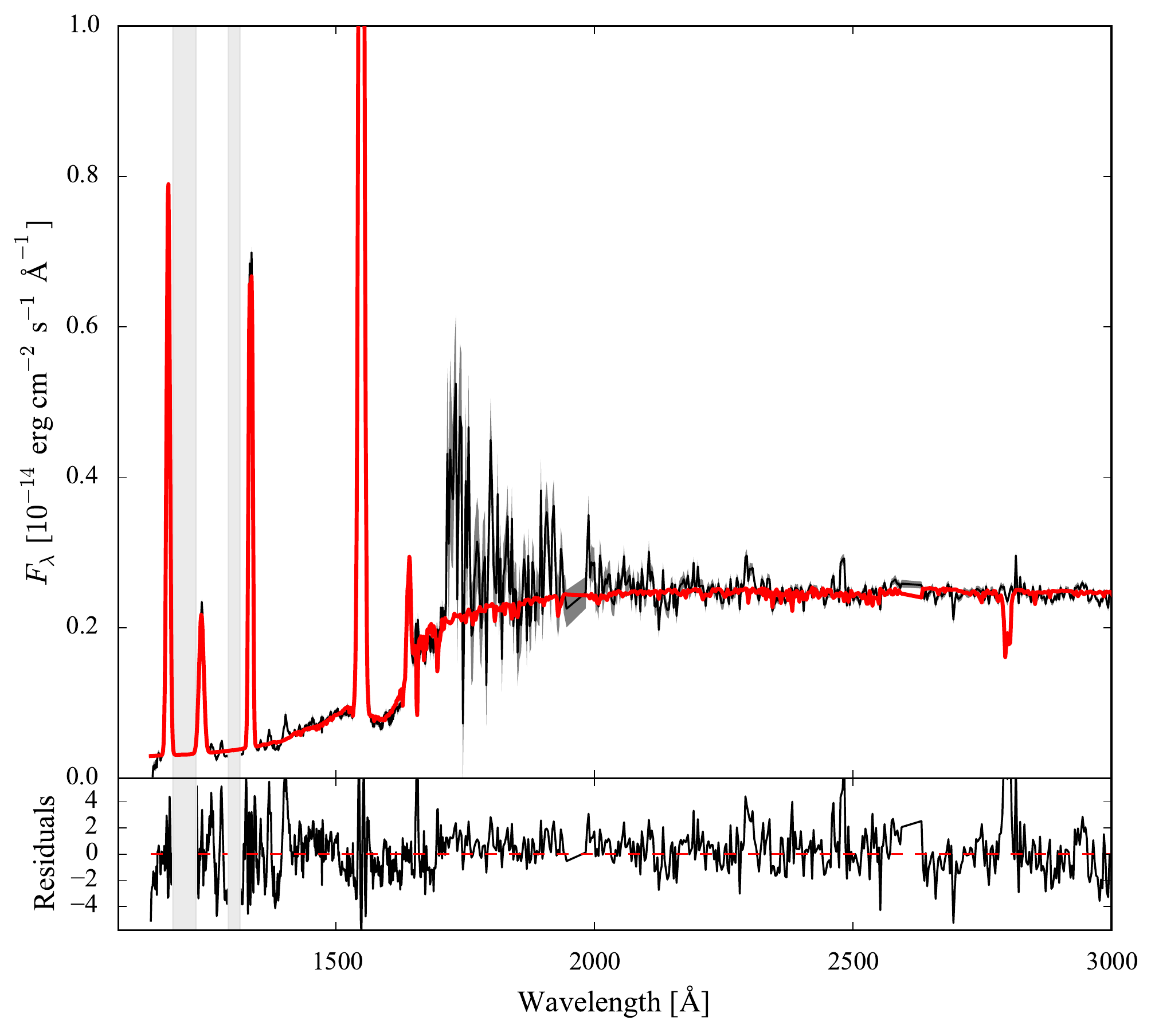}
\put (15,84) {GD\,552}
\end{overpic}
  \caption{Ultraviolet \hst spectra (black) of  GD\,552 along with the best-fitting model (red), computed as described in Section~\ref{sec:uv_fit}.}
\end{figure*}
\newpage
\clearpage

\thispagestyle{empty}
\begin{table*}
  \centering
  \caption{Best-fitting parameters for IY\,UMa.}
  \begin{tabular}{ l r }
    \toprule
     Instrument & COS \\ [0.13cm]
     Observation date & 2013 Mar 30 \\ [0.13cm]
     $E(B-V)$ (mag) & $0.015^{+0}_{-0.015}$ \\ [0.13cm]
     $P_\mathrm{orb}$ (min) & 106.43 \\ [0.13cm]
     $\mathit{Gaia}$ EDR3 ID & 855167540988615296 \\ [0.13cm]
     $\varpi$ (mas) & $5.52\pm0.07$ \\ [0.13cm]
     $d$ (pc) & $181{\pm2}$ \\ [0.13cm]
     $Z$ (Z$_\mathrm{\odot}$) & 1.0 \\ [0.13cm]
     $T_\mathrm{eff}$ (K) & $17057^{+179}_{-79}$ \\ [0.13cm]
     $R_\mathrm{WD}$ ($0.01\,\mathrm{R}_\odot$) & $0.83^{+0.04}_{-0.05}$ \\ [0.13cm]
     $M_\mathrm{WD}$ ($\mathrm{M}_\odot$) & $0.99^{+0.04}_{-0.03}$ \\ [0.13cm]
     $\log(g)$ & $8.59^{+0.07}_{-0.05}$ \\ [0.13cm]
     $\langle \dot{M} \rangle$ $(10^{-10}\,\mathrm{M}_\odot$ yr$^\mathrm{-1}$) & $0.85 \pm 0.07$ \\ [0.13cm]
     WD contribution & 79\% \\ [0.13cm]
     \bottomrule
    \end{tabular}
\end{table*}
\vspace{0.8cm}
\begin{figure*}
    \centering
\begin{overpic}[width=0.8\textwidth]{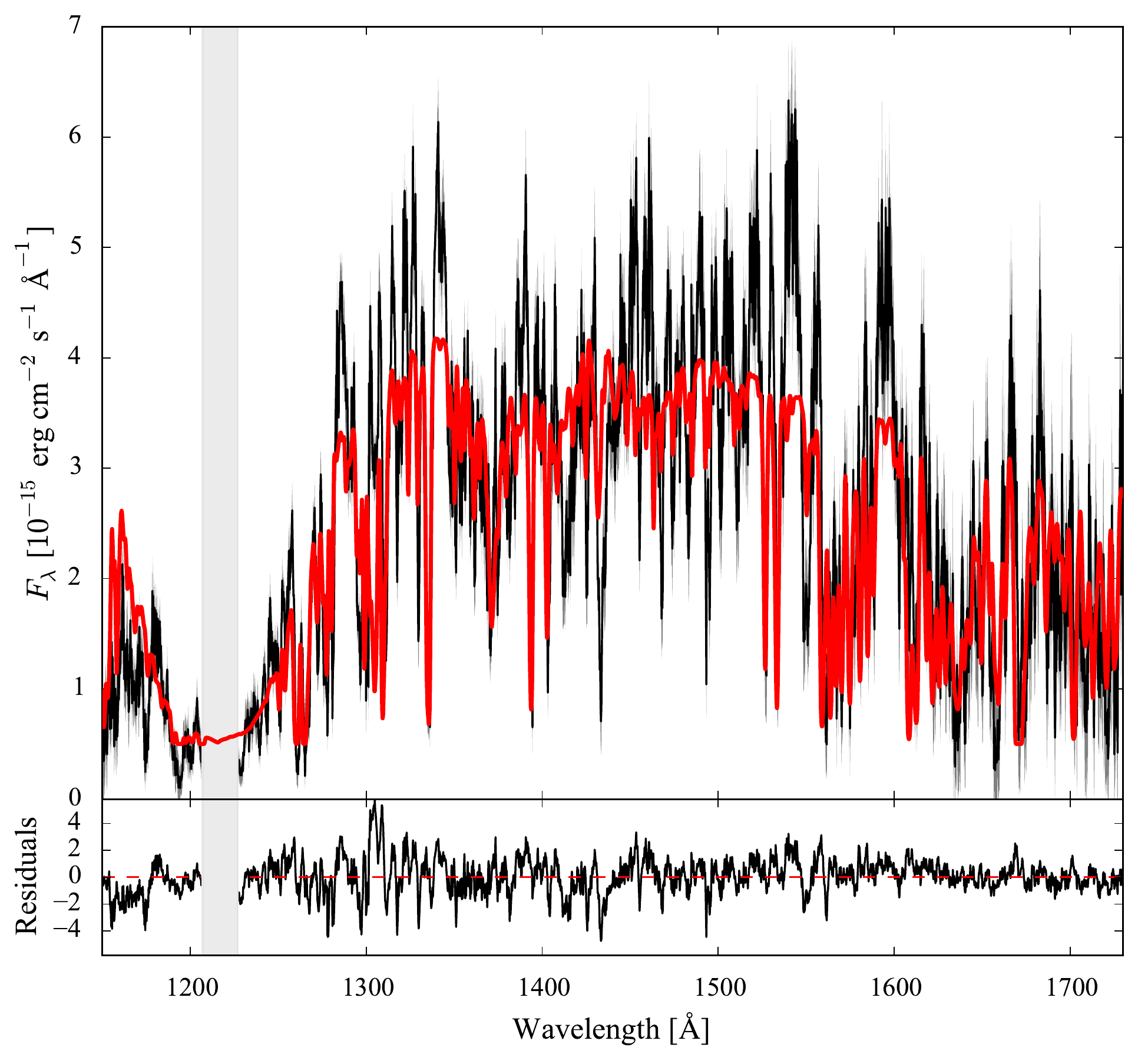}
\put (15,84) {IY\,UMa}
\end{overpic}
  \caption{Ultraviolet \hst spectra (black) of  IY\,UMa along with the best-fitting model (red), computed as described in Section~\ref{sec:uv_fit}.}
\end{figure*}
\newpage
\clearpage

\thispagestyle{empty}
\begin{table*}
  \centering
  \caption{Best-fitting parameters for SDSS\,J100515.38+191107.9.}
  \begin{tabular}{ l r }
    \toprule
     Instrument & COS \\ [0.13cm]
     Observation date & 2013 Jan 31 \\ [0.13cm]
     $E(B-V)$ (mag) & $0.021^{+0.011}_{-0.014}$ \\ [0.13cm]
     $P_\mathrm{orb}$ (min) & 107.6 \\ [0.13cm]
     $\mathit{Gaia}$ EDR3 ID & 626719772406892288 \\ [0.13cm]
     $\varpi$ (mas) & $2.95\pm0.17$ \\ [0.13cm]
     $d$ (pc) & $339^{+21}_{-19}$ \\ [0.13cm]
     $Z$ (Z$_\mathrm{\odot}$) & 0.2 \\ [0.13cm]
     $T_\mathrm{eff}$ (K) & $14483^{+520}_{-430}$ \\ [0.13cm]
     $R_\mathrm{WD}$ ($0.01\,\mathrm{R}_\odot$) & $1.7^{+0.4}_{-0.3}$ \\ [0.13cm]
     $M_\mathrm{WD}$ ($\mathrm{M}_\odot$) & $0.44^{+0.15}_{-0.09}$ \\ [0.13cm]
     $\log(g)$ & $7.6 \pm 0.3$ \\ [0.13cm]
     $\langle \dot{M} \rangle$ $(10^{-10}\,\mathrm{M}_\odot$ yr$^\mathrm{-1}$) & $2.4^{+1.2}_{-0.8}$ \\ [0.13cm]
     WD contribution & 76\% \\ [0.13cm]
     \bottomrule
    \end{tabular}
\end{table*}
\vspace{0.8cm}
\begin{figure*}
    \centering
\begin{overpic}[width=0.8\textwidth]{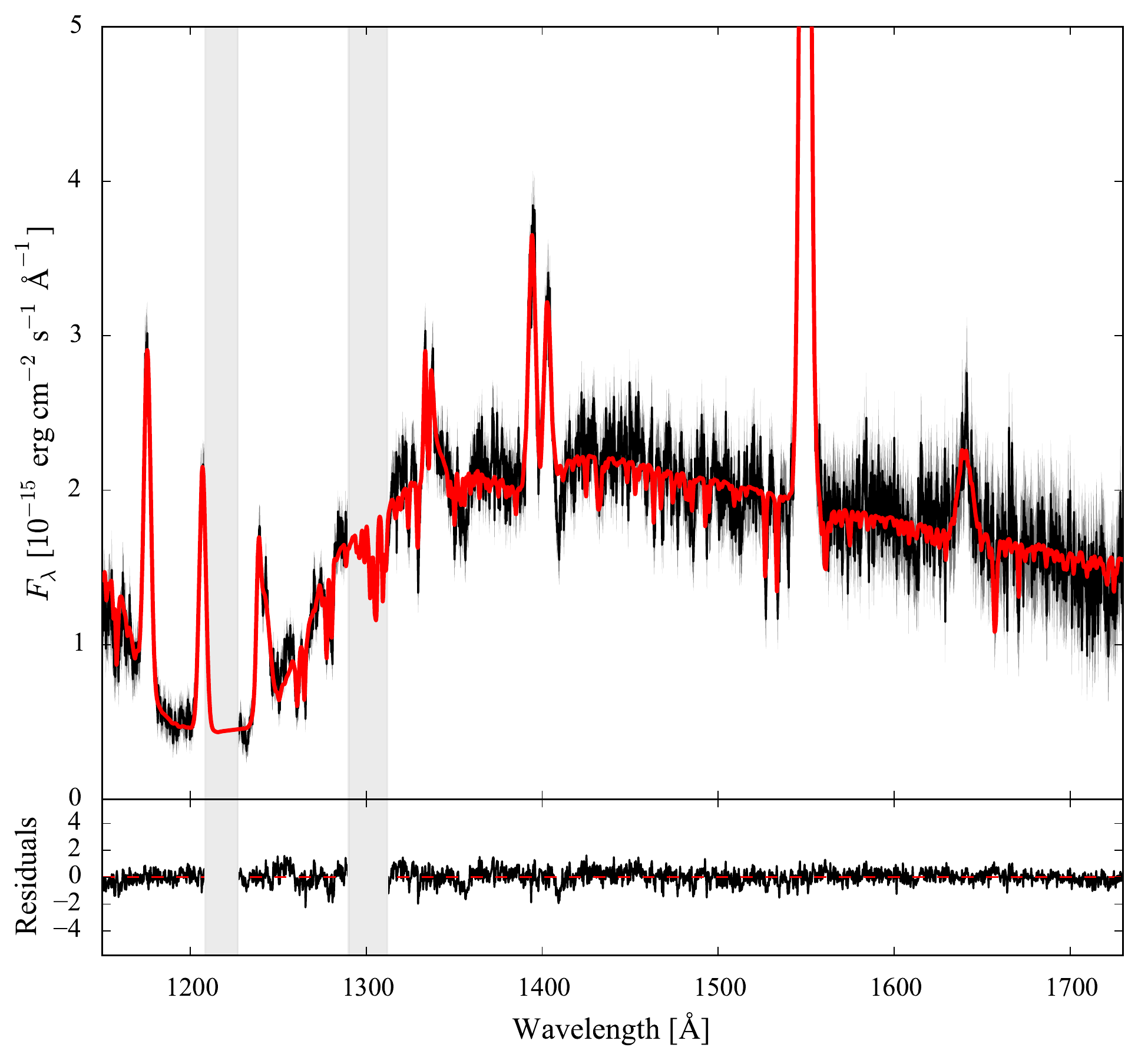}
\put (15,84) {SDSS\,J100515.38+191107.9}
\end{overpic}
  \caption{Ultraviolet \hst spectra (black) of  SDSS\,J100515.38+191107.9 along with the best-fitting model (red), computed as described in Section~\ref{sec:uv_fit}.}
\end{figure*}
\newpage
\clearpage

\thispagestyle{empty}
\begin{table*}
  \centering
  \caption{Best-fitting parameters for RZ\,Leo.}
  \begin{tabular}{ l r }
    \toprule
     Instrument & COS \\ [0.13cm]
     Observation date & 2013 Apr 11 \\ [0.13cm]
     $E(B-V)$ (mag) & $0.023^{+0.015}_{-0.023}$ \\ [0.13cm]
     $P_\mathrm{orb}$ (min) & 110.17 \\ [0.13cm]
     $\mathit{Gaia}$ EDR3 ID & 3799290858445023488 \\ [0.13cm]
     $\varpi$ (mas) & $3.59\pm0.15$ \\ [0.13cm]
     $d$ (pc) & $279^{+12}_{-11}$ \\ [0.13cm]
     $Z$ (Z$_\mathrm{\odot}$) & 0.5 \\ [0.13cm]
     $T_\mathrm{eff}$ (K) & $15573^{+437}_{-424}$ \\ [0.13cm]
     $R_\mathrm{WD}$ ($0.01\,\mathrm{R}_\odot$) & $0.85^{+0.2}_{-0.16}$ \\ [0.13cm]
     $M_\mathrm{WD}$ ($\mathrm{M}_\odot$) & $0.97^{+0.13}_{-0.16}$ \\ [0.13cm]
     $\log(g)$ & $8.6^{+0.2}_{-0.3}$ \\ [0.13cm]
     $\langle \dot{M} \rangle$ $(10^{-10}\,\mathrm{M}_\odot$ yr$^\mathrm{-1}$) & $0.62^{+0.3}_{-0.19}$ \\ [0.13cm]
     WD contribution & 81\% \\ [0.13cm]
     \bottomrule
    \end{tabular}
\end{table*}
\vspace{0.8cm}
\begin{figure*}
    \centering
\begin{overpic}[width=0.8\textwidth]{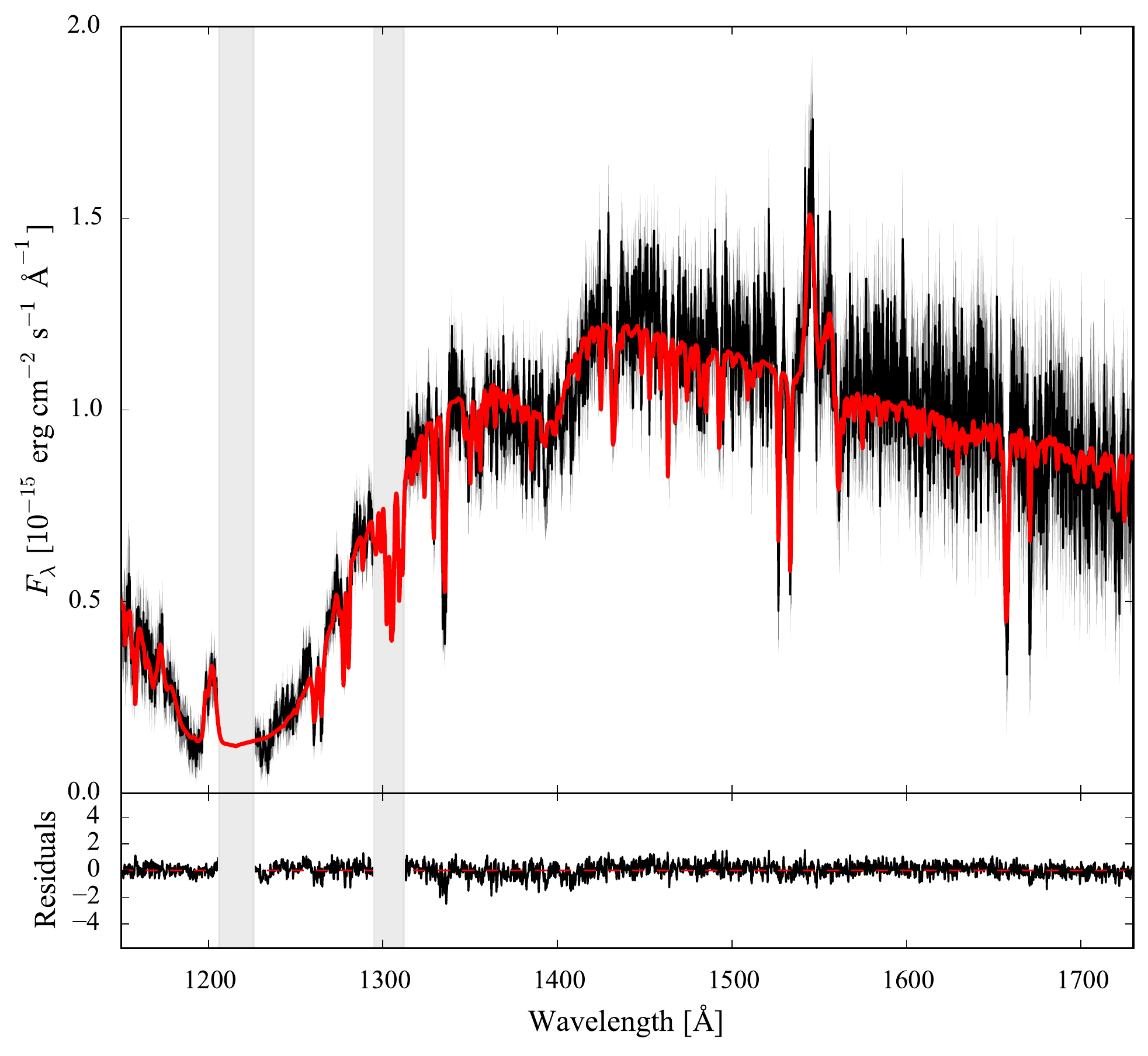}
\put (15,84) {RZ\,Leo}
\end{overpic}
  \caption{Ultraviolet \hst spectra (black) of  RZ\,Leo along with the best-fitting model (red), computed as described in Section~\ref{sec:uv_fit}.}
\end{figure*}
\newpage
\clearpage

\thispagestyle{empty}
\begin{table*}
  \centering
  \caption{Best-fitting parameters for AX\,For.}
  \begin{tabular}{ l r }
    \toprule
     Instrument & COS \\ [0.13cm]
     Observation date & 2013 Jul 11 \\ [0.13cm]
     $E(B-V)$ (mag) & $0.018\pm0.018$ \\ [0.13cm]
     $P_\mathrm{orb}$ (min) & 113.04 \\ [0.13cm]
     $\mathit{Gaia}$ EDR3 ID & 5067753236787919232 \\ [0.13cm]
     $\varpi$ (mas) & $2.86\pm0.08$ \\ [0.13cm]
     $d$ (pc) & $349\pm10$ \\ [0.13cm]
     $Z$ (Z$_\mathrm{\odot}$) & 1.0 \\ [0.13cm]
     $T_\mathrm{eff}$ (K) & -- \\ [0.13cm]
     $R_\mathrm{WD}$ ($0.01\,\mathrm{R}_\odot$) & $1.09^{+0.08}_{-0.09}$ \\ [0.13cm]
     $M_\mathrm{WD}$ ($\mathrm{M}_\odot$) & $0.76 \pm 0.07$ \\ [0.13cm]
     $\log(g)$ & $8.24 \pm 0.08$ \\ [0.13cm]
     $\langle \dot{M} \rangle$ $(10^{-10}\,\mathrm{M}_\odot$ yr$^\mathrm{-1}$) & -- \\ [0.13cm]
     WD contribution & --\% \\ [0.13cm]
     \bottomrule
    \end{tabular}
\end{table*}
\vspace{0.8cm}
\begin{figure*}
    \centering
\begin{overpic}[width=0.8\textwidth]{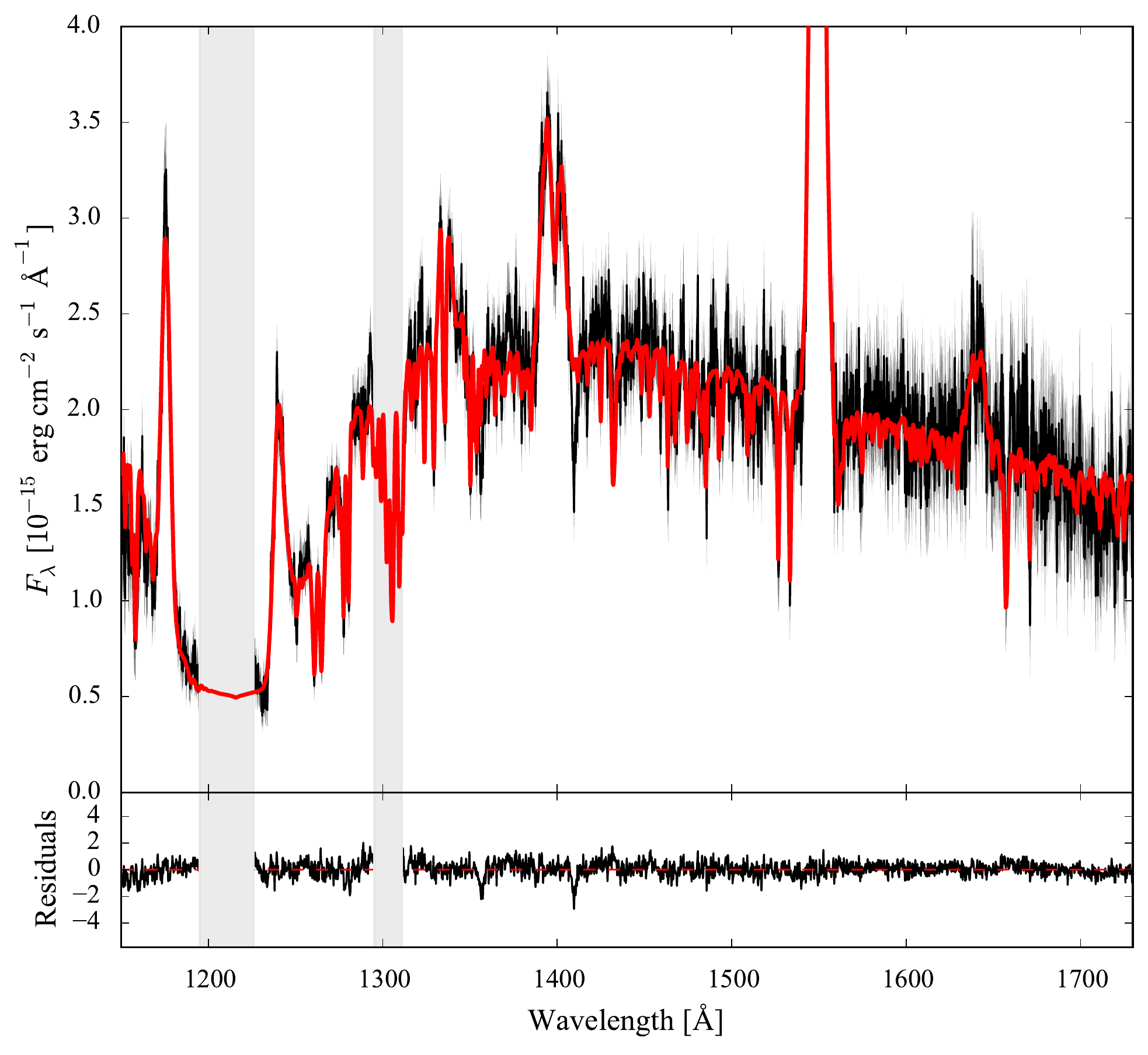}
\put (15,84) {AX\,For}
\end{overpic}
  \caption{Ultraviolet \hst spectra (black) of  AX\,For along with the best-fitting model (red), computed as described in Section~\ref{sec:uv_fit}.}
\end{figure*}
\newpage
\clearpage

\thispagestyle{empty}
\begin{table*}
  \centering
  \caption{Best-fitting parameters for CU\,Vel.}
  \begin{tabular}{ l r }
    \toprule
     Instrument & COS \\ [0.13cm]
     Observation date & 2013 Jan 18 \\ [0.13cm]
     $E(B-V)$ (mag) & $0.007^{+0.015}_{-0.007}$ \\ [0.13cm]
     $P_\mathrm{orb}$ (min) & 113.04 \\ [0.13cm]
     $\mathit{Gaia}$ EDR3 ID & 5524430207364715520 \\ [0.13cm]
     $\varpi$ (mas) & $6.31\pm0.04$ \\ [0.13cm]
     $d$ (pc) & $158{\pm1}$ \\ [0.13cm]
     $Z$ (Z$_\mathrm{\odot}$) & 0.1 \\ [0.13cm]
     $T_\mathrm{eff}$ (K) & $14174^{+117}_{-169}$ \\ [0.13cm]
     $R_\mathrm{WD}$ ($0.01\,\mathrm{R}_\odot$) & $1.58^{+0.11}_{-0.08}$ \\ [0.13cm]
     $M_\mathrm{WD}$ ($\mathrm{M}_\odot$) & $0.47^{+0.04}_{-0.05}$ \\ [0.13cm]
     $\log(g)$ & $7.71^{+0.08}_{-0.11}$ \\ [0.13cm]
     $\langle \dot{M} \rangle$ $(10^{-10}\,\mathrm{M}_\odot$ yr$^\mathrm{-1}$) & $1.97^{+0.28}_{-0.19}$ \\ [0.13cm]
     WD contribution & 90\% \\ [0.13cm]
     \bottomrule
    \end{tabular}
\end{table*}
\vspace{0.8cm}
\begin{figure*}
    \centering
\begin{overpic}[width=0.8\textwidth]{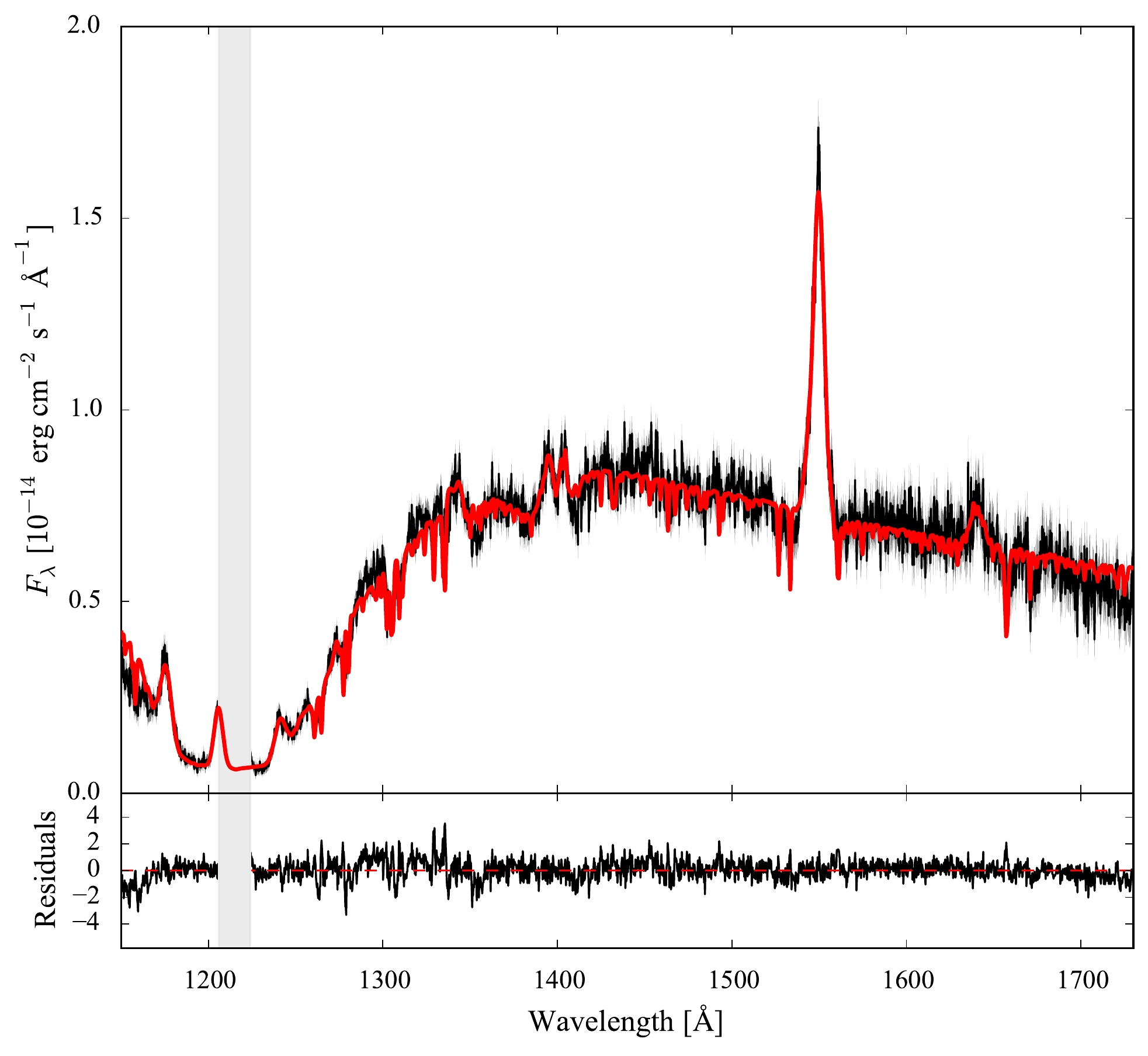}
\put (15,84) {CU\,Vel}
\end{overpic}
  \caption{Ultraviolet \hst spectra (black) of  CU\,Vel along with the best-fitting model (red), computed as described in Section~\ref{sec:uv_fit}.}
\end{figure*}
\newpage
\clearpage

\thispagestyle{empty}
\begin{table*}
  \centering
  \caption{Best-fitting parameters for EF\,Peg.}
  \begin{tabular}{ l r }
    \toprule
     Instrument & STIS \\ [0.13cm]
     Observation date & 2000 Jun 18 \\ [0.13cm]
     $E(B-V)$ (mag) & $0.048\pm0.019$ \\ [0.13cm]
     $P_\mathrm{orb}$ (min) & 120.53 \\ [0.13cm]
     $\mathit{Gaia}$ EDR3 ID & 1759321791033449472 \\ [0.13cm]
     $\varpi$ (mas) & $3.5\pm0.2$ \\ [0.13cm]
     $d$ (pc) & $288^{+21}_{-18}$ \\ [0.13cm]
     $Z$ (Z$_\mathrm{\odot}$) & 0.2 \\ [0.13cm]
     $T_\mathrm{eff}$ (K) & $16644^{+448}_{-570}$ \\ [0.13cm]
     $R_\mathrm{WD}$ ($0.01\,\mathrm{R}_\odot$) & $1.1 \pm 0.2$ \\ [0.13cm]
     $M_\mathrm{WD}$ ($\mathrm{M}_\odot$) & $0.8^{+0.16}_{-0.17}$ \\ [0.13cm]
     $\log(g)$ & $8.3 \pm 0.3$ \\ [0.13cm]
     $\langle \dot{M} \rangle$ $(10^{-10}\,\mathrm{M}_\odot$ yr$^\mathrm{-1}$) & $1.3^{+0.5}_{-0.4}$ \\ [0.13cm]
     WD contribution & 90\% \\ [0.13cm]
     \bottomrule
    \end{tabular}
\end{table*}
\vspace{0.8cm}
\begin{figure*}
    \centering
\begin{overpic}[width=0.8\textwidth]{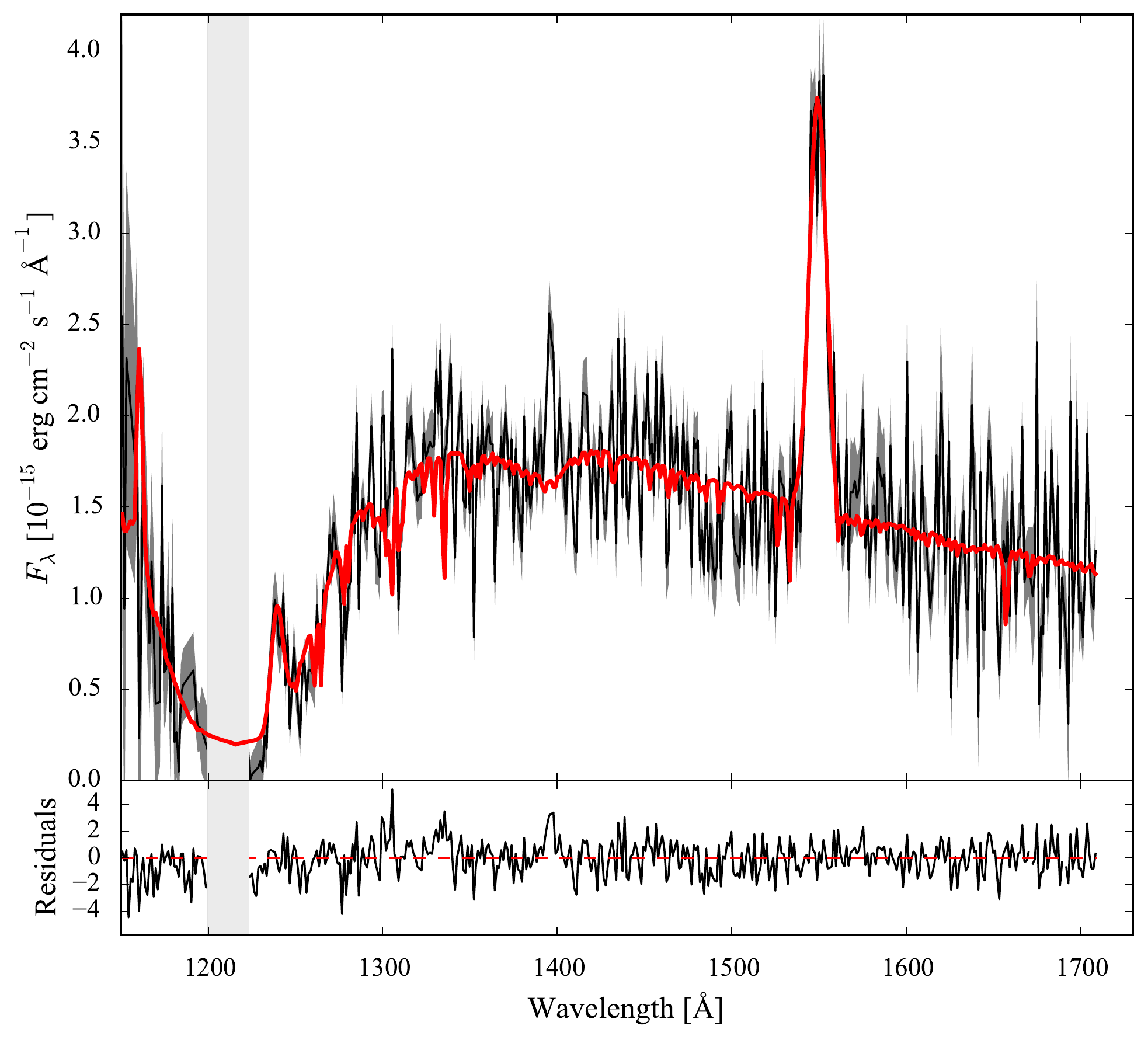}
\put (15,84) {EF\,Peg}
\end{overpic}
  \caption{Ultraviolet \hst spectra (black) of  EF\,Peg along with the best-fitting model (red), computed as described in Section~\ref{sec:uv_fit}.}
\end{figure*}
\newpage
\clearpage

\thispagestyle{empty}
\begin{table*}
  \centering
  \caption{Best-fitting parameters for DV\,UMa.}
  \begin{tabular}{ l r }
    \toprule
     Instrument & STIS \\ [0.13cm]
     Observation date & 2004 Feb 08 \\ [0.13cm]
     $E(B-V)$ (mag) & $0.02^{+0}_{-0.012}$ \\ [0.13cm]
     $P_\mathrm{orb}$ (min) & 123.62 \\ [0.13cm]
     $\mathit{Gaia}$ EDR3 ID & 820959638305816448 \\ [0.13cm]
     $\varpi$ (mas) & $2.60\pm0.15$ \\ [0.13cm]
     $d$ (pc) & $382^{+24}_{-21}$ \\ [0.13cm]
     $Z$ (Z$_\mathrm{\odot}$) & 1.0 \\ [0.13cm]
     $T_\mathrm{eff}$ (K) & $19410^{+244}_{-400}$ \\ [0.13cm]
     $R_\mathrm{WD}$ ($0.01\,\mathrm{R}_\odot$) & $0.86^{+0.12}_{-0.09}$ \\ [0.13cm]
     $M_\mathrm{WD}$ ($\mathrm{M}_\odot$) & $0.96^{+0.07}_{-0.1}$ \\ [0.13cm]
     $\log(g)$ & $8.55^{+0.13}_{-0.16}$ \\ [0.13cm]
     $\langle \dot{M} \rangle$ $(10^{-10}\,\mathrm{M}_\odot$ yr$^\mathrm{-1}$) & $1.5^{+0.4}_{-0.3}$ \\ [0.13cm]
     WD contribution & 89\% \\ [0.13cm]
     \bottomrule
    \end{tabular}
\end{table*}
\vspace{0.8cm}
\begin{figure*}
    \centering
\begin{overpic}[width=0.8\textwidth]{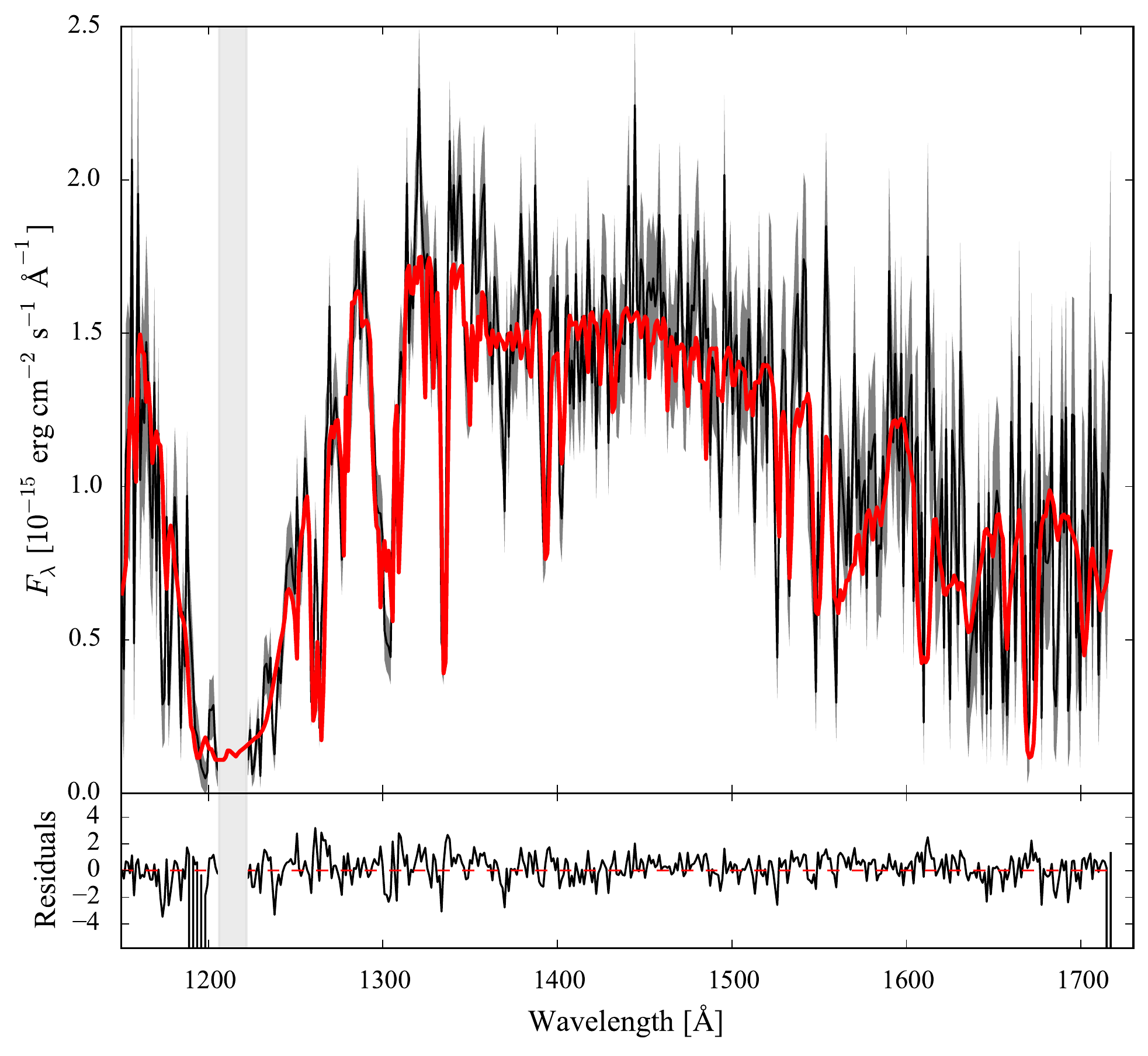}
\put (15,84) {DV\,UMa}
\end{overpic}
  \caption{Ultraviolet \hst spectra (black) of  DV\,UMa along with the best-fitting model (red), computed as described in Section~\ref{sec:uv_fit}.}
\end{figure*}
\newpage
\clearpage

\thispagestyle{empty}
\begin{table*}
  \centering
  \caption{Best-fitting parameters for IR\,Com.}
  \begin{tabular}{ l r }
    \toprule
     Instrument & COS \\ [0.13cm]
     Observation date & 2013 Jul 11 \\ [0.13cm]
     $E(B-V)$ (mag) & $0.019^{+0.021}_{-0.019}$ \\ [0.13cm]
     $P_\mathrm{orb}$ (min) & 125.34 \\ [0.13cm]
     $\mathit{Gaia}$ EDR3 ID & 3955313418148878080 \\ [0.13cm]
     $\varpi$ (mas) & $4.63\pm0.06$ \\ [0.13cm]
     $d$ (pc) & $216{\pm3}$ \\ [0.13cm]
     $Z$ (Z$_\mathrm{\odot}$) & 1.0 \\ [0.13cm]
     $T_\mathrm{eff}$ (K) & $17531^{+236}_{-271}$ \\ [0.13cm]
     $R_\mathrm{WD}$ ($0.01\,\mathrm{R}_\odot$) & $0.78^{+0.11}_{-0.08}$ \\ [0.13cm]
     $M_\mathrm{WD}$ ($\mathrm{M}_\odot$) & $1.03^{+0.07}_{-0.09}$ \\ [0.13cm]
     $\log(g)$ & $8.67^{+0.12}_{-0.15}$ \\ [0.13cm]
     $\langle \dot{M} \rangle$ $(10^{-10}\,\mathrm{M}_\odot$ yr$^\mathrm{-1}$) & $0.82^{+0.21}_{-0.15}$ \\ [0.13cm]
     WD contribution & 87\% \\ [0.13cm]
     \bottomrule
    \end{tabular}
\end{table*}
\vspace{0.8cm}
\begin{figure*}
    \centering
\begin{overpic}[width=0.8\textwidth]{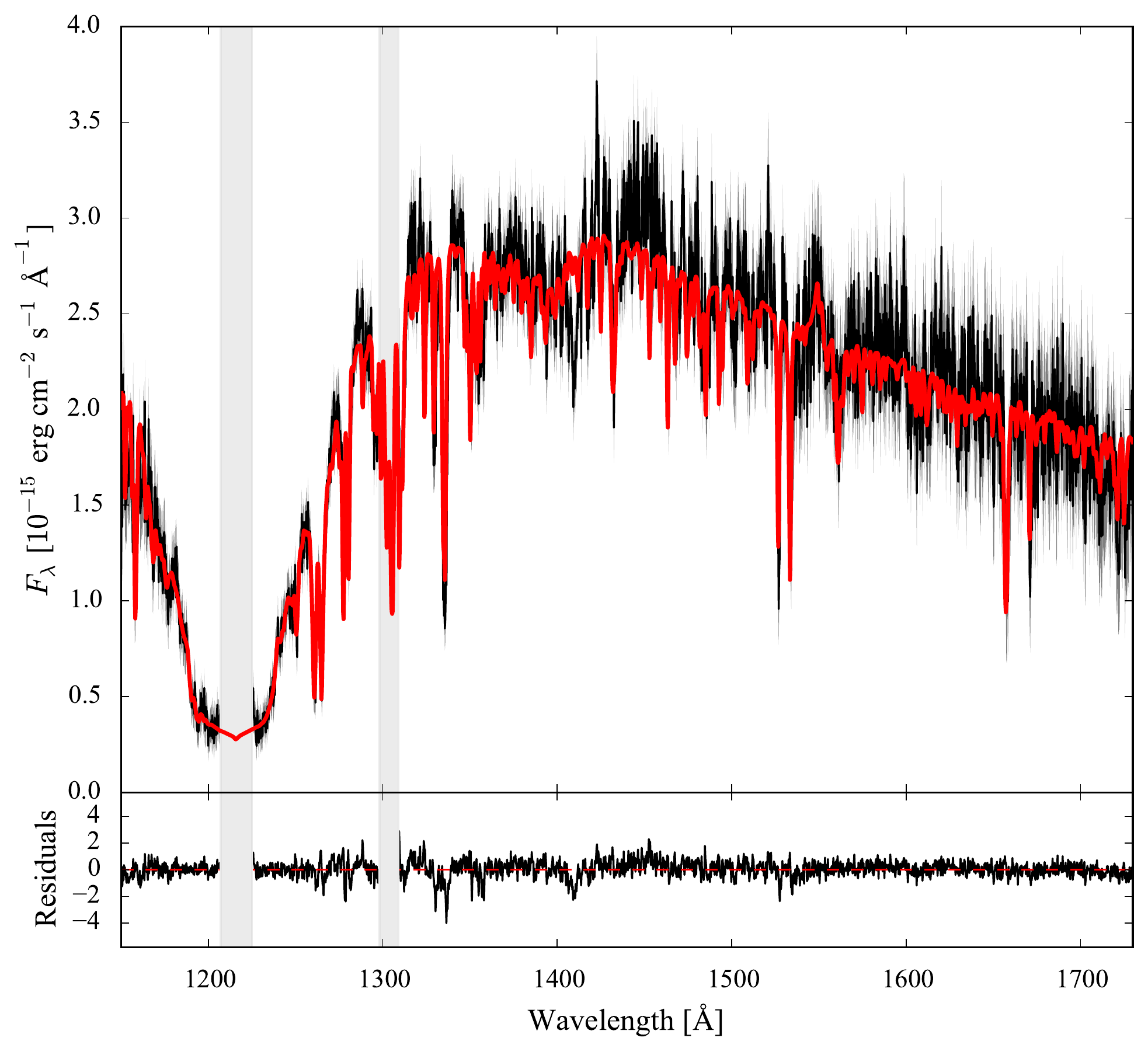}
\put (15,84) {IR\,Com}
\end{overpic}
  \caption{Ultraviolet \hst spectra (black) of  IR\,Com along with the best-fitting model (red), computed as described in Section~\ref{sec:uv_fit}.}
\end{figure*}
\newpage
\clearpage

\thispagestyle{empty}
\begin{table*}
  \centering
  \caption{Best-fitting parameters for AM\,Her.}
  \begin{tabular}{ l r }
    \toprule
     Instrument & STIS \\ [0.13cm]
     Observation date & 2002 Jul 11/12 \\ [0.13cm]
     $E(B-V)$ (mag) & $0.017\pm0.016$ \\ [0.13cm]
     $P_\mathrm{orb}$ (min) & 185.65 \\ [0.13cm]
     $\mathit{Gaia}$ EDR3 ID & 2123837555230207744 \\ [0.13cm]
     $\varpi$ (mas) & $11.37\pm0.03$ \\ [0.13cm]
     $d$ (pc) & $87.9{\pm0.2}$ \\ [0.13cm]
     $Z$ (Z$_\mathrm{\odot}$) & 0.001 \\ [0.13cm]
     $T_\mathrm{eff}$ (K) & $19248^{+486}_{-450}$ \\ [0.13cm]
     $R_\mathrm{WD}$ ($0.01\,\mathrm{R}_\odot$) & $1.3^{+0.17}_{-0.14}$ \\ [0.13cm]
     $M_\mathrm{WD}$ ($\mathrm{M}_\odot$) & $0.63^{+0.1}_{-0.08}$ \\ [0.13cm]
     $\log(g)$ & $8.01^{+0.17}_{-0.16}$ \\ [0.13cm]
     $\langle \dot{M} \rangle$ $(10^{-10}\,\mathrm{M}_\odot$ yr$^\mathrm{-1}$) & $4.0^{+0.9}_{-0.7}$ \\ [0.13cm]
     WD contribution & 92\% \\ [0.13cm]
     \bottomrule
    \end{tabular}
\end{table*}
\vspace{0.8cm}
\begin{figure*}
    \centering
\begin{overpic}[width=0.8\textwidth]{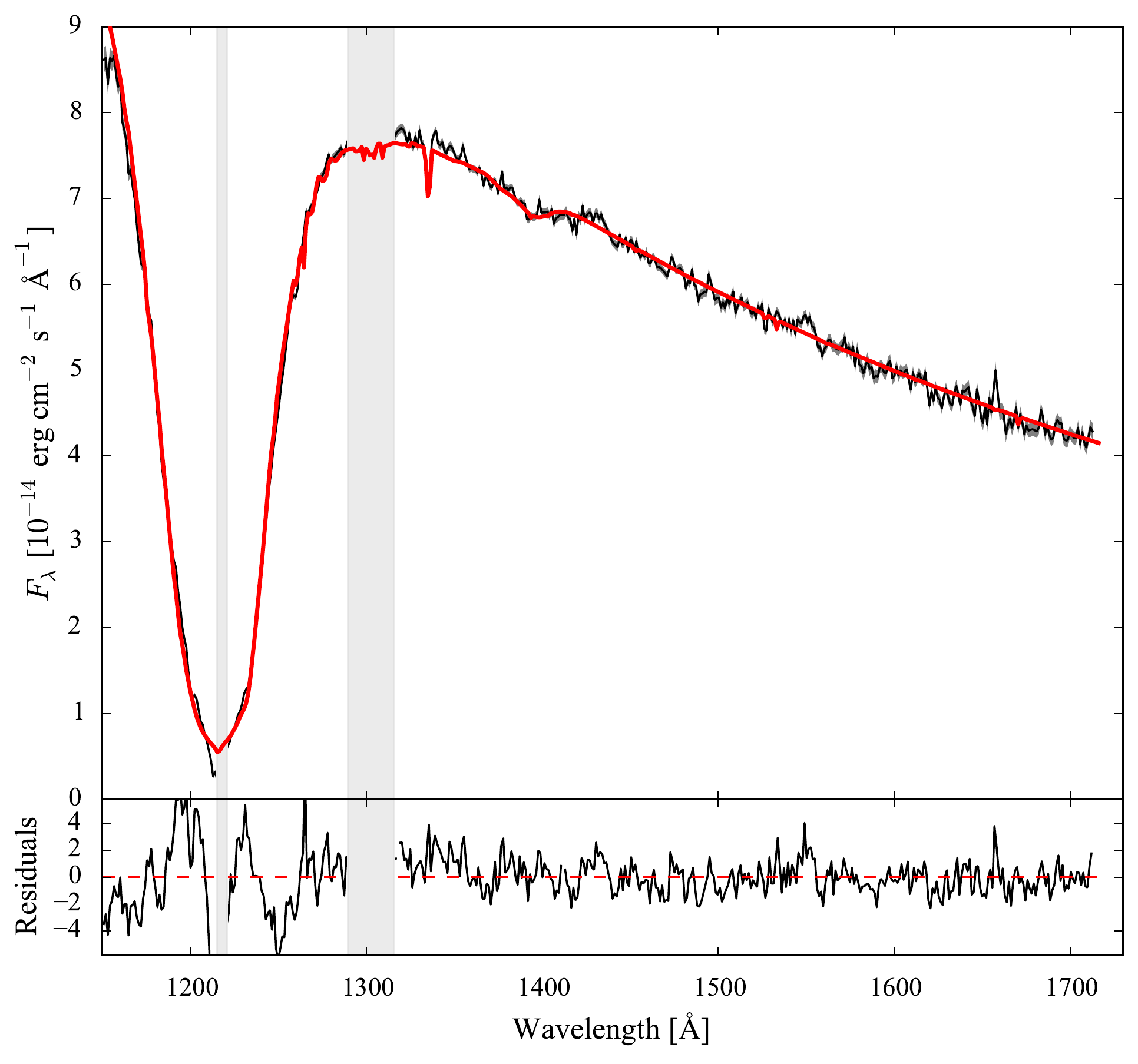}
\put (15,84) {AM\,Her}
\end{overpic}
  \caption{Ultraviolet \hst spectra (black) of  AM\,Her along with the best-fitting model (red), computed as described in Section~\ref{sec:uv_fit}.}
\end{figure*}
\newpage
\clearpage

\thispagestyle{empty}
\begin{table*}
  \centering
  \caption{Best-fitting parameters for DW\,UMa.}
  \begin{tabular}{ l r }
    \toprule
     Instrument & STIS \\ [0.13cm]
     Observation date & 1999 Jan 25 \\ [0.13cm]
     $E(B-V)$ (mag) & $0.009^{+0.018}_{-0.009}$ \\ [0.13cm]
     $P_\mathrm{orb}$ (min) & 196.71 \\ [0.13cm]
     $\mathit{Gaia}$ EDR3 ID & 855119196836523008 \\ [0.13cm]
     $\varpi$ (mas) & $1.73\pm0.02$ \\ [0.13cm]
     $d$ (pc) & $579^{+7}_{-6}$ \\ [0.13cm]
     $Z$ (Z$_\mathrm{\odot}$) & 0.71 \\ [0.13cm]
     $T_\mathrm{eff}$ (K) & $56760^{+146}_{-259}$ \\ [0.13cm]
     $R_\mathrm{WD}$ ($0.01\,\mathrm{R}_\odot$) & $1.19^{+0.07}_{-0.05}$ \\ [0.13cm]
     $M_\mathrm{WD}$ ($\mathrm{M}_\odot$) & $0.82^{+0.03}_{-0.04}$ \\ [0.13cm]
     $\log(g)$ & $8.2^{+0.05}_{-0.07}$ \\ [0.13cm]
     $\langle \dot{M} \rangle$ $(10^{-10}\,\mathrm{M}_\odot$ yr$^\mathrm{-1}$) & $228.0 \pm 24.0$ \\ [0.13cm]
     WD contribution & 97\% \\ [0.13cm]
     \bottomrule
    \end{tabular}
\end{table*}
\vspace{0.8cm}
\begin{figure*}
    \centering
\begin{overpic}[width=0.8\textwidth]{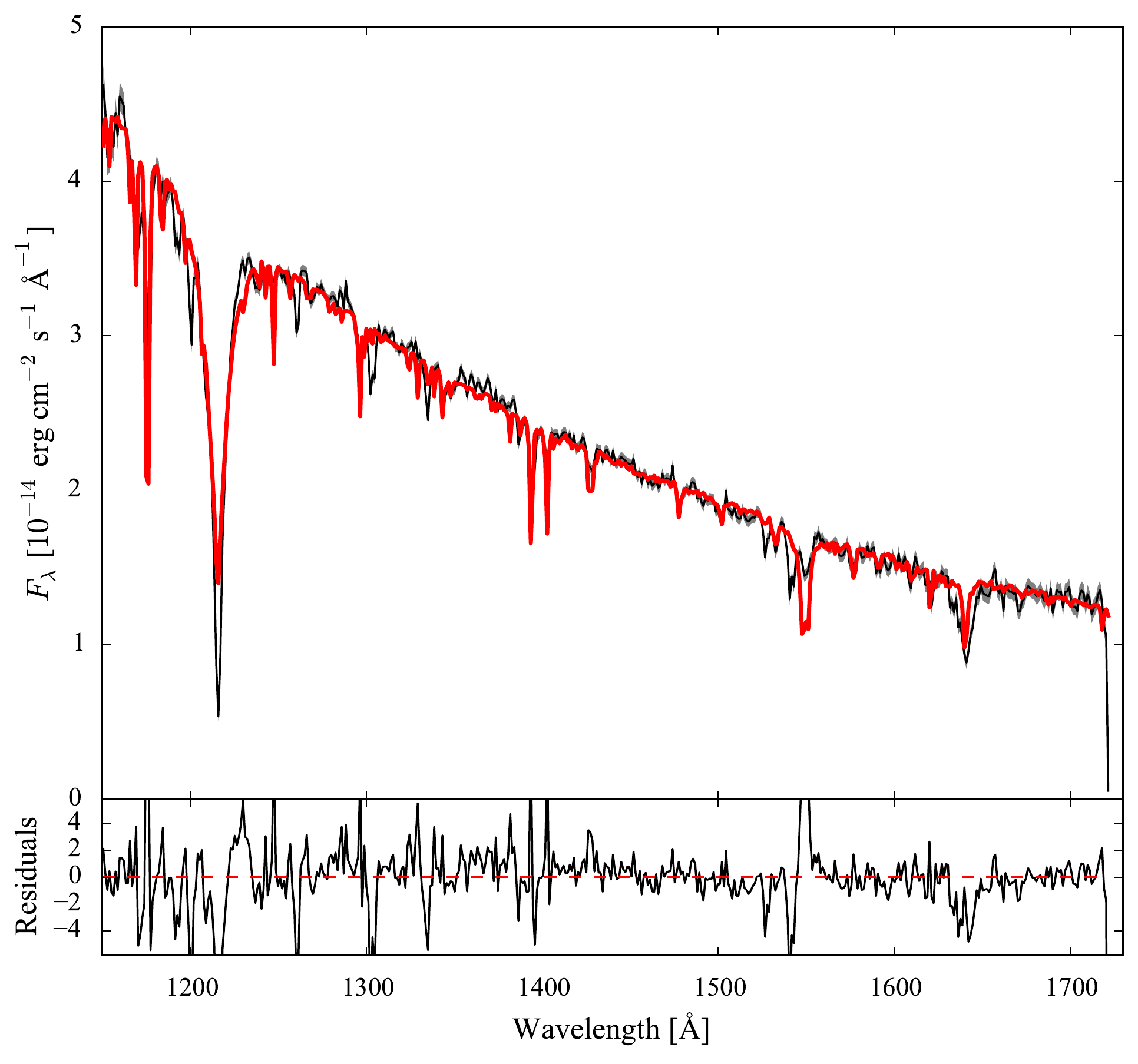}
\put (15,84) {DW\,UMa}
\end{overpic}
  \caption{Ultraviolet \hst spectra (black) of  DW\,UMa along with the best-fitting model (red), computed as described in Section~\ref{sec:uv_fit}.}
\end{figure*}
\newpage
\clearpage

\thispagestyle{empty}
\begin{table*}
  \centering
  \caption{Best-fitting parameters for U\,Gem.}
  \begin{tabular}{ l r }
    \toprule
     Instrument & FOS \\ [0.13cm]
     Observation date & 1992 Sep 28 \\ [0.13cm]
     $E(B-V)$ (mag) & $0.003^{+0.015}_{-0.003}$ \\ [0.13cm]
     $P_\mathrm{orb}$ (min) & 254.74 \\ [0.13cm]
     $\mathit{Gaia}$ EDR3 ID & 674214551557961984 \\ [0.13cm]
     $\varpi$ (mas) & $10.75\pm0.03$ \\ [0.13cm]
     $d$ (pc) & $93.0{\pm0.3}$ \\ [0.13cm]
     $Z$ (Z$_\mathrm{\odot}$) & 1.0 \\ [0.13cm]
     $T_\mathrm{eff}$ (K) & $33070^{+648}_{-616}$ \\ [0.13cm]
     $R_\mathrm{WD}$ ($0.01\,\mathrm{R}_\odot$) & $0.634 \pm 0.016$ \\ [0.13cm]
     $M_\mathrm{WD}$ ($\mathrm{M}_\odot$) & $1.160 \pm 0.013$ \\ [0.13cm]
     $\log(g)$ & $8.90 \pm 0.03$ \\ [0.13cm]
     $\langle \dot{M} \rangle$ $(10^{-10}\,\mathrm{M}_\odot$ yr$^\mathrm{-1}$) & $6.53^{+0.22}_{-0.17}$ \\ [0.13cm]
     WD contribution & 100\% \\ [0.13cm]
     \bottomrule
    \end{tabular}
\end{table*}
\vspace{0.8cm}
\begin{figure*}
    \centering
\begin{overpic}[width=0.8\textwidth]{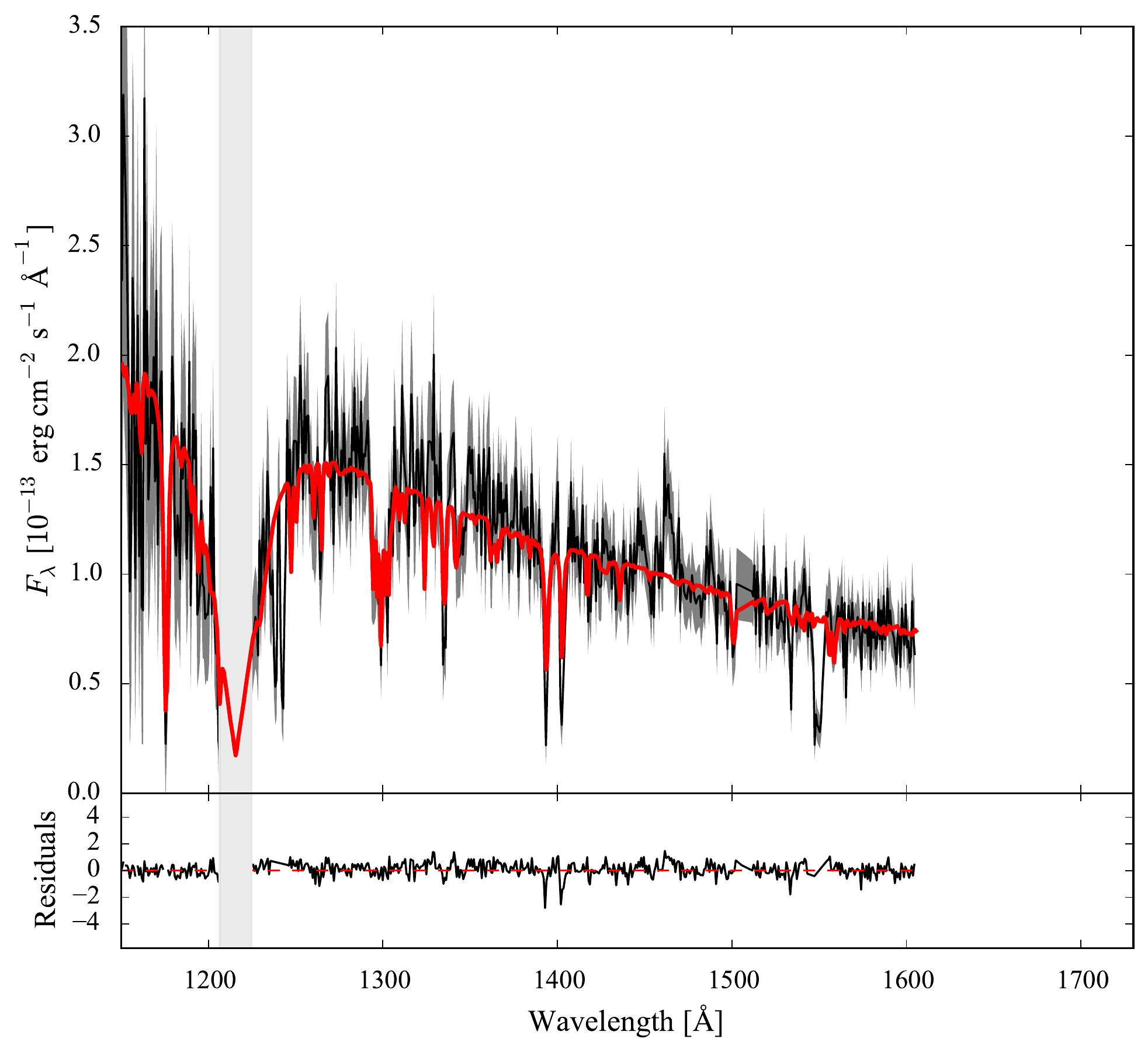}
\put (15,84) {U\,Gem}
\end{overpic}
  \caption{Ultraviolet \hst spectra (black) of  U\,Gem along with the best-fitting model (red), computed as described in Section~\ref{sec:uv_fit}.}
\end{figure*}
\newpage
\clearpage

\thispagestyle{empty}
\begin{table*}
  \centering
  \caption{Best-fitting parameters for SS\,Aur.}
  \begin{tabular}{ l r }
    \toprule
     Instrument & STIS \\ [0.13cm]
     Observation date & 2003 Mar 20 \\ [0.13cm]
     $E(B-V)$ (mag) & $0.047\pm0.033$ \\ [0.13cm]
     $P_\mathrm{orb}$ (min) & 263.23 \\ [0.13cm]
     $\mathit{Gaia}$ EDR3 ID & 968824328534823936 \\ [0.13cm]
     $\varpi$ (mas) & $4.02\pm0.03$ \\ [0.13cm]
     $d$ (pc) & $249{\pm2}$ \\ [0.13cm]
     $Z$ (Z$_\mathrm{\odot}$) & 0.1 \\ [0.13cm]
     $T_\mathrm{eff}$ (K) & $28627^{+190}_{-269}$ \\ [0.13cm]
     $R_\mathrm{WD}$ ($0.01\,\mathrm{R}_\odot$) & $0.86^{+0.15}_{-0.11}$ \\ [0.13cm]
     $M_\mathrm{WD}$ ($\mathrm{M}_\odot$) & $0.98^{+0.09}_{-0.11}$ \\ [0.13cm]
     $\log(g)$ & $8.56^{+0.16}_{-0.19}$ \\ [0.13cm]
     $\langle \dot{M} \rangle$ $(10^{-10}\,\mathrm{M}_\odot$ yr$^\mathrm{-1}$) & $7.3^{+3.0}_{-1.9}$ \\ [0.13cm]
     WD contribution & 100\% \\ [0.13cm]
     \bottomrule
    \end{tabular}
\end{table*}
\vspace{0.8cm}
\begin{figure*}
    \centering
\begin{overpic}[width=0.8\textwidth]{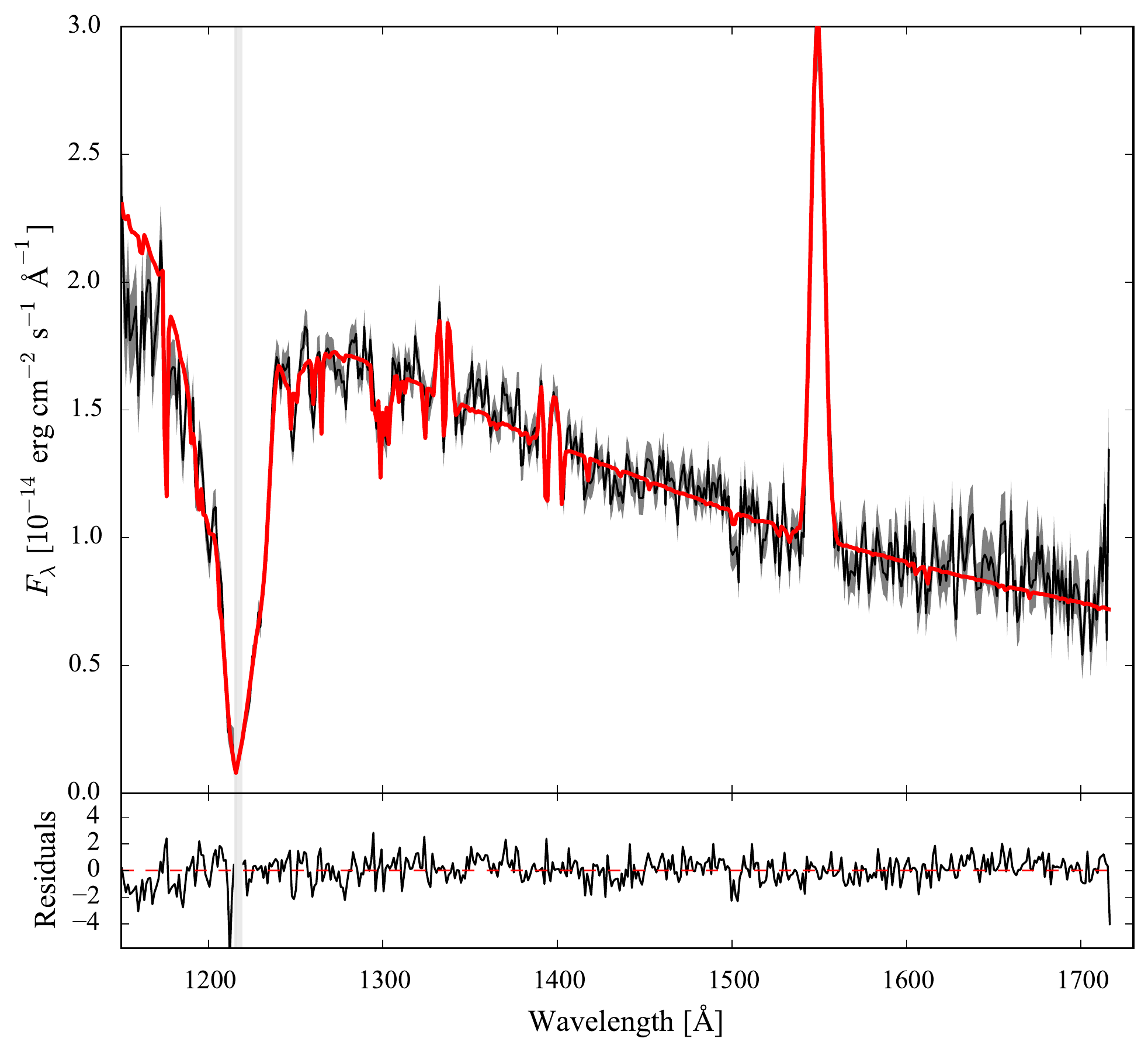}
\put (15,84) {SS\,Aur}
\end{overpic}
  \caption{Ultraviolet \hst spectra (black) of  SS\,Aur along with the best-fitting model (red), computed as described in Section~\ref{sec:uv_fit}.}
\end{figure*}
\newpage
\clearpage

\thispagestyle{empty}
\begin{table*}
  \centering
  \caption{Best-fitting parameters for RX\,And.}
  \begin{tabular}{ l r }
    \toprule
     Instrument & GHRS \\ [0.13cm]
     Observation date & 1996 Dec 22 \\ [0.13cm]
     $E(B-V)$ (mag) & $0.02\pm0.02$ \\ [0.13cm]
     $P_\mathrm{orb}$ (min) & 302.25 \\ [0.13cm]
     $\mathit{Gaia}$ EDR3 ID & 374510294830244992 \\ [0.13cm]
     $\varpi$ (mas) & $5.08\pm0.03$ \\ [0.13cm]
     $d$ (pc) & $197{\pm1}$ \\ [0.13cm]
     $Z$ (Z$_\mathrm{\odot}$) & 0.5 \\ [0.13cm]
     $T_\mathrm{eff}$ (K) & $33900^{+634}_{-995}$ \\ [0.13cm]
     $R_\mathrm{WD}$ ($0.01\,\mathrm{R}_\odot$) & $1.12^{+0.12}_{-0.08}$ \\ [0.13cm]
     $M_\mathrm{WD}$ ($\mathrm{M}_\odot$) & $0.81^{+0.06}_{-0.08}$ \\ [0.13cm]
     $\log(g)$ & $8.25^{+0.1}_{-0.13}$ \\ [0.13cm]
     $\langle \dot{M} \rangle$ $(10^{-10}\,\mathrm{M}_\odot$ yr$^\mathrm{-1}$) & $26.0^{+6.0}_{-4.0}$ \\ [0.13cm]
     WD contribution & 100\% \\ [0.13cm]
     \bottomrule
    \end{tabular}
\end{table*}
\vspace{0.8cm}
\begin{figure*}
    \centering
\begin{overpic}[width=0.8\textwidth]{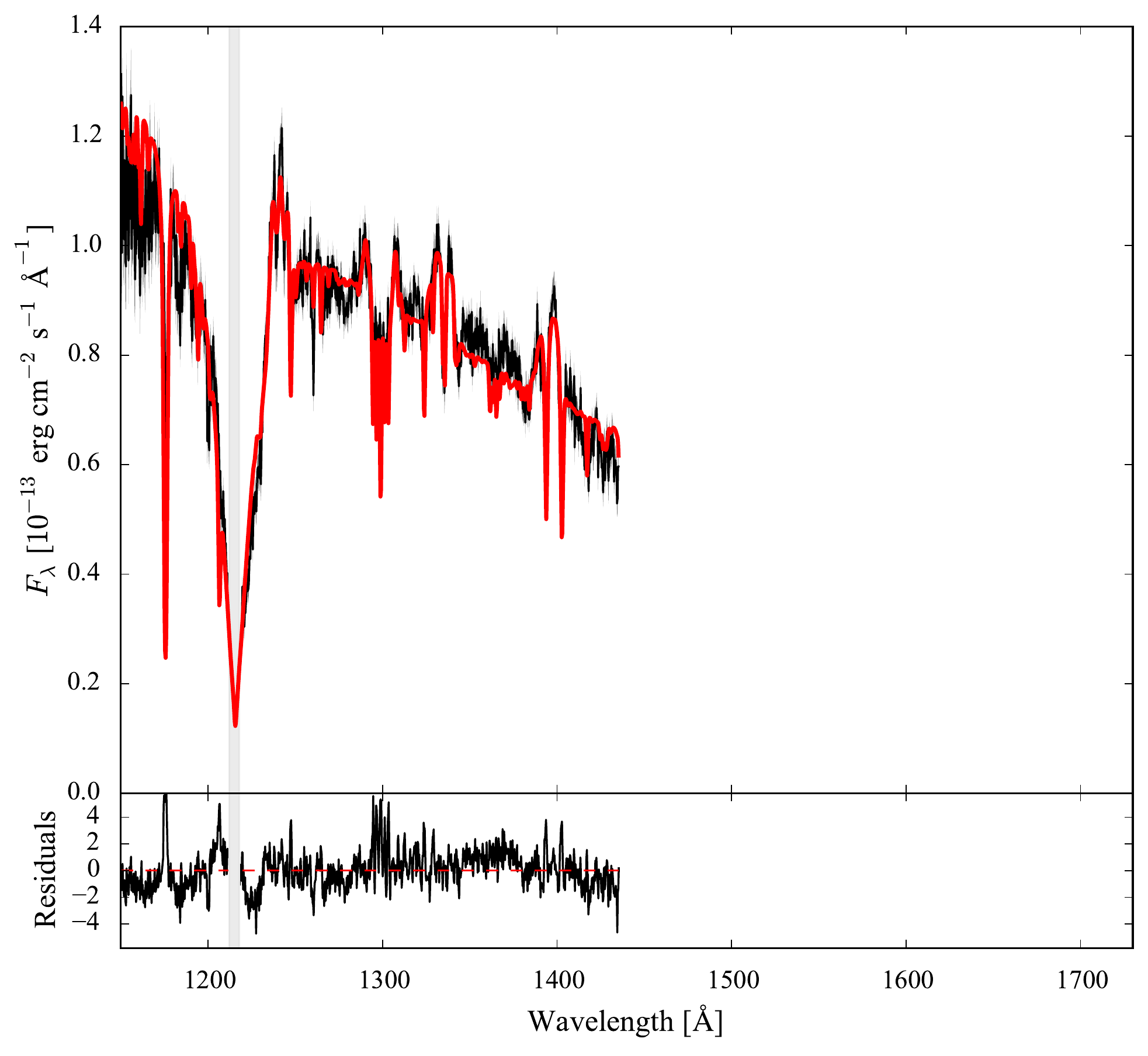}
\put (15,84) {RX\,And}
\end{overpic}
  \caption{Ultraviolet \hst spectra (black) of  RX\,And along with the best-fitting model (red), computed as described in Section~\ref{sec:uv_fit}.}
\end{figure*}
\newpage
\clearpage

\thispagestyle{empty}
\begin{table*}
  \centering
  \caption{Best-fitting parameters for V442\,Cen.}
  \begin{tabular}{ l r }
    \toprule
     Instrument & STIS \\ [0.13cm]
     Observation date & 2002 Dec 29 \\ [0.13cm]
     $E(B-V)$ (mag) & $0.048\pm0.015$ \\ [0.13cm]
     $P_\mathrm{orb}$ (min) & 662.4 \\ [0.13cm]
     $\mathit{Gaia}$ EDR3 ID & 5398867830598349952 \\ [0.13cm]
     $\varpi$ (mas) & $2.92\pm0.04$ \\ [0.13cm]
     $d$ (pc) & $343\pm5$ \\ [0.13cm]
     $Z$ (Z$_\mathrm{\odot}$) & 0.01 \\ [0.13cm]
     $T_\mathrm{eff}$ (K) & $29802^{+211}_{-247}$ \\ [0.13cm]
     $R_\mathrm{WD}$ ($0.01\,\mathrm{R}_\odot$) & $1.35^{+0.14}_{-0.12}$ \\ [0.13cm]
     $M_\mathrm{WD}$ ($\mathrm{M}_\odot$) & $0.64^{+0.06}_{-0.05}$ \\ [0.13cm]
     $\log(g)$ & $7.98 \pm 0.12$ \\ [0.13cm]
     $\langle \dot{M} \rangle$ $(10^{-10}\,\mathrm{M}_\odot$ yr$^\mathrm{-1}$) & $25.0^{+5.0}_{-4.0}$ \\ [0.13cm]
     WD contribution & 100\% \\ [0.13cm]
     \bottomrule
    \end{tabular}
\end{table*}
\vspace{0.8cm}
\begin{figure*}
    \centering
\begin{overpic}[width=0.8\textwidth]{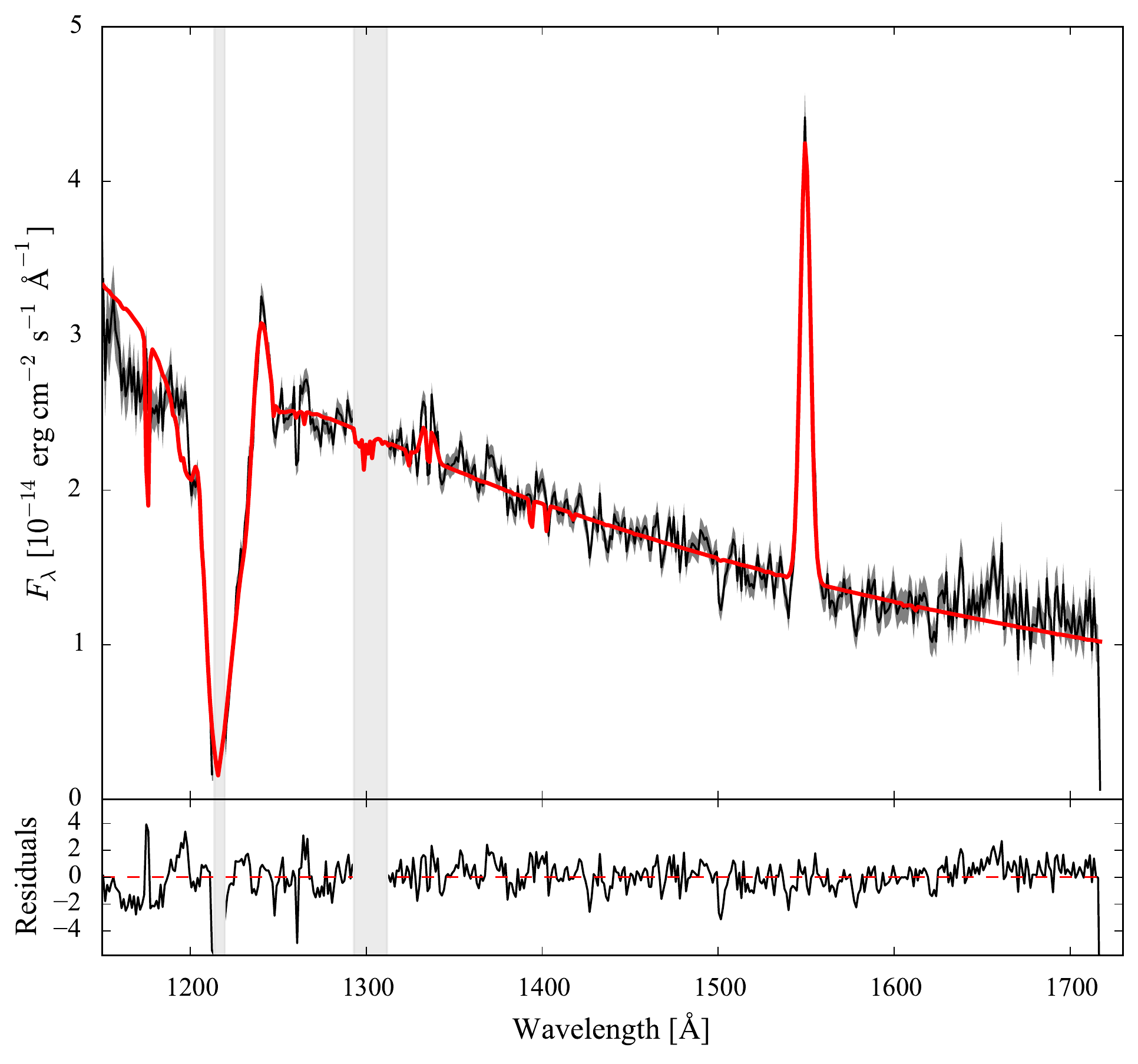}
\put (15,84) {V442\,Cen}
\end{overpic}
  \caption{Ultraviolet \hst spectra (black) of  V442\,Cen along with the best-fitting model (red), computed as described in Section~\ref{sec:uv_fit}.}
\end{figure*}
\newpage
\clearpage

\section{Correction applied to the $\log(\lowercase{g})$-distance relationships from the literature}\label{ap:correction}
\begin{figure}
\includegraphics[width=\columnwidth]{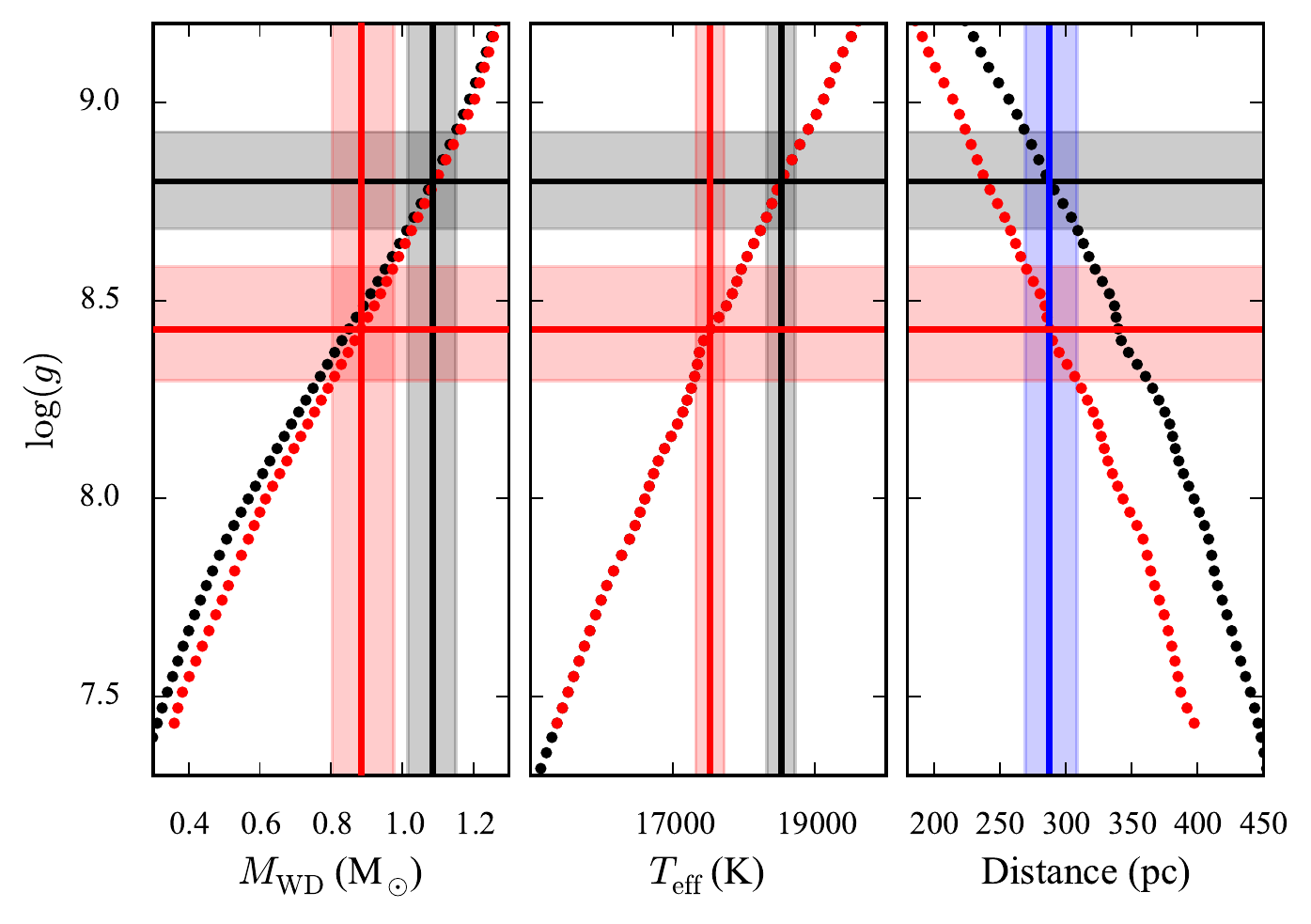}
\caption{$\log(g)$-$\mwd$ (left panel), $\log(g)$-$\teff$ (middle panel) and $\log(g)$-distance relationships (right panel) for EF\,Peg from \protect\cite{Howell+2002} (black dots) and their correction (red dots) accounting for the dependency of the mass-radius relationship on the white dwarf $\teff$ and the reddening. The blue line represents the distance to the system derived from its \gaia EDR3 parallax while the black and red lines correspond to the $\log(g)$, $\mwd$ and $\teff$ from \protect\cite{Howell+2002} and the procedure described in Appendix~\ref{ap:correction}, respectively. The shaded areas represent the $1\sigma$ statistical uncertainties.}\label{fig:correction_efpeg}
\end{figure}

The method described by \citet{Gaensicke+2005} allows us to investigate of the correlation between $\log(g)$ and the best-fit value for \teff, for different distances.
It consists of fitting the ultraviolet data by stepping through a grid of atmosphere models with fixed values for $\log(g)$, leaving the temperature and scaling factor free.
For each best-fitting model, the mass and radius of the white dwarf are then computed assuming a mass-radius relationship and the distance to the system is then estimated using Equation~\ref{eq:scaling}. 

\cite{Howell+2002,Szkody+2002a,Gaensicke+2005,Gaensicke+2006} used the above method to compute the $\log(g)$-distance relationship for AM\,Her, BC\,UMa, BW\,Scl, EF\,Peg, EG\,Cnc,  HV\,Vir, LL\,And and SW\,UMa. However, these works assumed a zero-temperature mass-radius relationship and, therefore, did not account for its dependency on the white dwarf $\teff$. Moreover, the authors did not apply any reddening correction in their spectral analysis. 
To take into account these effects, we corrected the $\log(g)$-distance relationship by applying the procedure described in the following, which we illustrate using EF\,Peg as an example:

\begin{enumerate}
\item we retrieved the $\log(g)$-$\mwd$, $\log(g)$-$\teff$ and $\log(g)$-distance relationships for EF\,Peg from the work by \cite{Howell+2002} (black dots Figure~\ref{fig:correction_efpeg}). For each couple of $\log(g)$-$\mwd$ values, we computed the corresponding $\rwd$. This, combined with the $\log(g)$-distance provide the scaling factor ($S$) for each best-fitting model identified by each couple of $\log(g)$-$\teff$ values.

\item To take into account the dependency of the mass-radius relationship on the white dwarf temperature, for each couple of values $\log(g)$-$\teff$, we computed the corresponding white dwarf $\mwd$ and $\rwd$ using the mass-radius relationship by \citet{Holberg+2006,Tremblay+2011} (red dots in the left panel Figure~\ref{fig:correction_efpeg}). The difference between the two mass-radius relationships is more important the lower the mass of the white dwarf. 

\item To correct the results by \cite{Howell+2002} for the reddening, we computed the scaling factor ($\mathrm{S}_\mathrm{corr}$)  corresponding to the best-fitting model reddened by $E(B-V)$ using the  following equation:
\begin{equation}
\mathrm{S}_\mathrm{corr} = S \times 10^{0.4 A_\lambda}
\end{equation}
where $A_\lambda$ is the extinction from \cite{Cardelli+1989} and $S$ the scaling factor computed at point~1.

\item Combining the $\rwd$ computed at point~2 and $\mathrm{S}_\mathrm{corr}$, we derived the distance to the system (red dots in the right panel in Figure~\ref{fig:correction_efpeg}). The non-null reddening of EF\,Peg ($E(B-V) = 0.048$) implies that the computation by \cite{Howell+2002} systematically overestimated the distance to the system.
\end{enumerate}

Assuming the distance to the system derived from its \gaia EDR3 parallax (vertical blue line), it is now possible to derive the mass of the white dwarf. In the case of EF\,Peg, $d= 288$\,pc, implies $\teff = 18\,522\,$K and $\mwd  = 1.08\,\msun$ for the original computation by \cite{Howell+2002} (black lines) and $\teff = 17\,526\,$K and $\mwd  = 0.88\,\msun$ (red lines) for the corrected relationship we computed. We applied this same procedure to AM\,Her, BC\,UMa, BW\,Scl, EG\,Cnc,  HV\,Vir, LL\,And and SW\,UMa and the corrected masses are reported in Table~\ref{tab:comparison}.

\section{White dwarf parameters from the literature}\label{ap:wd_par}
The following table lists the 54 reliable mass measurements of CV white dwarfs from the literature. A measurement is defined reliable when the white dwarf can be unambiguously detected either in the ultraviolet spectrum (from a broad Ly$\alpha$ absorption profile and, possibly, sharp absorption metal lines) or its ingress and egress can be seen in eclipse light curves. 

\begin{table*}
\vspace{-3.7cm}
\thisfloatpagestyle{empty}
\small
\caption{CVs with reliable measurements of the white dwarf quiescent effective temperature and mass, as compiled from the literature, sorted by their orbital periods. The method employed are as follows: eclipse, analysis of eclipse light curves; gr, gravitational redshift; rv, radial velocity curve; sp, spectrophotometric modelling. The mass accretion rates have been computed using eq.~1 from \protect\cite{Townsley+2009}. AE\,Aqr, KIC\,5608384 and V1640\,Her have not been considered in Section~\ref{subsec:evolution} as they host nuclear evolved donors.}\label{tab:masses_lit}
\begin{tabular}{@{} lcccccccc @{}}
\toprule
System & $P_\mathrm{orb}$ & $\teff$ & $\mwd$ & $\langle \dot{M} \rangle $ & Magnetic & Method & Reference \\
       &   (min)           &  (K)   & ($\mathrm{M}_\odot$) & $(\mathrm{M}_\odot$ yr$^\mathrm{-1}$) & &  &  & \\
\midrule
SDSS\,J150722.30+523039.8 & 66.61 & $11\,300\pm1000$ & $0.89\pm0.01$ & $(2.0^{+0.8}_{-0.6}) \times 10^{-11}$ & no & eclipse & 1 \\[0.08cm] 
PHL\,1445 & 76.3 & $13\,200\pm700$ & $0.73\pm0.03$ & $(8.0^{+3.0}_{-2.0}) \times 10^{-11}$ & no & eclipse & 2 \\[0.08cm] 
SDSS\,J143317.78+101123.3 & 78.11 & $12\,700\pm1500$ & $0.86\pm0.01$ & $(4.0^{+2.0}_{-1.0}) \times 10^{-11}$ & no & eclipse & 1 \\[0.08cm] 
SDSS\,J123813.73--033932.9 & 80.52 & -- & $0.98\pm0.06$ & -- & no & gr & 3 \\[0.08cm] 
V455\,And & 81.08 & $11\,300 \pm 1000$ & $0.87 \pm 0.08$ & $(2.2^{+1.0}_{-0.7}) \times 10^{-11}$ & no & sp & 35 \\[0.08cm]
WZ\,Sge & 81.6 & $14\,900\pm250$ & $0.85\pm0.04$ & $(7.0^{+2.0}_{-1.0}) \times 10^{-11}$ & no & rv,gr,sp & 4,5 \\[0.08cm] 
ASASSN-17jf & 81.78 & -- & $0.669 \pm 0.031$ & -- & no & eclipse & 36\\[0.08cm]
SDSS\,J150137.22+550123.4 & 81.85 & $13\,400\pm1100$ & $0.72^{+0.02}_{-0.01}$ & $(9.0\pm3.0)\times 10^{-11}$ & no & eclipse & 6 \\[0.08cm] 
SDSS\,J103533.02+055158.4 & 82.09 & $10\,000\pm1100$ & $0.83\pm0.01$ & $(1.6^{+0.8}_{-0.6}) \times 10^{-11}$ & no & eclipse & 1 \\[0.08cm] 
SSS100615\,J200331--284941 & 84.53 & $13\,600\pm1500$ & $0.88\pm0.03$ & $(4.0\pm2.0)\times 10^{-11}$ & no & eclipse & 6 \\[0.08cm] 
SDSS\,J150240.98+333423.9 & 84.83 & $11\,800\pm1200$ & $0.71\pm0.0$ & $(6.0^{+3.0}_{-2.0}) \times 10^{-11}$ & no & eclipse & 1 \\[0.08cm] 
SDSS\,J090350.73+330036.1 & 85.07 & $13\,300\pm1700$ & $0.87\pm0.01$ & $(4.0^{+3.0}_{-2.0}) \times 10^{-11}$ & no & eclipse & 1 \\[0.08cm] 
CSS080623\,J140454--102702 & 85.79 & $15\,500\pm1700$ & $0.71\pm0.02$ & $(1.7^{+0.9}_{-0.6}) \times 10^{-10}$ & no & eclipse & 6 \\[0.08cm] 
ASASSN--14ag & 86.85 & $14\,000^{+2200}_{-2000}$ & $0.63\pm0.04$ & $(2.0^{+2.0}_{-1.0}) \times 10^{-10}$ & no & eclipse & 7 \\[0.08cm] 
XZ\,Eri & 88.07 & $15\,300\pm2400$ & $0.77\pm0.02$ & $(1.2^{+1.0}_{-0.6}) \times 10^{-10}$ & no & eclipse & 1 \\[0.08cm] 
ASASSN-16kr & 88.17 & -- & $0.952 \pm 0.018$ & -- & no & eclipse & 36\\[0.08cm]
CRTS\,SSS11126\,J052210--350530 & 89.56 & -- & $0.760 \pm 0.023$ & -- & no & eclipse & 36\\[0.08cm]
SDSS\,J105754.25+275947.5 & 90.42 & $13\,300\pm1100$ & $0.8\pm0.01$ & $(6.0\pm2.0)\times 10^{-11}$ & no & eclipse & 8 \\[0.08cm] 
SDSS\,J122740.83+513925.0 & 90.66 & $15\,900\pm1400$ & $0.8\pm0.02$ & $(1.3^{+0.5}_{-0.4}) \times 10^{-10}$ & no & eclipse & 1 \\[0.08cm] 
OY\,Car & 90.89 & $18\,600^{+2800}_{-1600}$ & $0.88\pm0.01$ & $(1.6^{+1.3}_{-0.8}) \times 10^{-10}$ & no & eclipse & 6 \\[0.08cm] 
CTCV\,J2354--4700 & 94.39 & $14\,800\pm700$ & $0.94\pm0.03$ & $(4.9^{+1.3}_{-1.0}) \times 10^{-11}$ & no & eclipse & 1 \\[0.08cm] 
LSQ1725 & 94.67 & -- & $0.97\pm0.03$ & -- & yes & eclipse & 9 \\[0.08cm] 
SSS130413\,J094551--194402 & 94.71 & -- & $0.84\pm0.03$ & -- & no & eclipse & 6 \\[0.08cm]
CSS110113\,J043112--031452 & 95.11 & $14\,500\pm2200$ & $1.0^{+0.04}_{-0.01}$ & $(4.0^{+3.0}_{-2.0}) \times 10^{-11}$ & no & eclipse & 6 \\[0.08cm] 
SDSS\,J115207.00+404947.8 & 97.56 & $15\,900\pm2000$ & $0.62\pm0.04$ & $(3.0\pm2.0)\times 10^{-10}$ & no & eclipse & 6 \\[0.08cm] 
EX\,Hya & 98.25 & -- & $0.78\pm0.03$ & -- & yes & rv & 28 \\[0.08cm] 
OU\,Vir & 104.7 & -- & $0.7\pm0.01$ & -- & no & eclipse & 1 \\[0.08cm] 
HT\,Cas & 106.1 & $14\,000\pm1000$ & $0.61\pm0.04$ & $(2.2^{+1.0}_{-0.7}) \times 10^{-10}$ & no & eclipse & 10 \\[0.08cm] 
IY\,UMa & 106.43 & $17\,750\pm1000$ & $0.95^{+0.01}_{-0.03}$ & $(9.0^{+3.0}_{-2.0}) \times 10^{-11}$ & no & eclipse & 6 \\[0.08cm] 
VW\,Hyi & 107.0 & $19\,300\pm250$ & $0.7\pm0.2$ & $(3.3^{+0.6}_{-0.4}) \times 10^{-10}$ & no & gr & 11,12,13 \\[0.08cm] 
Z\,Cha & 107.28 & $16\,300\pm1400$ & $0.8\pm0.01$ & $(1.3^{+0.5}_{-0.4}) \times 10^{-10}$ & no & eclipse & 6 \\[0.08cm] 
SDSS\,J090103.94+480911.0 & 112.15 & $14\,900\pm2000$ & $0.75\pm0.02$ & $(1.3^{+0.8}_{-0.6}) \times 10^{-10}$ & no & eclipse & 6 \\[0.08cm] 
V713\,Cep & 123.0 & $17\,000^{+6000}_{-3000}$ & $0.7\pm0.01$ & $(3.0^{+6.0}_{-2.0}) \times 10^{-10}$ & no & eclipse & 6 \\[0.08cm] 
DV\,UMa & 123.63 & $17\,400\pm1900$ & $1.09\pm0.03$ & $(5.0^{+3.0}_{-2.0}) \times 10^{-11}$ & no & eclipse & 6 \\[0.08cm] 
CTCV\,J1300--3052 & 128.07 & $11\,000\pm1000$ & $0.72\pm0.02$ & $(4.0\pm2.0)\times 10^{-11}$ & no & eclipse & 6 \\[0.08cm] 
SDSS\,J170213.26+322954.1 & 144.12 & $15200\pm1200$ & $0.91\pm0.03$ & $(6.0\pm2.0)\times 10^{-11}$ & no & eclipse & 1 \\[0.08cm] 
HY\,Eri & 171.32 & -- & $0.42\pm0.05$ & -- & yes & eclipse,rv & 14 \\[0.08cm] 
AM\,Her & 185.7 & $19\,800\pm700$ & $0.78\pm0.15$ & $(3.0^{+5.0}_{-1.0}) \times 10^{-10}$ & yes & sp & 15 \\[0.08cm] 
DW\,UMa & 196.7 & $50\,000\pm1000$ & $0.77\pm0.07$ & $(1.4^{+0.6}_{-0.4}) \times 10^{-8}$ & no & eclipse & 16 \\[0.08cm] 
HS\,0220+0603 & 214.86 & $30\,000\pm5000$ & $0.87\pm0.09$ & $(1.1^{+1.3}_{-0.6}) \times 10^{-9}$ & no? & eclipse & 17 \\[0.08cm] 
IP\,Peg & 227.8 & -- & $1.16\pm0.02$ & -- & no & eclipse & 18 \\[0.08cm] 
KIS\,J1927+4447 & 238.04 & $23\,000\pm3000$ & $0.69\pm0.07$ & $(1.0^{+1.0}_{-0.5}) \times 10^{-9}$ & no & eclipse & 19 \\[0.08cm] 
GY\,Cnc & 252.64 & $25\,900\pm2300$ & $0.88\pm0.02$ & $(6.0^{+3.0}_{-2.0}) \times 10^{-10}$ & no & eclipse & 6 \\[0.08cm] 
U\,Gem & 254.7 & $30\,000\pm1000$ & $1.2\pm0.05$ & $(3.1^{+1.2}_{-0.8}) \times 10^{-10}$ & no & rv,gr,sp & 20,21,22,23,24,25 \\[0.08cm] 
SDSS\,J100658.40+233724.4 & 267.71 & $16\,000\pm1000$ & $0.82\pm0.11$ & $(1.1^{+1.0}_{-0.5}) \times 10^{-10}$ & no & eclipse & 6 \\[0.08cm] 
DQ\,Her & 278.8 & -- & $0.6\pm0.07$ & -- & yes & rv & 26 \\[0.08cm] 
V1640\,Her & 299.31 & -- & $0.87\pm0.01$ & -- & no & rv & 27 \\[0.08cm] 
V347\,Pup & 334.0 & -- & $0.63\pm0.04$ & -- & no & rv & 29 \\[0.08cm] 
1RXS\,J064434.5+334451 & 387.9 & -- & $0.73\pm0.07$ & -- & no & eclipse & 30 \\[0.08cm] 
EM\,Cyg & 418.9 & -- & $1.0\pm0.06$ & -- & no & rv & 31 \\[0.08cm] 
AC\,Cnc & 432.7 & -- & $0.76\pm0.03$ & -- & no & rv & 32 \\[0.08cm] 
V363\,Aur & 462.6 & -- & $0.9\pm0.06$ & -- & no? & rv & 32 \\[0.08cm] 
KIC\,5608384 & 524.32 & -- & $0.46\pm0.02$ & -- & no & rv & 33 \\[0.08cm] 
AE\,Aqr & 592.8 & -- & $0.63\pm0.05$ & -- & yes & rv & 34 \\[0.08cm] 
\bottomrule
\end{tabular}
\begin{tablenotes}
\item \textbf{Notes.} For OU\,Vir and SSS130413\,J094551--194402, the effective temperatures reported by \cite{Savoury+2011} and \cite{McAllister+2019} have been obtained during descent from superoutburst and within a month from the last outburst, respectively. We do not include their $\teff$ in our analysis since only represent an upper limit for their quiescent temperatures.
\item \textbf{References.}
(1) \cite{Savoury+2011}, (2) \cite{McAllister+2015}, (3) \cite{Pala+2019}, (4) \cite{Long+2004}, (5) \cite{Steeghs+2007}, (6) \cite{McAllister+2019}, (7) \cite{McAllister+2017a}, (8) \cite{McAllister+2017}, (9) \cite{Fuchs+2016}, (10) \cite{Feline+2005}, (11) \cite{Gaensicke+2006}, (12) \cite{Sion+1997}, (13) \cite{Smith+2006}, (14) \cite{Beuermann+2020}, (15) \cite{Gaensicke+2006}, (16) \cite{Araujo+2003}, (17) \cite{Pablo+2015}, (18) \cite{Copperwheat+2010}, (19) \cite{Littlefair+2014}, (20) \cite{Zhang+1987}, (21) \cite{Sion+1998}, (22) \cite{Long+1999}, (23) \cite{Naylor+2005}, (24) \cite{Long+2006}, (25) \cite{Echevarria+2007}, (26) \cite{Horne+1993}, (27) \cite{Ashley+2020}, (28) \cite{Echevarria+2016}, (29) \cite{Thoroughgood+2005}, (30) \cite{Hernandez+2017}, (31) \cite{Welsh+2007}, (32) \cite{Thoroughgood+2004}, (32) \cite{Thoroughgood+2004}, (33) \cite{Yu+2019}, (34) \cite{Echevarria+2008}, (35) Pala et al.,in preparation, (36) \cite{Wild+2021}.
\end{tablenotes}
\end{table*}

\section{CVs observed with \hst}\label{ap:hst_archive}
\subsubsection{Notes on individual objects}\label{subsubsec:notes} 
\begin{figure}
\includegraphics[width=\columnwidth]{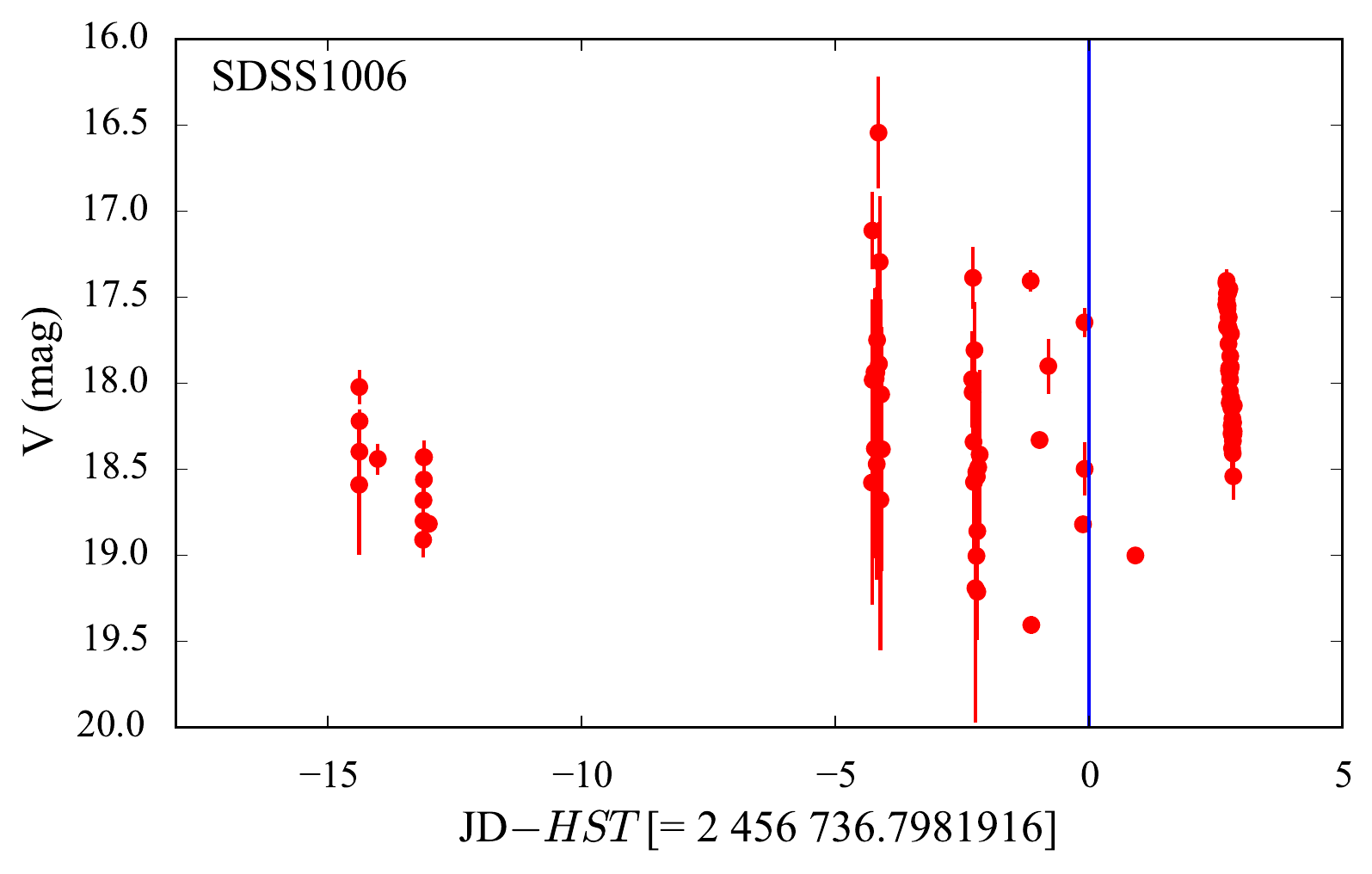}
\caption{Light curve of SDSS1006 as retrieved from the AAVSO database. The blue line represents the epoch of the \hst observation, which has been subtracted from the Julian date on the x--axis.}\label{fig:lc_sdss1006}
\end{figure}
\paragraph*{Systems observed after an outbursts}\mbox{}\\
\citet{Pala+2017} reported that IY\,UMa went into outburst about 25 and five days before the \hst observations. Nonetheless, we can conclude that the white dwarf emission has returned to the quiescent level, as our white dwarf parameters are consistent (see Section~\ref{subsec:comparison_eclipsing}) with the independent measurement by \cite{McAllister+2019}.
For SDSS\,J100658.41+233724.4 (SDSS1006), SDSS\,J040714.78-064425.1 (SDSS0407) and V844\,Her, \citet{Pala+2017} reported that the \hst observations have been acquired during the quiescent state. However, from an inspection of the light curve retrieved from the American Association of Variable star Observations (AAVSO) database\footnote{\url{https://www.aavso.org/data-download}}, we noticed that SDSS0407 and V844\,Her underwent an outburst within two months before the \hst observations, while SDSS1006 system showed some level of variability just before the \hst data were acquired (Figure~\ref{fig:lc_sdss1006}). We therefore did not include these systems in our sample since the analysis of their ultraviolet data would provide only a lower limit on their $\mwd$.

\newgeometry{right=0.8cm,left=0.8cm,bottom=1cm}
\onecolumn
\captionsetup{width=18cm}
\setlength{\tabcolsep}{0.1cm}
\begin{longtable}{lccccccl}
\caption{CVs with ultraviolet \hst observations providing (at least) the full coverage of the Ly$\alpha$ ($1100-1600\,$\AA) from the white dwarf photosphere with a resolution $R \simeq 1000-3000$.  Systems observed in the quiescent phase and in which the white dwarf signature is clearly detected in the spectra have been analysed in this work. For the remaining objects, we specify here why the corresponding data were not suitable for our analysis. The systems are sorted by their orbital periods.}\label{tab:cvs_with_hst_obs}\\
 \toprule
 System         & $\alpha$ & $\delta$ & Period & Instrument & Grating & Central         & Comment  \\ 
                     &               &                & (min)  &                 &             &  wavelength &                 \\ \midrule
 \endfirsthead
\multicolumn{8}{c}%
{{\tablename\ \thetable{} -- continued from previous page}} \\
\toprule
 System         & $\alpha$ & $\delta$ & Period & Instrument & Grating & Central         & Comment  \\ 
                     &               &                & (min)  &                 &             &  wavelength &                 \\ \midrule\endhead
\bottomrule
\endfoot
\bottomrule
\endlastfoot
\multirow{2}{*}{CSS\,120422:111127+571239} & \multirow{2}{*}{11:11:26.83} & \multirow{2}{*}{+57:12:38.6} & \multirow{2}{*}{55.37} & COS & G140L & 1105 & \multirow{2}{*}{Evolved donor.} \\
           &                     &                     &           & COS & G140L & 1280  & \\[0.1cm]
\multirow{2}{*}{V485\,Cen} & \multirow{2}{*}{12:57:23.12} & \multirow{2}{*}{--33:12:05.64} & \multirow{2}{*}{59.03} & STIS & G140L & 1425 & \multirow{2}{*}{Evolved donor.} \\
           &                     &                     &           & COS & G140L & 1105 & \\[0.1cm]
EI\,Psc & 23:29:54.22 & +06:28:11.5 & 64.18 & STIS & G140L & 1425 & Evolved donor. \\[0.1cm]
\multirow{2}{*}{SDSS\,J150722.30+523039.8} & \multirow{2}{*}{15:07:22.11} & \multirow{2}{*}{+52:30:40.46} & \multirow{2}{*}{66.61} & COS & G140L & 1230 & \multirow{2}{*}{This work.}\\
           &                     &                     &           & STIS & G230L & 2376 & \\[0.1cm]
SDSS\,J074531.91+453829.5 & 07:45:31.90 & +45:38:29.16 & 76.0 & COS & G140L & 1105 & This work. \\[0.1cm]
GW\,Lib & 15:19:55.26 & --25:00:24.24 & 76.78 & COS & G140L & 1105 & This work. \\[0.1cm]
SDSS\,J143544.02+233638.7 & 14:35:44.01 & +23:36:38.84 & 78.0 & COS & G140L & 1105 & This work. \\[0.1cm]
OT\,J213806.6+261957 & 21:38:06.71 & +26:19:57.03 & 78.1 & COS & G140L & 1105 & This work. \\[0.1cm]
BW\,Scl & 23:53:00.99 & --38:51:47.67 & 78.23 & STIS & E140M & 1425 & This work. \\[0.1cm]
V844\,Her & 16:25:01.74 & +39:09:26.2 & 78.69 & STIS & G140L & 1425 & Obtained after to an outburst. \\[0.1cm]
LL\,And & 00:41:51.51 & +26:37:20.75 & 79.28 & STIS & G140L & 1425 & This work. \\[0.1cm]
GZ\,Cet & 01:37:01.06 & --09:12:34.4 & 79.69 & COS & G140L & 1105 & Evolved donor. \\[0.1cm]
SDSS\,J013701.06--091234.8 & 01:37:01.04 & --09:12:35.65 & 79.71 & COS & G140L & 1105 & Evolved donor. \\[0.1cm]
V386\,Ser & 16:10:33.61 & --01:02:23.0 & 80.52 & COS & G140L & 1105 & Obtained after an outburst. \\[0.1cm]
\multirow{2}{*}{SDSS\,J123813.73-033933.0} & \multirow{2}{*}{12:38:13.73} & \multirow{2}{*}{--03:39:32.8} & \multirow{2}{*}{80.52} & \multirow{2}{*}{COS} & \multirow{2}{*}{G140L} & \multirow{2}{*}{1105} & UV data analysed in combination \\
& & & & & & & with the Gaia parallaxes by \citet{Pala+2019}. \\[0.1cm]
\multirow{3}{*}{V455\,And} & \multirow{3}{*}{23:34:01.37} & \multirow{3}{*}{+39:21:38.64} & \multirow{3}{*}{81.08} & STIS & G140L & 1425 & Weak detection of the white dwarf.\\
           &                     &                     &           &  COS & G140L & 1105 & Obtained after an outburst.\\
           &                     &                     &           & COS & G160M & 1577 & Obtained after an outburst. \\[0.1cm]
AL\,Com & 12:32:25.77 & +14:20:41.78 & 81.6 & STIS & G140L & 1425 & This work. \\[0.1cm]
WZ\,Sge & 20:07:36.58 & +17:42:14.34 & 81.63 & FOS & G160L	 & 2440 & This work. \\[0.1cm]
PU\,CMa & 06:40:47.65 & --24:23:13.51 & 81.63 & COS & G140L & 1105 & Obtained after an outburst. \\[0.1cm]
SW\,UMa & 08:36:42.70 & +53:28:38.07 & 81.81 & STIS & G140L & 1425 & This work. \\[0.1cm]
V1108\,Her & 18:39:26.14 & +26:04:09.97 & 81.87 & COS & G140L & 1105 & This work. \\[0.1cm]
ASAS\,J002511+1217.2 & 00:25:11.04 & +12:17:11.79 & 82.0 & COS & G140L & 1105 & This work. \\[0.1cm]
HV\,Vir & 13:21:03.18 & +01:53:28.92 & 82.18 & STIS & G140L & 1425 & This work. \\[0.1cm]
SDSS\,J103533.02+055158.4 & 10:35:33.01 & +05:51:59.00 & 82.22 & COS & G140L & 1105 & This work. \\[0.1cm]
WX\,Cet & 01:17:04.17 & --17:56:22.38 & 83.9 & STIS & E140M & 1425 & This work. \\[0.1cm]
CC\,Scl & 23:15:31.80 & --30:48:48.57 & 84.1 & COS & G140L & 1105 & Obtained after an outburst. \\[0.1cm]
KV\,Dra & 14:50:38.31 & +64:03:28.6 & 84.61 & STIS & G140L & 1425 & White dwarf not detected. \\[0.1cm]
SDSS\,J075507.70+143547.6 & 07:55:07.68 & +14:35:47.26 & 84.76 & COS & G140L & 1105 & This work. \\[0.1cm]
SDSS\,J080434.20+510349.2 & 08:04:34.15 & +51:03:48.71 & 84.97 & COS & G140L & 1105 & This work. \\[0.1cm]
GQ\,Mus & 11:52:02.42 & --67:12:20.6 & 85.49 & FOS & G160L & 2440 &  Classical nova, white dwarf not detected. \\[0.1cm]
FS\,Aur & 05:47:48.36 & +28:35:11.2 & 85.8 & STIS & G140L & 1425 & White dwarf not detected. \\[0.1cm]
EK\,TrA & 15:14:00.09 & --65:05:36.68 & 86.36 & STIS & E140M & 1425 & This work. \\[0.1cm]
EG\,Cnc & 08:43:03.97 & +27:51:49.71 & 86.36 & STIS & G140L & 1425 & This work. \\[0.1cm]
AQ\,Eri & 05:06:13.11 & --04:08:07.3 & 87.75 & STIS & G140L & 1425 & White dwarf not detected. \\[0.1cm]
1RXS\,J105010.8--140431 & 10:50:10.59 & --14:04:36.95 & 88.56 & COS & G140L & 1105 & This work. \\[0.1cm]
\multirow{4}{*}{DP\,Leo} & \multirow{4}{*}{11:17:15.93} & \multirow{4}{*}{+17:57:42.0} & \multirow{4}{*}{89.8} & \multirow{4}{*}{FOS} & \multirow{4}{*}{G160L} & \multirow{4}{*}{2440} &  Magnetic, weak white dwarf signature in\\
& & & & & & & the UV data. Time-resolved observations \\
& & & & & & & available but of too low quality \\
& & & & & & & for spectral analysis. \\[0.1cm]
\multirow{2}{*}{V2051\,Oph} & \multirow{2}{*}{17:08:19.09} & \multirow{2}{*}{--25:48:31.7} & \multirow{2}{*}{89.9} & \multirow{2}{*}{FOS} & \multirow{2}{*}{G160L} & \multirow{2}{*}{2440} &  Weak white dwarf signature, data of too\\
& & & & & & & low quality for spectral analysis. \\[0.1cm]
V436\,Cen & 11:14:00.06 & --37:40:47.7 & 90.0 & STIS & G140L & 1425 & White dwarf not detected. \\[0.1cm]
V347\,Pav & 18:44:48.14 & --74:18:34.1 & 90.08 & STIS & G140L & 1425 &  Magnetic, no time-resolved observations available. \\[0.1cm]
\multirow{3}{*}{EU\,UMa} & \multirow{3}{*}{11:49:55.71} & \multirow{3}{*}{+28:45:07.4} & \multirow{3}{*}{90.14} & \multirow{3}{*}{FOS} & \multirow{3}{*}{G130H} & \multirow{3}{*}{1600} &  Magnetic. Time-resolved observations are \\
& & & & & & & available but the white dwarf signature is not\\ 
& & & & & & & detected in the UV spectrum. \\[0.1cm]
BC\,UMa & 11:52:15.79 & +49:14:41.68 & 90.16 & STIS & G140L & 1425 & This work. \\[0.1cm]
VY\,Aqr & 21:12:09.30 & --08:49:37.47 & 90.85 & STIS & E140M & 1425 & This work. \\[0.1cm]
\multirow{3}{*}{OY\,Car} & \multirow{3}{*}{10:06:22.10} & \multirow{3}{*}{--70:14:04.6} & \multirow{3}{*}{90.89} & \multirow{2}{*}{FOS} & \multirow{2}{*}{G160L} & \multirow{2}{*}{2440} & Obtained in 1991. White dwarf + iron curtain \\
& & & & & & & detected but too low quality for spectral analysis. \\
           &                     &                     &           & FOS & G160L & 2440 & Obtained in 1992, during an outburst. \\[0.1cm]
MR\,Uma & 11:31:22.38 & +43:22:38.66 & 91.17 & COS & G140L & 1105 & Obtained after an outburst. \\[0.1cm]
QZ\,Lib & 15:36:16.02 & --08:39:08.63 & 92.36 & COS & G140L & 1105 & This work. \\[0.1cm]
SDSS\,J153817.35+512338.0 & 15:38:17.35 & +51:23:37.98 & 93.11 & COS & G140L & 1105 & This work. \\[0.1cm]
IX\,Dra & 18:12:31.50 & +67:04:45.9 & 93.31 & STIS & G140L & 1425 & White dwarf not detected. \\[0.1cm]
UV\,Per & 02:10:08.34 & +57:11:21.00 & 93.44 & STIS & G140L & 1425 & This work. \\[0.1cm]
1RXS\,J023238.8--371812 & 02:32:38.23 & --37:17:54.65 & 95.04 & COS & G140L & 1105 & This work. \\[0.1cm]
SDSS\,J093249.57+472523.0 & 09:32:49.57 & +47:25:23.0 & 95.48 & COS & G140L & 1105 & Evolved donor. \\[0.1cm]
SS\,UMi & 15:51:22.33 & +71:45:11.9 & 97.6 & STIS & G140L & 1425 & White dwarf not detected. \\[0.1cm]
KS\,UMa & 10:20:26.54 & +53:04:33.1 & 97.86 & STIS & G140L & 1425 & White dwarf not detected. \\[0.1cm]
BZ\,UMa & 08:53:44.15 & +57:48:40.6 & 97.91 & STIS & E140M & 1425 & Evolved donor. \\[0.1cm]
\multirow{2}{*}{EX\,Hya} & \multirow{2}{*}{12:52:24.22} & \multirow{2}{*}{--29:14:56.0} & \multirow{2}{*}{98.26} & FOS & G130H & 1600 & Magnetic, time-resolved observations available\\
           &                     &                     &           & STIS & G140L & 1425 &  but the white dwarf is not detected  in the spectra. \\[0.1cm]
RZ\,Sge & 20:03:18.44 & +17:02:51.57 & 98.32 & STIS & G140L & 1425 & This work. \\[0.1cm]
TY\,Psc & 01:25:39.35 & +32:23:08.9 & 98.4 & STIS & E140M & 1425 & Obtained in outburst. \\[0.1cm]
CY\,UMa & 10:56:56.95 & +49:41:18.38 & 100.18 & STIS & G140L & 1425 & This work. \\[0.1cm]
VV\,Pup & 08:15:06.79 & --19:03:17.7 & 100.44 & STIS & G140L & 1425 &  Magnetic, no time-resolved observations available. \\[0.1cm]
BB\,Ari & 02:44:57.80 & +27:31:09.13 & 101.2 & COS & G140L & 1105 & Obtained after an outburst. \\[0.1cm]
V834\,Cen & 14:09:07.30 & --45:17:16.1 & 101.52 & STIS & G140L & 1425 &  Magnetic, no time-resolved observations available. \\[0.1cm]
\multirow{2}{*}{GD\,552} & \multirow{2}{*}{22:50:40.25} & \multirow{2}{*}{+63:28:37.66} & \multirow{2}{*}{102.73} & \multirow{2}{*}{STIS} & G230LB & 2375 & \multirow{2}{*}{This work.} \\
           &                     &                     &           & & G140L & 1425 & \\[0.1cm]
DT\,Oct & 18:40:52.39 & --83:43:10.59 & 104.54 & COS & G140L & 1105 & Obtained after an outburst. \\[0.1cm]
HT\,Cas & 01:10:12.96 & +60:04:36.3 & 106.05 & FOS & G160L & 2440 & Obtained in outburst. \\[0.1cm]
IY\,Uma & 10:43:56.63 & +58:07:31.60 & 106.43 & COS & G140L & 1105 & This work. \\[0.1cm]
\multirow{4}{*}{VW\,Hyi} & \multirow{4}{*}{04:09:11.28} & \multirow{4}{*}{--71:17:41.4} & \multirow{4}{*}{106.95} & FOS & G160L & 2440 & Obtained in 1993 after an outburst. \\
           &                     &                     &           & FOS & G160L & 2440 & Obtained in 1995 after an outburst. \\
           &                     &                     &           & STIS & E140M & 1425 & Obtained in 2000 after an outburst. \\
           &                     &                     &           & STIS & E140M & 1425 & Obtained in 2001 after an outburst. \\[0.1cm]
SDSS\,J100515.38+191107.9 & 10:05:15.41 & +19:11:07.5 & 107.6 & COS & G140L & 1105 & Evolved donor. \\[0.1cm]
WX\,Hyi & 02:09:50.83 & --63:18:39.9 & 107.73 & GHRS & G140L & 1425 & White dwarf not detected. \\[0.1cm]
BK\,Lyn & 09:20:11.20 & +33:56:42.4 & 107.97 & STIS & E140M & 1425 &  Nova-like, white dwarf not detected. \\[0.1cm]
V893\,Sco & 16:15:14.98 & --28:37:32.1 & 109.38 & STIS & G140L & 1425 & White dwarf not detected. \\[0.1cm]
\multirow{2}{*}{T\,Pyx} & \multirow{2}{*}{09:04:41.51} & \multirow{2}{*}{--32:22:47.6} & \multirow{2}{*}{109.77} & \multirow{2}{*}{COS} & G130M  & 1055       &  Recurrent nova,\\
           &                     &                     &           & & G140L  & 1105       &  white dwarf not detected. \\[0.1cm]
RZ\,Leo & 11:37:22.18 & +01:48:58.86 & 110.17 & COS & G140L & 1105 & This work. \\[0.1cm]
SDSS\,J173008.38+624754.7 & 17:30:08.36 & +62:47:54.5 & 110.22 & STIS & G140L & 1425 & White dwarf not detected. \\[0.1cm]
CD\,Ind & 21:15:40.98 & --58:40:53.6 & 110.89 & STIS & E140M & 1425 & White dwarf not detected. \\[0.1cm]
V503\,Cyg & 20:27:17.41 & +43:41:22.5 & 111.89 & STIS & G140L & 1425 & White dwarf not detected. \\[0.1cm]
V2301\,Oph & 18:00:35.53 & +08:10:13.9 & 112.97 & FOS & G160L & 2440 &  Nova like, white dwarf not detected. \\[0.1cm]
CU\,Vel & 08:58:33.09 & --41:47:52.54 & 113.04 & COS & G140L & 1105 & This work. \\[0.1cm]
AX\,For & 02:19:28.01 & --30:45:45.96 & 113.04 & COS & G140L & 1105 & This work. \\[0.1cm]
\multirow{2}{*}{MR\,Ser} & \multirow{2}{*}{15:52:47.34} & \multirow{2}{*}{+18:56:25.6} & \multirow{2}{*}{113.47} & \multirow{2}{*}{STIS} & \multirow{2}{*}{G140L} & \multirow{2}{*}{1425} &  Magnetic, no time-resolved \\
& & & & & & & observations available.\\[0.1cm]
SDSS\,J164248.52+134751.4 & 16:42:48.51 & +13:47:51.51 & 113.6 & COS & G140L & 11105 & Obtained close to an outburst. \\[0.1cm]
\multirow{2}{*}{BL\,Hyi}  & \multirow{2}{*}{01:41:00.41} & \multirow{2}{*}{--67:53:27.5} & \multirow{2}{*}{113.64} & \multirow{2}{*}{STIS} & \multirow{2}{*}{G140L} & \multirow{2}{*}{1425} &  Magnetic, no time-resolved \\
& & & & & & & observations available.\\[0.1cm]
\multirow{2}{*}{ST\,LMi} & \multirow{2}{*}{11:05:39.77} & \multirow{2}{*}{+25:06:28.6} & \multirow{2}{*}{113.89} & \multirow{2}{*}{STIS} & \multirow{2}{*}{G140L} & \multirow{2}{*}{1425} &  Magnetic, no time-resolved \\
& & & & & & & observations available.\\[0.1cm]
\multirow{2}{*}{AN\,UMa} & \multirow{2}{*}{11:04:25.65} & \multirow{2}{*}{+45:03:13.8} & \multirow{2}{*}{114.84} & \multirow{2}{*}{STIS} & \multirow{2}{*}{G140L} & \multirow{2}{*}{1425} &  Magnetic, no time-resolved \\
& & & & & & & observations available.\\[0.1cm]
AR\,UMa & 11:15:44.62 & +42:58:22.3 & 115.92 & STIS & G140L & 1425 & White dwarf not detected. \\[0.1cm]
V1974\,Cyg & 20:30:31.66 & +52:37:50.9 & 117.01 & STIS & E140M & 1425 &  Classical nova, white dwarf not detected. \\[0.1cm]
\multirow{2}{*}{QZ\,Ser} & \multirow{2}{*}{15:56:54.48} & \multirow{2}{*}{+21:07:18.93} & \multirow{2}{*}{119.75} & STIS & G140L & 1425 & \multirow{2}{*}{Evolved donor.} \\
 &   & & & COS & G140L & 1105 &\\[0.1cm]
EF\,Peg & 21:15:04.11 & +14:03:49.15 & 120.53 & STIS & E140M & 1425 & This work. \\[0.1cm]
DV\,UMa & 09:46:36.55 & +44:46:44.62 & 123.62 & STIS & G140L & 1425 & This work. \\[0.1cm]
\multirow{2}{*}{HU\,Aqr} & \multirow{2}{*}{21:07:58.19} & \multirow{2}{*}{--05:17:40.4} & \multirow{2}{*}{125.02} & \multirow{2}{*}{FOS} & \multirow{2}{*}{G160L} & \multirow{2}{*}{2440} &  Magnetic, time-resolved observations available \\
& & & & & & & but the white dwarf is not detected in the spectra.\\[0.1cm]
IR\,Com & 12:39:31.99 & +21:08:06.25 & 125.34 & COS & G140L & 1105 & This work. \\[0.1cm]
\multirow{2}{*}{EU\,Cnc} & \multirow{2}{*}{08:51:27.16} & \multirow{2}{*}{+11:46:57.4} & \multirow{2}{*}{125.42} & \multirow{2}{*}{GHRS} & \multirow{2}{*}{G140L} & \multirow{2}{*}{1425} &  Magnetic, no time-resolved \\
& & & & & & & observations available.\\[0.1cm]
\multirow{2}{*}{UZ\,For} & \multirow{2}{*}{03:35:28.64} & \multirow{2}{*}{--25:44:21.8} & \multirow{2}{*}{126.53} & \multirow{2}{*}{FOS} & \multirow{2}{*}{G160L} & \multirow{2}{*}{2440} &  Magnetic, time-resolved observations available \\
& & & & & & & but the white dwarf is not detected in the spectra.\\[0.1cm]
UW\,Pic & 05:31:35.66 & --46:24:04.5 & 133.4 & STIS & G140L & 1425 & White dwarf not detected. \\[0.1cm]
\multirow{2}{*}{QS\,Tel} & \multirow{2}{*}{19:38:35.81} & \multirow{2}{*}{--46:12:57.0} & \multirow{2}{*}{139.95} & \multirow{2}{*}{FOS} & \multirow{2}{*}{G160L} & \multirow{2}{*}{2440} &  Magnetic, time-resolved observations available \\
& & & & & & & but the white dwarf is not detected in the spectra.\\[0.1cm]
NY\,Ser & 15:13:02.29 & +23:15:08.4 & 140.4 & STIS & G140L & 1425 & White dwarf not detected. \\[0.1cm]
OR\,And & 23:04:37.43 & +49:27:23.82 & 144.4 & COS & G140L & 1105 & Nova like, white dwarf not detected. \\[0.1cm]
SDSS\,J001153.08-064739.1 & 00:11:53.08 & --06:47:39.2 & 144.41 & COS & G140L & 1105 & Evolved donor. \\[0.1cm]
V348\,Pup & 07:12:32.91 & --36:05:38.5 & 146.65 & FOS & G160L & 2440 &  Nova-like, white dwarf not detected. \\[0.1cm]
IM\,Nor & 15:39:26.35 & --52:19:18.8 & 147.79 & COS & G140L & 1105 & Recurrent nova. \\[0.1cm]
\multirow{2}{*}{RXJ1554.2+2721} & \multirow{2}{*}{15:54:12.40} & \multirow{2}{*}{+27:21:51.2} & \multirow{2}{*}{151.87} & \multirow{2}{*}{STIS} & \multirow{2}{*}{G140L} & \multirow{2}{*}{1425} &  Magnetic, no time-resolved \\
& & & & & & & observations available.\\[0.1cm]
V795\,Her & 17:12:56.18 & +33:31:19.3 & 155.9 & FOS & G130H & 1600 &  Nova-like, white dwarf not detected. \\[0.1cm]
TU\,Men & 04:41:40.75 & --76:36:45.52 & 168.77 & STIS & G140L & 1425 & Obtained after an outburst. \\[0.1cm]
AM\,Her & 18:16:13.18 & +49:52:05.21 & 185.65 & STIS & G140L & 1425 & This work. \\[0.1cm]
MV\,Lyr & 19:07:16.29 & +44:01:08.6 & 191.38 & STIS & G140L & 1425 & White dwarf not detected. \\[0.1cm]
V482\,Cam & 07:33:41.42 & +67:32:15.6 & 192.41 & STIS & G140L & 1425 &  Nova-like, white dwarf not detected. \\[0.1cm]
SDSS\,J080908.39+381406.2 & 08:09:08.40 & +38:14:06.1 & 193.01 & STIS & G140L & 1425 &  Nova-like, white dwarf not detected. \\[0.1cm]
DW\,UMa & 10:33:52.88 & +58:46:54.74 & 196.71 & STIS & G140L & 1425 & This work. \\[0.1cm]
TT\,Ari & 02:06:53.09 & +15:17:41.8 & 198.07 & FOS & G160L & 2440 &  Nova-like, white dwarf not detected. \\[0.1cm]
\multirow{2}{*}{V603\,Aql} & \multirow{2}{*}{18:48:54.64} & \multirow{2}{*}{+00:35:02.9} & \multirow{2}{*}{199.01} & GHRS & G140L & 1425 &  \multirow{2}{*}{Nova-like, white dwarf not detected.} \\[0.1cm]
 & & & & FOS & G160L & 2440 &  \\[0.1cm]
V1500\,Cyg & 21:11:36.61 & +48:09:02.4 & 201.04 & FOS & G160L & 2440 &  Classical nova, white dwarf not detected. \\[0.1cm]
\multirow{2}{*}{V1432\,Aql} & \multirow{2}{*}{19:40:11.42} & \multirow{2}{*}{--10:25:25.8} & \multirow{2}{*}{201.94} & \multirow{2}{*}{FOS} & \multirow{2}{*}{G160L} & \multirow{2}{*}{2440} &  Magnetic, time-resolved observations available \\
& & & & & & & but the white dwarf is not detected in the spectra.\\[0.1cm]
HV\,And & 00:40:55.37 & +43:24:59.6 & 202.03 & STIS & G140L & 1425 &  Nova-like, white dwarf not detected. \\[0.1cm]
1ES\,0439-68.2 & 04:39:49.7 & --68:09:02 & 202.03 & COS & G130M & 1291 & Super-Soft X-ray source. \\[0.1cm]
\multirow{2}{*}{MN\,Hya} & \multirow{2}{*}{09:29:07.04} & \multirow{2}{*}{--24:05:05.5} & \multirow{2}{*}{203.39} & \multirow{2}{*}{FOS} & \multirow{2}{*}{G160L} & \multirow{2}{*}{2440} &  Magnetic, time-resolved observations available \\
& & & & & & & but the white dwarf is not detected in the spectra.\\[0.1cm]
\multirow{2}{*}{LS\,Cam} & \multirow{2}{*}{05:57:23.97} & \multirow{2}{*}{+72:41:52.4} & \multirow{2}{*}{205.03} & \multirow{2}{*}{STIS} & \multirow{2}{*}{G140L} & \multirow{2}{*}{1425} &  Magnetic, no time-resolved \\
& & & & & & & observations available.\\[0.1cm]
V751\,Cyg & 20:52:12.80 & +44:19:26.0 & 208.03 & STIS & G140L & 1425 & White dwarf not detected. \\[0.1cm]
VZ\,Sex & 09:44:31.71 & +03:58:05.6 & 214.13 & STIS & G140L & 1425 & White dwarf not detected. \\[0.1cm]
HS\,0455+8315 & 05:06:48.28 & +83:19:23.3 & 214.16 & STIS & G140L & 1425 &  Nova-like, white dwarf not detected. \\[0.1cm]
V1294\,Tau & 04:00:37.26 & +06:22:45.9 & 214.59 & STIS & G140L & 1425 &  Nova-like, white dwarf not detected. \\[0.1cm]
AO\,Psc & 22:55:17.99 & --03:10:40.0 & 215.46 & STIS & E140M & 1425 &  Magnetic, time-tag, da analizzare?. \\[0.1cm]
BP\,Lyn & 09:03:08.89 & +41:17:47.6 & 220.05 & STIS & E140M & 1425 &  Nova-like, white dwarf not detected. \\[0.1cm]
V794\,Aql & 20:17:33.91 & --03:39:51.0 & 220.75 & STIS & G140L & 1425 &  Nova-like, white dwarf not detected. \\[0.1cm]
V380\,Oph & 17:50:13.63 & +06:05:29.3 & 221.01 & STIS & G140L & 1425 & White dwarf not detected. \\[0.1cm]
BZ\,Cam & 06:29:34.09 & +71:04:37.0 & 221.08 & GHRS & G140L & 1425 &  Nova-like, white dwarf not detected. \\[0.1cm]
BB\,Dor & 05:29:28.58 & --58:54:46.04 & 221.9 & COS & G140L & 1105 & Nova like, white dwarf not detected. \\[0.1cm]
GS\,Pav & 20:08:07.62 & --69:48:58.7 & 223.59 & STIS & G140L & 1425 & White dwarf not detected. \\[0.1cm]
BH\,Lyn & 08:22:36.11 & +51:05:25.0 & 224.46 & STIS & E140M & 1425 &  Nova-like, white dwarf not detected. \\[0.1cm]
V382\,Vel & 10:44:48.40 & --52:25:31.1 & 227.66 & STIS & E140M & 1425 &  Classical nova, white dwarf not detected. \\[0.1cm]
IP\,Peg & 23:23:08.60 & +18:24:59.6 & 227.82 & FOS & G160L & 2440 & White dwarf not detected. \\[0.1cm]
V842\,Cen & 14:35:52.58 & --57:37:35.1 & 236.16 & COS & G140L & 1105 &  Old nova, white dwarf not detected. \\[0.1cm]
\multirow{2}{*}{RX\,J0153.3+7446} & \multirow{2}{*}{01:53:20.99} & \multirow{2}{*}{+74:46:22.3} & \multirow{2}{*}{236.38} & \multirow{2}{*}{STIS} & \multirow{2}{*}{G140L} & \multirow{2}{*}{1425} & Magnetic, no time-resolved \\
& & & & & & & observations available.\\[0.1cm]
V1776\,Cyg & 20:23:30.55 & +46:31:29.8 & 237.22 & STIS & G140L & 1425 &  Nova-like, white dwarf not detected. \\[0.1cm]
& & & & & & & but the white dwarf is not detected in the spectra.\\[0.1cm]
SDSS\,J040714.78-064425.1 & 04:07:14.78 & --06:44:25.2 & 245.04 & COS & G140L & 1105 & Obtained after an outburst. \\[0.1cm]
\multirow{2}{*}{V405\,Aur} & \multirow{2}{*}{05:57:59.29} & \multirow{2}{*}{+53:53:44.9} & \multirow{2}{*}{248.58} & \multirow{2}{*}{STIS} & \multirow{2}{*}{G140L} & \multirow{2}{*}{1425} &  Magnetic, time-resolved observations available \\
& & & & & & & but the white dwarf is not detected in the spectra.\\[0.1cm]
\multirow{2}{*}{RXJ1313.2-3259} & \multirow{2}{*}{13:13:17.11} & \multirow{2}{*}{--32:59:12.4} & \multirow{2}{*}{251.41} & \multirow{2}{*}{STIS} & \multirow{2}{*}{G140L} & \multirow{2}{*}{1425} &  Magnetic, time-resolved observations available \\
& & & & & & & but they do not cover the minimum. \\[0.1cm]
CW\,Mon & 06:36:54.59 & +00:02:17.20 & 254.3 & COS & G140L & 1105 & Obtained after an outburst. \\[0.1cm]
U\,Gem & 07:55:05.20 & +22:00:04.40 & 254.74 & FOS & G130H & 1600 & This work. \\[0.1cm]
V405\,Peg & 23:09:49.10 & +21:35:16.26 & 255.81 & COS & G140L & 1105 & Obtained after an outburst. \\[0.1cm]
HS2214 & 22:16:31.15 & +29:00:19.82 & 258.02 & COS & G140L & 1105 & Obtained after an outburst. \\[0.1cm]
\multirow{2}{*}{BD\,Pav} & \multirow{2}{*}{18:43:11.96} & \multirow{2}{*}{--57:30:45.26} & \multirow{2}{*}{258.19} & STIS & G140L & 1425 & \multirow{2}{*}{Evolved donor.} \\
& & & & COS & G140L & 1105 & \\[0.1cm]
SS\,Aur & 06:13:22.44 & +47:44:24.97 & 263.23 & STIS & G140L & 1425 & This work. \\[0.1cm]
HM\,Leo & 09:38:36.99 & +07:14:54.61 & 267.71 & COS & G140L & 1105 & Obtained after an outburst. \\[0.1cm]
SDSS\,J100658.41+233724.4 & 10:06:58.42 & +23:37:24.6 & 267.72 & COS & G140L & 1105 & Obtained after an outburst. \\[0.1cm]
\multirow{2}{*}{V416\,Dra} & \multirow{2}{*}{18:57:20.36} & \multirow{2}{*}{+71:31:18.8} & \multirow{2}{*}{272.32} & \multirow{2}{*}{STIS} & \multirow{2}{*}{G140L} & \multirow{2}{*}{1425} & Magnetic, no time-resolved \\
& & & & & & & observations available.\\[0.1cm]
\multirow{2}{*}{DQ\,Her} & \multirow{2}{*}{18:07:30.26} & \multirow{2}{*}{+45:51:32.4} & \multirow{2}{*}{278.81} & \multirow{2}{*}{FOS} &  G130H &  1600     &  Magnetic, time-resolved observations available\\
& & & & & G160L & 2440 & but the white dwarf is not detected in the spectra.\\[0.1cm]
IX\,Vel & 08:15:18.97 & --49:13:20.7 & 279.25 & STIS & E140M & 1425 &  Nova-like, white dwarf not detected. \\[0.1cm]
\multirow{2}{*}{MU\,Cam} & \multirow{2}{*}{06:25:16.26} & \multirow{2}{*}{+73:34:39.1} & \multirow{2}{*}{283.17} & \multirow{2}{*}{STIS} & \multirow{2}{*}{G140L} & \multirow{2}{*}{1425} & Magnetic, no time-resolved \\
& & & & & & & observations available.\\[0.1cm]
UX\,UMa & 13:36:40.96 & +51:54:49.5 & 283.21 & STIS & E140M & 1425 &  Nova-like, white dwarf not detected. \\[0.1cm]
AY\,Sex & 10:23:47.68 & +00:38:41.3 & 285.26 & COS & G140L & 1105 & Evolved donor. \\[0.1cm]
\multirow{2}{*}{V895\,Cen} & \multirow{2}{*}{14:29:27.20} & \multirow{2}{*}{--38:04:09.6} & \multirow{2}{*}{285.92} & \multirow{2}{*}{STIS} & \multirow{2}{*}{G140L} & \multirow{2}{*}{1425} & Magnetic, no time-resolved \\
& & & & & & & observations available.\\[0.1cm]
\multirow{2}{*}{FO\,Aqr} & \multirow{2}{*}{22:17:55.38} & \multirow{2}{*}{--08:21:03.8} & \multirow{2}{*}{290.97} & \multirow{2}{*}{FOS} & \multirow{2}{*}{G160L} & \multirow{2}{*}{2440} & Magnetic, no time-resolved \\
& & & & & & & observations available.\\[0.1cm]
T\,Aur & 05:31:59.13 & +30:26:45.5 & 294.3 & STIS & G140L & 1425 & White dwarf not detected. \\[0.1cm]
V825\,Her & 17:18:37.02 & +41:15:50.9 & 296.64 & STIS & E140M & 1425 &  Classical nova, white dwarf not detected. \\[0.1cm]
V446\,Her & 18:57:21.60 & +13:14:29.0 & 298.08 & STIS & G140L & 1425 &  Classical nova, white dwarf not detected. \\[0.1cm]
\multirow{2}{*}{V3885\,Sgr} & \multirow{2}{*}{19:47:40.53} & \multirow{2}{*}{--42:00:26.4} & \multirow{2}{*}{298.31} & STIS & E140M & 1425 &  \multirow{2}{*}{Nova-like, white dwarf not detected.} \\
& & & & FOS & G160L & 2440 &  \\[0.1cm]
1SWASP\,J162117.36+441254.2 & 16:21:17.36 & +44:12:54.1 & 299.31 & COS & G140L & 1105 & Evolved donor. \\[0.1cm]
\multirow{3}{*}{RX\,And} & \multirow{3}{*}{01:04:35.54} & \multirow{3}{*}{+41:17:57.42} & \multirow{3}{*}{302.25} & \multirow{3}{*}{GHRS} & G140L & 1425 & This work. \\
& & & & & G140L & 1425 & Obtained in 1997 after an outburst. \\
& & & & & G140L & 1425 & Obtained in 1997 after an outburst. \\[0.1cm]
EX\,Dra & 18:04:14.11 & +67:54:12.2 & 302.31 & FOS & G160L & 2440 & Obtained in outburst. \\[0.1cm]
\multirow{2}{*}{PQ\,Gem} & \multirow{2}{*}{07:51:17.33} & \multirow{2}{*}{+14:44:23.9} & \multirow{2}{*}{311.56} & \multirow{2}{*}{FOS} & \multirow{2}{*}{G160L} & \multirow{2}{*}{2440} &  Magnetic, time-resolved observations available \\
& & & & & & & but the white dwarf is not detected in the spectra.\\[0.1cm]
\multirow{2}{*}{V0709\,Cas} & \multirow{2}{*}{00:28:48.80} & \multirow{2}{*}{+59:17:22.2} & \multirow{2}{*}{319.97} & \multirow{2}{*}{STIS} & \multirow{2}{*}{G140L} & \multirow{2}{*}{1425} &  Magnetic, time-resolved observations available \\
& & & & & & & but the white dwarf is not detected in the spectra.\\[0.1cm]
RW\,Tri & 02:25:36.16 & +28:05:50.9 & 333.91 & STIS & G140L & 1425 &  Nova-like, white dwarf not detected. \\[0.1cm]
V347\,Pup & 06:10:33.66 & --48:44:25.4 & 333.99 & FOS & G130H & 1600 &  Nova-like, white dwarf not detected. \\[0.1cm]
VY\,Scl & 23:29:00.48 & --29:46:45.9 & 334.08 & STIS & G140L & 1425 &  Nova-like, white dwarf not detected. \\[0.1cm]
\multirow{2}{*}{RW\,Sex} & \multirow{2}{*}{10:19:56.62} & \multirow{2}{*}{--08:41:56.1} & \multirow{2}{*}{353.01} & GGHRS & G140L & 1425 &  \multirow{2}{*}{Nova-like, white dwarf not detected.} \\[0.1cm]
 &  &  &  & STIS & E140M & 1425 &  \\[0.1cm]
SDSS\,J154453.60+255348.8 & 15:44:53.61 & +25:53:48.9 & 361.85 & COS & G140L & 1105 & No valid data. \\[0.1cm]
TZ\,Per & 02:13:50.96 & +58:22:52.3 & 378.58 & STIS & G140L & 1425 & Obtained in outburst. \\[0.1cm]
TT\,Crt & 11:34:47.26 & --11:45:31.0 & 386.52 & STIS & G140L & 1425 & Obtained after an outburst. \\[0.1cm]
LY\,UMa & 10:48:18.01 & +52:18:29.8 & 390.64 & STIS & G140L & 1425 & White dwarf not detected. \\[0.1cm]
SS\,Cyg & 21:42:42.79 & +43:35:09.9 & 396.19 & GHRS & G140L & 1425 & Obtained during or close to an outburst. \\[0.1cm]
EM\,Cyg & 19:38:40.11 & +30:30:28.4 & 418.91 & STIS & G140L & 1425 & White dwarf not detected. \\[0.1cm]
V959\,Mon & 06:39:38.60 & +05:53:53.0 & 426.96 & STIS & E140M & 1425 &  Classical nova, white dwarf not detected. \\[0.1cm]
HS\,0218+3229 & 02:21:33.48 & +32:43:23.8 & 428.01 & COS & G140L & 1105 & Evolved donor. \\[0.1cm]
AF\,Cam & 03:32:15.48 & +58:47:22.1 & 466.67 & STIS & G140L & 1425 & White dwarf not detected. \\[0.1cm]
\multirow{2}{*}{V1309\,Ori} & \multirow{2}{*}{05:15:41.41} & \multirow{2}{*}{+01:04:40.5} & \multirow{2}{*}{478.96} & \multirow{2}{*}{FOS} & \multirow{2}{*}{G160L} & \multirow{2}{*}{2440} &  Magnetic, time-resolved observations available \\
& & & & & & & but the white dwarf is not detected in the spectra.\\[0.1cm]
HS1055 & 10:57:56.29 & +09:23:14.87 & 541.88 & COS & G140L & 1105 & Obtained after an outburst. \\[0.1cm]
\multirow{2}{*}{AE\,Aqr} & \multirow{2}{*}{20:40:09.16} & \multirow{2}{*}{--00:52:15.1} & \multirow{2}{*}{592.78} & \multirow{2}{*}{FOS} & \multirow{2}{*}{G160L} & \multirow{2}{*}{2440} & Evolved donor. Magnetic, time-resolved observations available \\
& & & & & & & but the white dwarf is not detected in the spectra.\\[0.1cm]
EY\,Cyg & 19:54:36.71 & +32:21:55.1 & 661.43 & STIS & G140L & 1425 & Evolved donor. \\[0.1cm]
V442\,Cen & 11:24:51.87 & --35:54:36.69 & 662.4 & STIS & G140L & 1425 & This work. \\[0.1cm]
NGC\,6397-CV1 & 17:40:41.6 & --53:40:19 & 678.53 & FOS & G160L & 2440 & White dwarf not detected. \\[0.1cm]
\multirow{2}{*}{IGR\,J17014-4306} & \multirow{2}{*}{17:01:28.15} & \multirow{2}{*}{--43:06:12.2} & \multirow{2}{*}{768.97} & \multirow{2}{*}{COS} & \multirow{2}{*}{G140L} & \multirow{2}{*}{1105} & Magnetic, time-resolved observations available \\
& & & & & & & but the white dwarf is not detected in the spectra.\\[0.1cm]
DI\,Lac & 22:35:48.58 & +52:42:59.5 & 783.03 & STIS & E140M & 1425 & Old nova. \\[0.1cm]
CI\,Aql & 18:52:03.59 & --01:28:39.3 & 890.44 & COS & G140L & 1105 &  Recurrent nova, white dwarf not detected. \\[0.1cm]
QR\,And & 00:19:49.93 & +21:56:52.1 & 951.05 & STIS & E140M & 1425 & Super-Soft X-ray source. \\[0.1cm]
U\,Sco & 16:22:30.78 & --17:52:42.8 & 1772.0 & STIS & G140L & 1425 & Recurrent nova. \\[0.1cm]
NOVA\,LMC 2000 & 05:25:01.60 & --70:14:17.3 & -- & STIS & E140M & 1425 &  Classical nova, white dwarf not detected. \\[0.1cm]
V339\,Del & 20:23:30.68 & +20:46:03.8 & -- & STIS & E140M & 1425 &  Classical nova, white dwarf not detected. \\[0.1cm]
QU\,Car & 11:05:42.48 & --68:37:58.3 & -- & STIS & E140M & 1425 &  Nova-like, white dwarf not detected. \\[0.1cm]
M31N\,2008-12a & 00:45:28.81 & +41:54:09.9 & -- & STIS & G140L & 1425 &  Classical nova, white dwarf not detected. \\[0.1cm]
SDSS\,J074716.81+424849.0 & 07:47:16.80 & +42:48:49.0 & -- & STIS & G140L & 1425 &  Nova-like, white dwarf not detected. \\[0.1cm]
V1369\,Cen & 13:54:45.35 & --59:09:04.2 & -- & STIS & E140M & 1425 &  Classical nova, white dwarf not detected. \\[0.1cm]
V5668\,Sgr & 18:36:56.83 & --28:55:40.0 & -- & STIS & E140M & 1425 &  Classical nova, white dwarf not detected. \\[0.1cm]
NOVA\,LMC 2012 & 04:54:56.84 & --70:26:56.1 & -- & STIS & E140M & 1425 &  Classical nova, white dwarf not detected. \\[0.1cm]
\bottomrule
\end{longtable}


\bsp	
\label{lastpage}
\end{document}